\documentclass[10pt,twocolumn]{article} %acmtog
\pdfoutput=1
\newcommand{\anon}[1]{}
\newcommand{\known}[1]{#1}

\usepackage{epsf,amssymb,algorithm,algorithmic,psfrag}
\usepackage{amsmath}
\usepackage{amsfonts,amsmath,amsthm} % is incompatile with tog format
\usepackage{graphicx}
\usepackage{array}
\usepackage{psfrag}
\usepackage{subfigure}
\usepackage{bm}
\usepackage{color}
\usepackage{multirow}
\usepackage{booktabs}
\usepackage{threeparttable}

\usepackage[normalem]{ulem}
\usepackage{hyperref}
\hypersetup{
    colorlinks,%
    citecolor=black,%
    filecolor=black,%
    linkcolor=black,%
    urlcolor=black
}

\newcommand{\kflat}{\mbox{\ensuremath{k}-d~flat}} % use \darttext\ with trailing slash for a space
\newcommand{\kflats}{\mbox{\ensuremath{k}-d~flats}} % use \darttext\ with trailing slash for a space

\newcommand{\dart}{\mbox{\ensuremath{k}-d~dart}} % use \darttext\ with trailing slash for a space
\newcommand{\darts}{\mbox{\ensuremath{k}-d~darts}} % use \darttext\ with trailing slash for a space
\newcommand{\Dart}{\mbox{\ensuremath{k}-d~dart}} %this should still be lower-case k, since k is a math variable.
\newcommand{\dchoosek}{\ensuremath{\binom{d}{k}}}
\newcommand{\ddim}{\ensuremath{d}} %dimension of space
\newcommand{\adart}{\ensuremath{s}} %a dart
\newcommand{\kdim}{\ensuremath{k}} % dimension of dart
\newcommand{\samp}{\ensuremath{p}} % the sample point defining the dart
\newcommand{\linedart}{\ensuremath{\adart^1}} % one-dimensional line dart
\newcommand{\linedarti}{\ensuremath{\linedart_i}} % one of the search lines of  a line dart, the ith search, for i = 1..d
\newcommand{\disk}[1]{\ensuremath{D(#1)}}

\newcommand{\func}{\ensuremath{f}}
\newcommand{\domain}{\ensuremath{\mathcal{X}}}
\newcommand{\Flat}{\ensuremath{F}}
\newcommand{\figref}[1]{Figure~\ref{#1}} %use this to reference figures
\newcommand{\secref}[1]{Section~\ref{#1}} % use this to reference sections
\newcommand{\tabref}[1]{Table~\ref{#1}} % use this to reference tables
\newcommand{\equref}[1]{Equation~(\ref{#1})} % use this for equation
\newcommand{\equrefs}[2]{Equations~(\ref{#1}) and (\ref{#2})} % use this for two equations

\newcommand{\shortcite}[1]{\cite{#1}} 

\graphicspath{{./fig/}{./fig/proj-experiments/}{.}}
\usepackage{color}
\definecolor{darkred}{rgb}{0.7,0,0.0}

\newcommand{\mytitle}{$k$-d Darts: Sampling by $k$-Dimensional Flat Searches}
\begin{document}

%\markboth{\myauthors}{\mytitle}

\title{\mytitle}

%\known{
 % this looks silly. too long. but it appears to be required
% author order: best guess as of Jan 13. re-arrange as we see fit.
%  \author{ %
%  MOHAMED S. EBEIDA \affil{Sandia National Laboratories} %
%  \and ANJUL PATNEY \affil{University of California, Davis} %
%  \and SCOTT A. MITCHELL \affil{Sandia National Laboratories} %
%  \and KEITH R. DALBEY \affil{Sandia National Laboratories} % 
%  \and ANDREW A. DAVIDSON \affil{University of California, Davis} %
%  \and JOHN D. OWENS \affil{University of California, Davis} %
%  }

%\numberofauthors{6}
%\alignauthor
%\begin{tabular}{c@{\,}c@{\,}c@{\,}c@\,}c@{\,}c}

% need to include this
\author{ %
%\begin{center}
%\newlength{\oldtabcolsep}
%\setlength{\oldtabcolsep}{\tabcolsep}
%\setlength{\tabcolsep}{3pt}
\begin{tabular}{ccc}
  {Mohamed S.~Ebeida$^{\ast}$} &
  {Anjul Patney$^{\dagger}$} &
  {Scott A.~Mitchell$^{\ast}$}\\ 
  {Keith R.~Dalbey$^{\ast}$} &
  {Andrew A.~Davidson$^{\dagger}$} &
  {John D.~Owens$^{\dagger}$}\\ 
\end{tabular}
\\
\vspace{-6pt}
\\
\begin{tabular}{cc}
  {$^{\ast}$Sandia National Laboratories} &
  {$^{\dagger}$University of California, Davis}\\ 
\end{tabular}
%\end{center}
}

%\author{ %
%  {Mohamed S.~Ebeida$^{\ast}$} \and
%  {Anjul Patney$^{\dagger}$} \and
%  {Scott A.~Mitchell$^{\ast}$} \and
%  {Keith R.~Dalbey$^{\ast}$} \and
%  {Andrew A.~Davidson$^{\dagger}$} \and
%  {John D.~Owens$^{\dagger}$}
%}

%\category{I.3.5}%{Computing Methodologies}%
%{Computer Graphics}{Computational Geometry and Object Modeling}

%\terms{Sampling, Dimension}

%\keywords{Line search, thin regions, rendering, depth of field, {P}oisson-disk sampling, {M}onte {C}arlo integration, {L}atin hypercube sampling, uncertainty quantification}

% todo, update
%\known
%{
%\acmformat{Ebeida, M. S., Patney, A., Mitchell, S. A., Dalbey, K. R., Davidson, A. A., and Owens, J. D. 2013. \mytitle. % 
%{ACM Trans. Graph.} \todo{update this} 28, 4, Article 106 (August 2012), 12 pages.\newline  DOI $=$
%10.1145/1559755.1559763\newline http://doi.acm.org/10.1145/1559755.1559763}

%\acmformat{Ebeida, M. S., Patney, A., Mitchell, S. A., Dalbey, K. R., Davidson, A. A., and Owens, J. D. 2013. \mytitle. % 
%{ACM Trans. Graph.} 
%TOG-12-0088, request reviewer continuity from SIGGRAPH-ASIA 2012 submission papers\_0450,
%15 pages.}
%}

\maketitle

%\begin{bottomstuff} 
%\known{Mohamed S. Ebeida: msebeid@sandia.gov}
%\end{bottomstuff}

\begin{abstract}

We formalize the notion of sampling a function using \dart{}s.
A \dart{} is a set of independent, mutually orthogonal, $k$-dimensional subspaces called \kflats{}.
Each dart has $d$ choose $k$ flats, aligned with the coordinate axes for efficiency. 
We show that \dart{}s are useful for exploring a function's properties, 
such as estimating its integral, or finding an exemplar above a threshold.
We describe a recipe for converting an algorithm from point sampling to \dart{} sampling,
assuming the function can be evaluated along a \kflat{}.

  We demonstrate that \dart{}s
  are more efficient than point-wise samples in high dimensions, depending on
  the characteristics of the sampling domain: e.g.\ the subregion
  of interest has small volume and evaluating the function along a flat is not too expensive.
We present three concrete
  applications using line darts (1-d darts): 
  relaxed maximal Poisson-disk sampling, high-quality
  rasterization of depth-of-field blur, and estimation of the
  probability of failure from a response surface for  uncertainty quantification. 
In these applications, line darts achieve the same fidelity output as point darts in less time.
We also demonstrate the accuracy of higher dimensional darts for a volume estimation problem.
For Poisson-disk sampling, we use significantly less memory, enabling the generation of larger point clouds in higher dimensions.
% to do, add a statement about UQ

%We
%illustrate the utility of \dart s with three concrete applications.
%These applications span computer graphics and uncertainty
%quantification, but have in common that the space is of high $\ddim$
%dimensions.  Specifically, we consider ray tracing with depth of field
%blur, completing a maximal Poisson-disk sampling, and estimating the
%probability of failure from an emulator.  

%We consider
%the following motivating class of problems. Given a function over
%abstract $\ddim$-dimensional space, we seek to find the function's
%properties such as its integral or an examplar whose function value is
%above a threshold.   For joined kd-darts, we sample a point from the
%space, then extend this point in axis-aligned directions with
%$\kdim$-dimensional flats.  (For split kd-darts, the flats are chosen
%independently and do not have a common point.) For each flat, we
%construct the imprint of the function along it.  The function
%properties are estimated from this imprint.  Particularly important
%instances are indicator functions, and line-darts which consist of
%$\ddim$ one-dimensional line searches.

\end{abstract}

\section{Introduction}
\label{sec:intro}
%
%%\begin{figure}[!ht]
%%       \begin{center}
%%%   \includegraphics[width=6.5truein]{fig/TeaserPlot.png}
%%    \includegraphics[width=0.32\columnwidth]{fig/TeaserPoint.pdf} %\hspace{0.09in}
%%    \includegraphics[width=0.32\columnwidth]{fig/TeaserLine.pdf}  %\hspace{0.09in} 
%%%    \includegraphics[width=0.32\columnwidth]{fig/TeaserPlane.pdf}
%%%    \includegraphics[width=0.32\columnwidth]{fig/paperTeaserTest.jpeg}
%%   \includegraphics[width=0.32\columnwidth]{fig/paperTeaserPlane.jpeg}
%%   \caption{Sampling long and thin subregions (gray) using points
%%     (left), lines (center), and planes (right). Point samples may be cheap to generate and
%%     evaluate, but they contribute nothing to the final result if they miss the region of interest. 
%%     Misses (blue) are frequent for regions with a small volume.
%%     Samples of higher dimensions, or \dart{}s, often intersect (red) the region of interest, especially if the region is long and thin.
%%     A \dart's  greater expense is offset
%%     by it providing more information. \label{fig:teaser}}
%%     \end{center}
%%\end{figure}
%
%\begin{figure}[ht]
%\begin{minipage}[b]{0.45\linewidth}
%\centering
%\includegraphics[width=\textwidth]{filename1}
%\caption{default}
%\label{fig:figure1}
%\end{minipage}
%\hspace{0.5cm}
%\begin{minipage}[b]{0.45\linewidth}
%\centering
%\includegraphics[width=\textwidth]{filename2}
%\caption{default}
%\label{fig:figure2}
%\end{minipage}
%\end{figure}
%
%\fontsize{10}{10}\selectfont
% could also use subfigure environment, to put caption at bottom. But I like the titles at the top.
\begin{figure}[!ht]
\begin{minipage}[b]{0.32\columnwidth}
\centering
\scriptsize 99 Point Darts
\includegraphics[width=\textwidth]{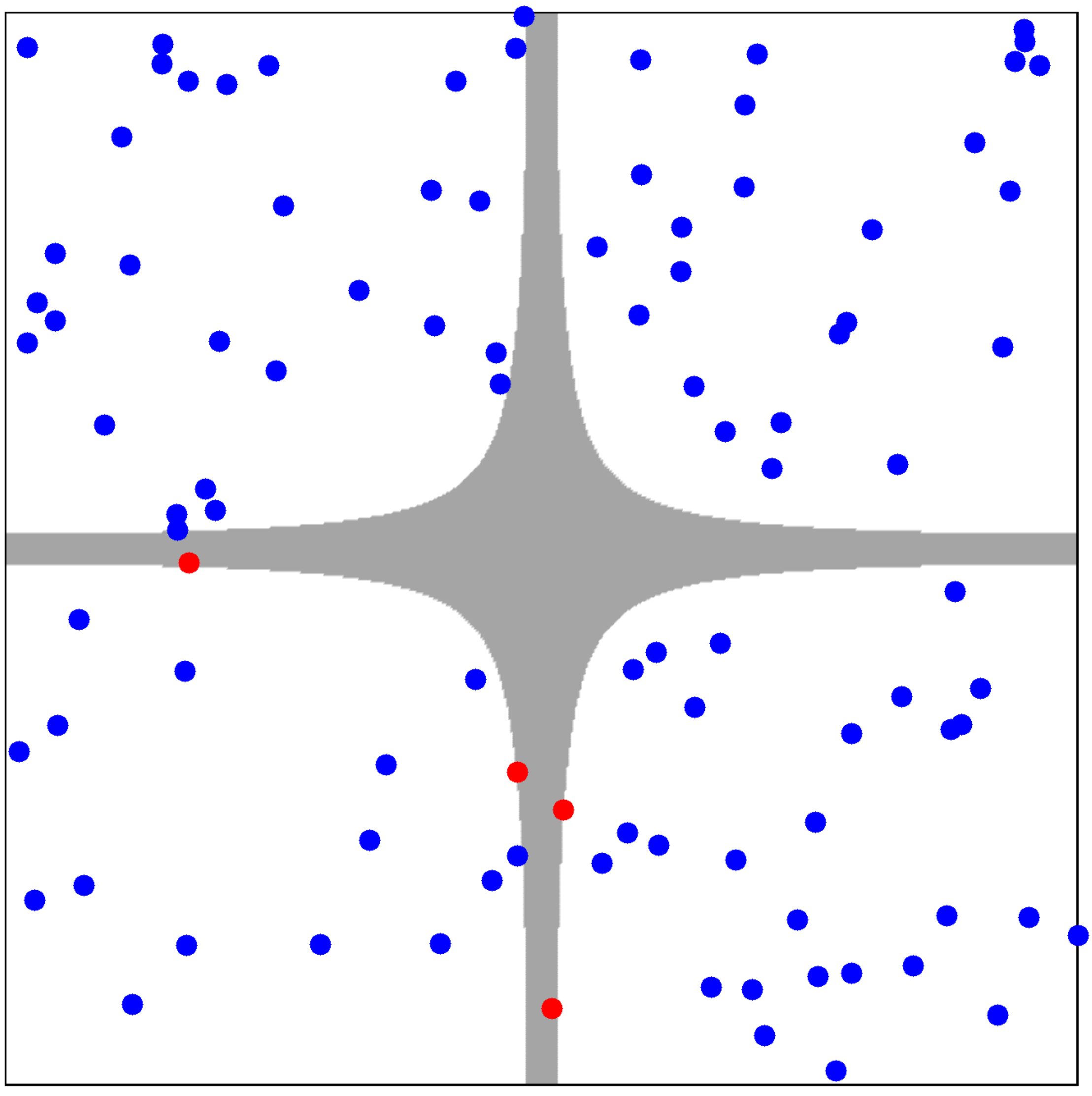}
\end{minipage}
\begin{minipage}[b]{0.32\columnwidth}
\centering
\scriptsize 6 Line Darts
\includegraphics[width=\textwidth]{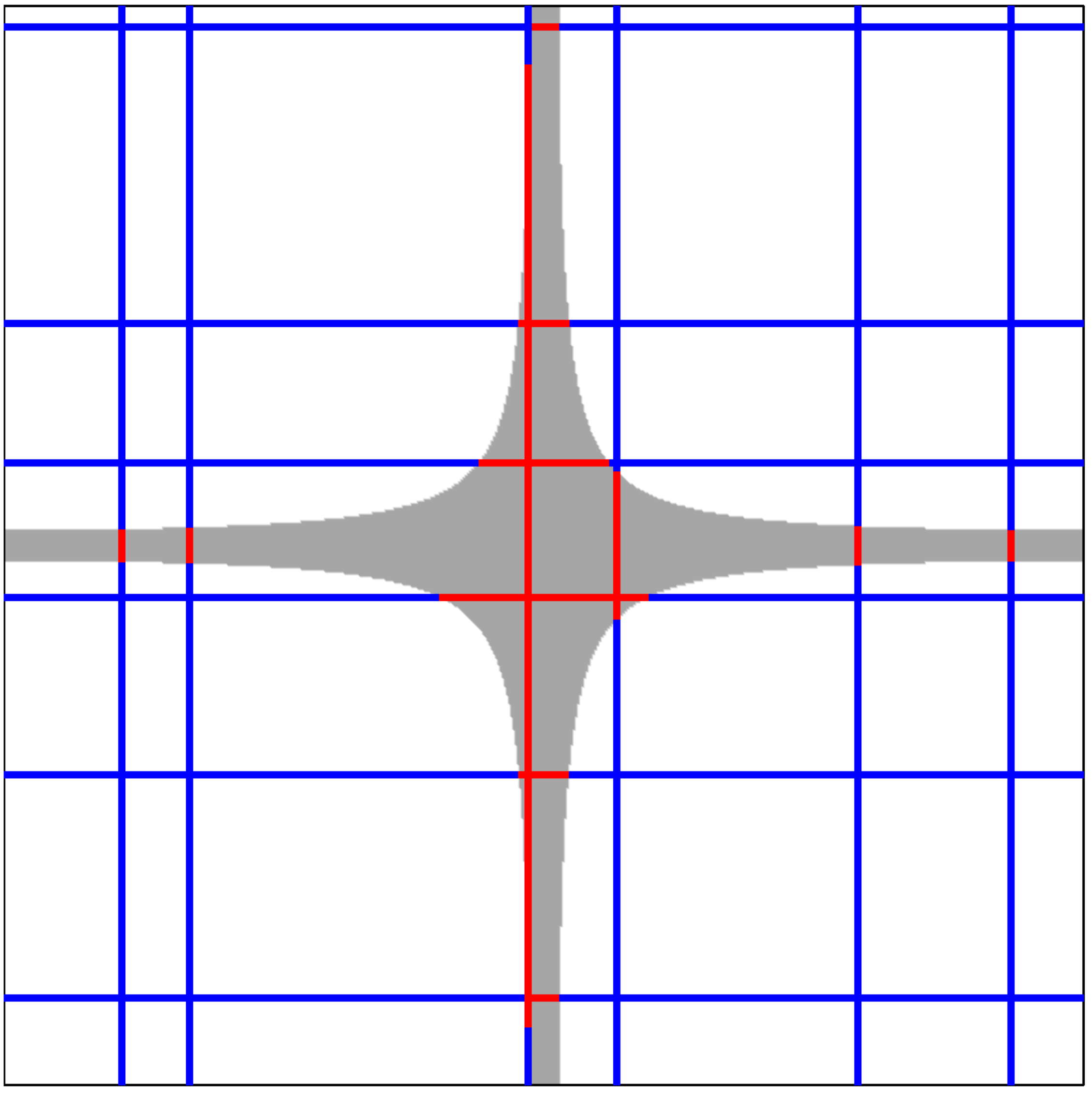}
\end{minipage}
\begin{minipage}[b]{0.32\columnwidth}
\centering
\scriptsize 1 Plane Dart
\includegraphics[width=\textwidth]{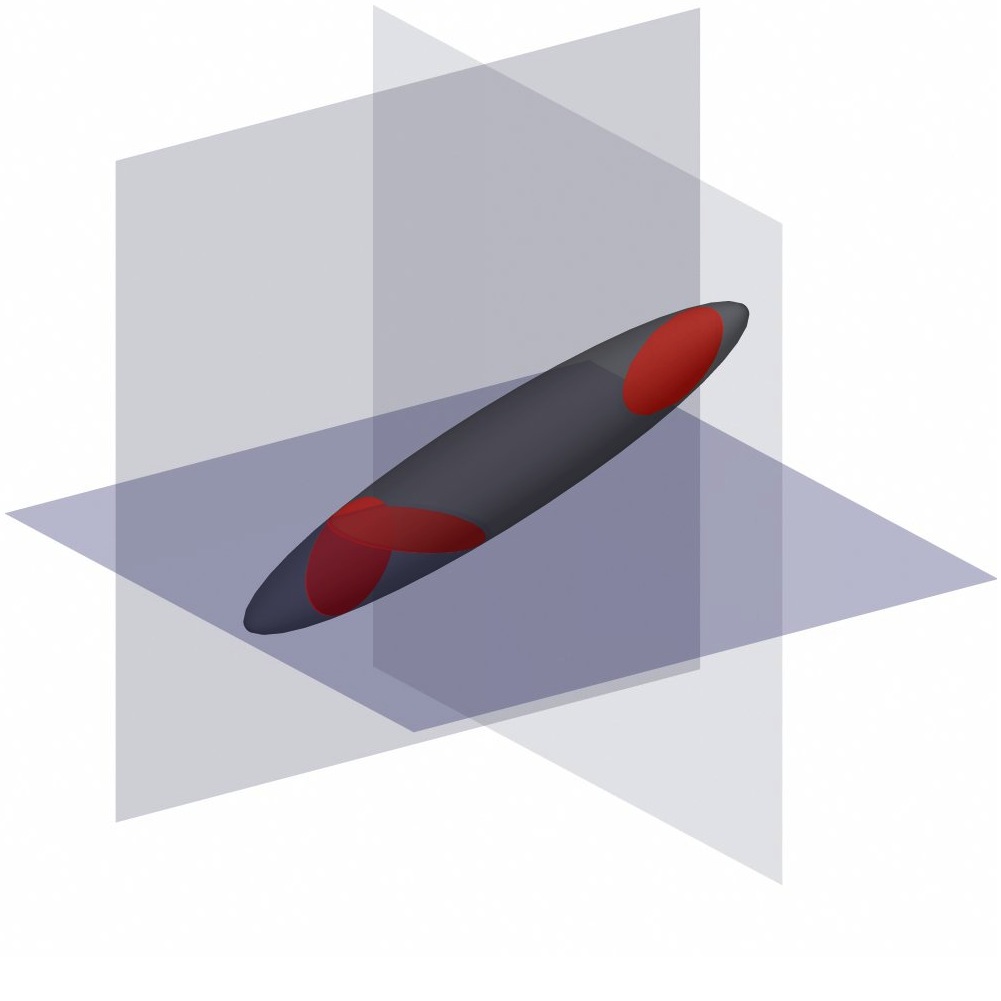}
\end{minipage}
   \caption{Sampling long and thin subregions (gray) using points
     (left), lines (center), and planes (right). Point samples may be cheap to generate and
     evaluate, but they contribute nothing to the final result if they miss the region of interest. 
     Misses (blue) are frequent for regions with a small volume.
     Samples of higher dimensions, or \dart{}s, often intersect (red) the region of interest, especially if the region is long and thin.
     A \dart's  greater expense is offset
     by it providing more information. \label{fig:teaser}}
\end{figure}

%\todo{scott}
%Few sentence description of technique
%Challenges in above domains addressed by the new technique
%description of 3 example applications

% \subsection{Motivation and Problem Definition}
In many applications we are interested in estimating some global property of a function because it is difficult to calculate that property exactly. 
\emph{Sampling} is the process of randomly selecting \emph{samples}, subsets of a domain.
The function is evaluated at these subsets, and the global property is estimated based on those values.

In typical sampling processes, the samples are points. However,  a recurring challenge is to deal efficiently with the case that the interesting part of the domain is very small compared to the entire domain. For example, suppose we have a function over a domain, and we are interested in estimating the volume of the subdomain where the function is negative. If this subdomain has a very small volume, only a correspondingly very small fraction of uniform sample points will land in it; see \figref{fig:teaser}. Consequently, point sampling will require a large number of samples to get any estimate, and will be inefficient since most samples will not contribute. 

We propose the \dart{} to address this problem. One key idea is that rather than evaluate the function at a single point, we evaluate it in a higher-dimensional region. 
For each sample, we evaluate the function along a set of higher-dimensional flats (i.e.\ lines, planes \ldots\ hyperplanes).  
The second key idea is to use a set of mutually orthogonal flats, aligned with the coordinate axes; a \dart\ denotes this set of flats.
Randomly oriented flats have been considered before, but orthogonal flats are more efficient and have better worst-case performance when probing high aspect-ratio settings.
To ensure that the expected mean of the function estimates is correct, each of these flats is chosen independently. An important case of flats are one-dimensional lines. Using our previous example, we may find the points along the line where the function value is zero, then partition the line into segments where the function value $f$ is strictly positive or negative, and finally estimate the volume where $f<0$ from the negative-interval lengths. 
While these samples are more expensive to compute, they are more powerful; depending on the function they can generate better results for the same amount of effort.

\begin{figure}[!ht]
       \begin{center}
       \subfigure[Monte Carlo Sampling (MC)]{\includegraphics[width=1.00\columnwidth]{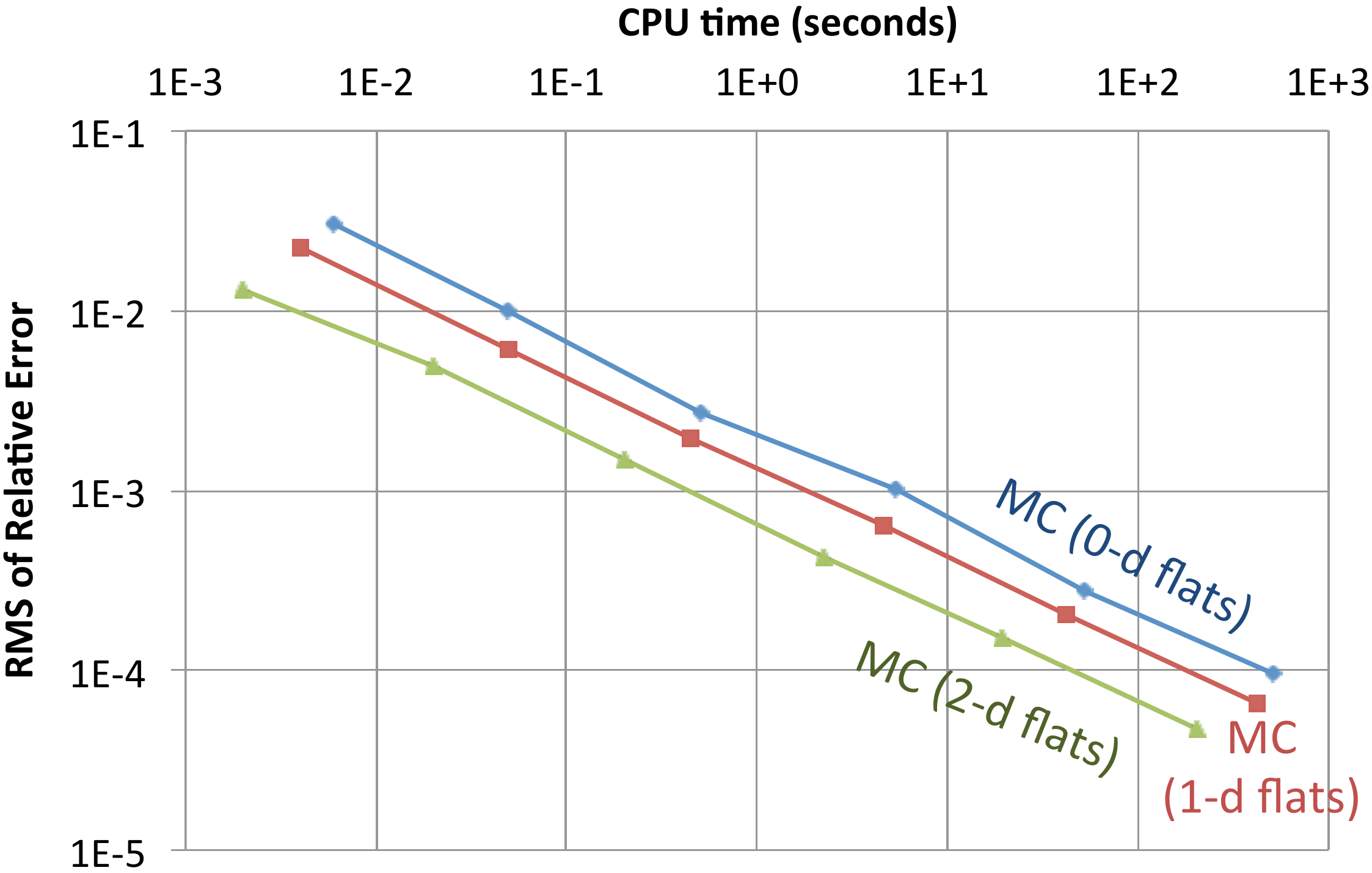}}   
       \subfigure[Latin Hypercube Sampling (LHS)]{\includegraphics[width=1.00\columnwidth]{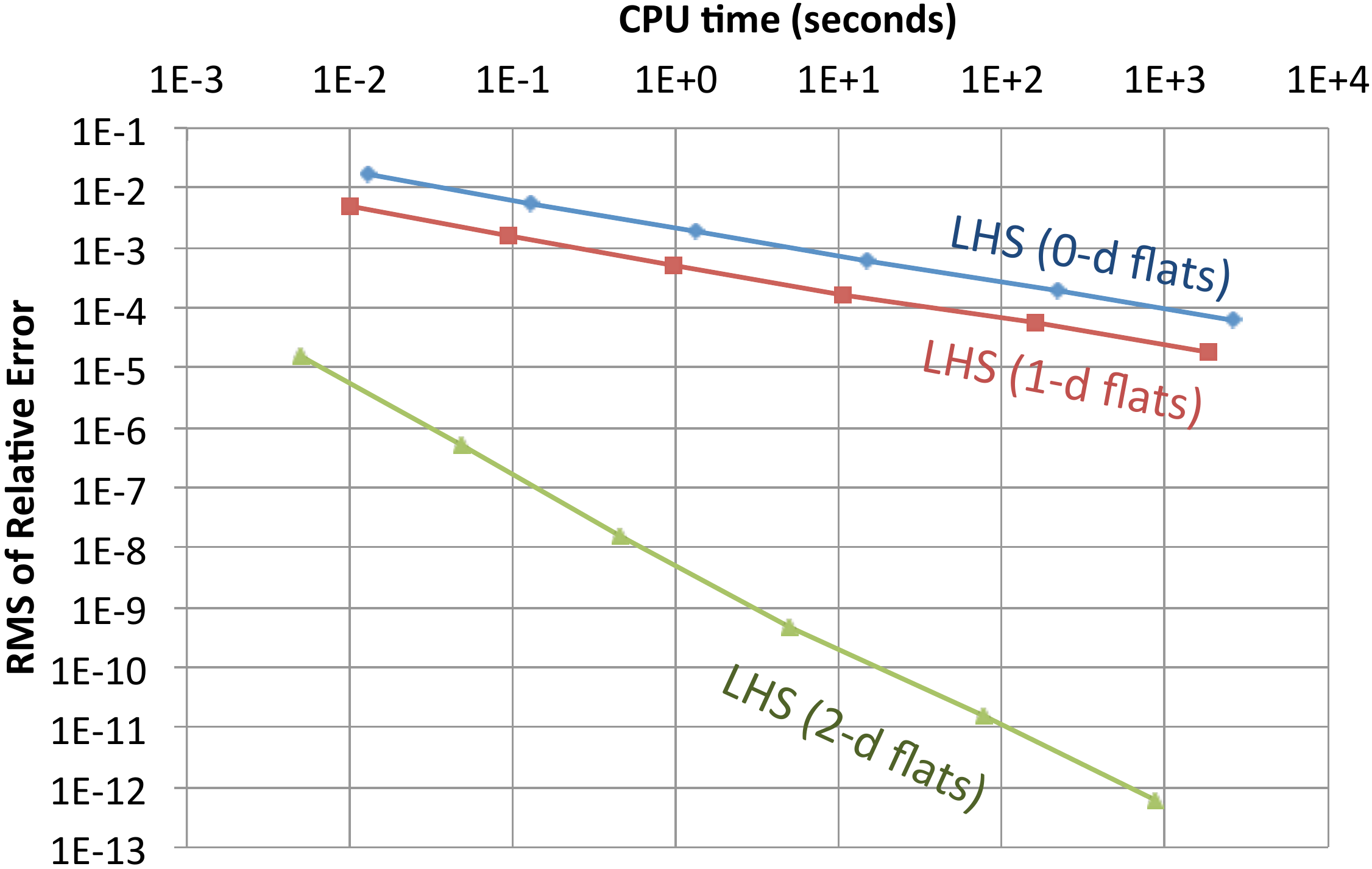}} 
       \caption{Estimating the volume of a sphere using random sampling via \kflats, k = 0, 1, 2. For each sample size, we performed 100 experiments and calculated the RMS error. The reported CPU time is the total time consumed by these experiments. For MC sampling (a) plane samples consumed an order of magnitude less time to achieve the same error as point samples. The savings were even more for LHS (b).} \label{fig:sphere_volume}
       \end{center}
\end{figure}

\begin{figure}[!ht]
       \begin{center}
        \subfigure[0-d flats]{\includegraphics[width=0.31\columnwidth]{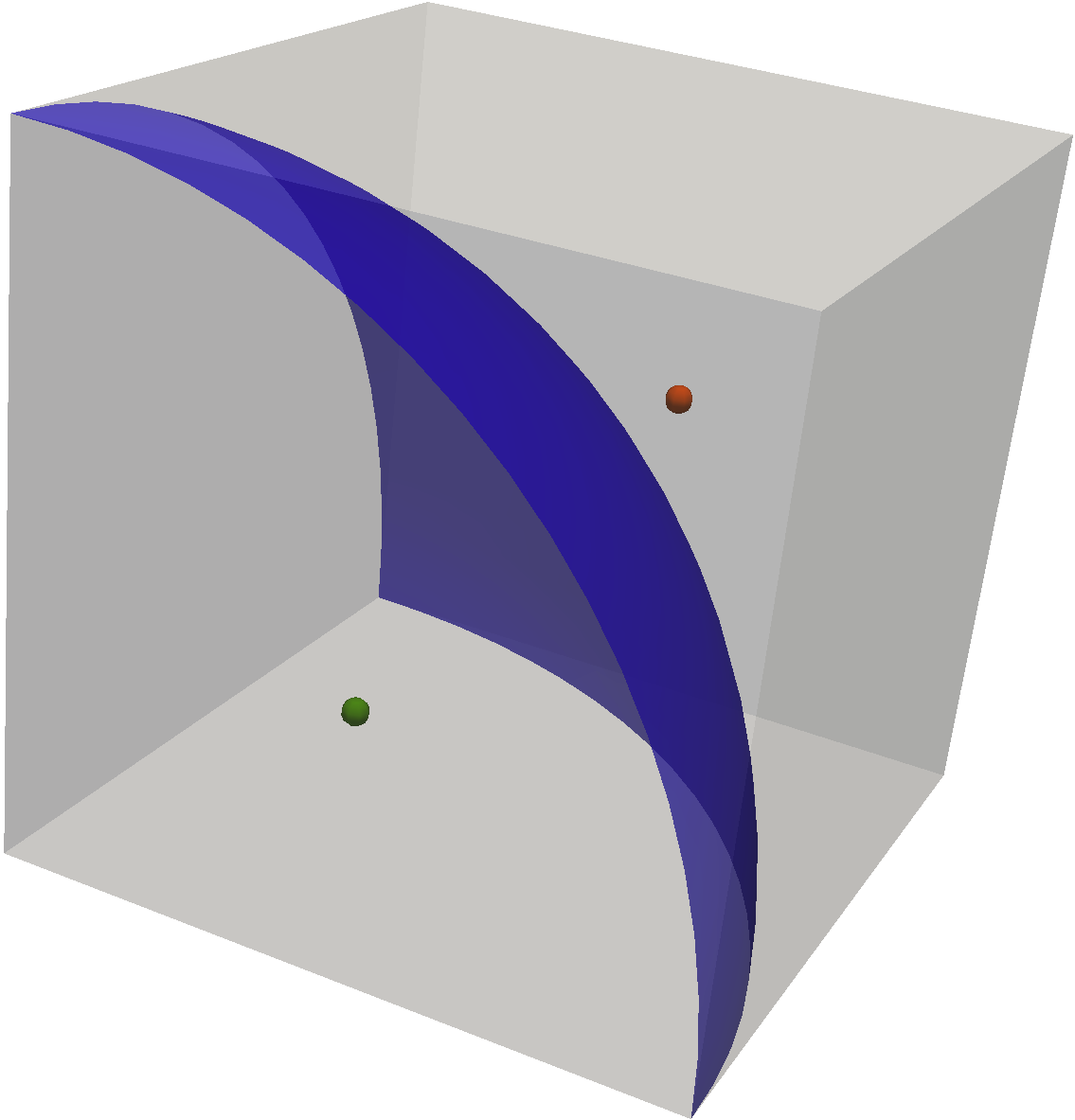}} $\,$
        \subfigure[1-d flat]{\includegraphics[width=0.31\columnwidth]{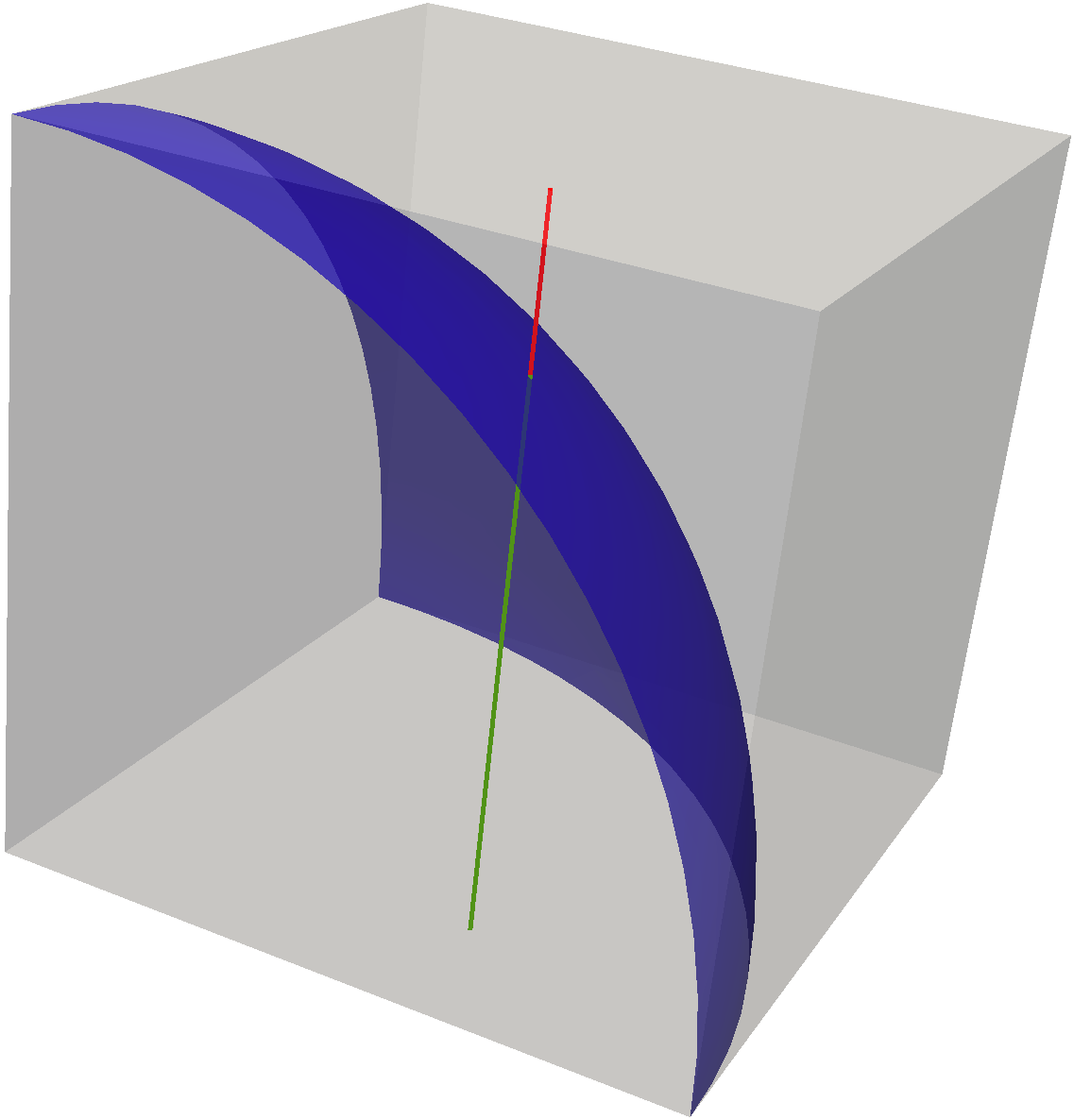}}  $\,$
        \subfigure[2-d flat]{\includegraphics[width=0.31\columnwidth]{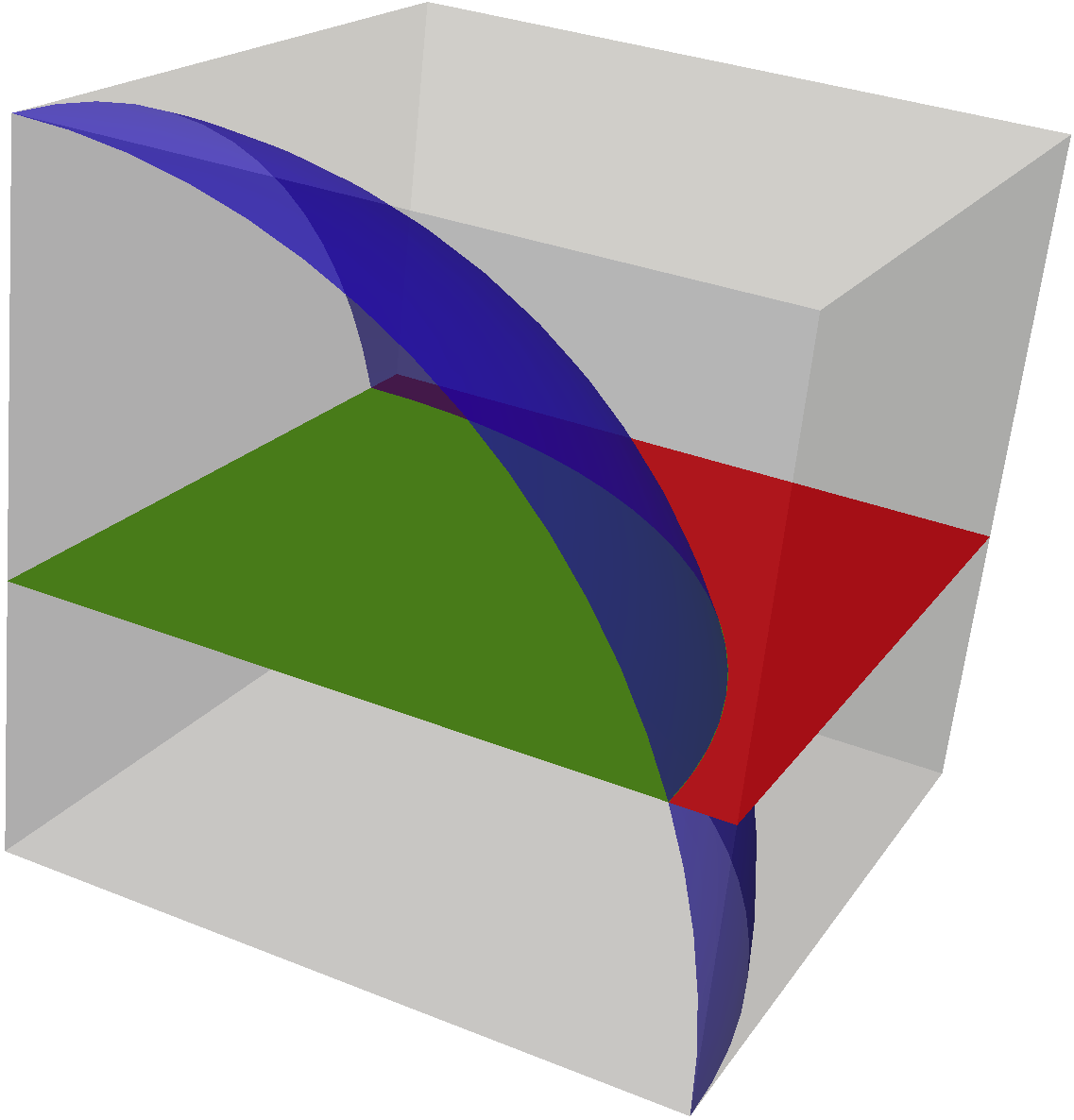}}
    \caption{\kflats{} used to estimate the volume of a sphere. The fraction of the sub-flats inside the sphere estimates the function average. } \label{fig:sphere_flats}
       \end{center}
\end{figure}

A simple example that demonstrates this concept is estimating the volume of a unit sphere by sampling from its bounding box:
$f=-1$ inside the sphere and $0$ outside it, and we seek an estimate of $\int_{f<0}1$.
\figref{fig:sphere_volume} shows the relation between error and time as the size of the sample increases. 
We sampled using \kflats{} of dimension $k = 0, 1, 2$.
For a point sample, we checked if the point was inside the sphere. 
For higher dimensions, we calculated the fraction of the flat inside the sphere; see \figref{fig:sphere_flats}.
We performed both Monte Carlo (MC) and Latin Hypercube Sampling (LHS).
For each sample size we ran 100 experiments and calculated the error in the volume estimate.
Plane samples consumed less CPU time than point samples for the same RMS error.
For MC sampling the payoff was about a factor of 5, and for LHS sampling the payoff was 3 to 8 orders of magnitude! 
The reasons behind these gains are the following:

\begin{itemize}
\item Evaluating $f$ along \kflats{} is cheap; in this case we exploited the analytic function of the sphere. 
\item A \kflat{} gives more information as $k$ increases.
\item A flat is cheap to generate. Each \kflat{} requires $d - k$ random numbers; here $d = 3$.
\end{itemize}

\begin{figure}[!hb]
       \begin{center}
       \subfigure[Thin object]{\hspace{0.2in} \includegraphics[width=0.3\columnwidth]{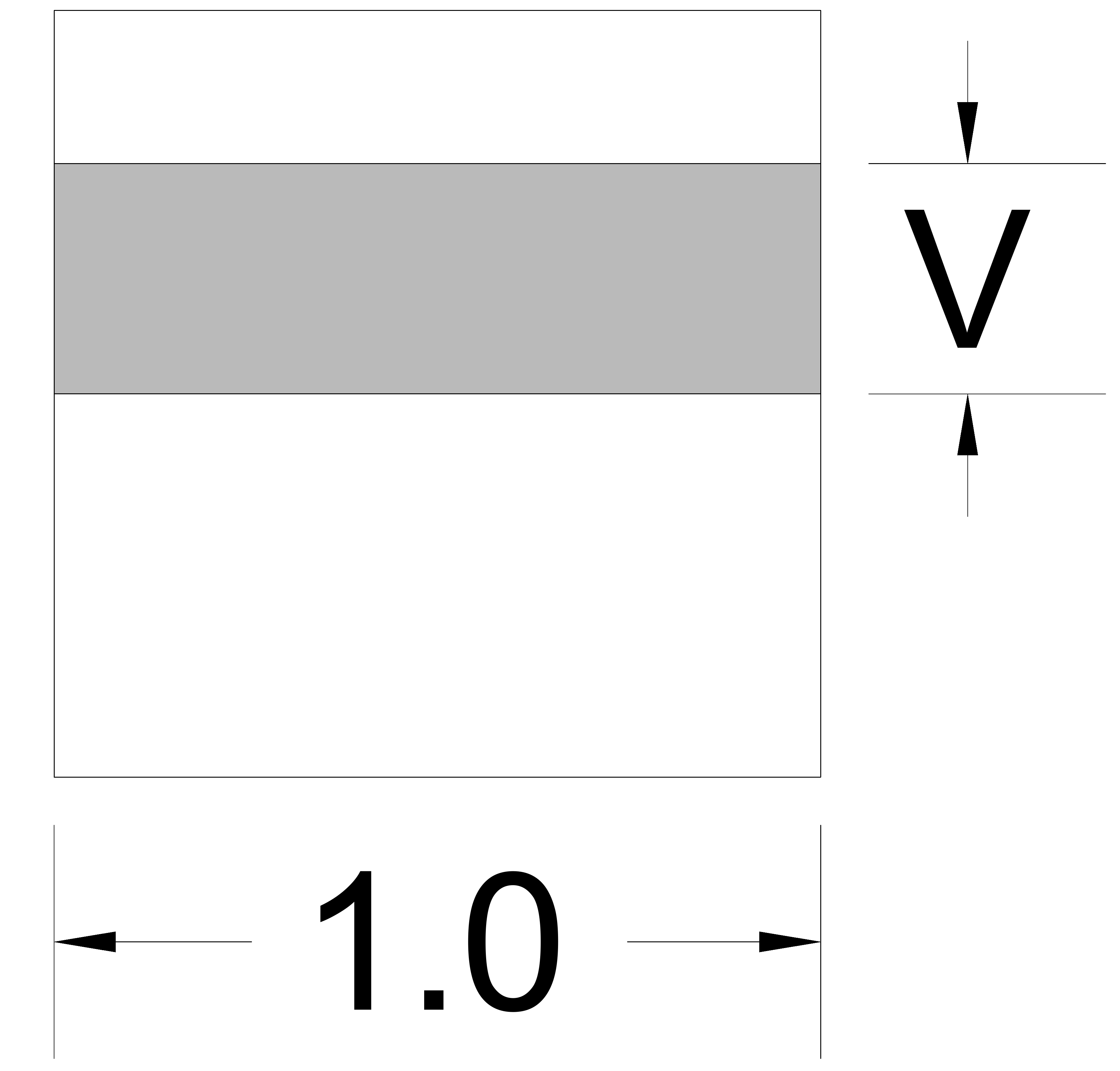}\label{fig:extreme_thin}} \hspace{0.06\columnwidth}
       \subfigure[Cubic object]{\hspace{.26in} \includegraphics[width=0.3\columnwidth]{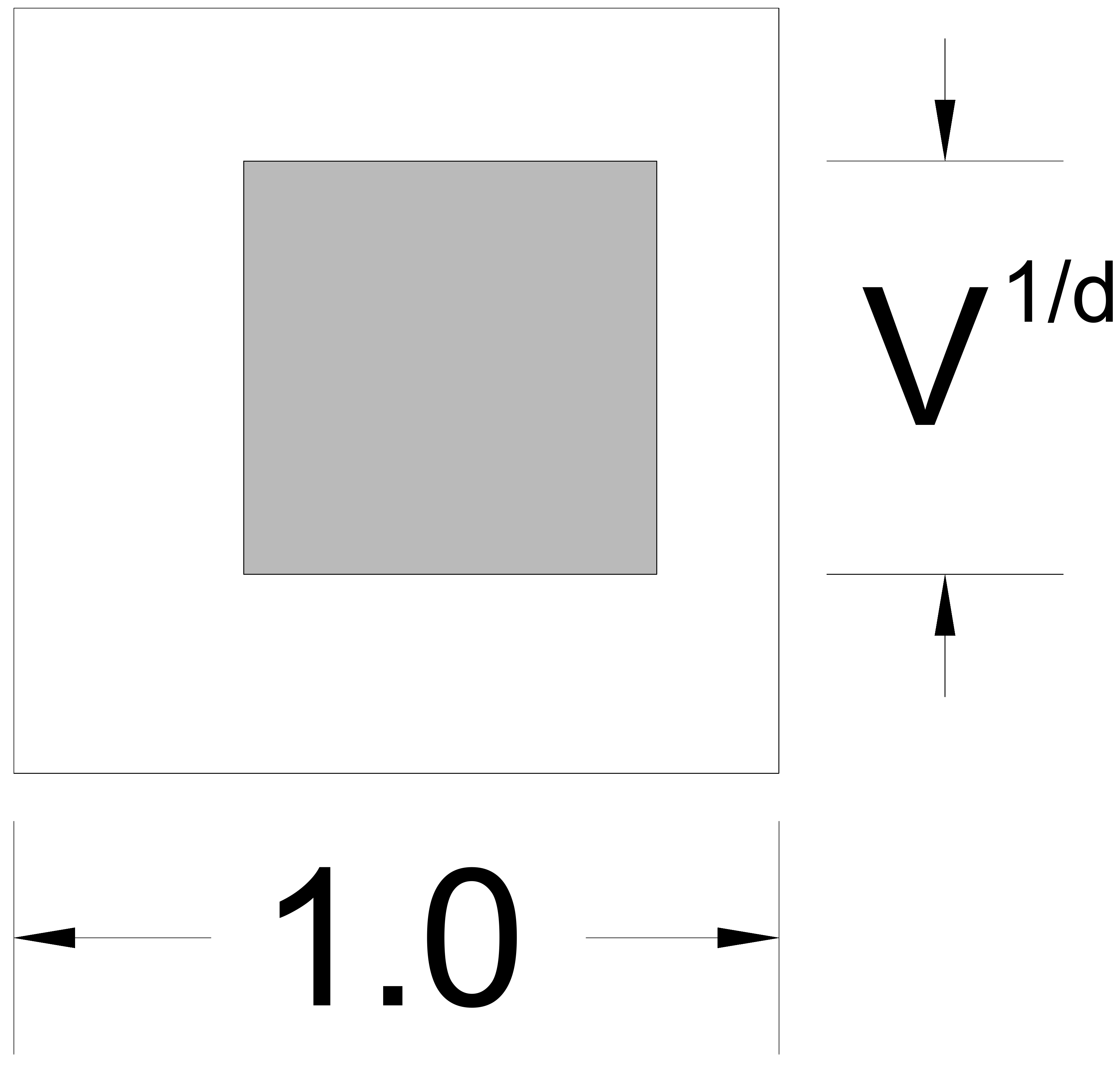}\label{fig:extreme_fat}}
        \caption{Extreme subregion shapes. In general, a \kflat{} has a better chance to intersect a region of interest as $k$ increases. For a given region volume, the advantage is higher for stretched regions than for square ones.}       
        \label{fig:extreme_cases}
       \end{center}
\end{figure}

In general, evaluating the integration function along a \kflat{} costs more than at a single point. However, for many problems, this extra cost is offset by the superior capability of a \kflat{} to capture narrow regions. For instance, consider  \figref{fig:extreme_thin}, where a line flat perpendicular to that narrow region of interest will capture it regardless of its thickness. On the other hand, the probability of a point sample landing in the region approaches zero as the thickness decreases.

The purpose of this paper is to formalize and demonstrate the \dart{} approach.
In \secref{sec:applications}, we show three practical applications:
completing a relaxed maximal Poisson-disk sampling, in dimensions 4--30;
rendering depth-of-field blur in four dimensions; 
and estimating the probability of failure
from a response surface for uncertainty quantification, where the probability is small, e.g.\ 1e-5, and the dimension is large, e.g.\ 15.
In each of these three applications the space is of moderate dimension, and line darts are particularly effective.
The experiments in \secref{sec:dart_experiments} verify the accuracy of higher-dimensional darts, using a volume estimation application.

%\todo{ the uq example we use is only two dimensions. it would be best to have a higher dimensional example. Otherwise, change this to "emulator in two dimensions".}
%\todo{we need to do an adequate job of describing sampling techniques. I think we should include the relevant ones during each of the following application contexts}

%{Note citation style: A random citation~\cite{Ebeida11} using cite.
%a short citation~\shortcite{Ebeida11} using shortcite.
%Both appear to be used parenthetically.}

% motivation
% gentler version of the abstract

\section{Previous Work}
\label{sec:prev-work}

Our work generalizes sampling.
There are many patterns for generating samples.
For graphics, maximal Poisson-disk sampling (\secref{sec:prev-work-mps}) is common. 
Much prior work focuses on dimensions 2 and 3, but our interests extend beyond.
Rendering applications use sampling in many forms (\secref{sec:prev-work-rendering}), often with high dimension. 
The field of uncertainty quantification (non-graphics, \secref{sec:prev-work-uq}) attempts to quantify the range of a computation, typically by performing the computation many times with a well-distributed selection of input parameters.  This is important in computational science for predictions and reliability.

%\todo{we use $N$ for number of samples in UQ, and $n$ for number of samples for the other apps. Make this consistent.}
Our \dart\ is a particular conception of high dimensional sampling.
Line sampling has been explored in varied contexts already, within graphics and other domains.
Unfortunately much of this work is isolated and does not consider dimension as a free parameter;
we hope to help provide a unified view of these approaches.

% scott for Keith
Lines were used early in the history of Monte Carlo sampling. For instance, ``Buffon's needle problem'' was published in 1777. %the 18th century. 
It considers the number of intersections of randomly-oriented short line segments with axis-aligned regularly spaced infinite lines.
Its solution involves geometric probability, either through integral geometry~\cite{Ramaley} or MC experiments~\cite{Hall1873edp}, and gives an estimate of the value of $\pi$.  Both the needle problem and volume estimation by line darts use finite objects, varying orientations, and infinite probes; but which things are known, measured, and estimated are different, as are which quantities are uniform-random and which are uniform-deterministic.  For example, using sets of orthogonal flats means the indices of the fixed dimensions of our flats are evenly spaced deterministically, whereas in Buffon's problem the geometry of the rule lines are evenly spaced.

In neutron transport physics simulations, a class of Monte Carlo algorithms known as ``track length estimators'' essentially performs Monte Carlo estimation using line segments~\cite{Spanier66}. This is in contrast to the ``collision estimators'' class that estimates using point samples.
In graphics, these collision estimators correspond to standard volumetric photon mapping, estimation using photon scattering locations.
Also track-length estimators (using line samples) correspond to photon beams; estimation uses entire random-walk path segments~\cite{Jarosz:2011:ACT}.
In surface reconstruction and CAD modeling, we can count the intersections of unoriented line samples with the surface, then make use of the integral geometry
Cauchy-Crofton formula to estimate integration quantities such as surface area or enclosed volume~\cite{Liu:2006:QCM:1649589.1649888}.
To get the right estimate with low variance, it is crucial to select the sample lines using the right probability model.
For example, models for bundles of uniformly-spaced parallel lines, reminiscent of both \darts\ and the regularly spaced lines in Buffon's needle problem; models for chordal lines; and pseudo-random and other numerical sequences have all been studied in the context of sampling surfaces~\cite{Rovira:2005:PSU:2386366.2386385}.

\subsection{Relaxed Maximal Poisson-Disk Sampling}
\label{sec:mps-intro}
\label{sec:prev-work-mps}
%literature survey of MPS. Highlight challenge of completing voids.
%
Maximal Poisson-disk Sampling (MPS) is a popular graphics technique to distribute a set of points in a domain. 
The points are random and have a blue-noise spectrum, which is well-suited to the human visual system and helps avoid visual artifacts. 
The points have a minimum distance between them, $r_f$, which helps to use the point budget efficiently. 
We denote the maximum distance from a domain point to its nearest sample by $r_c$. In a maximal sample, the $r_f$ disks around the points overlap to cover the whole domain, leaving no room to add another point, and $r_c \le r_f$. 
Otherwise, maximality is \emph{relaxed}, and we measure how far the geometry is from maximality by the distribution aspect 
ratio $\epsilon_r = r_c/r_f \ge 1$.
Point clouds with a meaningful upper bound on $\epsilon_r$ are 
separated-yet-dense, also known as 
``well-spaced.''

MPS algorithms abound, and often achieving maximality is the most challenging part. 
In some applications maximality is not required~\cite{Lagae08}.
The acceptable relaxation of maximality depends on the application. 
For example, in Voronoi mesh generation~\cite{ebeida_mitchell_vor}, 
the cells have an aspect ratio bound that varies smoothly with the relaxation, $2\epsilon_r$.
Some methods sacrifice maximality to terminate more quickly, but explicit statements about the achieved $\epsilon_r$ are rare. 
%While achieving strict maximality is often not required, it is useful to know how far a set is from it~\cite{Lagae08}. Here, we use a relaxed version of the maximal condition based on a probabilistic bound.

Many methods use some form of a background grid for point location and proximity queries~\cite{Ebeida11,Jones11,Wei08}.
The background grid may be refined, as in a quadtree, to track the remaining voids (uncovered regions)~\cite{Gamito09,White07}. 
This can be made efficient in dimensions up to about 5~\cite{EbeidaMPSEurographics2012}. 
However, even in dimensions below 5, refinement methods can run out of memory as the sample size increases. 
Memory problems are exacerbated on a GPU\@.

Uncertainty quantification motivates MPS sampling in higher dimensions, e.g.\ 10--30. 
No MPS methods in the literature scale to these dimensions due to the so-called ``curse of dimensionality.''
Classical dart throwing~\cite{Cook86,Dippe85} is not strongly dependent on dimension, but as the number of accepted samples increases, the runtime for the next sample becomes prohibitive and the algorithm must terminate well before maximality. 

Consider sampling a unit box with disk radius $r$ in dimension $d$.
Some issues are fundamental to the problem, independent of any specific algorithm or application.
The size of a maximal sampling $n$ is indeterminate, but its lower and upper bounds %$n_{\min}$ $n_{\max}$ 
grow exponentially in $1/r$ and $d$.
The kissing number, the number of disks that can touch another disk, grows exponentially in $d$.
(The geometry literature discusses these issues extensively. The densest packings and largest kissing numbers by dimension are summarized in
``A Catalogue of Lattices''~\cite{lattice}.)

The curses of dimensionality for grid-based methods go beyond those unavoidable issues:
\begin{enumerate}
\item The base grid size grows exponentially and faster than the output point set. \label{it:base_grid}
\item A grid cell is refined into $2^d$ subcells. \label{it:refine}
\item The number of nearby cells that might contain a conflicting sample grows faster than the kissing number.\label{it:kissing}
\item The ratio of misses/hits grows exponentially with the dimension~\cite{EbeidaMPSEurographics2012}. \label{it:miss_rate}
\end{enumerate}

Item~\ref{it:base_grid} arises because 
the size of the base grid is usually chosen such that each cell can accommodate at most one point: base squares have diagonal length $r$ and edge length $r/\sqrt{d}$. Worse, at maximality, the number of empty base cells grows exponentially with the dimension. These empty cells increase the time, and especially the memory requirements, to prohibitive levels. One possible solution is to choose a base grid level with \emph{edge} length $2d$, so that every square must contain one point. However, due to Item~\ref{it:refine}, this approach just defers the problem until cells are required to be refined a couple of times to represent voids. Because of the kissing number, representing voids through geometric constructions~\cite{Ebeida11} or explicit arrival times~\cite{Jones11} do not appear to be viable solutions at high dimension.

Some approximate methods~\cite{Wei08} put an upper bound on the number of misses per cell, which partly addresses Item~\ref{it:miss_rate}. The drawback is that the sampling is not maximal. How far it is from maximality has not been analyzed, but  volume arguments suggest that for a fixed box size, the number of allowed misses must grow exponentially in $d$ to bound the linear distance between an uncovered point and a sample's disk.

While some of these issues affect runtime, the real curse is the memory requirements for quadtrees.
Simple MPS~\cite{EbeidaMPSEurographics2012} is the quadtree method with the best memory scaling by dimension.
It seems unlikely to extend to even $d=10$ in the near future. %The consequence of the lack of computation and memory scalability is that 
Simple MPS uses a flat quadtree of same-sized squares, periodically refining all remaining squares uniformly.
We compare our results to Simple MPS in \secref{sec:experimental_results}.
%Today's maximal-Poisson-disk point sampling techniques are insufficient for high dimensions. %sampling functions of large dimensionality. 

\subsection{Motivation for High-Dimensional Point Clouds}
In the design of computer experiments, we generate points in parameter space, then evaluate a function at the sample points.  
We often build a surrogate model based on those points' values, e.g.\ Kriging models.
The dimension is equal to the number of parameters, so can be very high. The time it takes to generate the points is often very small compared to evaluating the function, so it is worthwhile to spend the time to find a set of points that span the space efficiently.
Well-spaced points use the point budget efficiently, and provide bounds on the condition number and interpolation error.

For some applications, if the function is not too expensive, it may make sense to sample the function directly using \dart{}s.
For example, if the integral of a function over a flat is available analytically, then sampling it directly using \dart{}s would be very efficient.
Otherwise, \dart{}s can provide well-spaced points (see \secref{sec:mps}); we can build a surrogate model over those points; then \dart{}s can integrate the \emph{surrogate model} analytically.

Another application where sample points are required is finite element simulation, where we need a computational mesh of the points.
%Most surrogate models are based on function values at a set of discrete points.

%to approximate this integral numerically.

%Even when the analytic integral is not available on the true model, a surrogate model (based on the available realization of the true model) 
%can be constructed and utilized to give a good evaluations for these integrals.

%\todo{Mohamed add some more motivation}.

\subsection{Rendering}
\label{sec:prev-work-rendering}

%\todo{I'd still rather see the examples here briefly touch on each of
%  the following: (a) very brief summary of what the previous work
%  does; (b) how that previous work is actually a higher-dimension k-d
%  dart; (c) why a higher-dimension k-d dart is better than taking
%  point samples; and (d) how you might make the previous work better
%  now that we actually understand k-d darts. Largely, missing c and
%  d.--JDO}

High-quality rendering is an important application of
multi-dimensional sampling. Photorealistic effects like
motion blur, depth-of-field (defocus) and soft shadows can be expressed as
integrals over multiple dimensions. Classical techniques often employ
stochastic point sampling to estimate these integrals. Noise-free
rendering using point sampling can require a large number of
samples, which can be extremely expensive for complicated scenes.

% ++ Anjul, include "The authors may find it helpful to refer
% to the work of Don Mitchell (variance in the context of antialiasing)
% and Eric Veach et al. (on Monte Carlo for low-variance estimation in
% rendering), and place their work within that context and the work that
% follows from it."
Thus a long history of research has targeted choosing samples wisely.
Mitchell's classic antialiasing paper~\shortcite{Mitchell:1987:GAI},
for instance, ``focus[es] on reducing sampling density while still
producing an image of high quality,'' primarily in the context of ray
tracing. Metropolis light transport~\cite{Veach:1997:MLT} ``performs
especially well on problems that are usually considered difficult'' by
using mutations to preferentially (but in an unbiased way) sample
light paths in interesting regions of the path space.

Besides choosing samples carefully, another approach toward the same
goal is to reduce the number of required samples through 
techniques such as sample reuse and/or caching. For example, a notable
recent advance in this area, by Lehtinen et
al.~\shortcite{Lehtinen:2011:TLF}, specifically notes ``a clear need
for methods that maximize the image quality obtainable from a given
set of samples'' and exploits anisotropy in the temporal light field
to reuse samples between pixels. Our work has a similar goal---making
the best use of a limited number of samples---but instead reduces the
number of necessary samples by increasing their dimensionality.
% ++ Anjul,
% I feel the similarity to the clever use of samples in Lehtinen et al.
% 2011 for constructing smooth depth of field images from very low
% numbers of samples is appropriate given the primary application of
% this papers contribution to rendering.

% ++ Anjul, we should include the analogy, with references, that one
% reviewer wrote: "Just like an accumulation buffer samples 2-d images
% in a 3-d spatio-temporal space in order to more efficiently integrate
% the result, the proposed method generalizes to sampling lines, "
\dart{}s formalize a general way of multi-dimensional sampling.
Several early graphics applications use multi-dimensional samples in
limited and specific scenarios. The OpenGL accumulation
buffer~\cite{Haeberli:1990:TAB} uses a form of \dart{}s for motion
blur: each output pixel is an aggregate of several input pixels, each
of which may span a 2-d region (pixel area) for a constant shutter
time. Essentially, the accumulation buffer samples 2-d $x$-$y$ images
in a 3-d $x$-$y$-time space. Nelson Max~\shortcite{Max:1990:ASD} uses a
scan-line visible surface algorithm that generates line samples,
describing how to use the information in these samples to create
antialiased images.
% ++ Anjul, include "The authors should reference Nelson Max 1990,
% Antialiasing Scan-Line Data, an early application of line samples."

Recent research has also shown promise in rendering high-quality
motion blur using multi-dimensional samples. Gribel et
al.~\shortcite{Gribel:2010:AMB,Gribel:2011:HQS} present the use of
line samples (our ``1-d flats''). In their 3-d domain ($x,y,t$) they
fix $x, y$ and perform a 1-d flat in the $t$ domain\footnote{In their
  2011 paper, Gribel et al.\ use 2-d darts but call them line samples
  because they are lines in $x$-$y$ space.}. They also extend their
implementation to render motion-correct ambient occlusion.

More recently, line samples have proven useful in the representation
of light. Sun et al.~\shortcite{Sun:2010:LSG} represent lighting and
viewing rays directly in a 6-d Pl\"ucker space, which allows an
efficient formulation of finding nearby lighting rays. This, in turn,
allows accurate, fast rendering of large scenes with single scattering
even in the presence of occlusions and specular bounces. The
previously mentioned work of Jarosz et al.~\shortcite{Jarosz:2011:ACT}
concentrates on better representations for light paths in photon
tracing for the purposes of rendering participating media in light
interaction; previous methods had used photon particles. Instead, they
represent and store full light paths (samples), resulting in a more
compact and expressive lighting representation with corresponding
performance benefits.
% ++ Anjul: add
% Line samples have been used in scaterring media rendering algorithms, e.g.: 
% * Xin Sun, Kun Zhou, Stephen Lin, Baining Guo: Line space gathering for single scattering in large scenes. ACM Trans. Graph. 29(4): (2010) ~\cite{Sun:2010:LSG}
% * Wojciech Jarosz, Derek Nowrouzezahrai, Iman Sadeghi, Henrik Wann Jensen: A comprehensive theory of volumetric radiance estimation using photon points and beams. ACM Trans. Graph. 30(1): 5 (2011) ~\cite{Jarosz:2011:ACT}
% The lines samples used in these algorithms are not axis-aligned and thus more general than the 1-d darts introduced in this paper. 

Jones and Perry~\shortcite{Jones:2000:AWL} experimented with using analytical
line sampling for anti-aliased polygon rendering. They shoot
single-dimensional darts across a pixel's surface, analytically
compute triangle coverage for each of them, and then average them to
obtain pixel colors. 

Our paper generalizes the idea of multi-dimensional darts for
sampling, and it is this generalization that helps us design renderers
that use \dart{}s in different configurations. 
For rendering depth-of-field effects with added anti-aliasing, we 
present a generalized
configuration using a full 1-d dart in \secref{sec:rendering}.
That is, we use a set of orthogonal flats rather than just a single flat as in prior work.
We compute  high-quality depth-of-field images efficiently.

While we demonstrate just one configuration of \dart{}s, several different
strategies are possible for depth-of-field and other
effects. In general, \dart{}s offer a sampling process that converges
faster and has lower noise than point sampling; the use of \dart{}s
offers the potential benefit of faster convergence but must be weighed
against the higher complexity per \dart. 

% The orientation and density of the flats, and the weighting of the integral along a flat,
% may have to be carefully designed to 
% achieve the correct expected values, or at least to ensure that the deviation is not noticeable in the final rendering.

%\todo{Not sure if we are doing two versions.} We use two varieties of
%line darts in our implementation: .

%With all triangle contributions accounted for in a sample, it then
%sorts samples by starting time, and adds contributions (considering
%depth for transparency and order) until all these segments are
%processed.Though each line sample (1-d dart) is significantly more
%work than a point sample($0$d dart), a few line sample give a much
%better image than many point samples.  Importance
%sampling~\cite{importance_sampling} targets the interesting
%subdomain, but requires some (a priori or learned) knowledge about
%the function.  \todo{note: maybe less description here? andrew}

%% previous work does not consider split or joined darts, since they
%% stick to just one dimension.

%All of these graphics techniques are examples of \emph{joined} \dart{}s
%rather than split \dart{}s. The distinction between these two types of
%\dart{}s is explained in \secref{sec:overview}.

\subsection{Monte Carlo Sampling}
\label{sec:prev-work-uq}

Random sampling is one of the oldest~\cite{Hall1873edp} and most robust methods for uncertainty quantification. 
% a reviewer wanted this tidbit of history cut:
% Although it existed long before then, UQ first became a useful tool for solving problems with the advent of computers, specifically during the Manhattan Project of World War II\@.  At that time, Nicholas Metropolis and Stan Ulam named it ``Monte Carlo'' (MC) sampling as a reference to Monaco's famous casino.
%
The principal use of Monte Carlo (MC) sampling is to approximate a high-dimensional integral with 
a sample mean.  
%A relevant example is to compute the probability, i.e.\ the 
%mean rate of occurrence, that a system's output will lie in a ``failure region.''
The primary drawback of MC is the slow rate at which the sample mean converges 
to its true value. 
%Let ${X_i}$ be a 
%sequence of independent identically distributed random variables (i.e.\
%independent draws from the same distribution) with finite mean $\mu$ and 
%variance $\sigma^2$.
%The Central Limit Theorem states that 
%as the number of samples, $n$, approaches infinity 
%the sum of the sequence mean converges to the normal distribution, 
%$\mathcal{N}\left(n\mu,n\sigma^2\right)$.  With minor manipulation it follows 
%that
%\begin{eqnarray*}
%\lim_{n\to\infty}\frac{\sum_{i=1}^n\left(X_i-\mu\right)}{\sqrt{n}}&\sim&\mathcal{N}\left(0,\sigma^2\right),\\
%\lim_{n\to\infty}\frac{\sum_{i=1}^n\left(X_i-\mu\right)}{n}&\sim&\mathcal{N}\left(0,\frac{\sigma^2}{n}\right).
%\end{eqnarray*}
% In the limit of an infinite number of samples, the %standard deviation of the error (a.k.a.\ 
The 
standard error in the {\it computed}
mean for $n$ samples is
\begin{equation}
\sigma_\mathrm{err}=\frac{\sigma}{\sqrt{n}}. %\text replaced by mathrm
\label{eqn:mcstderr}
\end{equation}
Although this rate of convergence in $n$ is very slow, the number of dimensions, 
\ddim, does not appear in \equref{eqn:mcstderr}.  MC's primary advantage is that
it is not subject to the curse of dimensionality.
%: 
%to represent an unknown function to a given fidelity,
%the number of samples required is exponential in \ddim\@. 
Significant effort has been 
invested in developing variants of MC with faster rates of convergence; 
a full review is out of scope.
Latin Hypercube Sampling (LHS)~\cite{McKay1979ctm,Owen1992clt} is
known as ``N-rooks sampling'' in the graphics community.
%Jittered Sampling (JS)~\cite{liu1965ebj} is a 
%space-filling form of MC that randomly perturbs nodes of a tensor product 
%grid.
%Multi-Jittered Sampling (MJS)~\cite{chiu1994mjs} is  a two-dimensional combination of JS and LHS.
%Binning Optimal Symmetric Latin Hypercube Sampling (BOSLHS)~\cite{dalbey2010fast} generalized MJS to 
% higher dimensions and a number of samples that is not exponential in \ddim. 
Importance sampling~\cite{glynn1989importance} involves preferentially sampling rare cases and compensating 
by reducing their weight.
% Markov Chain Monte Carlo (MCMC)~\cite{gilks1996markov} is a random or drunkard's-walk MC with a carefully constructed path preference. 
%The goal of all these methods and our \dart{}s is to
%accelerate the convergence of a MC sample mean for certain types of problems

%{all, please remember the submission needs to be anonymous. Use \\ anon\{\} for what we put in this version and \\ known\{\} for what goes in the final version.}

%{all, see notation.tex for newcommand definitions of frequently occurring symbols, like $\kdim$ and $\ddim$ and $\dart$ and $\kdart$ and $\samp$ and $\linedart$ and $\linedarti$.}

%\section{\DART\ Framework}
%\section{{$k$-d} Dart Framework} %math string not allowed
\section{Dart Framework} %math string not allowed
\label{sec:overview}
Given the ideas of the method (\secref{sec:intro}) and where it might
be useful (\secref{sec:prev-work}), we now define it formally and give
a general recipe for converting a point sampling algorithm to a
\dart{} sampling algorithm. 
We consider two application scenarios.
The first is sampling to evaluate a function, as in numerical
integration; see \secref{sec:framework:f-over-domain}. 
The second is sampling to find locations in the domain where the function has
a particular value; see \secref{sec:framework:find-value}.

We begin with terminology. 
Function $\func$ is defined over domain $\domain \subset \mathbb{R}^\ddim$. 
Often $\domain = [0,1]^\ddim$. 
A \emph{flat} $\Flat$ is a
$\kdim$-dimensional subspace of the domain, the vector space defined
by fixing $d-k$ coordinates. A \dart{} $\adart^k$ is a sample
defined by a set of $I=\dchoosek$ $k$-dimensional flats, $\adart^k_i$,
one for each combination of fixed coordinates. So in $\mathbb{R}^2$, a
1-d dart might consist of the two 1-d flats at $x=0.5$ and $y=0.5$. 

%% slow formal description of the method
%We are interested in analyzing a function $\func$, defined over the
%domain $\domain \subset \mathbb{R}^\ddim$. Often $\domain =
%[0,1]^\ddim$. To evaluate $\func$ on $\domain$ with traditional point
%sampling techniques, we take multiple 0-dimensional point samples from
%this domain, evaluate $\func$ at each of those points, and average the
%results, weighting the samples as appropriate. Here, we wish to
%perform the same analysis and achieve the same results, but use
%\dart{}s rather than point samples.

\subsection{Evaluating a Function over a Domain}
\label{sec:framework:f-over-domain}
Consider an algorithm that receives a single value, $y=\func(x)$, from
evaluating a function $\func$ at a uniform random point $x$. To get an
analogous single value $y^{\prime}$ from a \dart, we calculate the
average of $\func$ over its flats, weighted by the relative measure of
the flats. We generate a \dart{}'s flats one at a time. For each flat
$\Flat$ we choose its fixed coordinates uniformly at random, then:
\begin{enumerate}
\item{Clip $\Flat$ at the boundary of the domain and estimate its relative volume $|\Flat|=\int_{\Flat}1$. For the unit box this is trivial.}\label{step:clip}
\item{Integrate the function along the clipped flat: $G = \int_{\Flat}\func$.}\label{step:int}
\item{Now the weighted average of $f$ over the flat is simply the
    ratio of these two computations: $H = G/|\Flat|$.}\label{step:weight}
\end{enumerate}
Then $y^{\prime}$ is simply the average of $H$ over the flats:
$y^{\prime}=\sum_{i=1}^{I} H_i / I$. We can combine multiple \dart{}
$y^{\prime}$ measurements in the same way that we might combine
measurements made at multiple point samples. Thus we can convert
point-sampling function evaluation to \dart{} function evaluation.

A generic approach (for any $k$-dimensional flat) is to use numerical integration over a discretization of the flat.
Represent a flat with a uniform grid (mesh). 
(We assume it is too expensive to mesh the entire domain.)
In (\ref{step:clip}), clip the grid elements at the domain boundary, generating simplices. 
In (\ref{step:int}), evaluate $f$ at the grid points and perform standard numerical integration over the grid simplices. 

Our depth of field, probability of failure, and volume estimation applications follow this recipe,
because they use Monte Carlo integration, but the integration along the line
flats can be made faster and more accurate given application-specific
knowledge about the function. For probability of failure, the
function we integrate is the 0--1 indicator function of failure,
rather than the continuous value of the response function. We improve
accuracy by finding the roots of a single-variable equation to find
the boundary points where the indicator function switches its value.
For depth of field, $\func$ is the contribution of one ray (photon) to
a pixel, and we seek to estimate the contribution of all photons over
all focal depths for each pixel. For efficiency we use discrete
algorithms to find the occlusion boundaries of triangles, then
integrate a continuous function over each non-occluded segment of a
triangle.
% anything else to say about integration?

\subsection{Identifying Points in the Domain with a Particular
  Function Value}
\label{sec:framework:find-value}
MPS represents a different category of application. In this
application, instead of integrating $f(x)$, we are looking for a
random $x$ that satisfies $\func(x)>0$, i.e.\ finding a point outside
all prior disks. Point-sampling techniques would sample the domain
until such a point was found, which is expensive if the region of
interest is small compared to the size of the domain. Instead, we can
use \dart{}s for sampling using the following generic recipe:
\begin{enumerate}
\item{Clip $\Flat$ at the boundary of the domain. Retain $g$, a
    representation of the clipped flat. (For MPS, we clip a line by the
    unit cube.)}\label{step:clipflat}
\item{Retain the portions of $g$ where the function has the particular
    value of interest, e.g.\ $\func>0$. (For MPS, these are the
    subsegments outside all prior disks.)}
\item{Return a random point from $g$.}
\end{enumerate}

\section{Applying the Framework}
\label{sec:applications}
We now describe how to apply the \dart{} framework to our representative applications in more detail.

\subsection{Relaxed MPS}
\label{sec:mps}
%
% moved to introduction
%Maximal Poisson-disk sampling generates a point cloud that is separated-yet-dense, a.k.a.\  well-spaced.  The disk-free radius, $r_f$, specifies the minimum distance between any two points in the sample. The coverage radius, $r_c$, indicates the maximum distance between a sample point and any point (not necessarily a sample) of the domain. 
%If the sample is maximal then the radii are the same, $r_f = r_c$. 
%In relaxed MPS (RMPS) the coverage radius is greater than the disk-free radius to some extent.
%We define the \emph{distribution aspect ratio} as $\epsilon_r = {r_c}/{r_f}$ as a measure of maximality.
%
%In some graphics applications achieving maximality is not required~\cite{Lagae08}.
%An acceptable upper bound for $\epsilon_r$ depends on the application. For example, in Voronoi mesh generation~\cite{ebeida_mitchell_vor}, the cells of a relaxed MPS point cloud have aspect ratio below $2\epsilon_r$, which varies smoothly with $\epsilon_r$. 
%
%Some maximal Poisson-disk sampling algorithms suffer from the curse-of-dimensionality (\secref{sec:prev-work-mps}). 
%Others sacrifice maximality to terminate more quickly; explicit statements about the achieved distribution aspect ratio are rare. 
%
\begin{algorithm}
\begin{center}
\begin{algorithmic}
\label{alg:MPS}
\WHILE {maximality estimates are inadequate}
   \STATE generate a line dart \linedart
    \FORALL {$i=$permutation$(1..d)$}
      \STATE generate line segments $g = \linedarti \cap $ domain
        \FORALL {samples $\samp$}
          \STATE subdivide $g = g \setminus \disk{\samp}$
         \ENDFOR
         \IF {$g \neq \emptyset$}
           \STATE count \linedart\ as a hit
           \STATE select a sample point uniformly from $g$
           \STATE skip to next line dart
         \ENDIF
    \ENDFOR
   \STATE count \linedart\ as a miss
\ENDWHILE
\end{algorithmic}
\end{center}
\caption{A classical dart throwing algorithm using line darts.}
\label{alg:mps}
%\vspace{-0.1in}
\end{algorithm}

% our method
%Using \dart{}s, we provide a practical solution for the relaxed MPS problem in higher dimensions.
% reduced redundancy with intro
Our relaxed maximal Poisson-disk sampling algorithm is a variant of
traditional dart-throwing, throwing
point darts and keeping those that hit an uncovered region (void). 
Algorithm~\ref{alg:mps} specifies
our implementation in detail; in brief, we cast a line dart into the
domain and intersect it with the disks of previously accepted samples
to generate a set of uncovered segments,
then uniformly sample from those segments with a point dart. 
The user specifies an acceptable void volume $V$,
the fraction of the domain uncovered by sample disks.
We have a conservative stopping criteria based on the number of successive misses that usually achieves a smaller void volume. 
\figref{fig:mps_aspect_ratio} can be used as a guide for selecting $V$ in four dimensions.

\subsubsection{Complexity}
The memory requirements are only $O(nd)$, which is what is required to
represent the output point cloud because each sample has $d$
coordinates. Only $2n+d$ floats are needed for scratch space for line segments $g$.
We may generate and store only one flat of one dart at a time.
The runtime is $O(dn\log n + n d^2)$ per dart throw. 
% and this is even improved by employing a $k$-d tree to retrieve nearest neighbors rather than the lazy retrieval described in Algorithm~\ref{alg:mps}. 
The most significant feature is that the complexity does not suffer from a curse of dimensionality: there are no exponents containing $d$.
The number of throws is a function of $V$ and the miss rate; the community appears to lack the tools to analytically bound the miss rate, but we show that for line darts it can be made to be reasonable, or at least more reasonable than the alternatives.

Our approach is efficient because a line dart is
more likely to intersect an uncovered region than a point dart.
Only extremely simple and one-dimensional data structures are needed.
The cost of throwing a line dart is nearly the same as a point dart.

\subsubsection{Output and Process}
The main drawback is that the output is not maximal, but its deviation can be estimated. 
A second potential drawback is that the process is not identical to MPS\@.
There is no proof that the expected outputs are the same.
Indeed, for a non-maximal sample, the probability of inserting the next sample point in a given disk-free subregion depends only on the subregion's area in MPS; but for our variant the probability depends also on how much prior disks cover the axis-aligned lines through the subregion.
The main effect appears to be the order in which points are introduced, rather than their spatial distribution near maximality.
Choosing the position of the next sample is dependent on many prior random decisions, so there are few noticeable patterns between one run and another with a different random number seed.  

MPS and Algorithm~\ref{alg:mps} follow a different \emph{process}.
The quality of the \emph{outcome}, the point positions, does not appear to be sensitive to this, and we see little distinction between the outputs using the standard measures of FFT spectrum, power, and anisotropy. 
Additionally, we are unaware of any formal definition of an ideal point cloud nor any proof that MPS produces it, so exactly matching the MPS process and its output are not strict requirements. The conventional goal in the graphics literature is ``blue noise,'' meaning no discernible axis or boundary aligned patterns, and something resembling a Heaviside step function for the mean radial power, as in \figref{fig:blue_noise_freq} right.

\subsubsection{Implementation Details}

%\paragraph{Line Darts}
%\todo{implement and test mps using split darts, show results.}
% the efficiency differences between split and line darts are not mps specific, so were moved to the overview.tex file. It doesn't belong here anymore.
%The maximality estimates for coverage are better for split darts because they are independent. 
%However, the experimental results for joined darts showed that these, too, produced acceptable output. 
%At the time of the submission, we only implemented and tested the method using joined darts. 

\paragraph{$k$-d Tree}
To speed up the iteration over prior samples when generating segments $g$ we use a $k$-d tree.
This allows us to prune samples whose disks are too far away to intersect the line of $\linedarti$.
A $k$-d tree requires little memory regardless of the number of dimensions.
%The tree requires little memory, as every accepted sample is a single tree node. Non-leaves store a branching-dimension index and pointers to two child nodes.  samitch: nothing new here, this level of detail is not needed.

\paragraph{Line Segments $g$}
Beyond the $k$-d tree and storing the accepted samples, our only data structure is $g$, an array of segments in the line of $\linedarti$.
Each stored segment is inside the domain but outside the sample disks processed so far.
The set $g$ may be implemented as a 
one-dimensional linked list storing the $i$th coordinate of the starts and ends of segments: $g=[a_0b_0a_1b_1\ldots a_qb_q]$ where $q\le n$ if the domain is convex. 

\newcommand{\gp}{g^{\prime}}
\newcommand{\ap}{a^{\prime}}
\newcommand{\bp}{b^{\prime}}
\newcommand{\cp}{c^{\prime}}
To update $g$ for a new disk $\disk{c}$ centered at sample $c$, we first compute $\gp=\disk{c} \cap l = [\bp,\ap]$.
%
% Mohamed, uncomment the following to get all the details
%
%The closest point of $l$ to $c$ is $\cp$: replace all but the $i$th coordinate of $c$ by the coordinates of $l$. 
%The distance $h$ between $\cp$ and $c$ is $\lVert c-\cp \rVert_2$, computable in $d$ time.
%If $h$ is greater than $r$, the disk does not intersect the line. Otherwise march distance $v=\sqrt{r^2-h^2}$ along $l$ from $\cp$ to find $\ap$ and $\bp$.
%
Using binary search in $O(\log n)$ time, we find the position in $g$ where $\bp$ and $\ap$ should appear.
If they are beyond $a_0$ or $b_q$, the disk does not intersect $g$.
If $\bp$ and $\ap$ both lie between $a_j$ and $b_j$, the latter segment is split into two: $a_j \bp \ap b_j$.
Otherwise, if only $\bp$ lies between $a_j$ and $b_j$, then a segment is trimmed by replacing $b_j$ by $\bp$; a similar step applies for $\ap$.
Any endpoints between $\bp$ and $\ap$ are covered by the disk and discarded.
The updates are $O(1)$ time for a linked list.
%more details: (Since the list is at most length $2n$, no dynamic memory is required.)
%The list may be implemented using a fixed length array.) or randomized $O(1)$ time using an array with enough scratch space and considering the 

To choose a point from $g$ uniformly, we sum the lengths of the segments $L$, choose a random number between $0$ and $L$, and use binary search to find the corresponding point on $g$.

\paragraph{Maximality Estimates}
The user sets the remaining void volume $V$ that is acceptable.
Here we show how to translate that into a conservative stopping criteria. %, the number of consecutive misses $m$.
Denote the probability of hitting the void with a \dart{} by $P_k$.  
%Then on average, one out of $1/P_k$ darts will hit the void.  
Given a lower bound on $P_k$, we
set $m\ge\lceil 1/P_k \rceil.$
We stop after $m$ consecutive misses.

%m=\text{ceil}\left(1/P_k\right)
We seek lower bounds on $P_k$ in terms of $V$ in order to find a sufficiently large value of $m$.
For point darts, $k=0$ and $P_0=V$ regardless of void shape.
For higher dimensional darts, the worst shape for a void is 
a hypercube with edge length $b=V^\frac{1}{d}$.  
By worst shape, we mean the shape that has the smallest $P_k$ for a fixed $V$, assuming the entire domain is a hypercube.
For a hypercube void, the probability of hitting that void with a \kflat{} is given by  $p_k = V^\frac{d-k}{d}$. A \dart{} contains $l_k = {d \choose k}$ \kflats{} and hence $P_k = 1 - \left(1 - p_k\right)^{l_k}$. This gives a lower bound on $P_k$, and a sufficient value of $m$. We may consider this as a MC estimation of the remaining void using a sample size of the last $m + 1$ \dart{}s. 
As in any MC process, there is some variance, which decreases as $k$ increases. 
Moreover, the remaining void is typically scattered throughout the domain; $r_c$
%, the farthest distance from a domain point to its nearest sample, 
is less than if the void volume was a single ball.

%\begin{equation}
%r_c = \max_i ||p_i - q_i||_2 
%%P_k=\sum_{i=0}^k {d \choose i} b^{d-i} (1-b)^i.  \label{eq:Pk} % amsmath doesn't like \choose
%%P_k=\sum_{i=0}^k \binom{d}{k} b^{d-i} (1-b)^i.  \label{eq:Pk}
%\end{equation}

%\begin{equation}
%P_k=\sum_{i=0}^k {d \choose i} b^{d-i} (1-b)^i.  \label{eq:Pk} % amsmath doesn't like \choose
%P_k=\sum_{i=0}^k \binom{d}{k} b^{d-i} (1-b)^i.  \label{eq:Pk}
%\end{equation}

%\figref{fig:mps_misses} shows the values of $m$ for different $k$
%values using $V = 10^{-5}$ and $V = 10^{-10}$. For
%lower values of $V$, using a higher-dimension dart might be more
%efficient. 
%
%For joined line darts, \equref{eq:Pk} simplifies to 
%$P_1=V+d\left(V^{1-1/d}-V\right)$.  
%$P_1 \rightarrow V$ as $d \rightarrow \infty$.
%For split line darts, we have
%$P_1=1-\left(1-V^{1-1/d}\right)^d \rightarrow 1-V^{V/(1-V)} \approx -V \ln V$. 
%This suggests that split darts are better than joined darts at finding small voids for high dimensions.
%
%The probability of ``finding'' the void with $m$ throws 
%is the probability of not getting $m$ misses, or
%\begin{equation}
%P_\text{find}=1-\left(1-P_k\right)^m.
%\label{eqn:mpsPfind}
%\end{equation}  
%
%
%
%\todo{What information does P-find give us? Is the value of $m$ in terms of $P_k$ and $V$ correct, or does the user specify P-find?
%You can't solve for $m$ given just P-find, since $m$ is still in the equation?}
%\todo{ should we cut the following paragraph: 
%
%As an alternative to $V$, the user may specify $P_{\text{find}}$ ???
%Then \equref{eqn:mpsPfind} can be solved for $m$ instead
%of $P_\text{find}$. }

\subsubsection{Experimental Results}
\label{sec:experimental_results}          

\paragraph{Distribution Aspect Ratio}
The free radius, $r_f$, is the disk radius, the minimum distance between any two samples. 
The coverage radius, $r_c$, is the maximum distance between a domain point and its nearest sample.
We define the \emph{distribution aspect ratio} as $\epsilon_r = {r_c}/{r_f} \ge 1.$
This is a measure of maximality.

To compute this for a point cloud, we used
Qhull~\cite{Qhull} to generate a Voronoi diagram.
For each Voronoi vertex we retrieved the distance to its closest sample point: $r_c$ is the maximum of these distances.

\begin{figure}[!ht]
  \begin{center} 
    \includegraphics[width=0.48\columnwidth]{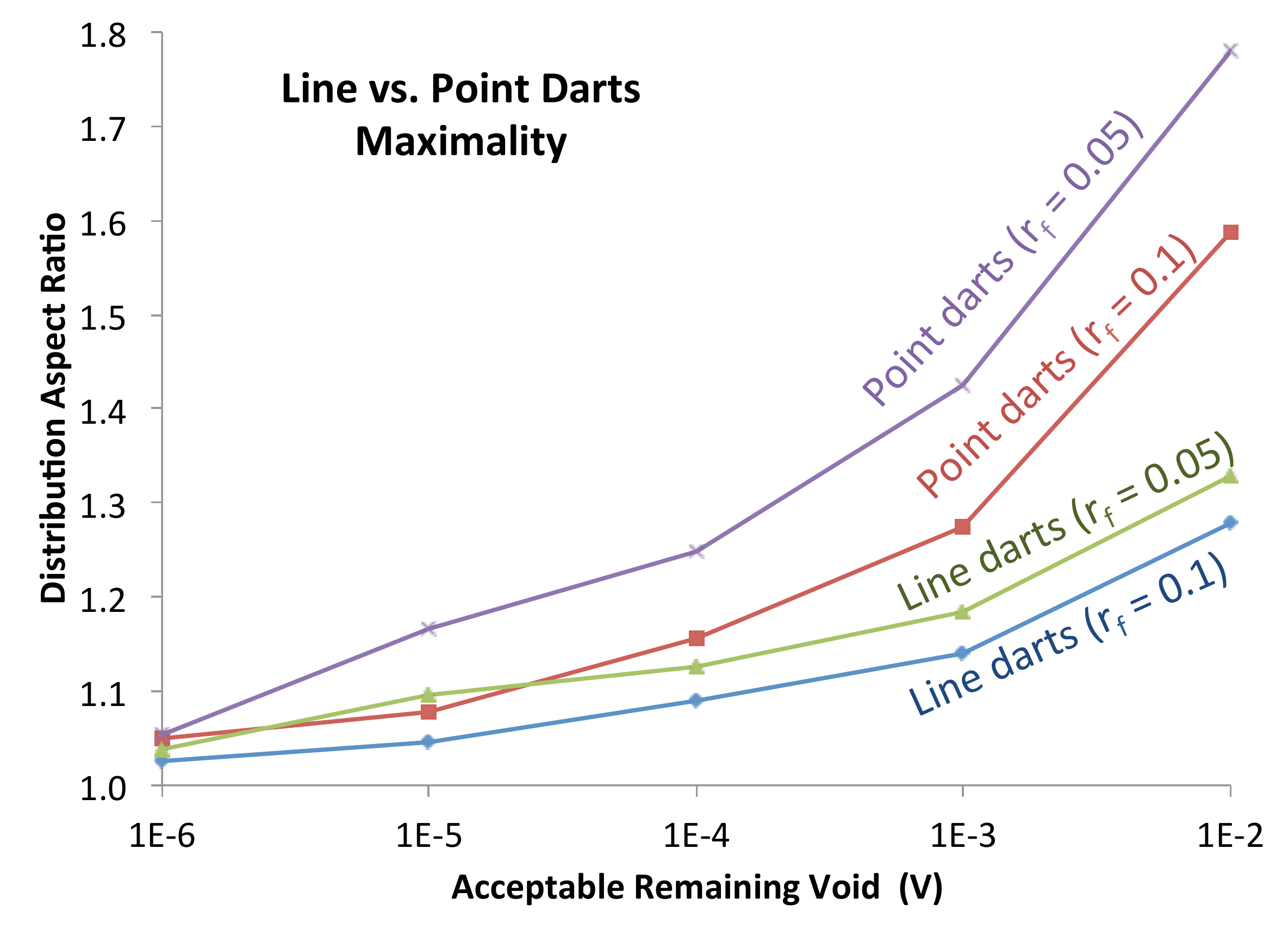}  \hspace{2pt}
    \includegraphics[width=0.48\columnwidth]{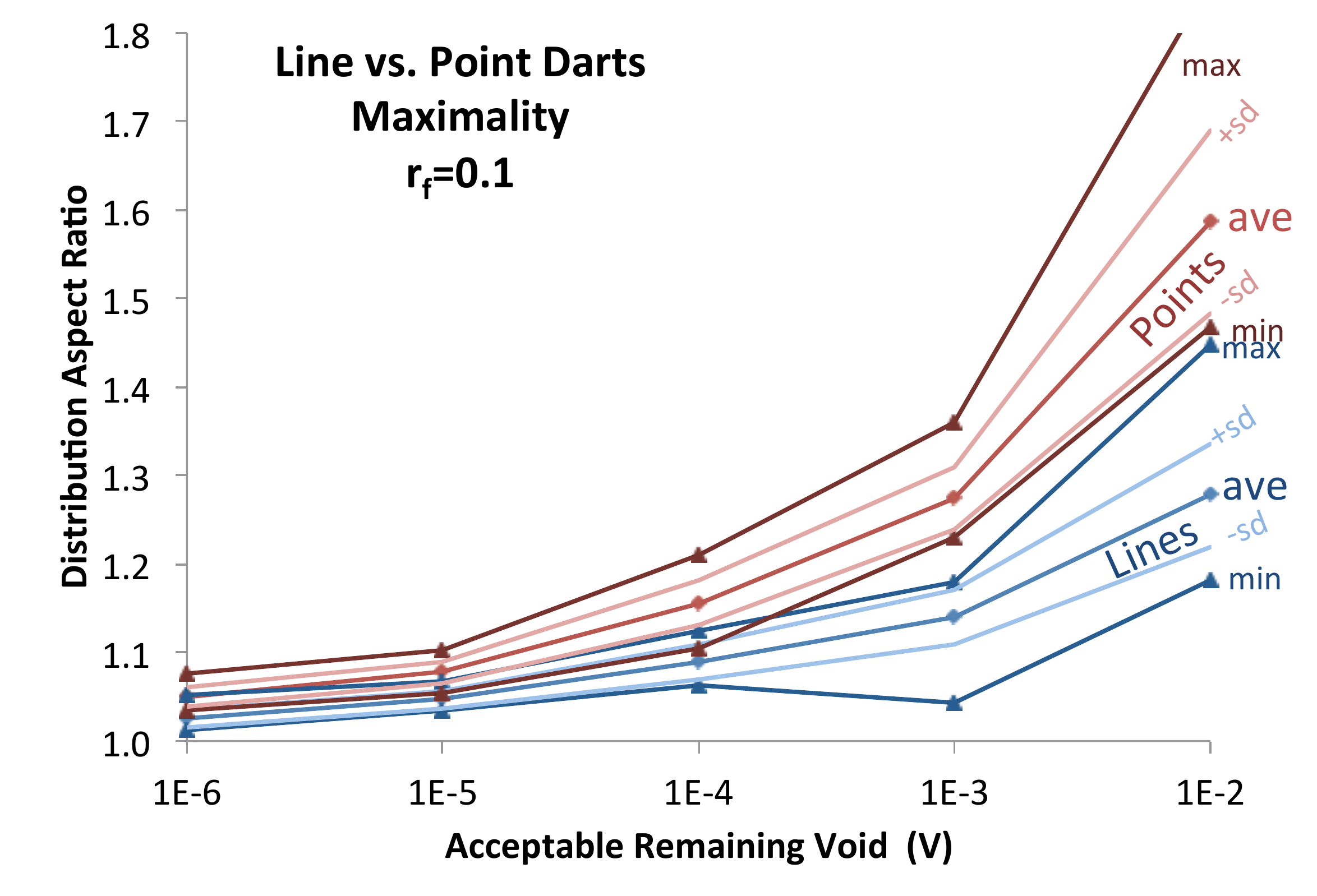} 
  \end{center}  
   \caption{Achieved distribution aspect ratio for void volume threshold $V$ in $d=4$. Data points are averages over 25 experiments. Right shows more statistics for $r_f=0.01$, including standard deviation, sd.} %See \tabref{tab:aspect_ratio} for more statistics.}
   \label{fig:mps_aspect_ratio}
\end{figure}

\begin{figure}[!ht]
 \begin{center}
   \includegraphics[width=0.78\columnwidth]{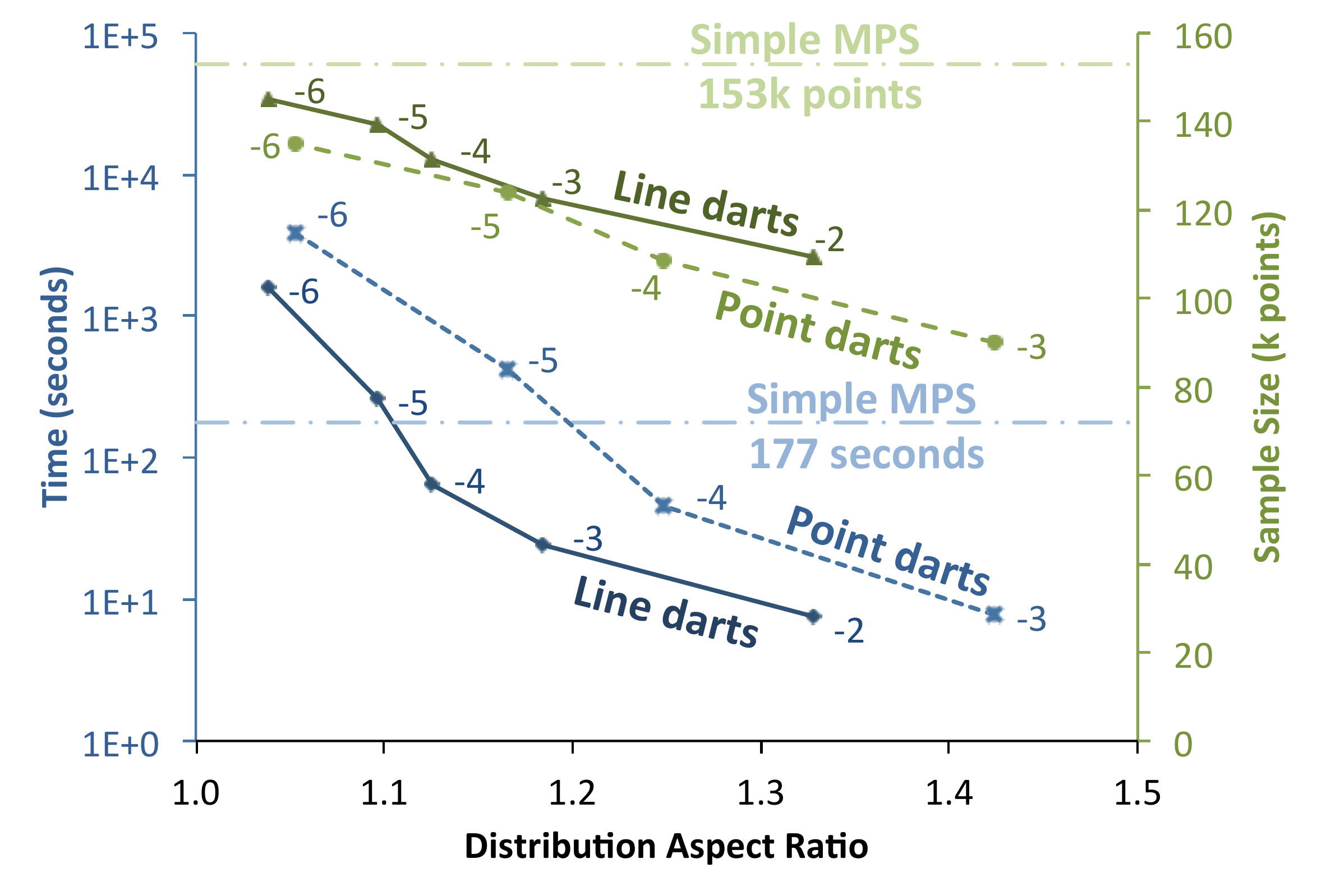}
  \end{center}
   \caption{$V$ threshold effects on time (blues) and sample size (greens) for line and point darts. Data points are labeled with $\log_{10}V$. As $V$ decreases, the sample approaches maximality and distribution aspect ratio 1.  Simple MPS~\cite{EbeidaMPSEurographics2012} % [5] 
   dashed lines are for a maximal distribution. \label{fig:mps_time}}
\end{figure}

\figref{fig:mps_aspect_ratio} shows the relation between the average distribution aspect ratio and the acceptable void volume over 
four-dimensional samples of two different disk-free radii: $r_f = 0.1$ and $r_f =0.05$. 
\figref{fig:mps_time} shows run-times for generating the point clouds. 
%Each data point in the figure represents an average value over 25 experiments using the same input parameters: $d, k, r_f,$ and $V$. 
%More statistical information for ($r_f = 0.1$) is given in \tabref{tab:aspect_ratio}. 
Line darts consistently produced better results.
%Combining the average value of the distribution aspect ratio and its standard deviation shows the advantage of line darts in relaxed MPS\@.
%
%
    
\begin{figure}[t]
  \begin{center}  
   \newcommand{\myfigheight}{0.37\columnwidth} %.41 for all on one horizontal row
  % without subfigure captions
    \subfigure   {\includegraphics[height=\myfigheight]{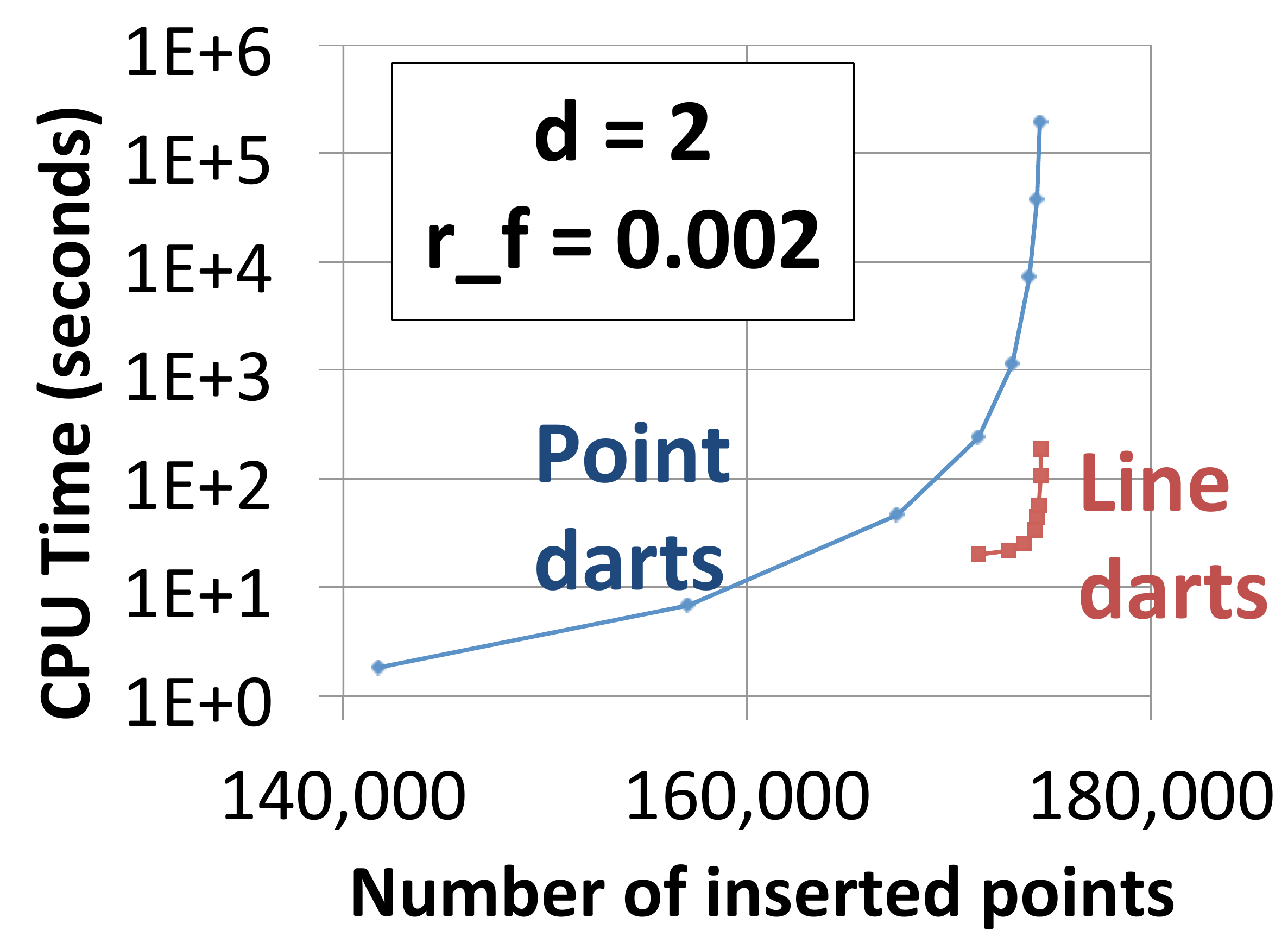}}
    \subfigure   {\includegraphics[height=\myfigheight]{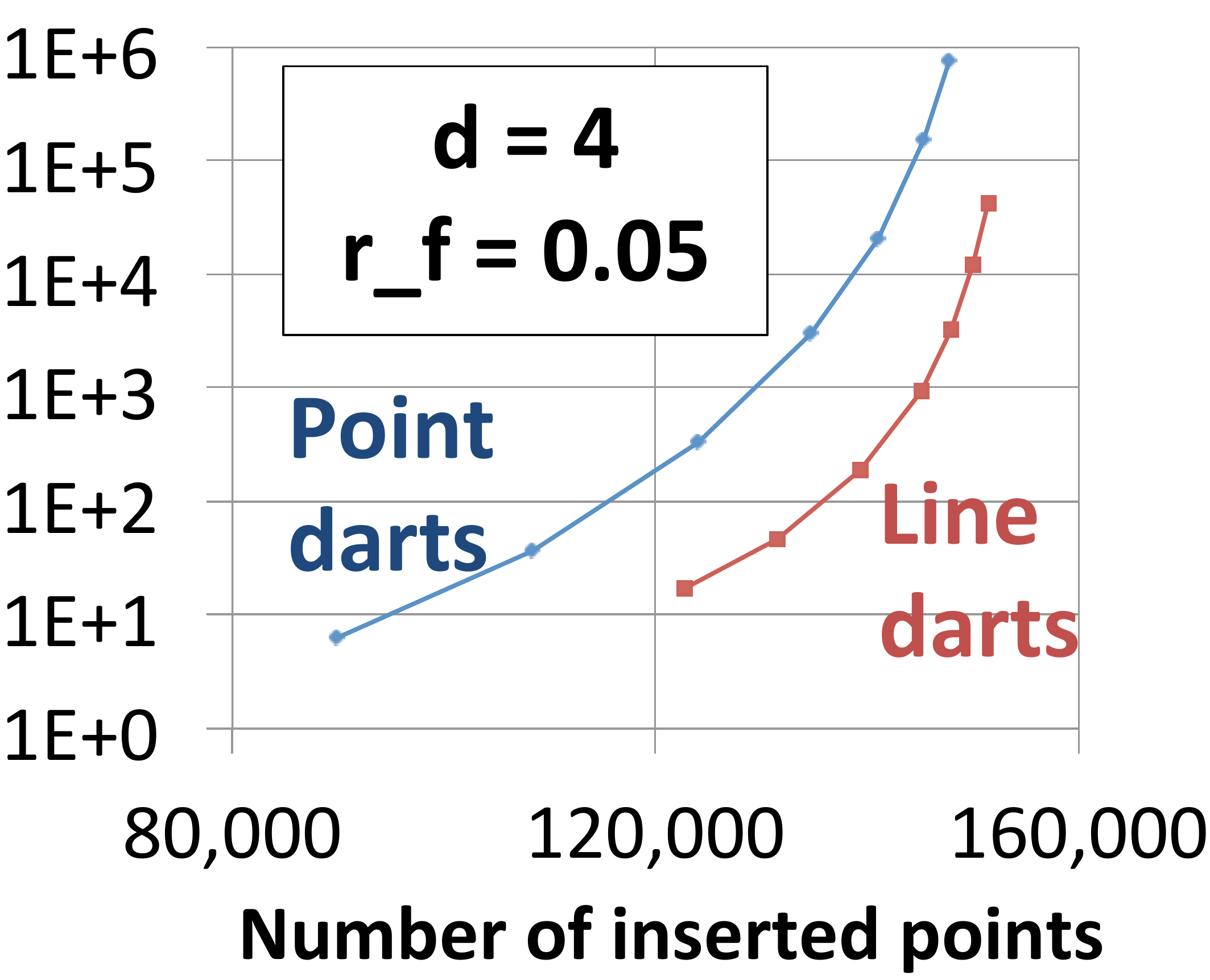}}
    \subfigure   {\includegraphics[height=\myfigheight]{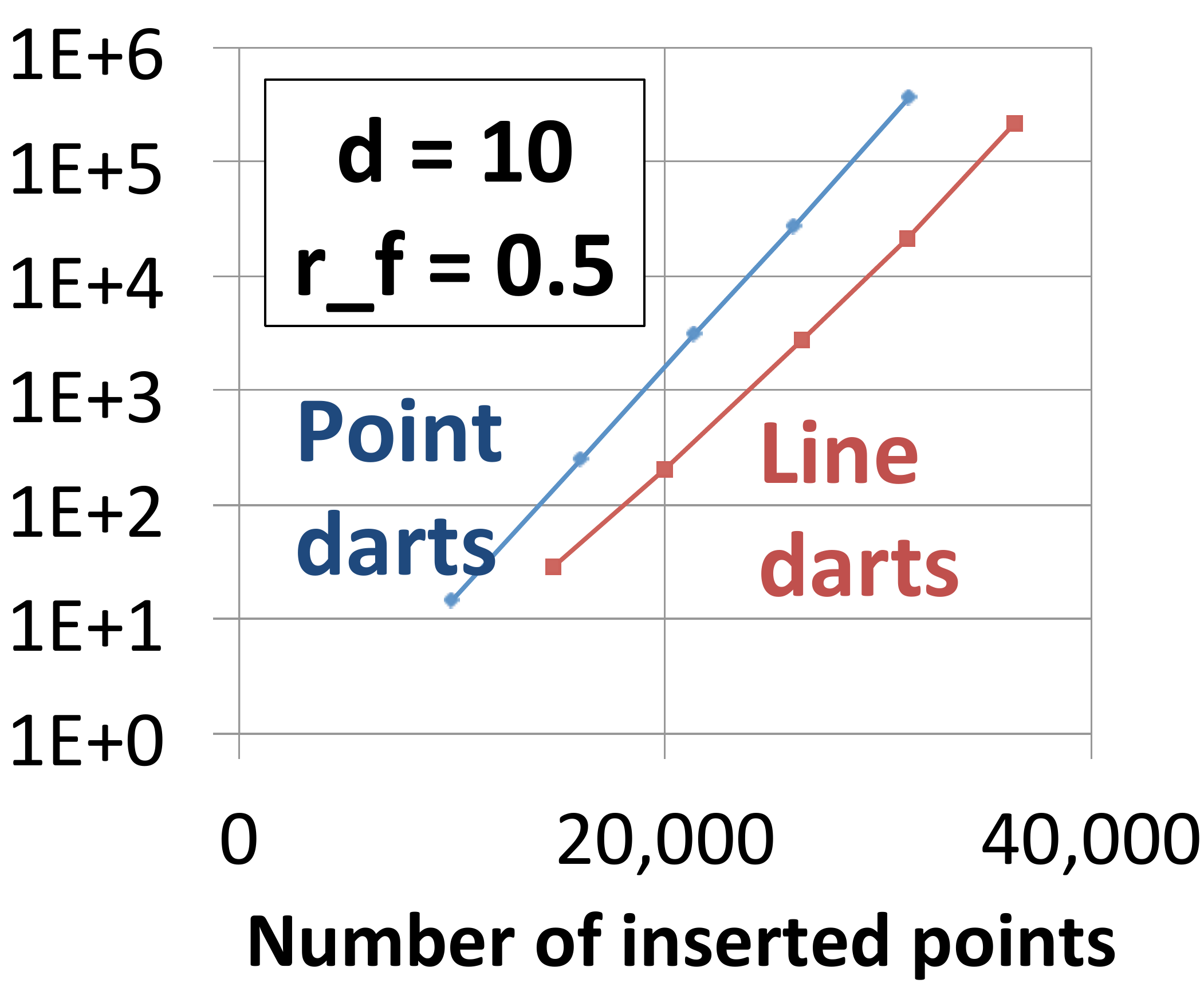}}
    \subfigure   {\includegraphics[height=\myfigheight]{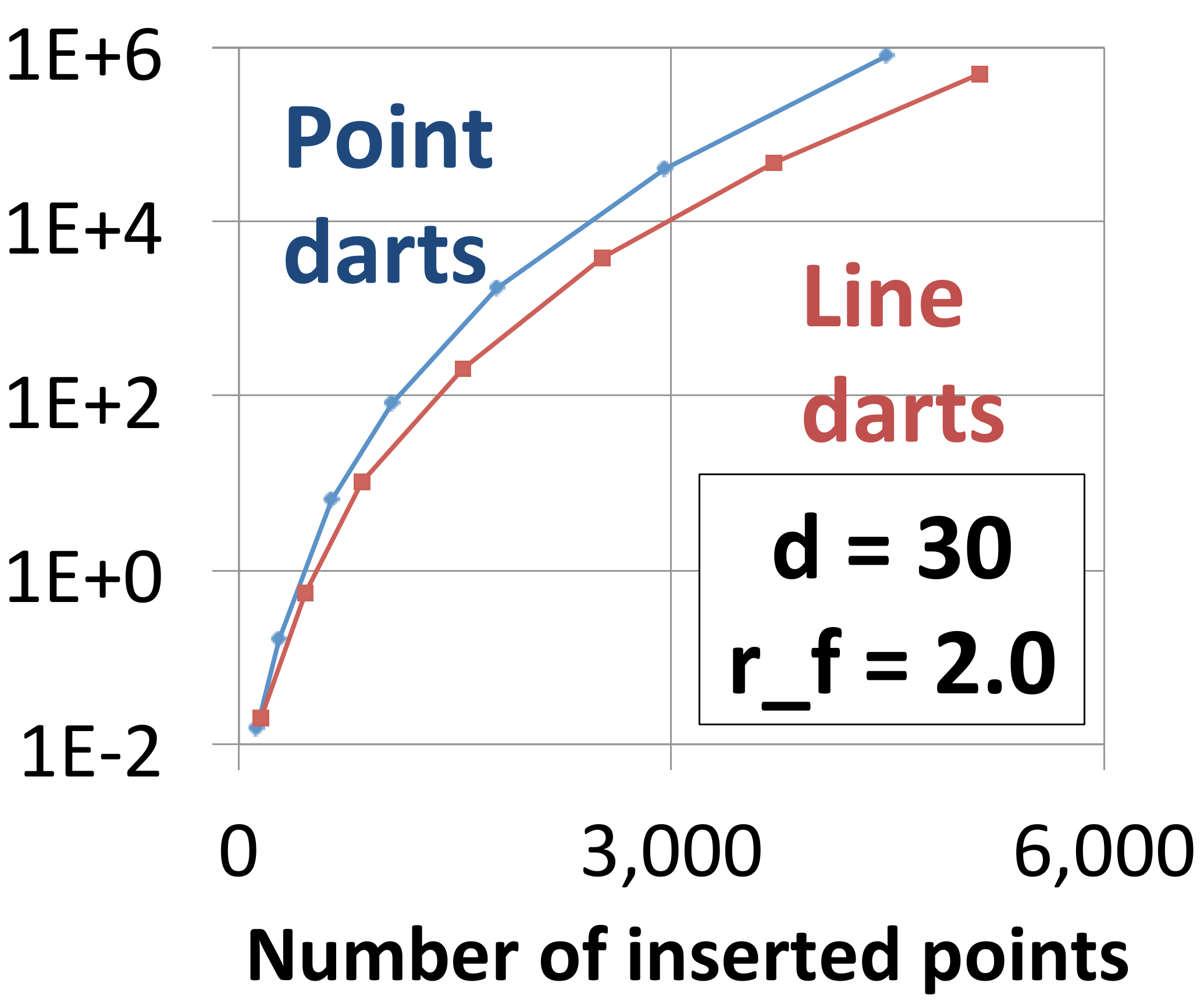}}
%    \subfigure   {\includegraphics[width=0.44\columnwidth]{2d_speed.pdf}}
%   \subfigure   {\includegraphics[width=0.44\columnwidth]{4d_speed.pdf}}
%    \subfigure   {\includegraphics[width=0.44\columnwidth]{10d_speed.pdf}}
%    \subfigure   {\includegraphics[width=0.44\columnwidth]{30d_speed.pdf}}
 % with subfigure captions, redundant with figure titles   
%    \subfigure [$\ddim=2$]  {\includegraphics[width=0.5\columnwidth]{mps_maximality_2d.pdf}}
%    \subfigure [$\ddim=4$]  {\includegraphics[width=0.5\columnwidth]{mps_maximality_4d.pdf}}
%    \subfigure [$\ddim=10$] {\includegraphics[width=0.5\columnwidth]{mps_maximality_10d.pdf}}
%    \subfigure [$\ddim=30$] {\includegraphics[width=0.5\columnwidth]{mps_maximality_30d.pdf}}
  \end{center}  
  \caption{Line darts approach maximality faster than point darts, as measured by the number of inserted points in a given run time. \label{fig:mps_maximality}}
\end{figure}

\paragraph{Speed of Approaching Maximality}
We tested our code over 2, 4, 10, and 30-dimensional domains using point darts and line darts. \figref{fig:mps_maximality} shows the number of points inserted over time. The expected void volume $V$ is related to the number of points; in practice the number of inserted points is a better indicator of maximality than our loose estimates of $V$ based on successive misses.
Line darts were able to generate larger samples for all $V$ and $d$.
\begin{figure}[!ht]
 \begin{center}
  \includegraphics[width=0.78\columnwidth]{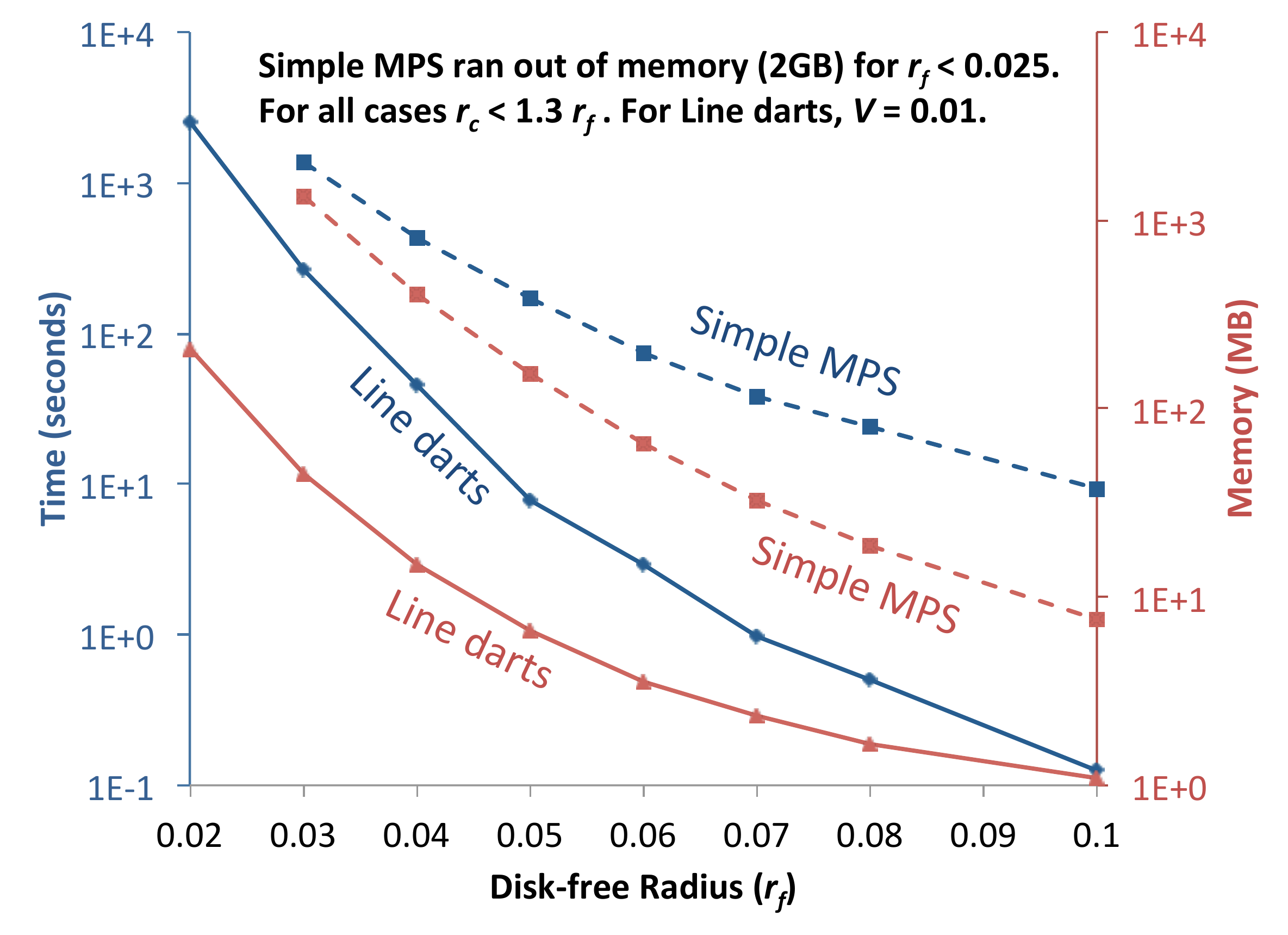} 
 \end{center}
   \caption{Time (blues) and memory (reds) for line darts compared to Simple MPS for the same acceptable void volume, $V$=1e-2.    \label{fig:mps_misses}}
\end{figure}

\paragraph{Efficiency by Method}
%See \tabref{tab:mps_comparison} for the efficiency of the method vs.\ the alternatives.
%citation inside caption doesn't work
\figref{fig:mps_misses} compares the performance of traditional
point darts and line darts (this work) to Simple MPS~\cite{EbeidaMPSEurographics2012}.
Recall Simple MPS is based on a flat quadtree, and is currently the fastest and
most memory-efficient of the provably-correct MPS methods. 
Our method is attractive at large values of $r_f$ for its
speed, and at low values of $r_f$ for its memory consumption. For
example, for $r_f < 0.025$, we generated 4M points in half an hour
using 107~MB of memory, while Simple MPS ran out of memory at 2~GB\@.

\begin{figure}[!ht]
  \begin{center}  
    \subfigure[Line darts]
    {
      \includegraphics[width=0.45\columnwidth]{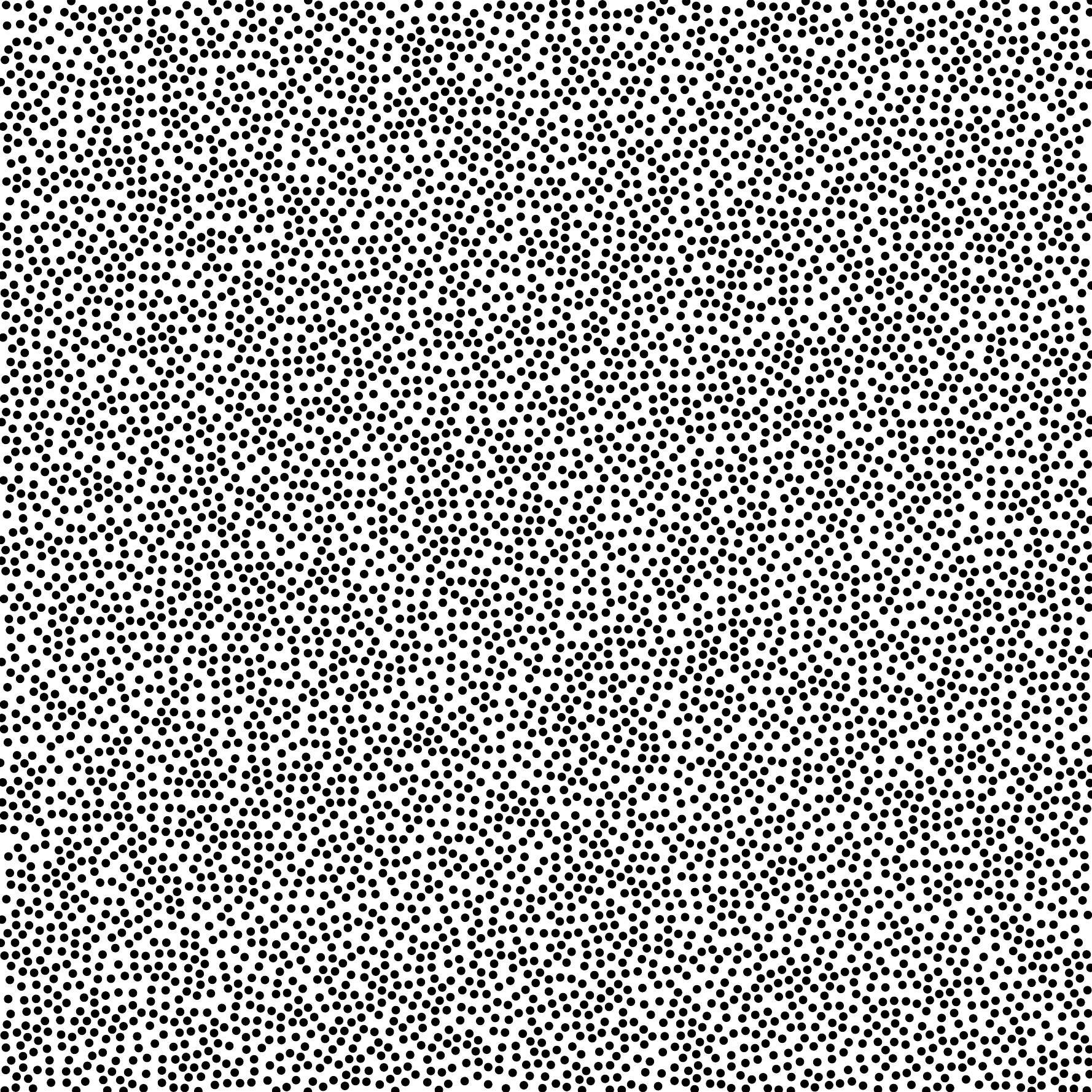} $\quad$
      \includegraphics[width=0.45\columnwidth]{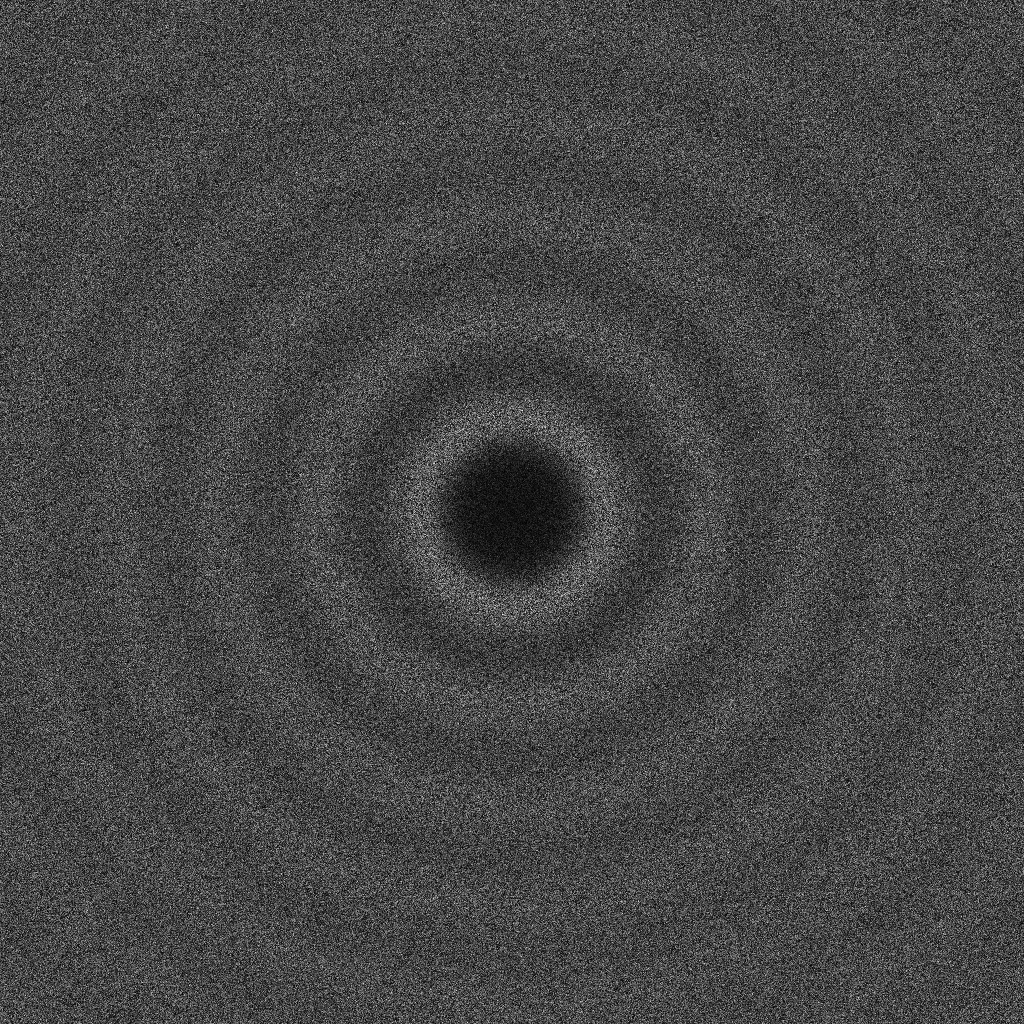}       
    } \\
    \subfigure[Point darts]
    {
        \includegraphics[width=0.45\columnwidth]{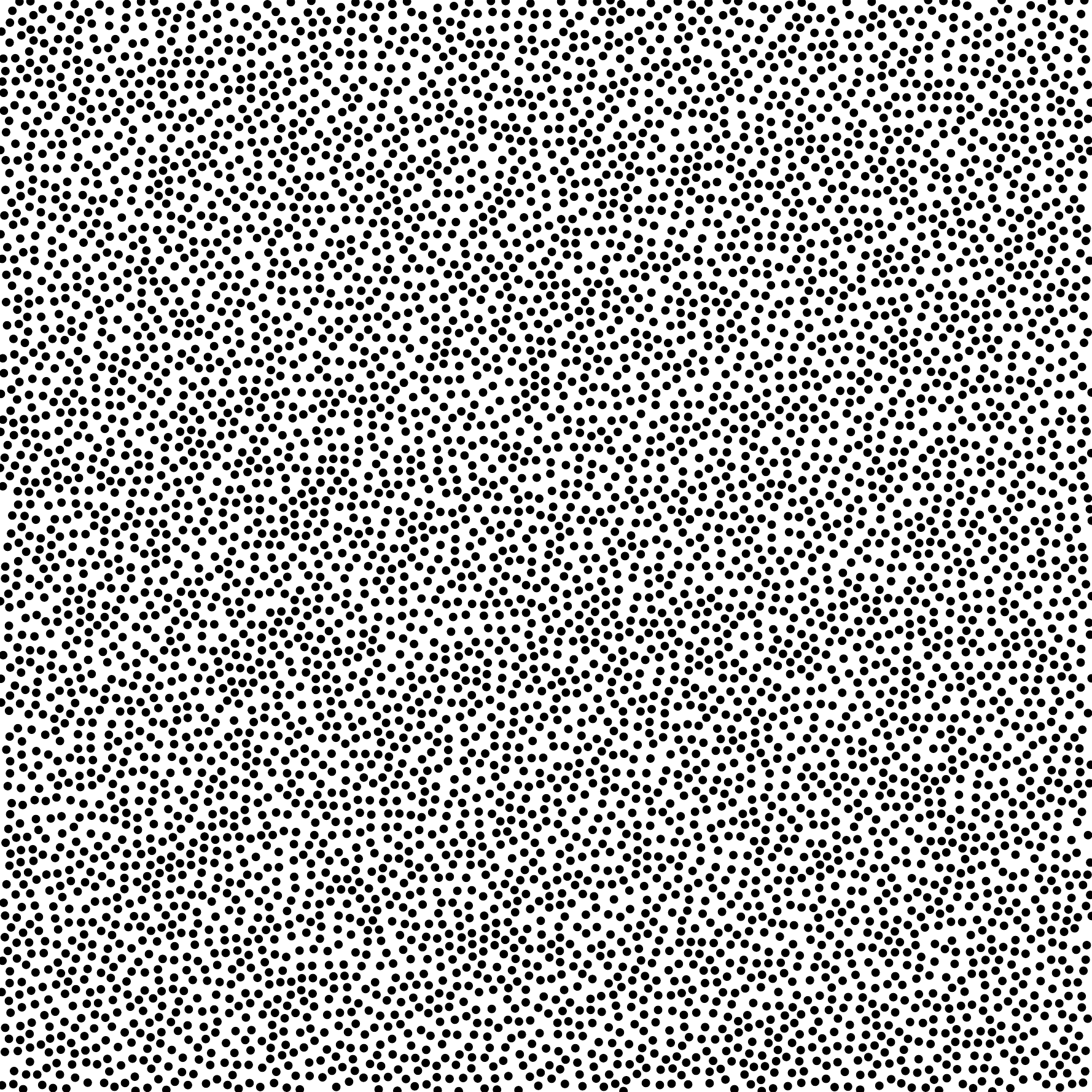} $\quad$
        \includegraphics[width=0.45\columnwidth]{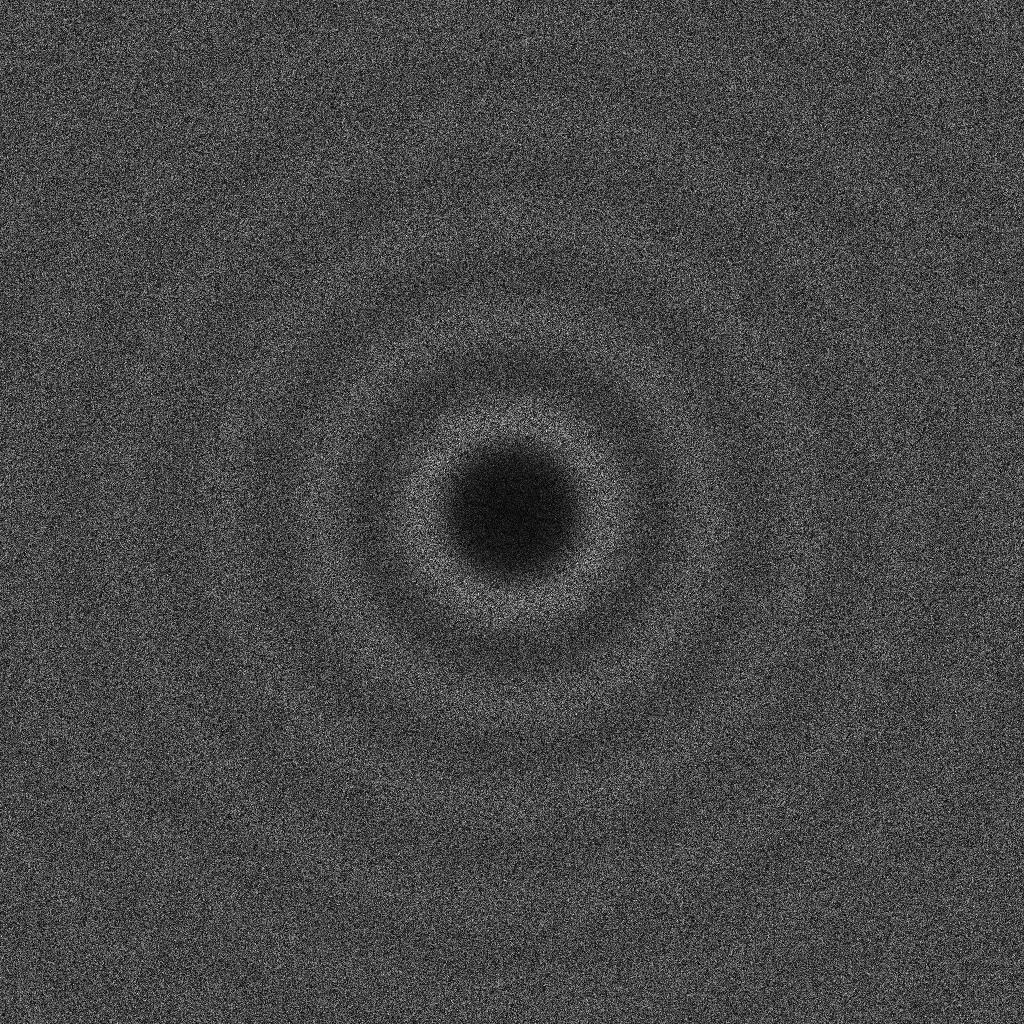}
    }
  \end{center}  
   \caption{FFT spectra (right) for the relaxed MPS point clouds (left)
     generated by line darts (top) and point darts (bottom).\label{fig:blue_noise_spec}}
\end{figure}

\begin{figure}[!ht]     
  \begin{center}  
    \subfigure[Line darts]
    {
      \includegraphics[width=0.45\columnwidth]{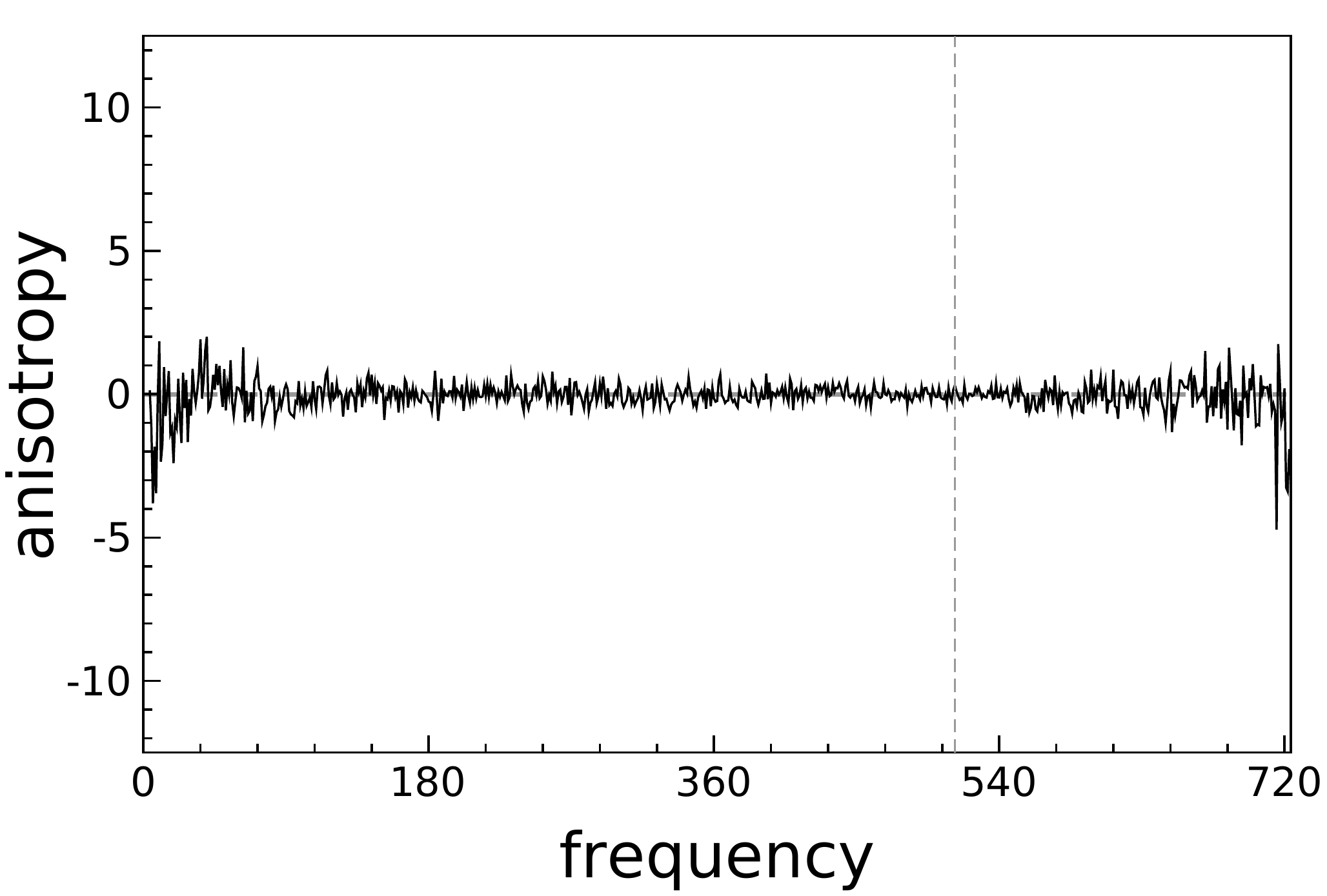} $\quad$
      \includegraphics[width=0.45\columnwidth]{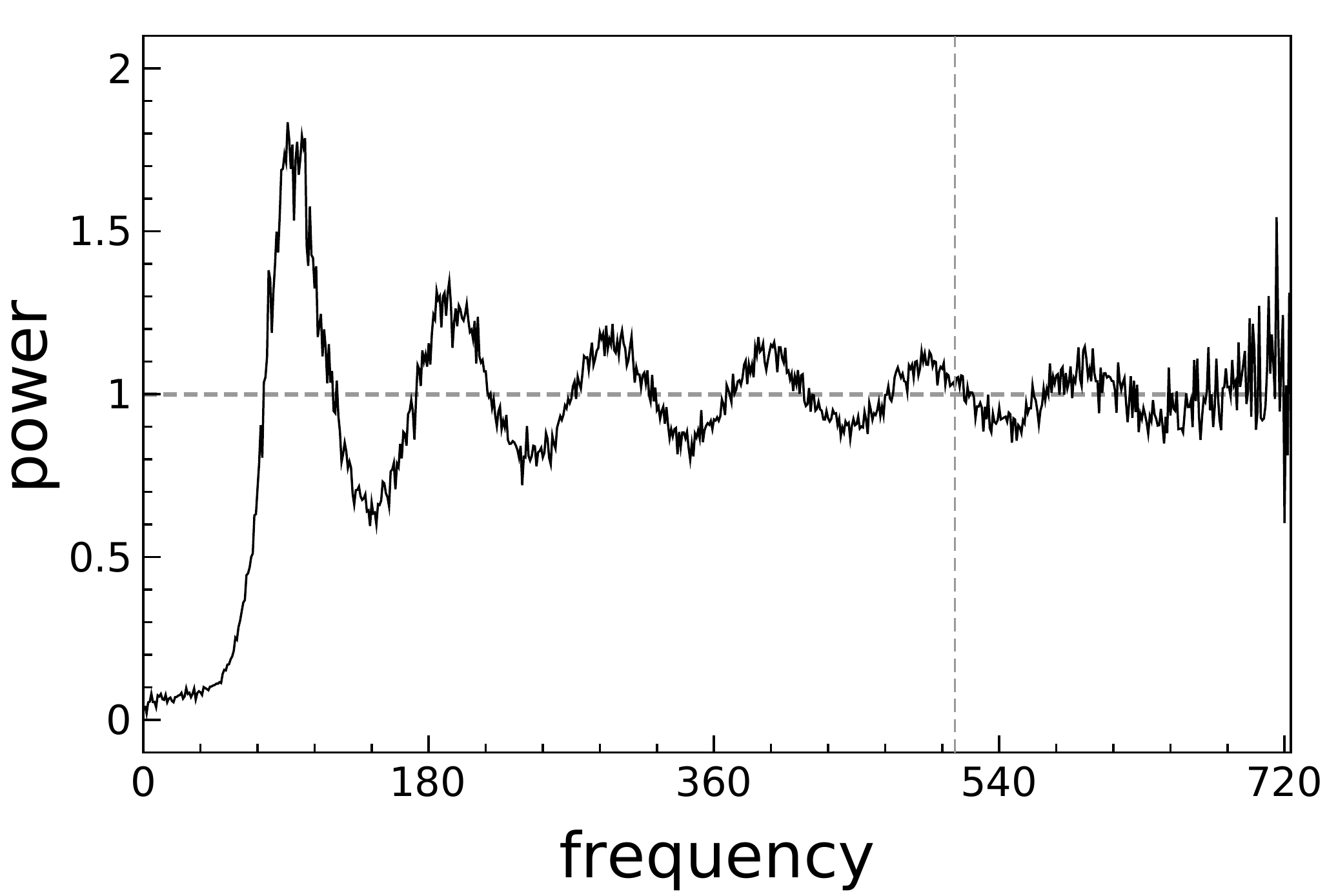}       
    } \\
    \subfigure[Point darts]
    {
        \includegraphics[width=0.45\columnwidth]{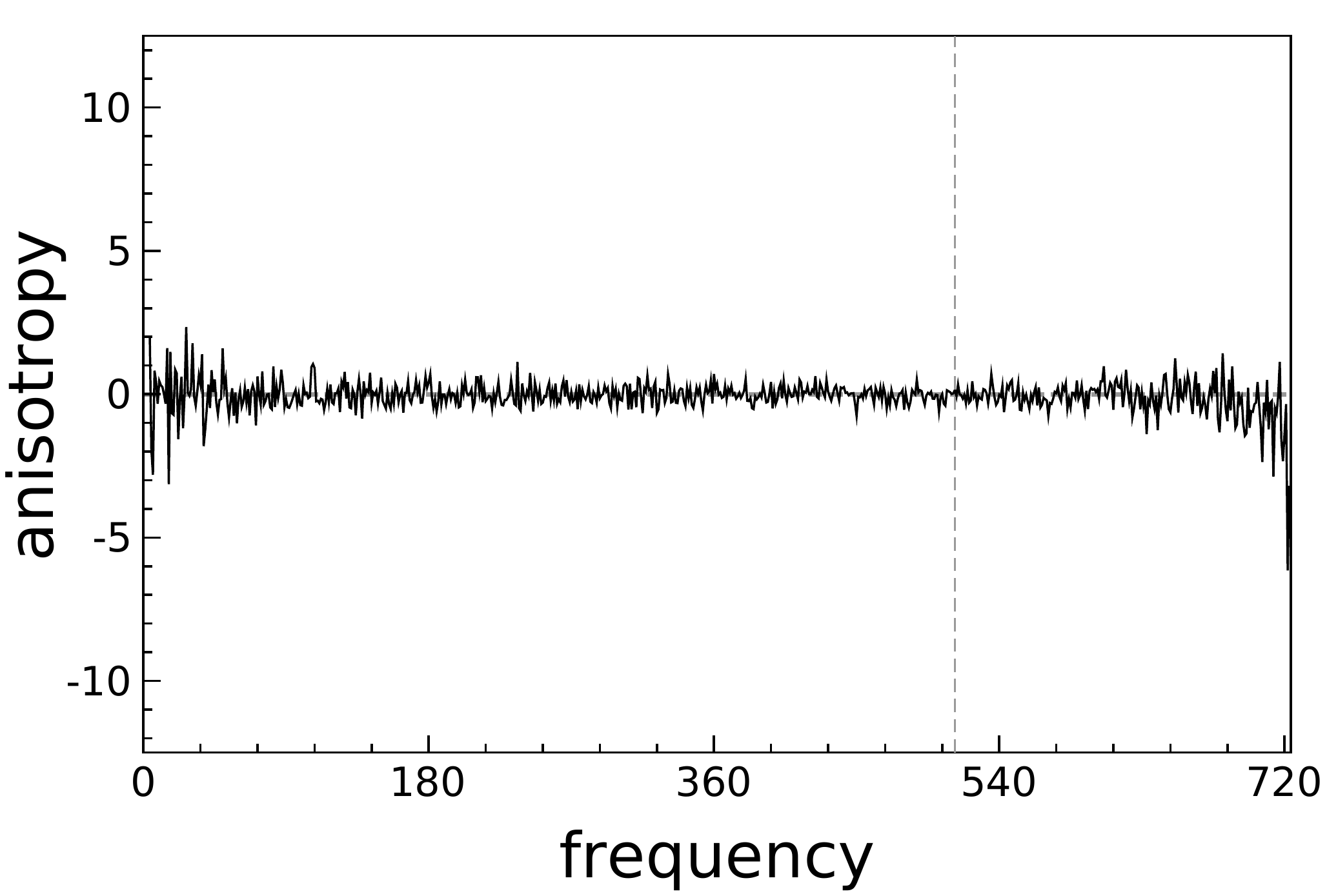} $\quad$
        \includegraphics[width=0.45\columnwidth]{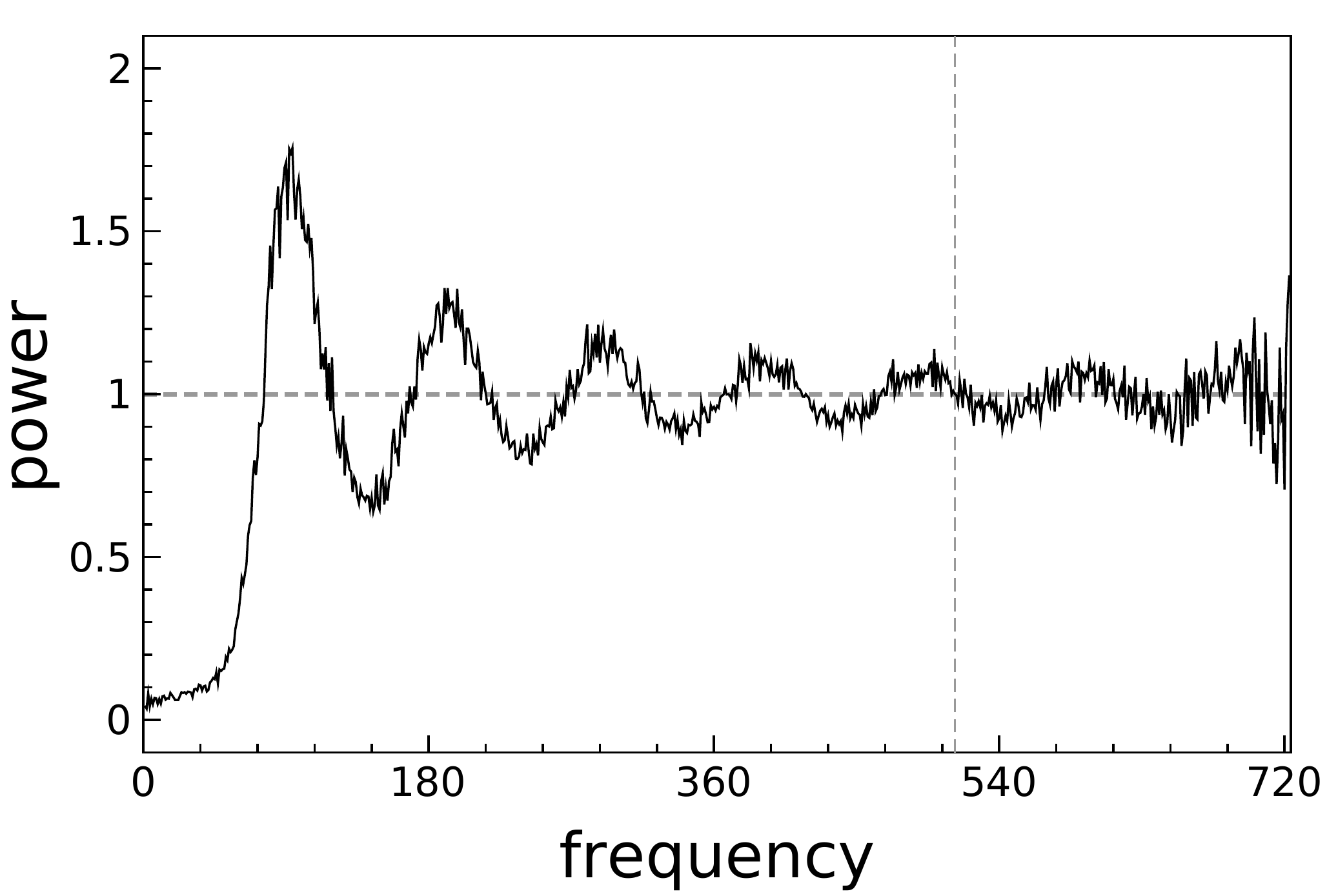}
    }
  \end{center}  
   \caption{Radial anisotropy and mean power estimates for line darts (top) and point darts (bottom), averaged over ten samplings.}
   \label{fig:blue_noise_freq}
\end{figure}

\paragraph{Output Quality}
We measure the quality of the distribution of 2-d output points using the PSA spectrum analysis tool~\cite{psa}. Using point darts, our process is the same as classic dart-throwing, so we it for our standard of correct output. Figures~\ref{fig:blue_noise_spec} and \ref{fig:blue_noise_freq} compare the outputs' blue-noise properties; the difference between point and line darts was insignificant, at least for these three metrics. %$V=10^{-6}$. 
%Line darts inserted 173,944 points in 33.241 seconds while point darts inserted 171,341 points in 235.947 seconds.

% figure moved to avoid breaking the above paragraph

%\todo{do experimental results to show the speed of the method as dimension increases. Find good biased-process and unbiased-outcome alternative to compare against.
%What we need are empirical studies showing that line darts are faster than the alternative point dart methods, e.g. our eurographics paper, for d>4.
%JDO's comment in the caption of the graph could be addressed this way.}

\subsection{Depth of Field With Antialiasing}
\label{sec:rendering}
{\Dart}s can be used for fast and
high-quality rendering of depth-of-field (DOF) effects in computer-synthesized
images.  
% reviewer thought this definition was unnecessary for Graphics (TOG) audience
%DOF causes objects in the
%ocal plane of the camera to appear sharp, and those farther or closer to be increasingly blurred.
%This is an optical effect caused
%by the large aperture of a physical lens, be it a camera lens or a
%human eye, in 
%contrast to a pinhole camera.
%People are used to seeing this limitation in the physical world, so reproducing it adds realism.
%
Mathematically, computing a pixel's color in the presence of
DOF can be expressed as a four-dimensional integral over
the pixel's spatial $(x,y)$ and lens aperture $(u,v)$ dimensions.
In most high-quality renderers, this is calculated using
Monte Carlo integration over many point samples. This
method suffers from a low rate of convergence; reducing noise for
a good quality image usually requires a very large number of samples per
pixel. \dart{}s offer the promise of faster convergence with low noise. 
Instead of using point samples for
reconstruction, we use 1-d (line) darts, thrown in the 4-d
$(x,y,u,v)$ space. 

We use Latin Hypercube Sampling (LHS) or
jittered sampling for each dimension~\cite{Cook86}. 
Given that our sample space is four-dimensional, each line dart consists of four line flats.
We select $n$ points, i.e.\ $4n$ flats. 
Each line requires a fixed location in 3-d and a variable fourth dimension. 

We compute coverage for these darts using a method inspired by Gribel
et al.'s work~\shortcite{Gribel:2010:AMB,Gribel:2011:HQS} on rendering
motion blur. 
% reviewer thought this was unnecessary for Graphics (TOG) audience
%(Motion adds blur from position changes over a time dimension.) 
Gribel et al.\ fix $x$ and $y$ and shoot line samples in the time
domain; instead, we use line darts that shoot line flats in all spatial dimensions
to compute depth-of-field blur. 
We also sample a higher-dimensional problem.

\subsubsection{Contrast to Tzeng et al.}
Tzeng et al.~\shortcite{Tzeng:2012:HPD} considered line sampling for DOF\@. 
They only consider the two $(u,v)$ dimensions, not the four $(u,v,x,y)$ dimensions as we do.
Further, they take advantage of the structure that the $(u,v)$ subspace of interest is uniformly circular.
This reduces 2-d $(u,v)$ sampling to 1-d sampling of lines through the origin, pinwheel sampling by angle. 
In their implementation, for each pixel, they fix $x$ and $y$ and use a pinwheel of line samples
to vary $u$ and $v$. The resulting DOF had high
performance when compared to point sampling strategies, with low
noise. However, due to both the pinwheel configuration and the fixing of $x$ and $y$ for each pixel's sample,
strobing artifacts tended to occur in regions with high frequency
changes. 
The formulation we present in this paper differs in the following ways: 

\begin{itemize}
\item Both implementations address DOF, but we also address
  antialiasing.
\item Tzeng et al.'s line darts are all radial and specified
  completely by their angle; they live in 1-d  $(\theta)$ space. In this
  paper, we use axis-aligned darts in 4-d $(x,y,u,v)$ space.
  The positive and negative consequences follow:
  \begin{itemize}
    \item Tzeng et al.\ exhibits screen-space aliasing because $x$ and $y$ are fixed.
    \item Tzeng et al.\ achieves higher quality DOF for the same number of samples, because 1-d spaces require fewer samples to cover than 4-d spaces.
   \end{itemize}
\item Tzeng et al.\ use non-random dart locations, but we
  randomly position darts. 
\item Tzeng et al.'s work specifically targets the GPU pipeline; 
  we do not discuss (or consider) implementation details.
\end{itemize}

% \todo{++ Anjul, Reference for comparison with experimental
%   results, "I'd like to see more analysis on the performance and
%   quality of the algorithm and comparisons with [Tzeng et al. 2012].
%   "}

% HPG Paper:
% ----------
% * Specific Application, one possible application of this paper
% * All line darts are radial and specified completely by theta, i.e.\ they belong
%   in 1-d space
% * Dart locations are non-random
% * Application specifically addresses DOF

% This Paper:
% -----------
% * General Method for k-d darts
% * Our graphics application uses axis-aligned darts in 4-d space (x,y,u,v)
% * Darts are randomly positioned
% * Application addresses both DOF and antialiasing

% Implementation:
% HPG paper only deals with GPU pipeline, our work does not talk about or consider implementation details.

% Quality:
% Quality of DOF is better for HPG paper because
% (a) HPG paper only considers the u,v dimension and not x,y dimensions. Thus the higher quality DOF comes at the expense of screen-space aliasing.
% (b) HPG solution is very specific to the problem because the area of interest is known to be uniformly circular. As a result it is able to reduce the problem to a 1-d domain. A hybrid is possibly conceivable I but I'd like to neither think nor talk about it :-)

\subsubsection{Triangle Edge Equations in 4-d}
\label{sec:uvtrans}
We now describe the triangle equations and how we create a line sample.
For a given triangle, we start by computing a signed radius of
Circle of Confusion (CoC) for each vertex, obtained using the following
expression~\cite{Hammon:2008:PPP}:
\begin{align*}
% \textnormal{CoC} = A\cdot \left( \frac{f(z-z_f)}{z(z_f-f)} \right )% is the extra parenthesis and dot traditional? If not, use:
\textnormal{CoC} = A \frac{f(z-z_f)}{z(z_f-f)}, 
\end{align*}
where $A$ and $f$ are the camera aperture and focal length,
respectively, and $z_f$ and $z$ indicate the respective depths of the
focal plane and the given vertex. 
Note $z$ is simply the $w$ coordinate of the vertex in clip space.
% The latter ... ? means z?
%%
%
Now that we have the circle of confusion for each triangle's vertex, we can begin formulating
a sampling strategy for each pixel. 

Given a set of coordinates on the lens $u$ and $v$, we assume a linear apparent motion
of the vertex screen coordinates. 
For a given screen space vertex $i \in \{0, 1, 2\}$, with coordinates $(x_{i},y_{i})$,
circle of confusion $c_{i}$,  and $u, v \in [-0.5,0.5]$, we have
\begin{align}
x_{i}^{u}=x_{i}+c_{i}u \label{eq:xcoc} \\
y_{i}^{v}=y_{i}+c_{i}v \label{eq:ycoc}
\end{align}

For each pixel we have a four-dimensional space $(x, y, u, v)$, where $x$ and $y$ are
the subpixel regions, and $u$ and $v$ are coordinates on the lens.  To get a 
realistic and noise-free image, we seek to sample this space uniformly in an effective manner.
Since we have four varying dimensions, we choose to use four different hypercubes for our Latin Hypercube Sampling (LHS)\@.
Each hypercube chooses three dimensions within which to sample, and one that will vary in our next stage. 

%For a given triangle, its $i$th oriented edge is between vertices $i$ and $j$.

For each triangle, we can consider the edge equations in this four-dimensional space to be the following:
we substitute \equrefs{eq:xcoc}{eq:ycoc} into the equations for testing whether a point $(x,y)$ lies within one of the triangle edges $i$.
%\todo{what does "covers" means? I understand a triangle covering a point. Is (x,y) necessarily on the boundary of the triangle? ---samitch}
Let $ES_i$ represent the edge-sum at point $(x,y)$ for the $i$th edge, between vertices $i$ and $j$,
\begin{eqnarray*}
ES_i(x,y,u,v) = (y - y^v_i)(x^u_j - x^u_i) - (x - x^u_i)(y^v_j - y^v_i).
\end{eqnarray*}
Expanding, we get equivalent right-hand sides
\begin{align*}
%ES_i(x,y,u,v) &
\Leftrightarrow & & (y - y_i - c_iv)\left(x_j - x_i + u(c_j-c_i)\right) & \\
& - & (x - x_i - c_iu)\left(y_j - y_i + v(c_j-c_i)\right)  \\
%\end{eqnarray*}
%and can be further simplified to get:
%\begin{eqnarray*}
%ES_i(x,y,u,v) &
\Leftrightarrow  & & ES_i(x,y,0,0)\\
& + & u\left(c_j(y-y_i) - c_i(y-y_j)\right) & \\
& - & v\left(c_j(x-x_i) - c_i(x-x_j)\right).
\end{align*}
This simplifies to
\begin{align*}   
ES_i(x,y,u,v) = C_i(x,y) + u A_i(y) - v B_i(x) \ge 0.
\end{align*}

Here we have four dimensions of variability ($x, y, u, v$). In conventional point sampling, one would randomly (or pseudo-randomly) select a set of points in this space.
We instead decide to employ line sampling using our Latin hypercubes. 
To illustrate this concept, consider fixing three of these dimensions with points on our hypercube from LHS: 
select $x_{1}, y_{1}$ and $u_{1}$ for $x, y$ and $u$. Now we have a line equation as follows:
\begin{eqnarray*}
ES_i(v) = C_i(x_{1},y_{1}) + u_{1} A_i(y_{1}) - v B_i(x_{1}),
\end{eqnarray*}
which simplifies to
\begin{eqnarray*}
ES_i(v) = P - vQ.
\end{eqnarray*}

This line equation is easy to analytically solve for each of our edge equations and determines
where line coverage exists. The same concept can be extended to our other hypercube setups, fixing three of the
dimensions and varying the last.

\begin{figure}[t]
\begin{center}
\subfigure[Two line darts in $(x,u)$]
{
\includegraphics[width=0.45\columnwidth]{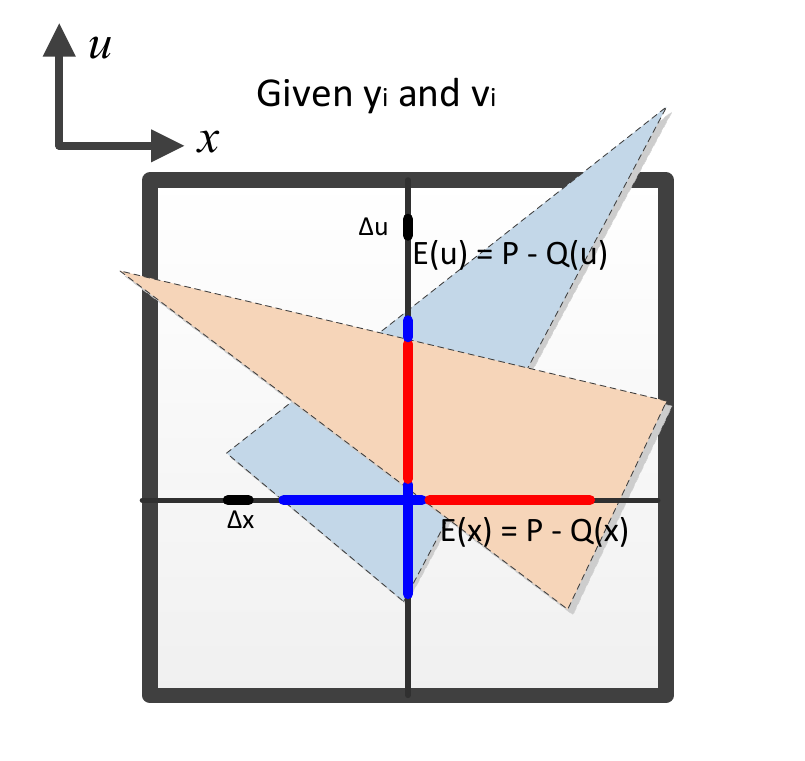}
}
\subfigure[Resolving occlusion depth]{
\includegraphics[width=0.45\columnwidth]{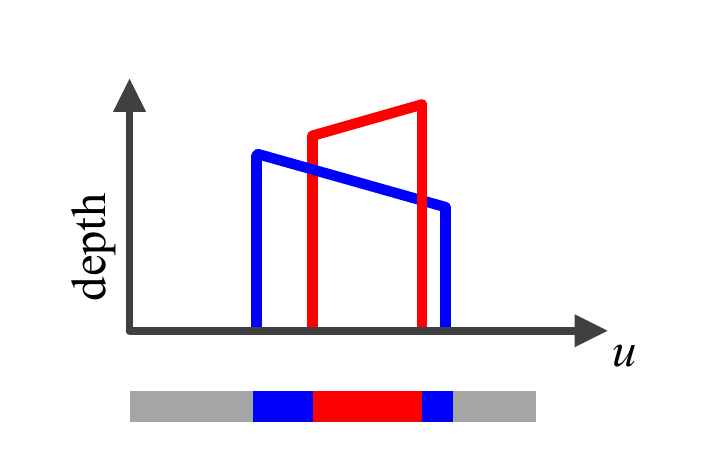}
}
\caption{Our technique for computing analytical
  coverage using line darts. In this example, consider two possible 
  line flats in the $x$ and $u$ directions. The domain can be transformed
  and we can test the triangles for occlusion along the line flat as shown in (a). Depth resolution per
  flat is shown in (b). 
  }
\label{fig:uvrast}
\end{center}
\end{figure}

\subsubsection{Analytical Coverage in the Hypercube Domain}
Knowing the edge equations in our 4-d domain, we can now compute
their coverage along a line sample.  \figref{fig:uvrast} summarizes
our technique for determining coverage.

We instantiate a set of line flats along each of our four hypercube configurations.
 A line dart is the combination of four different line flats (one in the direction of each of the four dimensions of the domain)
 with initial points chosen using our LHS\@. 

Rendering consists of testing each incoming triangle
against potentially covered line flats. For each pixel sample in the
triangle's bounding box, we use equations from \secref{sec:uvtrans}
to transform triangle edges to test for the correct hypercube domain. 

% Anjul, reviewer wants a summary of the steps:
To finish the calculation we follow Gribel et al.'s
approach~\shortcite{Gribel:2010:AMB} to construct and resolve line
darts. %; see their paper for the details.

For each line flat, we analytically compute its segment covered by the triangle. 
A per-line-sample queue stores the color and depth of covered segments. 
Once all triangles have been processed, we resolve
the final color for each sample. We sweep across the flat while
aggregating triangles closest in depth, and then use all pixel
samples to compute the final color for the pixel.

%\todo{Perhaps discuss how a wagon-wheel or purely $u,v$ space line-sampling might add bias? (i.e., compare vs hpg method)}
% \todo{cite samuli's paper for possible speedup using tighter bounds?}

\subsubsection{Implementation and Results}

To test our formulation, we built a simple CPU-based renderer.
It is capable of rendering scenes using traditional triangle
rasterization. We integrated two additional capabilities:
\begin{itemize}
    \item A stochastic sampler based on point darts.
    \item A sampler based on line darts.
\end{itemize}

\begin{table}[t]

    \begin{center}    

    \caption{Performance of our point vs.\ line darts.} {

    \begin{tabular}{l|cccc}
    \toprule
    		    Cessna   & 256 points & 1024 points  & 16 lines    & 30 lines \\
%              time (s)    &  125.01  & 493.69 & 97.47 & 176.44 \\                   
                  time (s)    &  125  & 494 & 97 & 176 \\                   
    \midrule
                 Teapot  & 256 points & 1024 points  & 16 lines  & 30 lines\\
%              time (s)  &  213.74 & 853.22 & 157.13  & 291.74 \\                 
                 time (s)  &  214       & 853  & 157    & 292  \\                 
    \bottomrule
    \end{tabular}
    }
   \label{tab:dofperf} 
    \end{center} 
\end{table}

The two noise artifacts that typically occur using a DOF
scheme are
noise in blurry regions and noisy aliasing artifacts
near the point in focus. For scenes far away from the focal plane, line flats
in the $x$ and $y$ direction become particularly \emph{narrow} when compared to the sampling space, and this
causes additional noise. In these regions the \emph{length} of flats in the $x$ and $y$ dimensions 
are significantly smaller than flats in either the $u$ or $v$ direction.

If we consider all such flats to have equal contributions, this results in the $x$ and $y$ samples adding a noticeable amount of noise into our system in blurry regions. 
To address this, rather than considering all line flats to have equal contribution, we select a weight constant ($\alpha=0.2$ in our case)
such that $x$ and $y$ line flats are scaled by $\alpha$ and contribute less to the scene.

This weight can be modified based on the aperture of the lens for the scene. In cases
where the aperture is small, contributions from $x$ and $y$ are deemed more
important (for antialiasing), and the weight is adjusted accordingly. A
more accurate dart sampling method might also consider the ratios of flat lengths
per tile (worst case) and decide on an $\alpha$ weight accordingly.
Research is needed to determine the optimal amount for each flat to contribute. 
Still, our simple heuristic
seems to be effective.

\begin{figure*}[!ht]
  \begin{center}  
  %\subfigure[Images rendered with 4 line darts ($n = 1$)]
    %{
      %\includegraphics[width=0.4\columnwidth]{dof-figs/cessna_4_735}
      %\includegraphics[width=0.4\columnwidth]{dof-figs/geo_4_214}       
      %\includegraphics[width=0.4\columnwidth]{dof-figs/hand_4_1336}
      %\includegraphics[width=0.4\columnwidth]{dof-figs/ogre_4_1423}       
      %\includegraphics[width=0.4\columnwidth]{dof-figs/teapot_4_1224}       
    %} \\
  \subfigure[Point darts, 256 per pixel.]
    {
      \includegraphics[width=0.33\columnwidth]{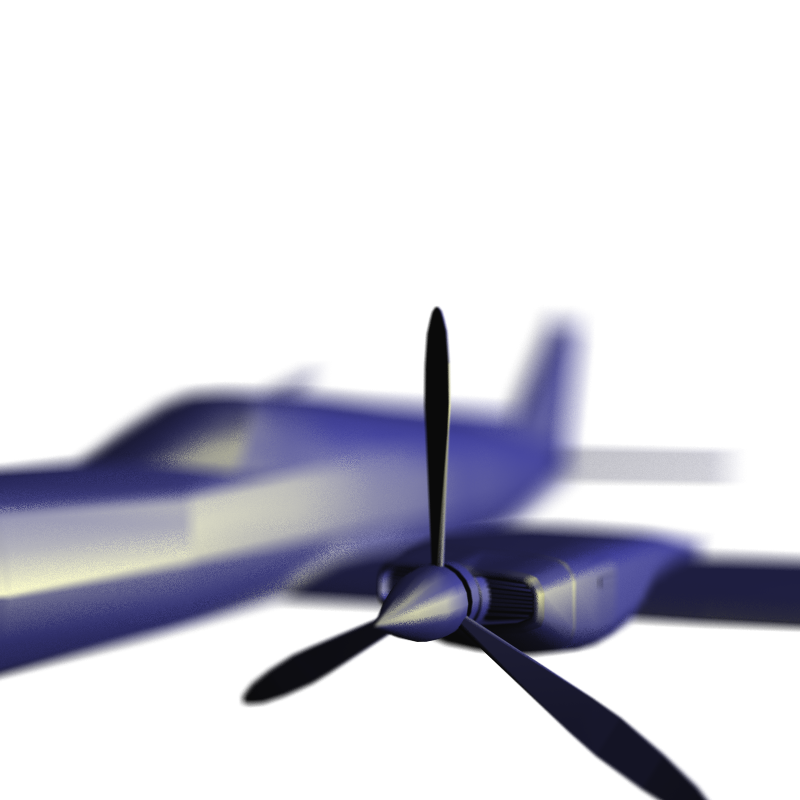}
      \includegraphics[width=0.33\columnwidth,  clip,trim=2.5in 1.5in 3.5in 4.5in]{dof-figs/cessna_p256_12501}            
      \includegraphics[width=0.33\columnwidth,  clip,trim=5.4in 3.3in 1.6in 3.7in]{dof-figs/cessna_p256_12501}            
      \includegraphics[width=0.33\columnwidth]{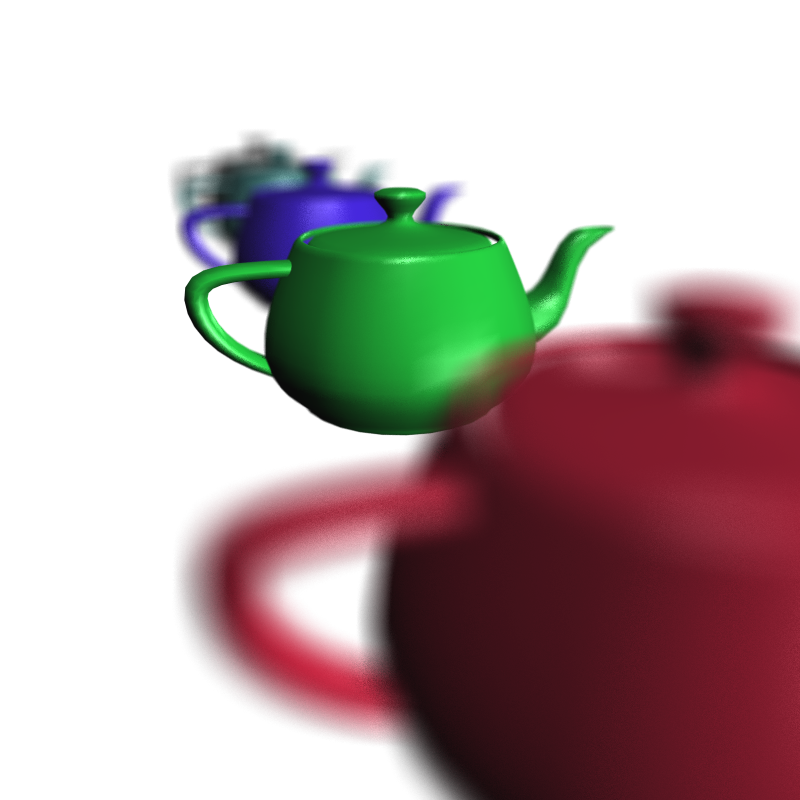}   
      \includegraphics[width=0.33\columnwidth,  clip,trim=5in 4.1in 1in 1.9in]{dof-figs/teapot_p256_21374}   
      \includegraphics[width=0.33\columnwidth, clip,trim=5.5in 4.6in 1.5in 2.4in]{dof-figs/teapot_p256_21374}
      
    }
    \subfigure[Point darts, 1024 per pixel.]
    {
      
      \includegraphics[width=0.33\columnwidth]{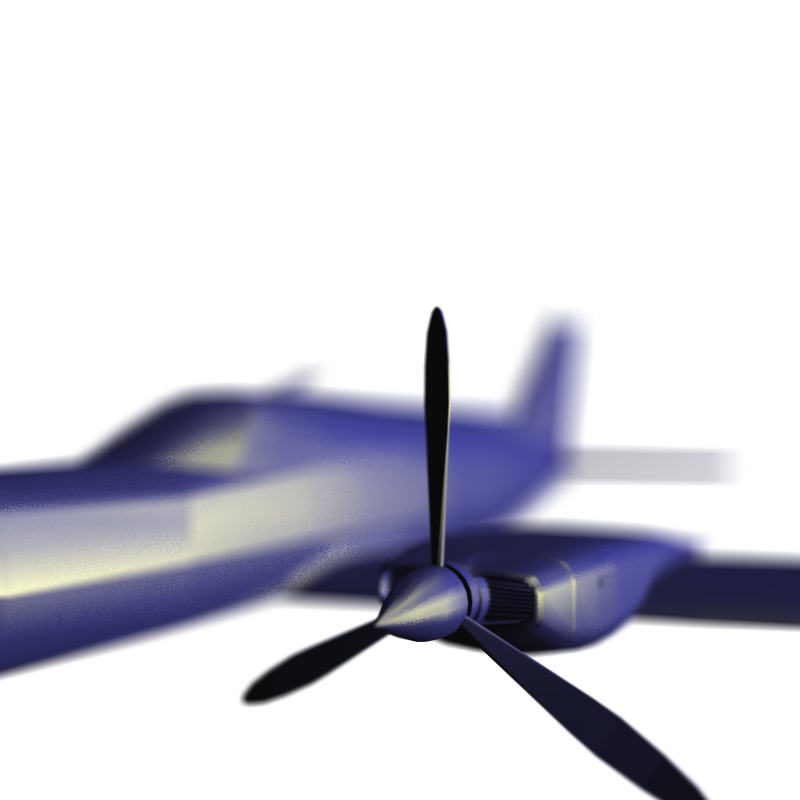}            
      \includegraphics[width=0.33\columnwidth,  clip,trim=2.5in 1.5in 3.5in 4.5in]{dof-figs/cessna_p1024_49369}     
      \includegraphics[width=0.33\columnwidth,  clip,trim=5.4in 3.3in 1.6in 3.7in]{dof-figs/cessna_p1024_49369}                   
      \includegraphics[width=0.33\columnwidth]{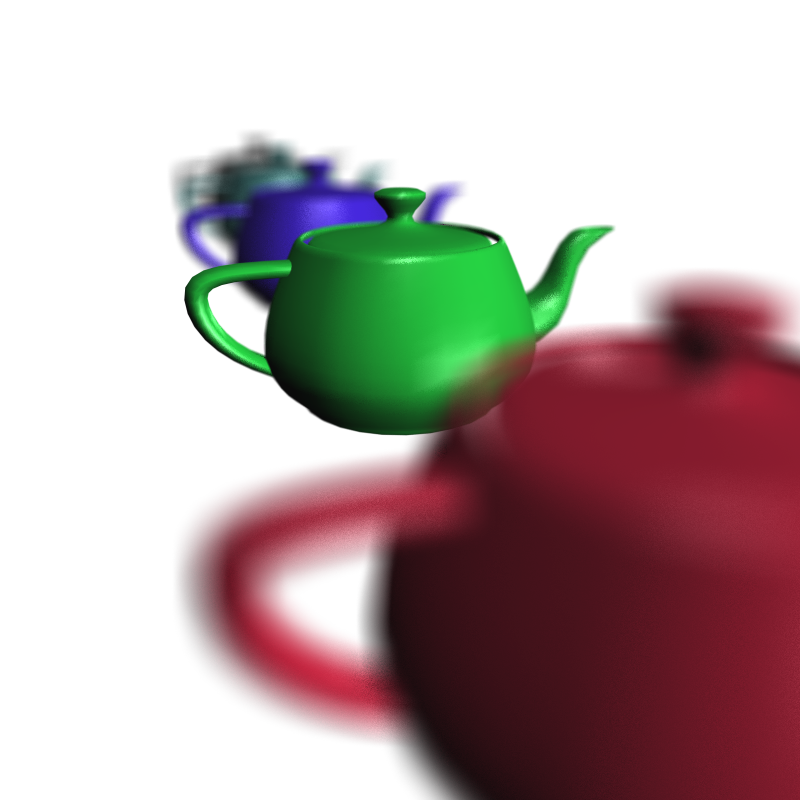}   
      \includegraphics[width=0.33\columnwidth, clip,trim=5in 4.1in 1in 1.9in]{dof-figs/teapot_p1024_85322}
      \includegraphics[width=0.33\columnwidth, clip,trim=5.5in 4.6in 1.5in 2.4in]{dof-figs/teapot_p1024_85322}
    }
  \subfigure[Line darts, 16 per pixel.]
    {
      \includegraphics[width=0.33\columnwidth]{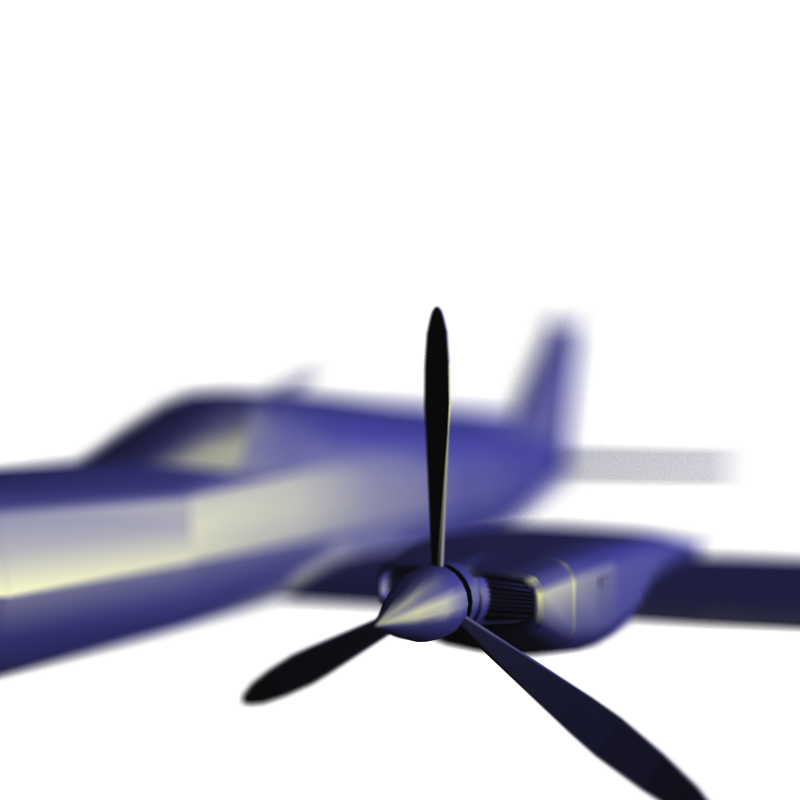}      
      \includegraphics[width=0.33\columnwidth,  clip,trim=2.5in 1.5in 3.5in 4.5in]{dof-figs/cessna_64_9747}       
      \includegraphics[width=0.33\columnwidth,  clip,trim=5.4in 3.3in 1.6in 3.7in]{dof-figs/cessna_64_9747}                 
      \includegraphics[width=0.33\columnwidth]{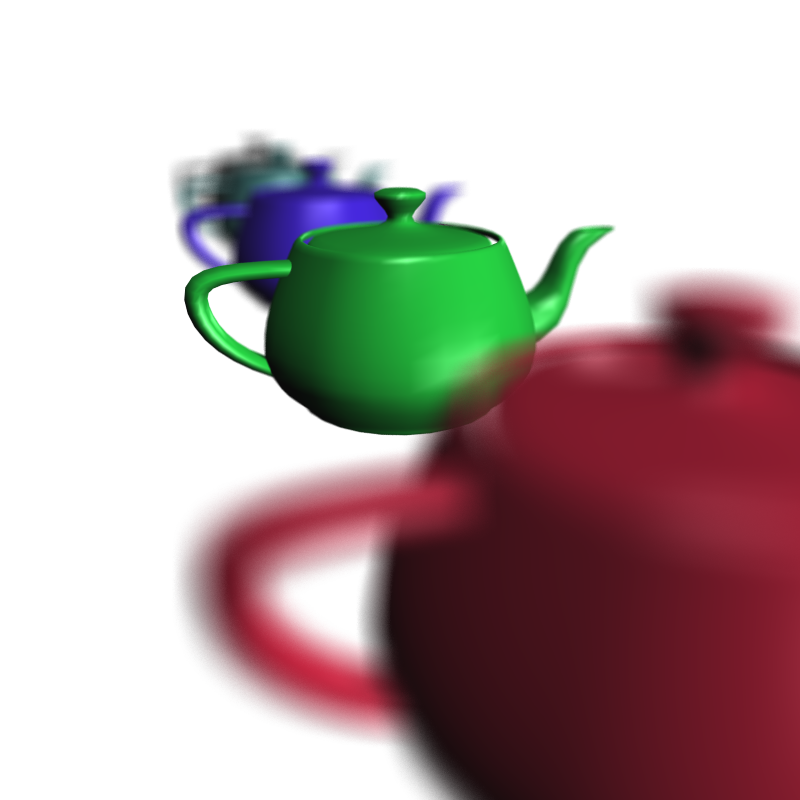}   
      \includegraphics[width=0.33\columnwidth, clip,trim=5in 4.1in 1in 1.9in]{dof-figs/teapot_64_15713}
      \includegraphics[width=0.33\columnwidth, clip,trim=5.5in 4.6in 1.5in 2.4in]{dof-figs/teapot_64_15713}
    }
    \subfigure[Line darts, 30 per pixel.]
    {
      
      \includegraphics[width=0.33\columnwidth]{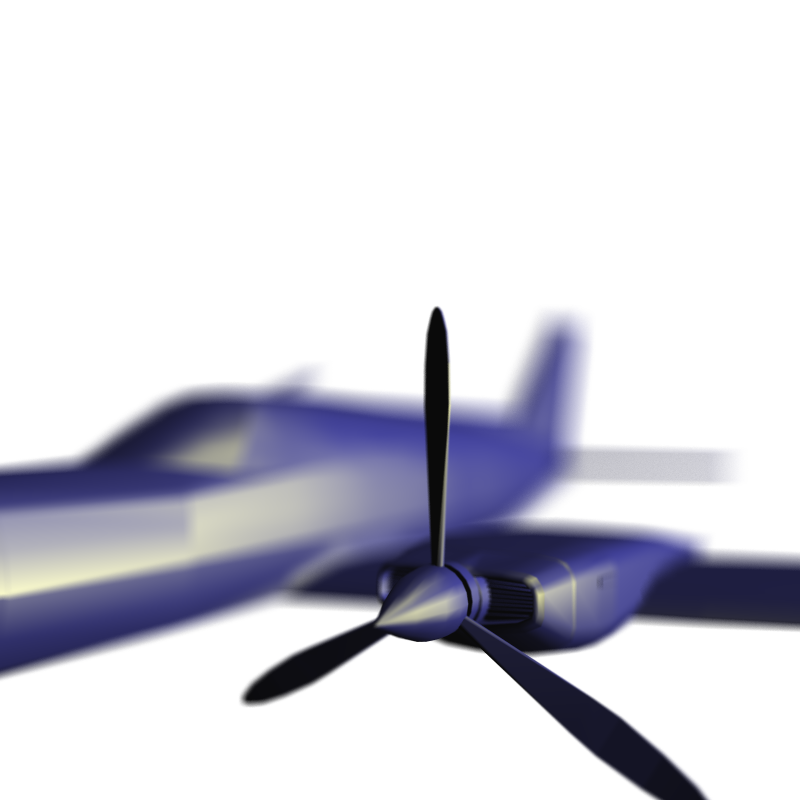}            
      \includegraphics[width=0.33\columnwidth,  clip,trim=2.5in 1.5in 3.5in 4.5in]{dof-figs/cessna_120_17644}     
      \includegraphics[width=0.33\columnwidth,  clip,trim=5.4in 3.3in 1.6in 3.7in]{dof-figs/cessna_120_17644}                   
      \includegraphics[width=0.33\columnwidth]{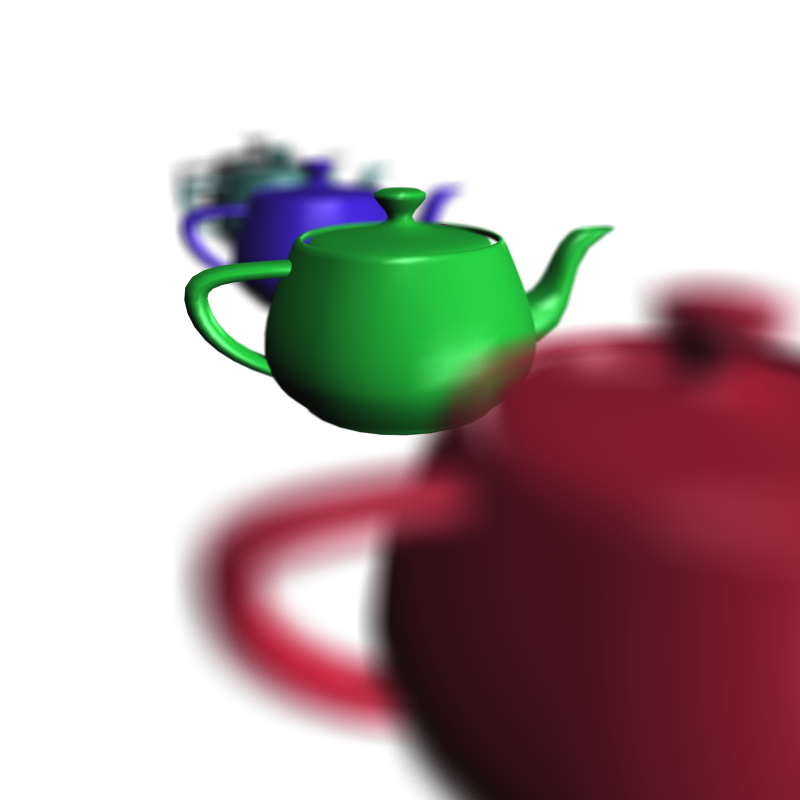}   
      \includegraphics[width=0.33\columnwidth, clip,trim=5in 4.1in 1in 1.9in]{dof-figs/teapot_120_29174}
      \includegraphics[width=0.33\columnwidth, clip,trim=5.5in 4.6in 1.5in 2.4in]{dof-figs/teapot_120_29174}
    }
    
    \end{center}  

    \caption{Depth Of Field (DOF) images using conventional point sampling (a--b) vs.\ our \dart{}s (c--d). \Dart{}s produce high-quality, antialiased images.    The third column from the left shows a close up of an extremely blurry region, the Cessna's tail.      In regions close to the focal plane we see some aliasing artifacts for point darts but not for line darts, e.g.\ the Cessna's prop cone, the junction of its body and wing, and the transition shades on the green teapot spout.   Furthermore, line darts tend to be faster for the same quality blur; see \tabref{tab:dofperf}. 
      \label{fig:dofimages}}

\end{figure*}

\figref{fig:dofimages} compares two scenes rendered with our point dart and line dart techniques.
Renderings based on {\dart}s are virtually free of noise and aliasing artifacts.
Point darts, however, retain noticeable aliasing artifacts even with 1024 darts.
Although 16 line darts produce a bit more noise in unfocused regions than 256 point darts,
30 line darts has better quality than 256 point darts and around the same quality as 1024 point darts in unfocused regions, and no
noticeable aliasing artifacts in focused areas.

\tabref{tab:dofperf} shows the performance of our samplers.
Clearly, throwing one line dart is more expensive than throwing one
point dart. However, fewer are needed, and correctly weighted line darts 
% improve bias significantly faster and  - samitch cut. I don't know the definition of bias here, so lets leave it out
converge more quickly to a less noisy image without aliasing artifacts.

%\todo{
%basic notes:

%1. we want to estimate monte carlo integral of scene over the lens
%position for each pixel

%2. DOF is complicated, noisy even for 100s of samples per
%pixel. kd darts should fix this.

%3. our method is similar to analytical antialiasing from hpg 2011

%4. we use semi-split line darts:

 %   4.1 for each pixel sample we shoot several line darts in the
  %  uv-domain

   % 4.2 show math: for given (x,y), each edge transforms to another in
    %u-v space. i.e. --> each triangle transforms to another for given
    %(x,y). the second set of line darts `rasterizes' triangles in
    %this domain (similar to jones et al.). we early exit to save time:

    %4.2.1 uv bounds from laine et al.

    %4.2.2 uv triangle bounding test

%5. final algorithm, implementation and results

%6. future work (gpu?)

%}

\subsection{Probability of Failure}
\label{sec:uq}
\newcommand{\myx}{\ensuremath{\underline{x}}}
\newcommand{\ythresh}{\ensuremath{y_{t}}}%{y^\ast}
Uncertainty quantification usually explores a vast high-dimensional space with a limited budget of sample points. Efficiency is crucial because typically the function evaluation is expensive and more sample points are desired than we can afford. Surrogate models create a  cheaper response surface that is evaluated instead. 
Even for very cheap response surfaces, if the failure region is small enough Monte Carlo (MC) sampling will not estimate it accurately.
Here we show that \dart{}s can improve MC efficiency.

\begin{figure}[!ht]
\begin{center}
\subfigure [Circular Parabola]  {\includegraphics[width=0.45\columnwidth]{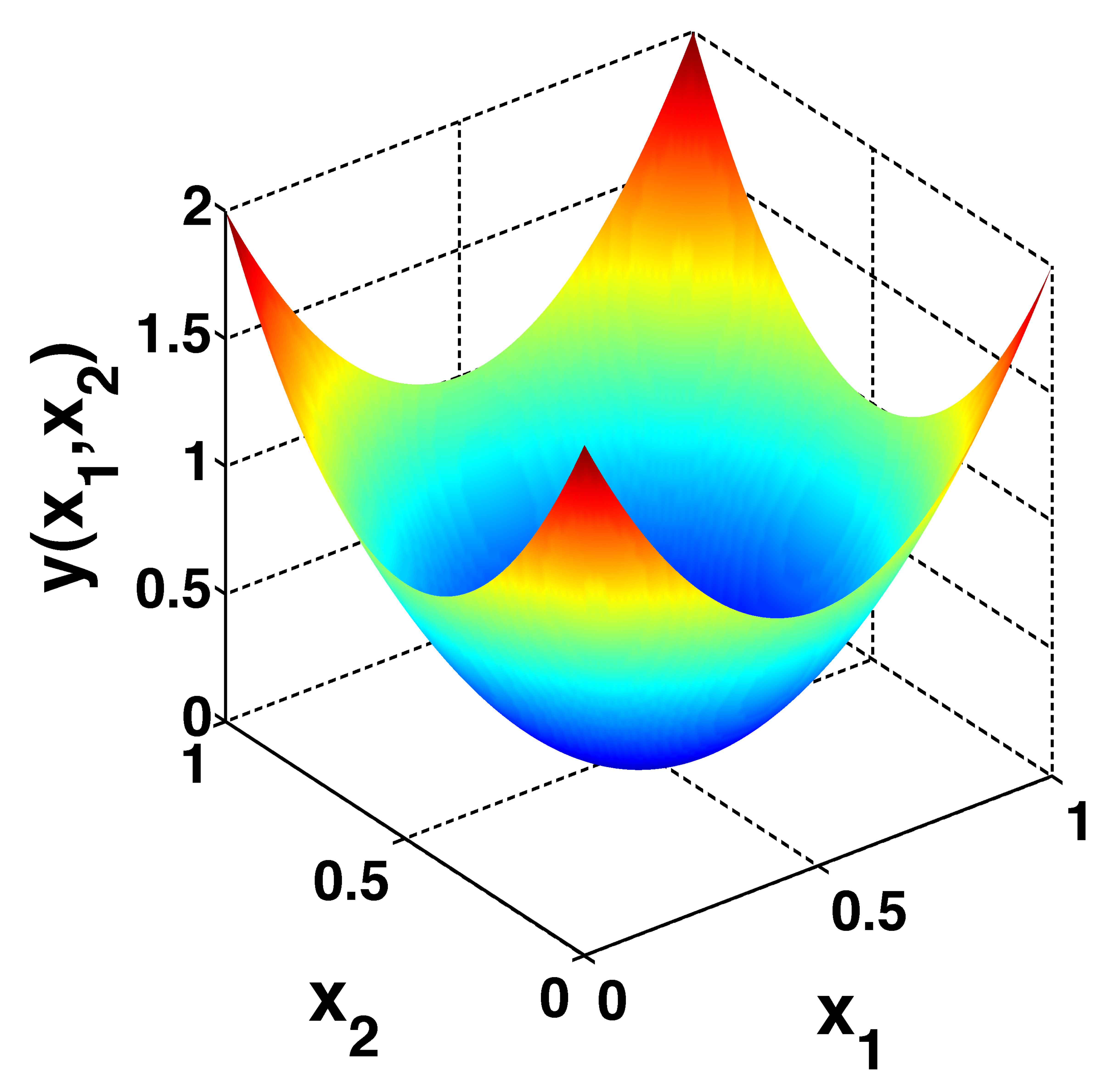}} $\quad$
\subfigure  [Planar Cross] {\includegraphics[width=0.45\columnwidth]{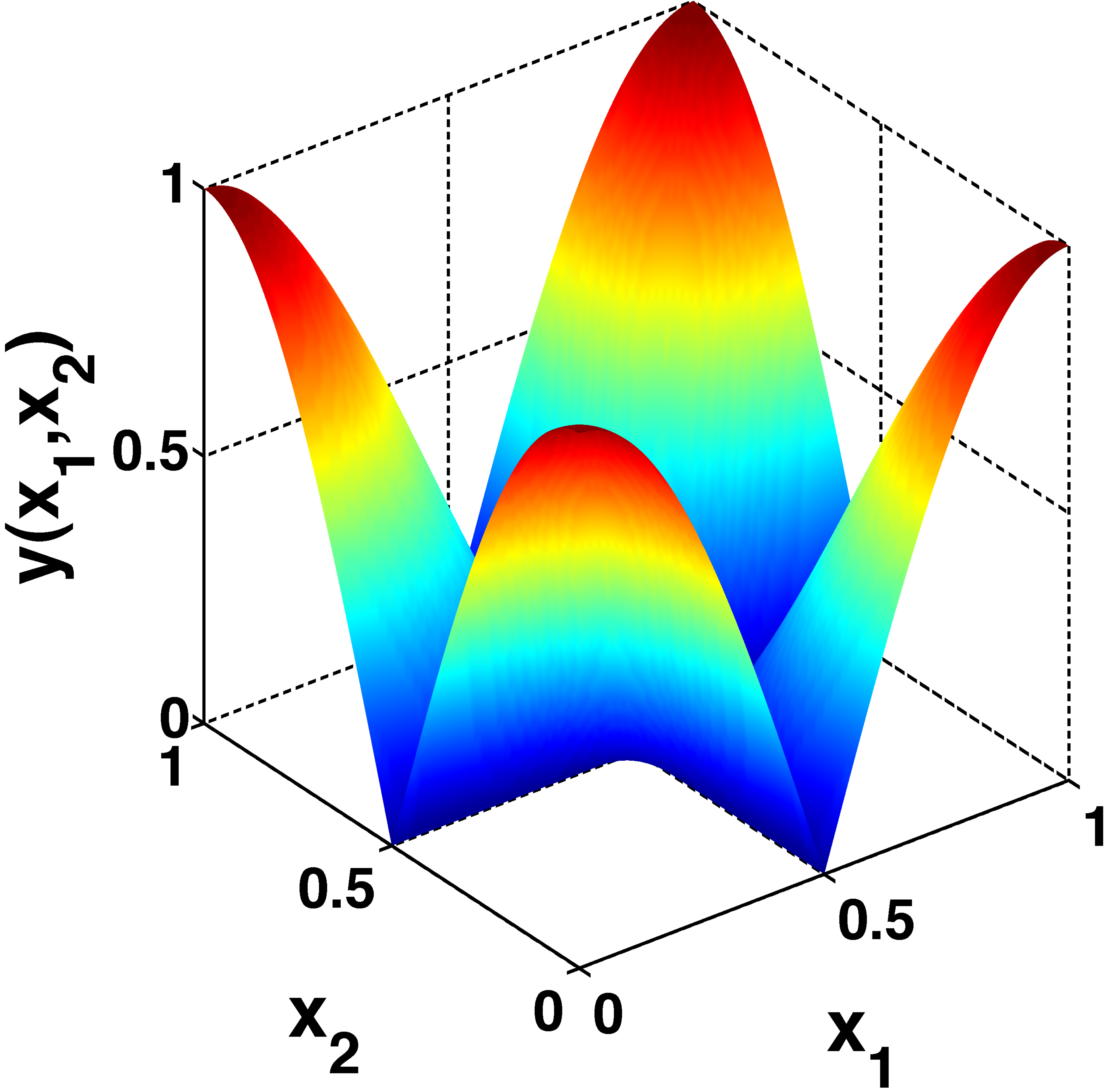}}
\end{center}    
\caption{Plots of the two surrogate model  test functions.}
\label{fig:test_problems}
\end{figure}

We test the ``circular parabola'' (\equref{eqn:circularparabola}) and ``planar cross'' (\equref{eqn:planar_cross}) surrogate models; see \figref{fig:test_problems}.
\begin{equation}
y(\myx)=\sum_{i=1}^d(2x_i-1)^2,\ 0<x_i<1.
\label{eqn:circularparabola}
\end{equation}
\begin{equation}
y(\myx)= \left[ \prod_{i=1}^d \frac{\left(1+\cos(2 \pi x_i)\right)}{2} \right]^{1/d},\ 0<x_i<1.
\label{eqn:planar_cross}
\end{equation}

\begin{figure}[!ht]
\begin{center}
\subfigure {\includegraphics[width=0.45\columnwidth]{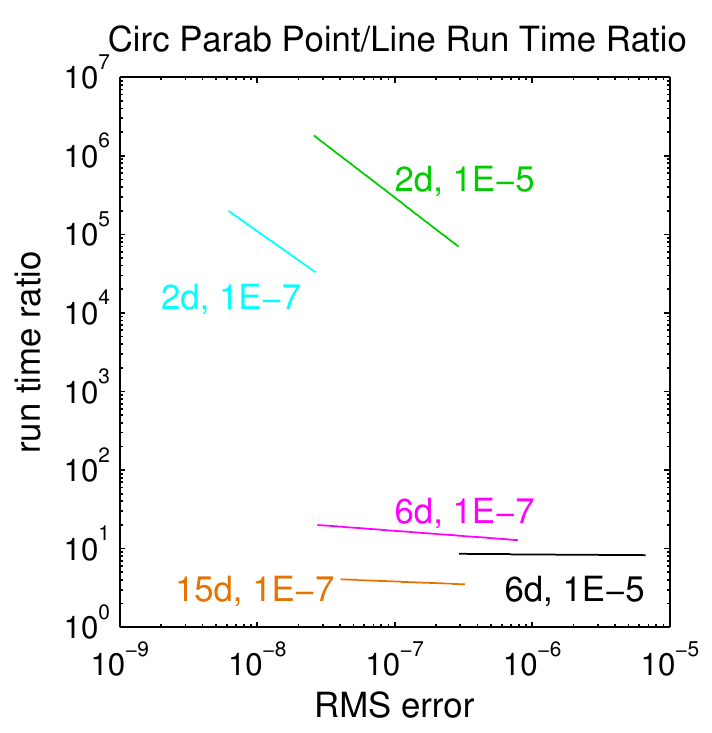}} $\quad$
\subfigure {\includegraphics[width=0.45\columnwidth]{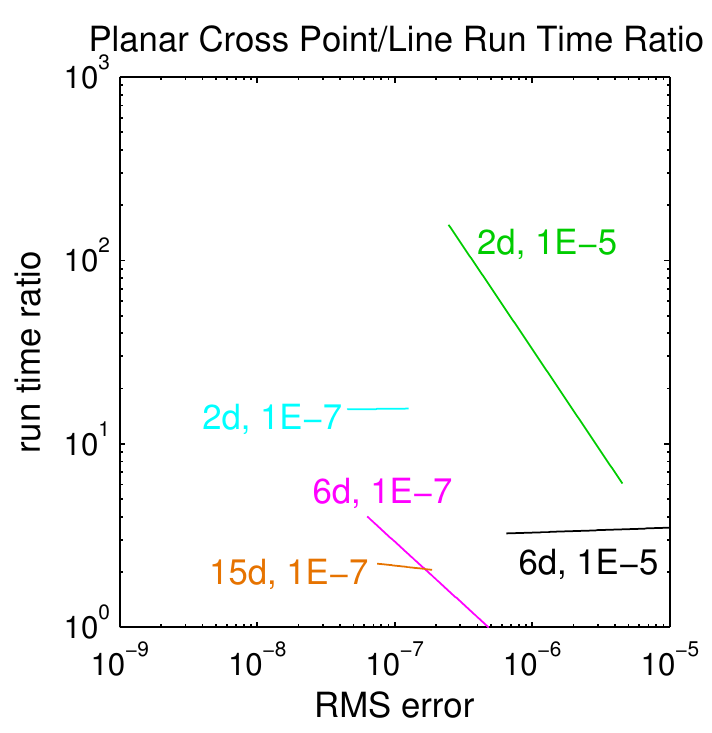}}
\end{center}    
\caption{Speedup of line darts over point sampling for estimating the probability of failure of two analytic response surfaces.}
\label{fig:uq_results}
\end{figure}

Failure is defined as the function value below some constant threshold: $y(\myx) < \ythresh$.
The shape of the failure region is different for the two test functions: a disk for the parabola and a fattened plus-sign for the planar cross; see \figref{fig:test_problems} and also \figref{fig:teaser}.
For uniform distributions, the probability of failure is the fraction of the domain volume where $y(\myx) < \ythresh$.
We choose $\ythresh$ so the probability of failure is exactly $10^{-5}$ or $10^{-7}$.
We estimate the failure volume using line darts.
For a line flat, we find the roots of the single variable equation $y(x_i) = \ythresh$. The length of the line segment between the two roots (if real roots exist) is used to estimate the volume of the failure region. 
\figref{fig:uq_results} demonstrates the benefit of line darts over conventional point sampling in reducing the time required to achieve a given accuracy level. The complexity of root-finding reduces the performance of line darts. However, for both test functions, line darts were better than point darts in dimensions up to 15.

\section{Accuracy Experiments}
\label{sec:dart_experiments}
\subsection{Problem Motivation}
\label{sec:exp_motivation}
We provide some experimental results on the accuracy of darts 
for the canonical Monte Carlo problem of estimating the volume of an object in high dimensions.
(Volume estimation is in the same category as the probability-of-failure problem in \secref{sec:uq}.)
In particular, we seek to show that the method produces good estimates, regardless of the size, shape, dimension and orientation of the object, and regardless of the dimension and orientation of the darts. The average estimate should be close to the true estimate, and the higher moments of the estimates should be low. We design our experiments to show the effects (if any) of the following factors.
\begin{itemize}
  \item $d$, the dimension of the object.
  \item $k$, the dimension of the dart. Of particular interest is comparing our results to standard MC point sampling, $k=0.$
  \item $s$, the squish factor of the object, which controls its aspect ratio.
  \item $r$, the number of rotations of the object. This allows us to compare axis-aligned objects to unaligned ones.
  \item Axis-aligned darts vs.\ unaligned darts.
\end{itemize}
We perform $N$ experiments of $n$ flats over the prior parameters, as  described in \tabref{tab:dart_experiments}.
Note that we keep the number of flats constant, rather than the number of darts, because the computational expense is more closely tied to the number of flats for objects with an analytic expression, and because middle-dimensional darts have many more flats than high and low dimensional ones.

%\subsection{Regression}

\begin{table}[!ht]
\tabcolsep6pt
\caption{\Dart\ parameter study} {%
\begin{tabular}{@{}ccccc@{}}\hline
%\begin{tabular}{@{}lllll@{}}\hline
{$d$} &{$k$} & $s$ & $r$ & $n$\\\hline\\ [-2.3ex]
%figs use d (k) s r n N
%  2  & 0--1 & {0.1, 0.5, 1.0, 2.0,  10.0}           & 10                        & $10^2$--$10^6$ \\
  2  & 0--1 & ${\frac{1}{10}, \frac{1}{2}, 1, 2,  10}$           & 10                        & $10^2$--$10^6$ \\
 2   & 0--1 & 0.5                                              & {0, 1, 5, 10, 20}   & $10^2$--$10^6$ \\
%  3 & 0--2 &  {0.1, 0.5, 1.0, 1.41,  3.16}        & 10                         & $10^2$--$10^6$\\
  3 & 0--2 &  ${\frac{1}{10}, \frac{1}{2}, 1, \sqrt{2}, \sqrt{10}}$        & 10                         & $10^2$--$10^6$\\
  3 & 0--2 & 0.5                                               & {0, 1, 5, 10, 20}   & $10^2$--$10^6$\\
% 10 & 0--9 & {0.1, 0.5, 1.0,  1.08,    1.29}     &  10                       & $10^2$--$10^6$ \\
 10 & 0--9 & ${\frac{1}{10}, \frac{1}{2}, 1,  \sqrt[9]{2},  \sqrt[9]{10}} $    &  10                       & $10^2$--$10^6$ \\
 10 & 0--9 & 0.5                                             & {0, 1, 5, 10, 20}   & $10^2$--$10^6$ \\\hline
\end{tabular}}
\label{tab:dart_experiments}
%\begin{tabnote}
The darts are aligned with the coordinate axis except for some $d=2$ experiments.
We repeat each parameter combination 1000 times, $N=1000$.
Squish parameters $s$ are symmetric around 1 with respect to ellipse volume: 
e.g.\ for $d=2$, $s=1/2$ and $s=2$ define ellipses with the same volume.
%\end{tabnote}
\end{table}

\subsection{Object Generation}
Instead of a spherical object, we estimate the volume of an ellipse (a.k.a.\ ellipsoid), randomly oriented and squished.
An ellipse provides enough generality to test the factors in \secref{sec:exp_motivation}, but enough simplicity to isolate numerical from methodology issues.
In particular, we choose an ellipse because it is possible to analytically calculate the volume of an ellipse's intersection with a \dart.

Our object is a $d$-dimensional ellipse centered at the origin. We construct it as follows.
We start with a $d$-sphere centered at the origin with radius 1. 
This fits in an origin-centered cube with side length 2, the two-cube.
We ensure the final ellipse also lies in the two-cube.

\begin{itemize}

  \item $s$: squish factor. We scale the ellipse along the $x$-axis by multiplying its $x$-extent by $s$.   The sphere has $s=1.$
  Note $s<1$ gives thin, coin-shaped objects.
  For $s>1$, we then shrink the ellipse so it fits in the two-cube: multiply all coordinates by a factor of $1/s$.
  The net effect is keeping the $x$-coordinate fixed and scaling the other axes by $1/s$. This gives needle-shaped objects.

  \item $r$: number of rotations. 
  We perform $r$ random rotations in sequence. Each is a Givens rotation with a random pair of coordinate indices $i$ and $j$, and a random angle $\theta \in [0,\pi]$:  multiply coordinates by the identity matrix with the $2 \times 2$,  $\{i,j\}$ submatrix  [1 0; 0 1] replaced by [$\cos{\theta}$  -$\sin{\theta}$; $\sin{\theta}$ $\cos{\theta}$].
  
\end{itemize}

\subsection{Dart Generation}

%\subsubsection{Unaligned Darts}
Most implementers will choose axis-aligned darts for three reasons.
First, it is easy to distribute aligned darts uniformly, which ensures that the expected mean of the function estimates is accurate.
Second, it is easiest to implement aligned darts, since it involves simply fixing coordinate values.
Third, in many cases it is most efficient because we may obtain an expression for the underlying function along a dart by substituting in the fixed coordinate values.
However, for completeness we provide some experimental results on the accuracy of unaligned darts.
%
%\subsection{Dart Implementation}
We shoot \darts\ into the two-cube as follows:
\begin{itemize}
  \item A point dart ($k=0$) is generated by selecting a random point. 
  Each of the $d$ coordinates is chosen independently, and uniformly in $[0,1]$.

% this is flat generation, not dart
%  \item Aligned darts are generated by selecting $d-k$ fixed coordinate indices for each flat;
%  the remaining $k$ coordinates are allowed to vary.
%  The fixed indices are chosen uniformly.
%  The coordinates for the fixed indices are chosen independently and uniformly as for point darts.

  \item Aligned darts are generated by their flats. Each flat has a unique combination of $d-k$ fixed coordinate indices;
  the remaining $k$ coordinates are allowed to vary.
%  The fixed indices are chosen uniformly.
  The coordinates for the fixed indices are chosen independently and uniformly as for point darts.

  \item Unaligned flats are generated so that the orientation of the flats is uniformly random. The only experimental setting where we generate unaligned flats is for  $k=1$ and $d=2$, line darts in the plane.
  We choose angle $\theta \in [0,\pi]$, which determines the orientation of the flat. Any line that intersects the square crosses one of its main diagonals. (It is guaranteed to cross the diagonal to which it is more perpendicular, which depends only on $\theta$.) We pick a point $p$ uniformly at random along the appropriate diagonal. We now have a point and an angle, which defines a line flat.  
  For random darts, the second flat is a line perpendicular to the first line, passing through some random point $q$ of the other diagonal.
\end{itemize}

  Aligned 1-d darts are labeled ``k=1a,''
  random flats are labeled ``k=1r,'' 
  and random darts, pairs of orthogonal flats, are labeled ``k=1o''
   in the top two rows of Figures~\ref{fig:mean_error} and \ref{fig:histogram_estimates}.

\subsection{Object-Dart Intersection}
For point darts, the volume estimation is the fraction of darts that landed inside the ellipse, multiplied by the volume of the two-cube sampling domain, $2^d$.
For $k>0$ darts, instead of this discrete ratio we average the geometric fraction of each dart inside the ellipse object. The details of these calculations follow.

%\begin{itemize}
%  \item 
For point darts, we simply back-project the points to the domain of the
  sphere: apply the inverse Givens rotations to the dart's point in reverse order; then scale the $x$-coordinate by $1/s$
  (or, for $s>1$, all the other coordinates by $s$). 
  If the distance from the transformed dart  to the origin is less than 1 it is inside the sphere, and the original dart is inside the ellipse.
  
%  \item 
For \darts, we back-project their hyperplanes into the sphere domain, where we can calculate the volume of intersection analytically, and then forward-weight it by the scaling. 
  
  %The details follow. 
  To back-project a flat, we back-project $k+1$ points spanning the flat.
  Each dart has $d-k$ fixed coordinates and $k$ free coordinates. 
  We pick spanning point $p_0$ with $0$ for all of its free coordinates, and spanning point $p_{i>0}$ with 1 for its $i$th free coordinate and 0 for its other free coordinates. Each $p_i$ is back-projected to $q_i$ using the same procedure as for a point dart. The $k$ vectors from $q_0$ to $\{q_{i>0}\}$ span the transformed flat, but are no longer orthonormal because of the final scaling step, so we must reconstruct an orthonormal basis. Now we are ready to calculate volumes, using forward transformations. We calculate the distance from the flat to the origin. This tells us the radius of the $k$-dimensional subsphere that is the intersection of the flat with the $d$-sphere. We compute the volume of this subsphere. We multiply this volume by the sum of the $x$-components of the orthonormal basis, which gives the volume of the (unrotated) ellipse of intersection. The (forward) rotations do not affect the volume so are skipped.

%  \item 
For unaligned line-darts in the plane, the distance
  from the origin is easy to measure. We use a process similar to the prior paragraph but it is a little easier because
  the $1$-sphere is simply a line segment.
%\end{itemize}

%\newcommand{\myspacing}{0.9\columnwidth}
%\newcommand{\myspacing}{0.195\textwidth}
%    \includegraphics[width=\myspacing]{row5_2_10_0p100000.pdf}
\newcommand{\myspacing}{0.13\textwidth}
\begin{figure*}[!ht]
  \begin{center} 
\fontsize{8}{8}\selectfont%fontsize, spacing after each line
$d=2$, varying squish\\
    \includegraphics[height=\myspacing]{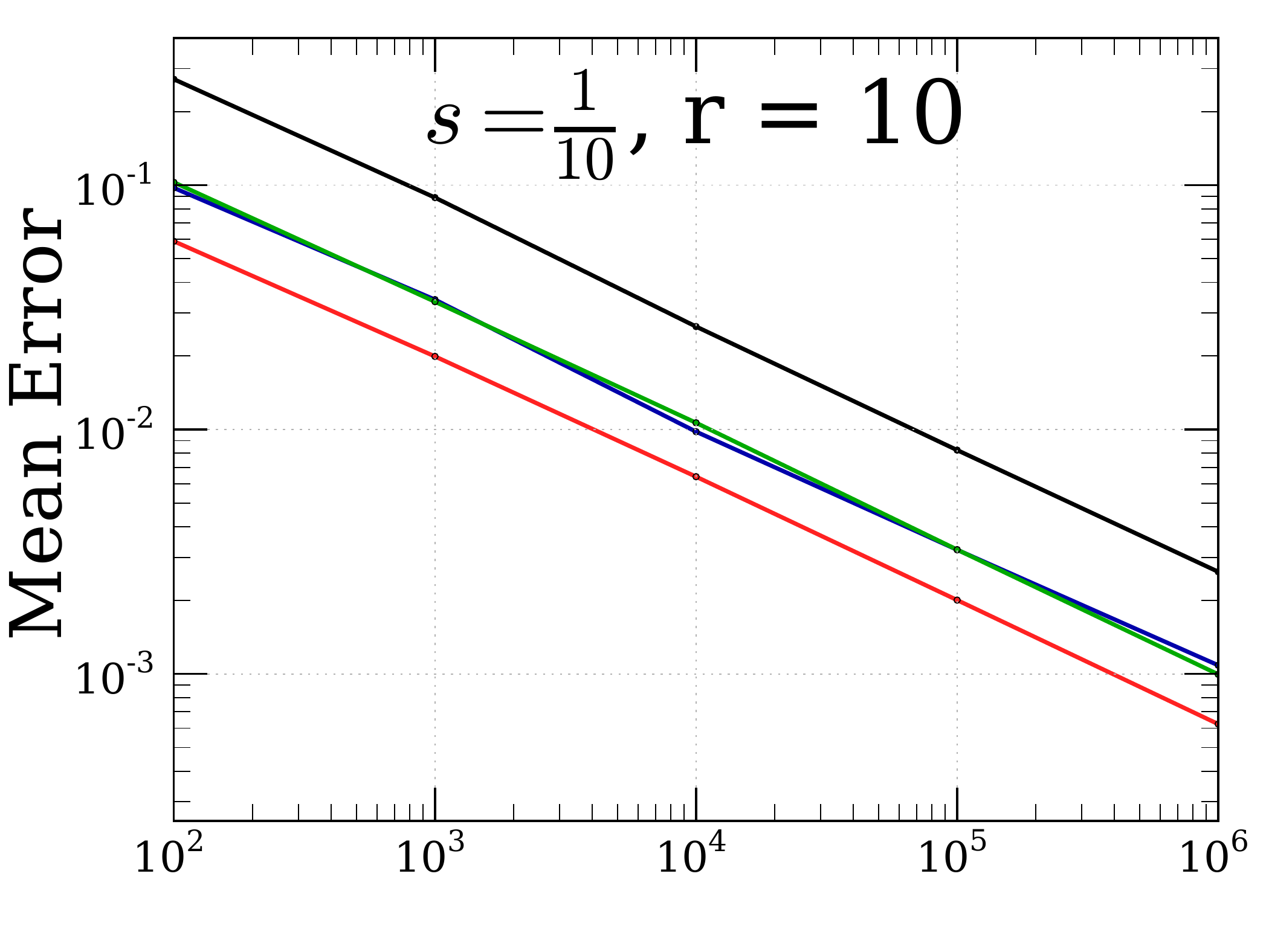}
    \includegraphics[height=\myspacing]{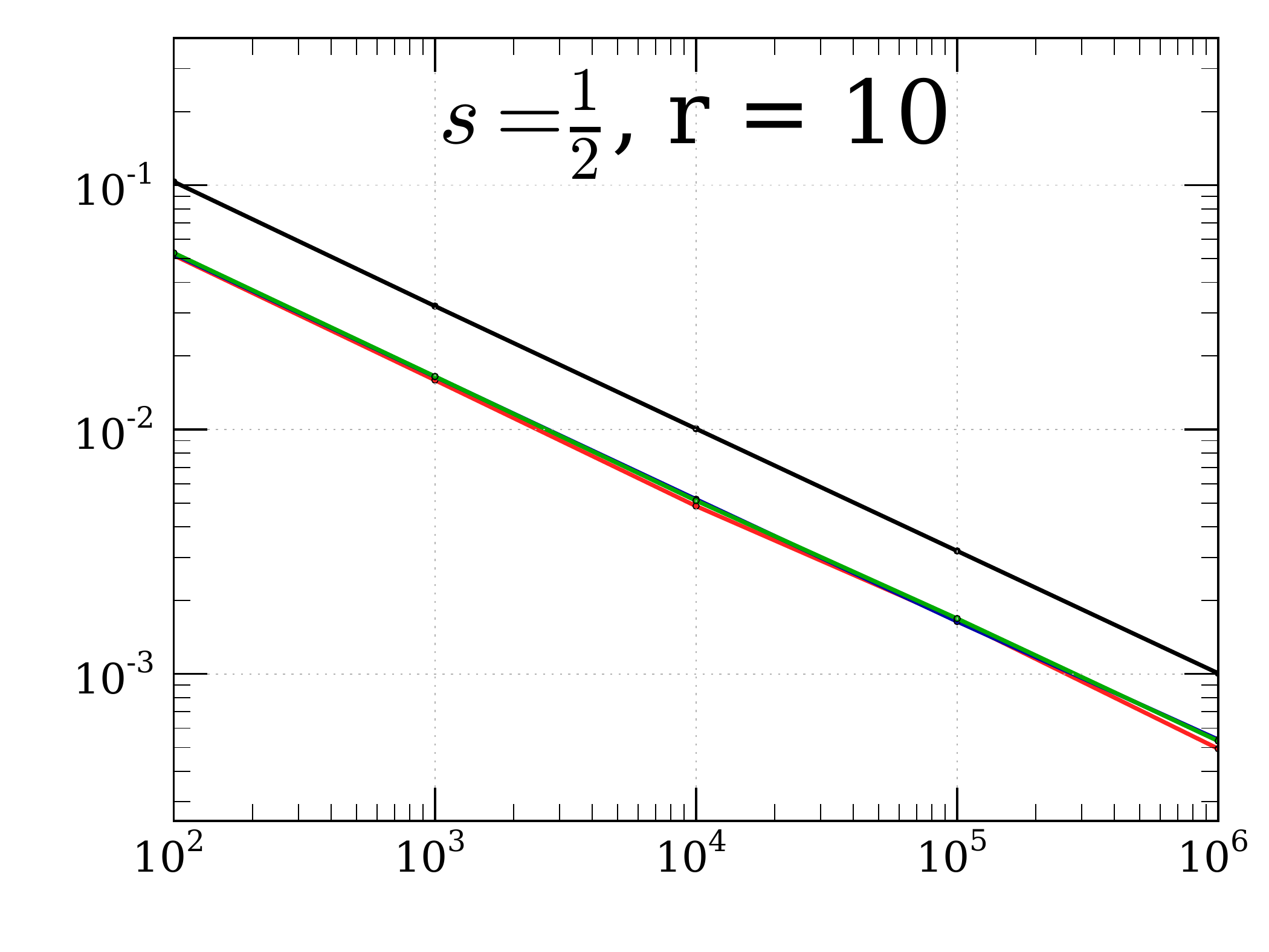}
    \includegraphics[height=\myspacing]{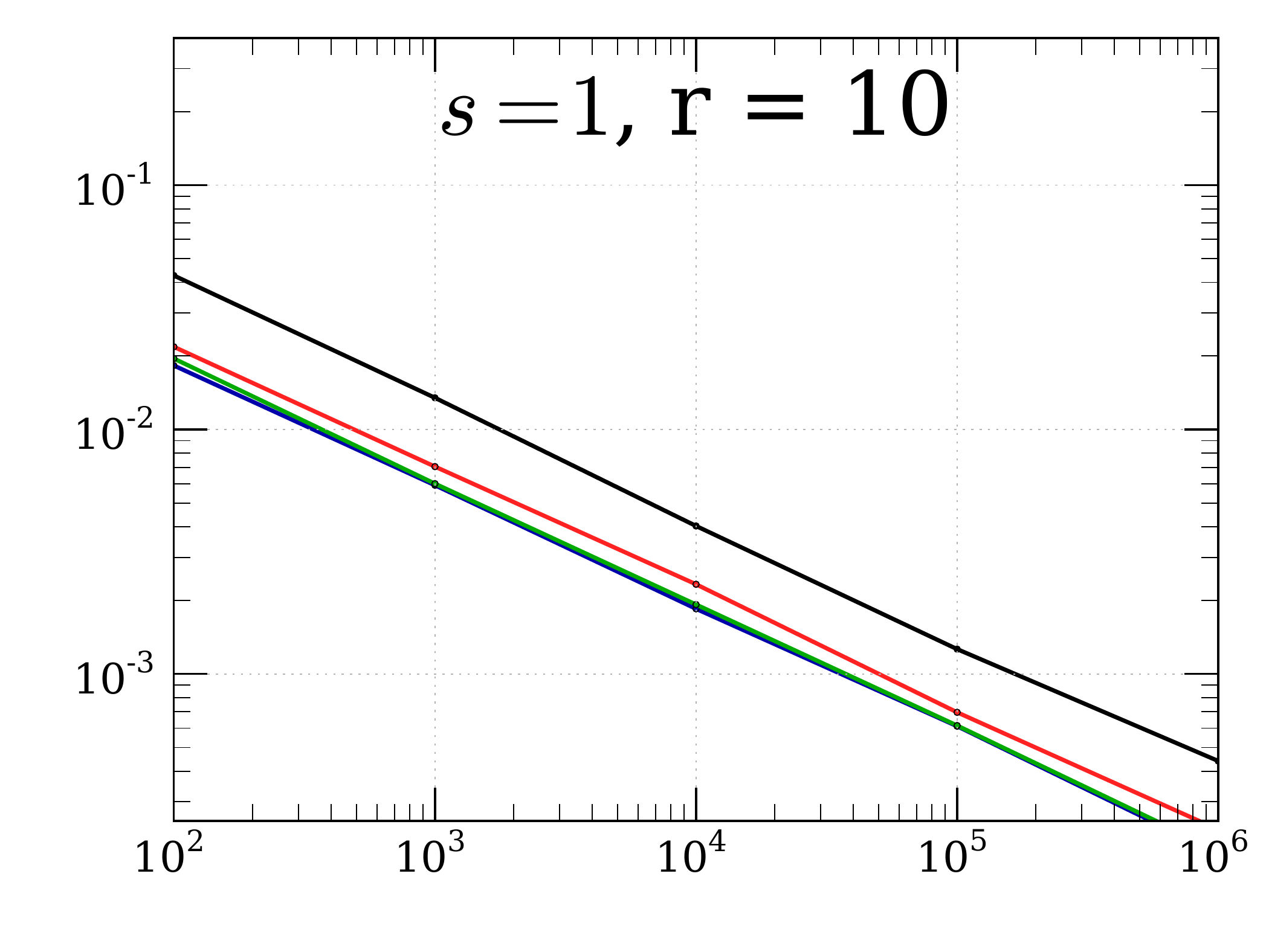}
    \includegraphics[height=\myspacing]{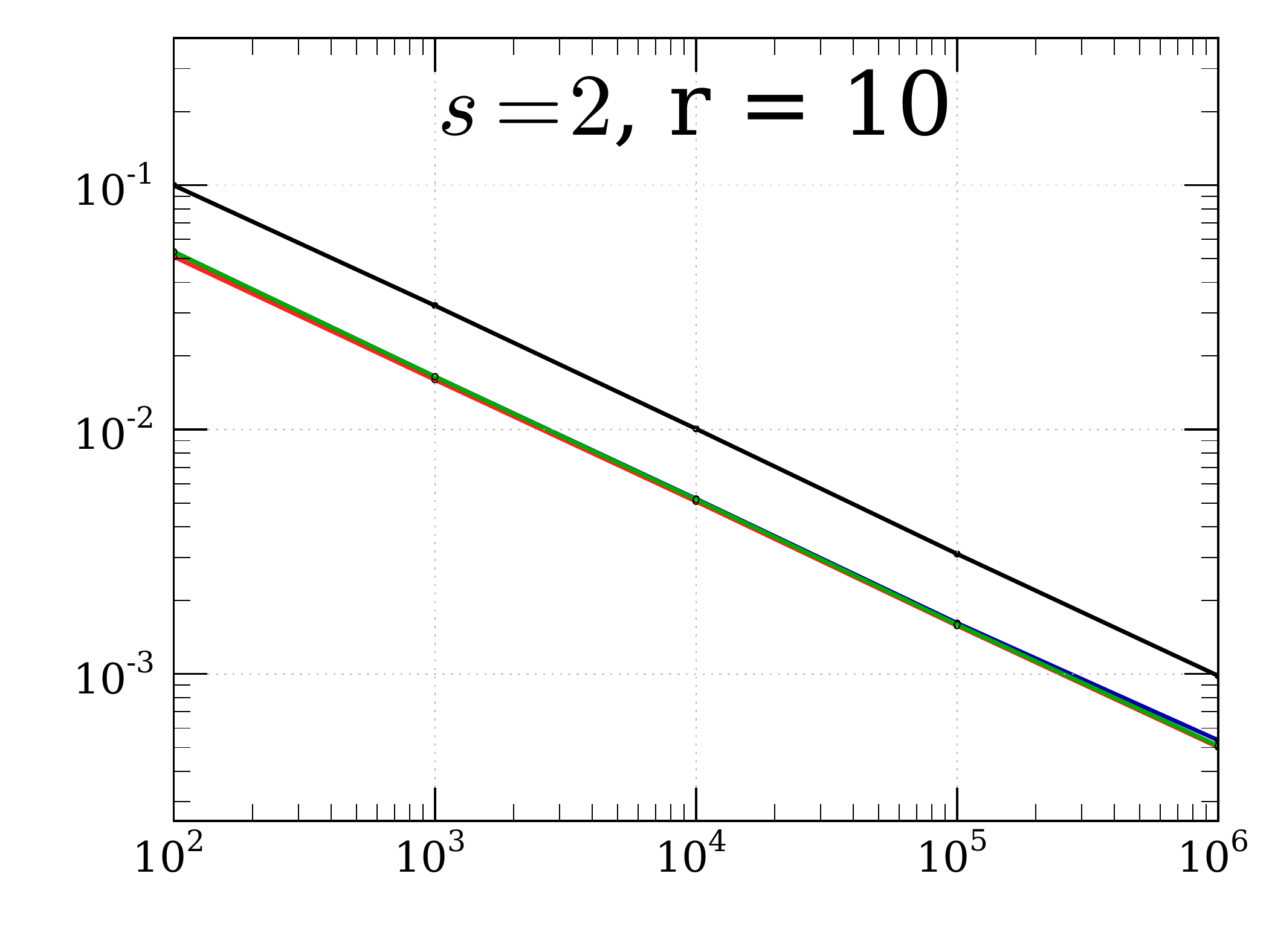} 
    \includegraphics[height=\myspacing]{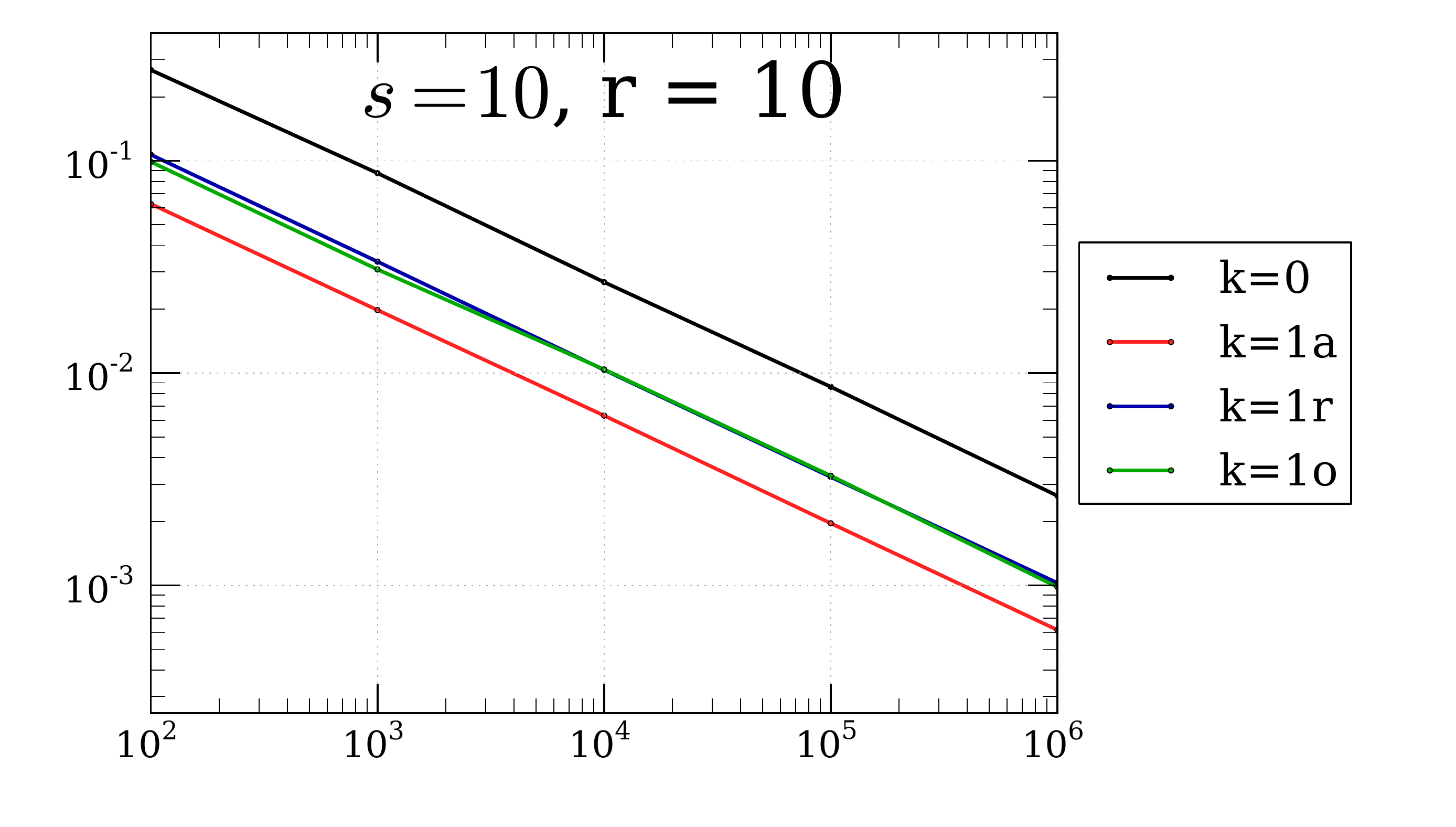}\\
 $d=2$, varying rotation\\
    \includegraphics[height=\myspacing]{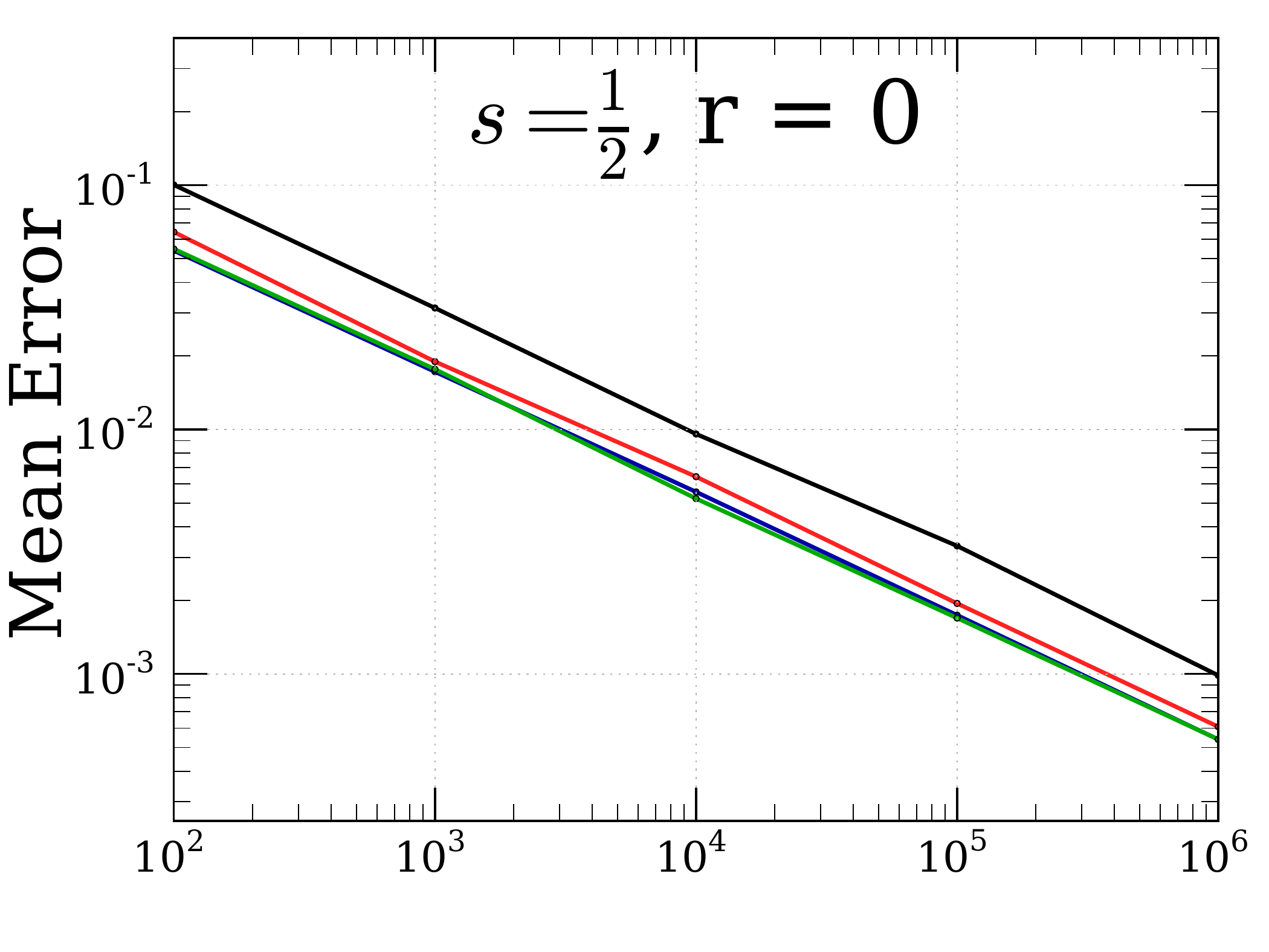}
    \includegraphics[height=\myspacing]{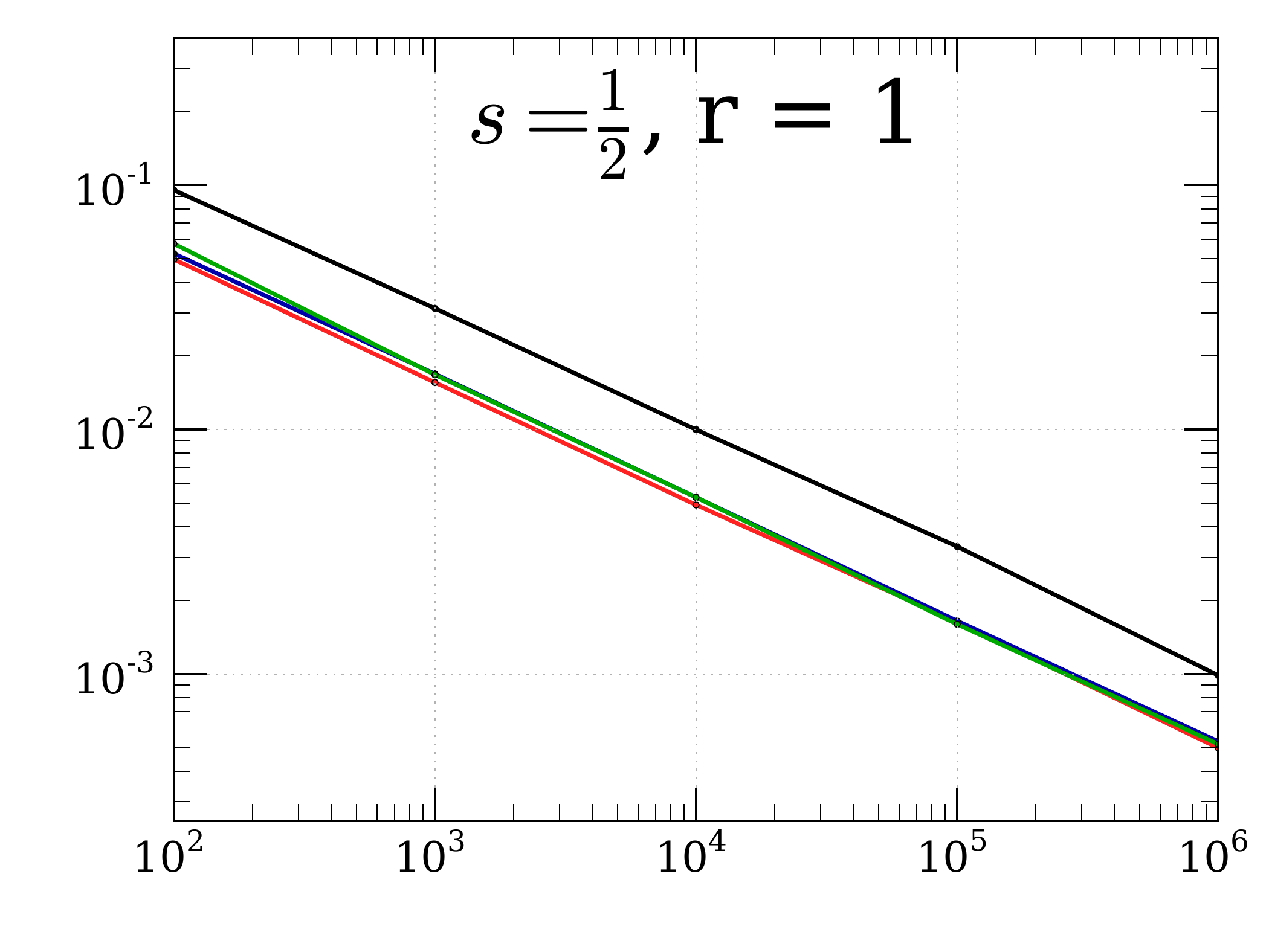}
    \includegraphics[height=\myspacing]{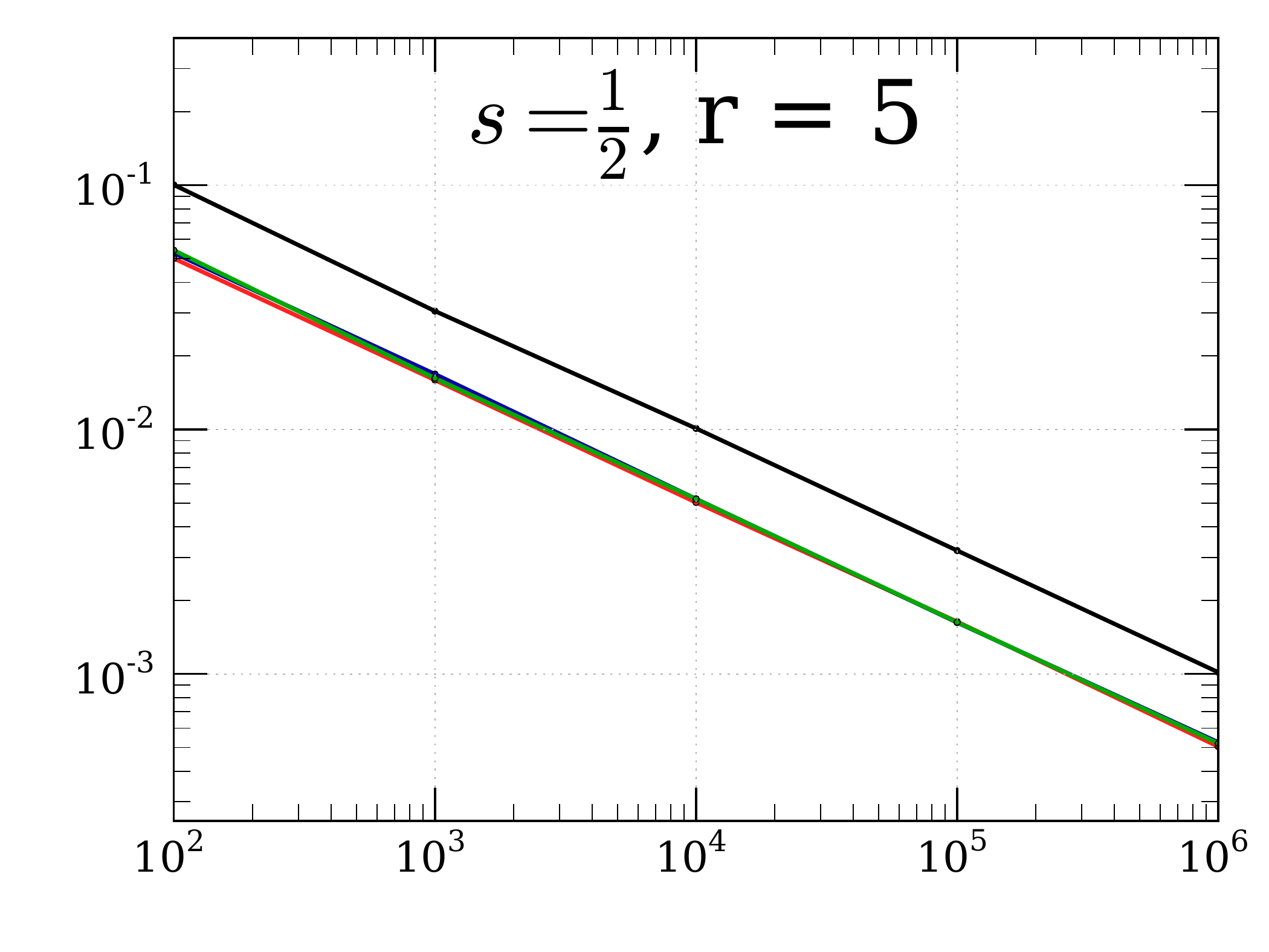} 
    \includegraphics[height=\myspacing]{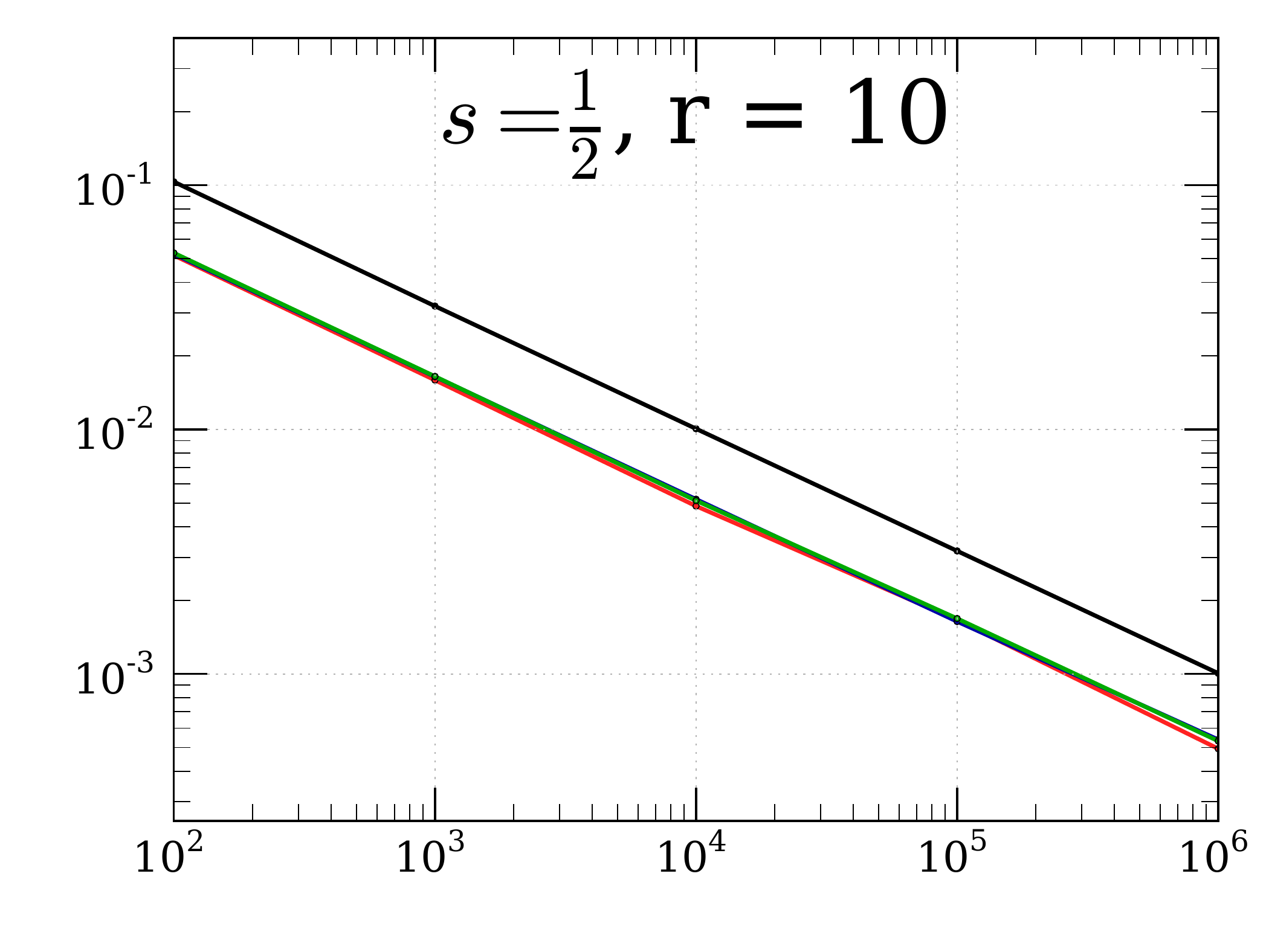}
    \includegraphics[height=\myspacing]{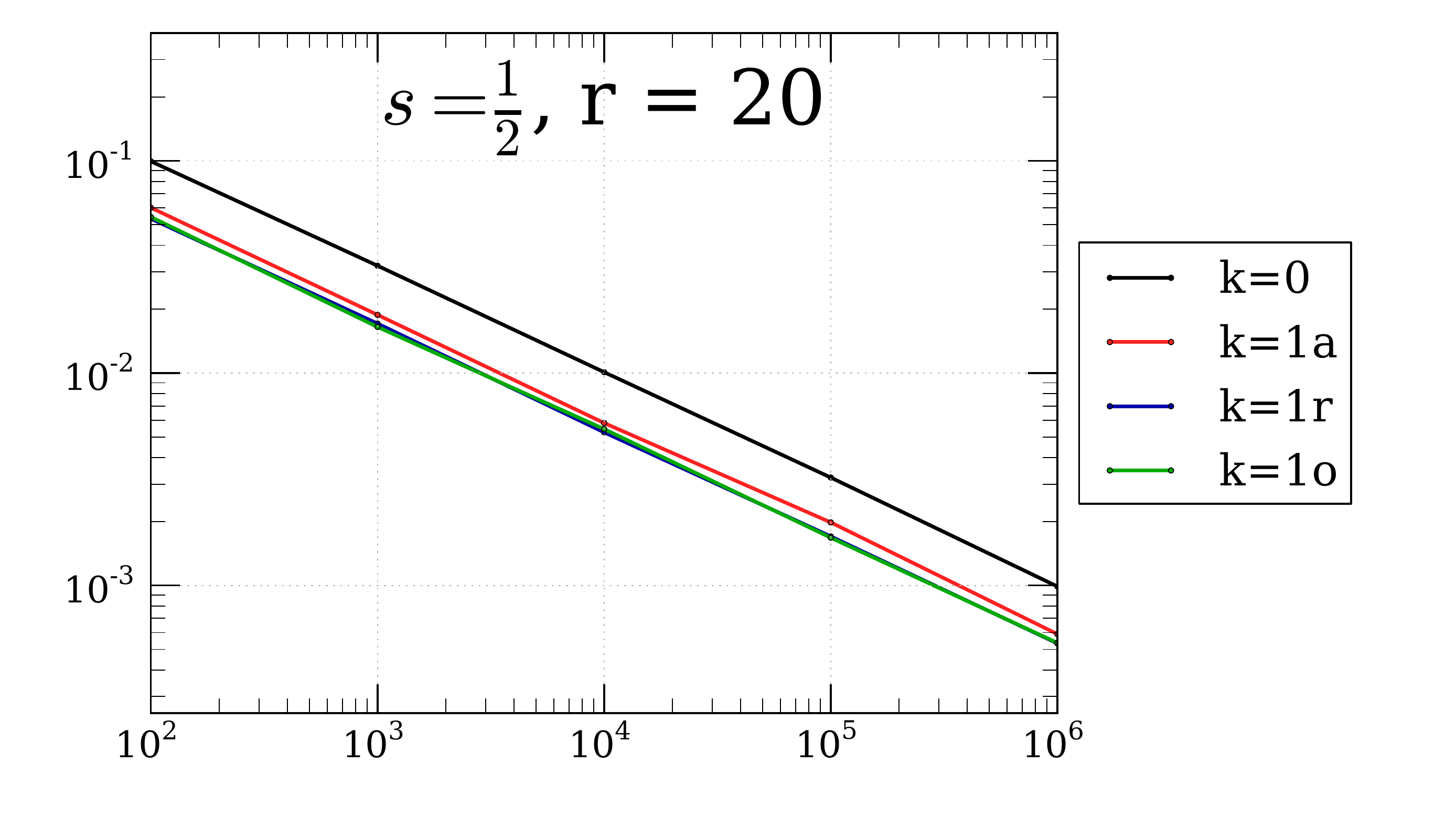}\\
$d=3$, varying squish\\
     \includegraphics[height=\myspacing]{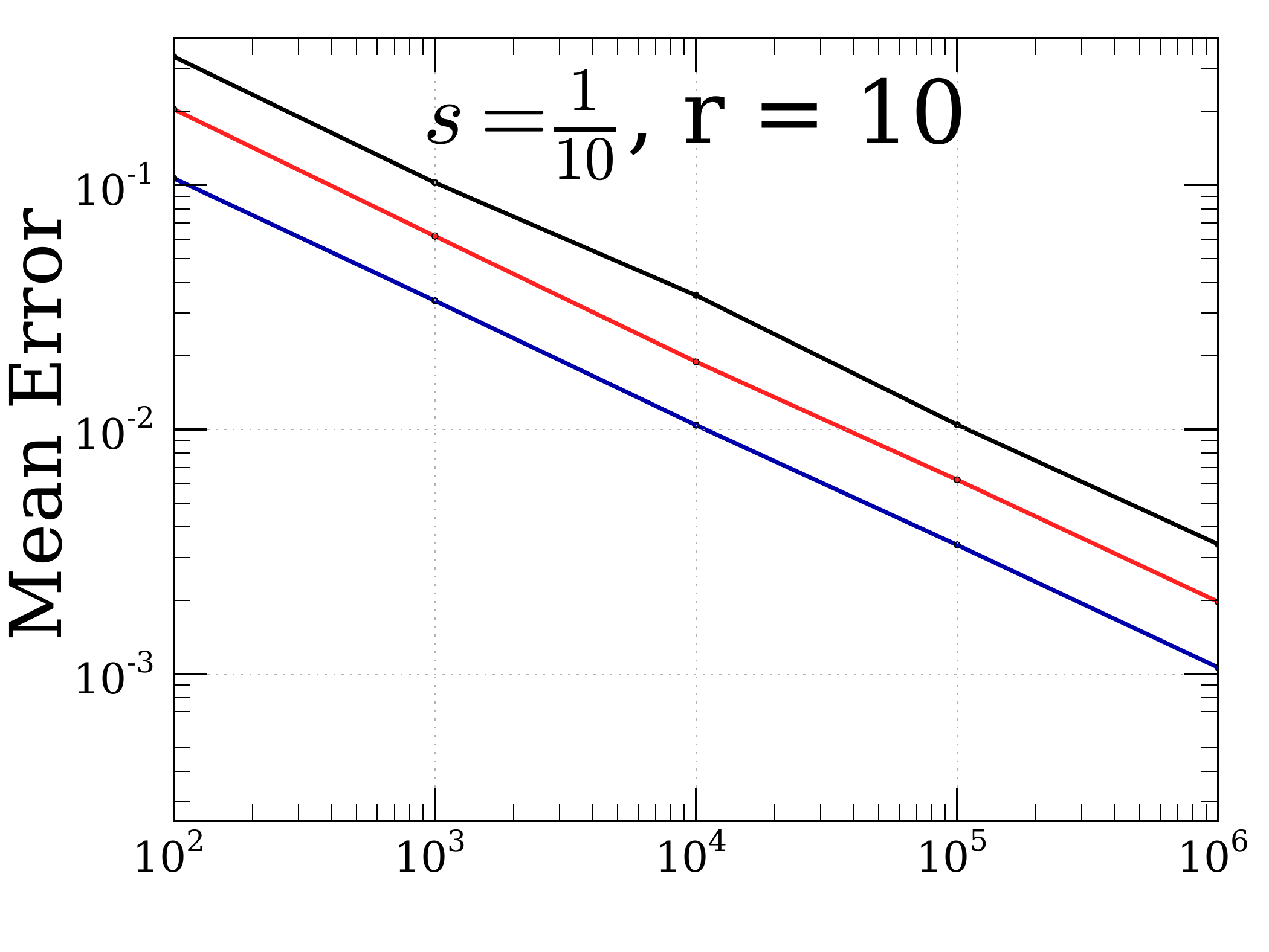}
    \includegraphics[height=\myspacing]{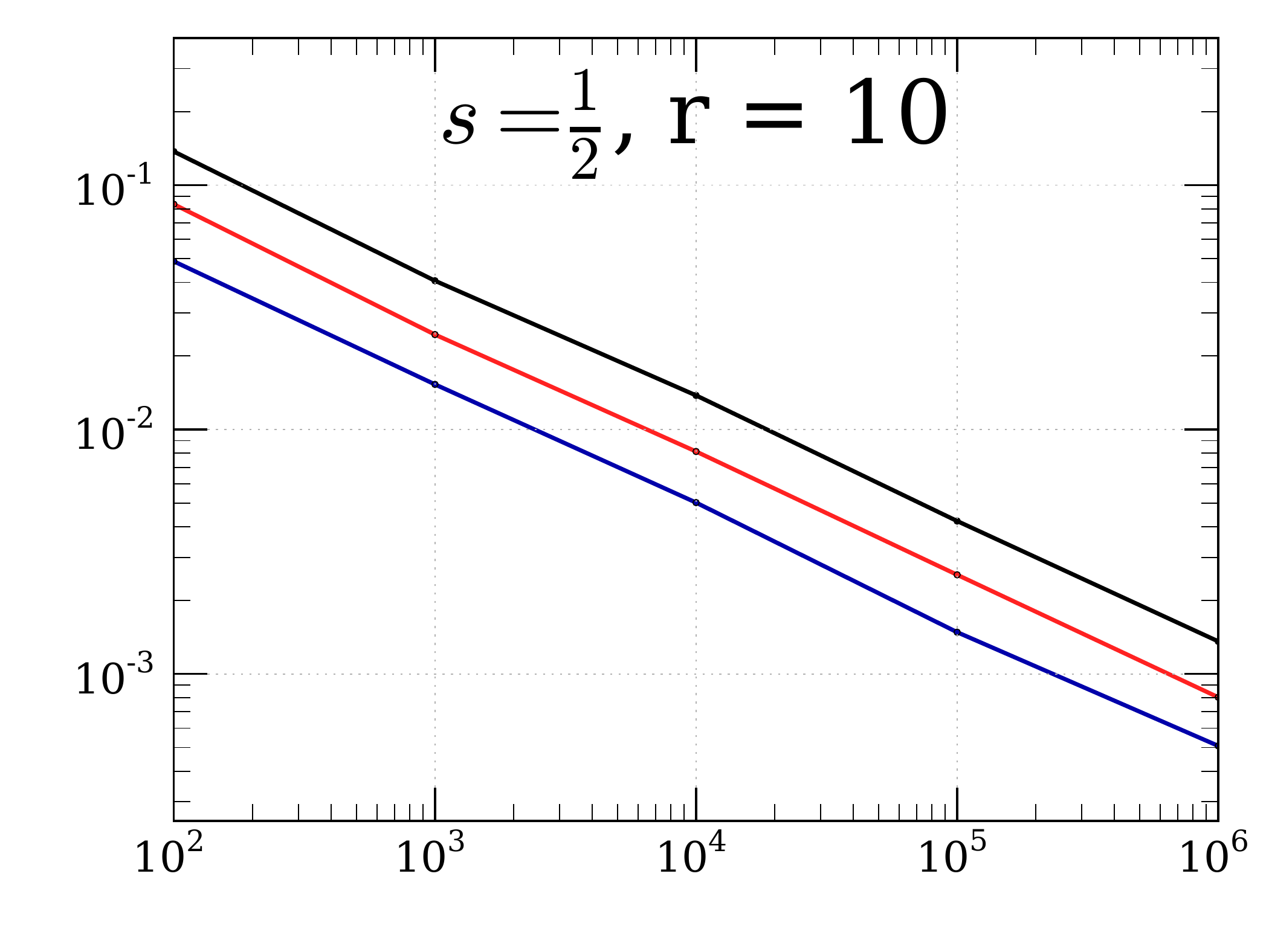}
    \includegraphics[height=\myspacing]{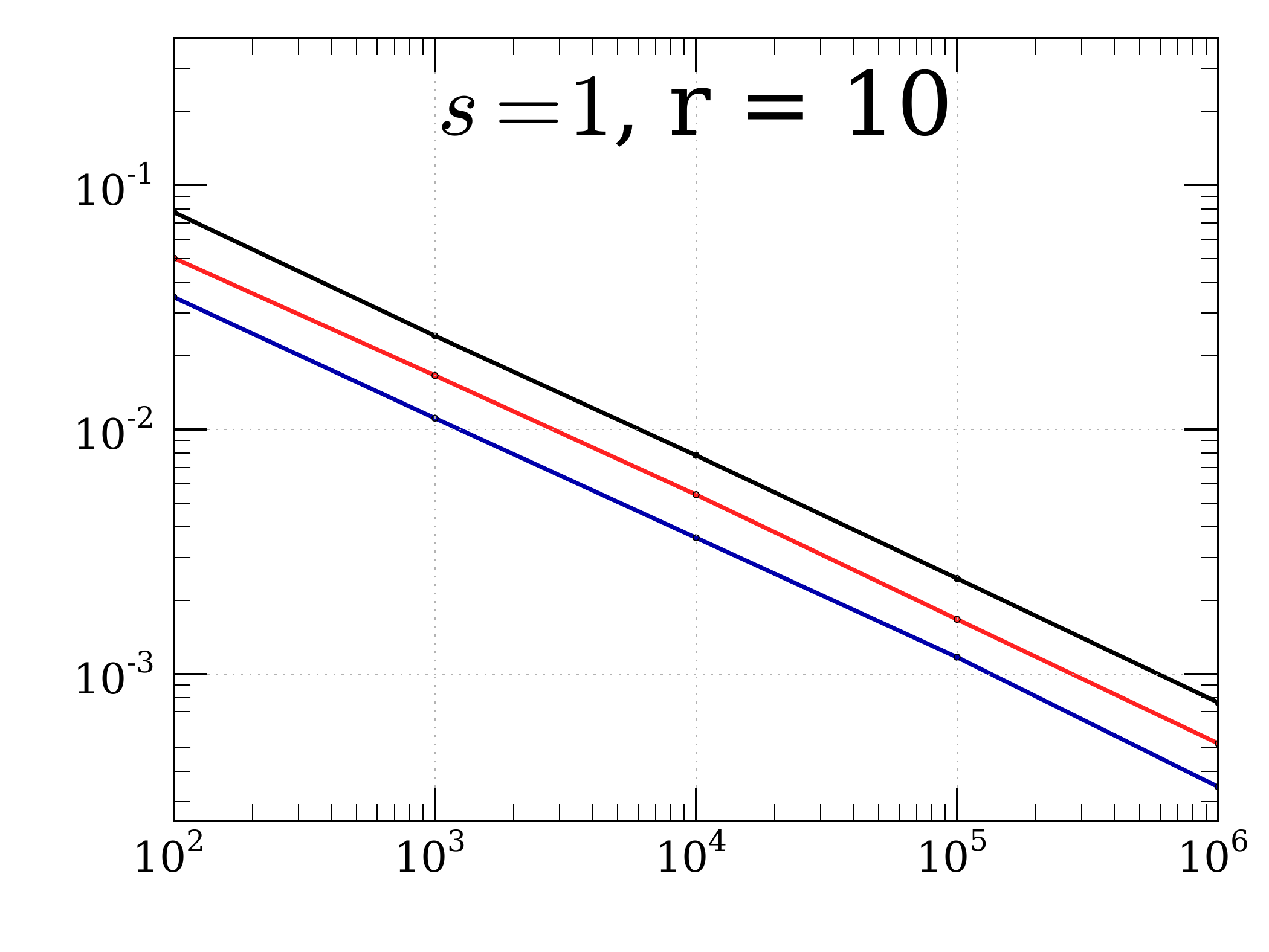}
    \includegraphics[height=\myspacing]{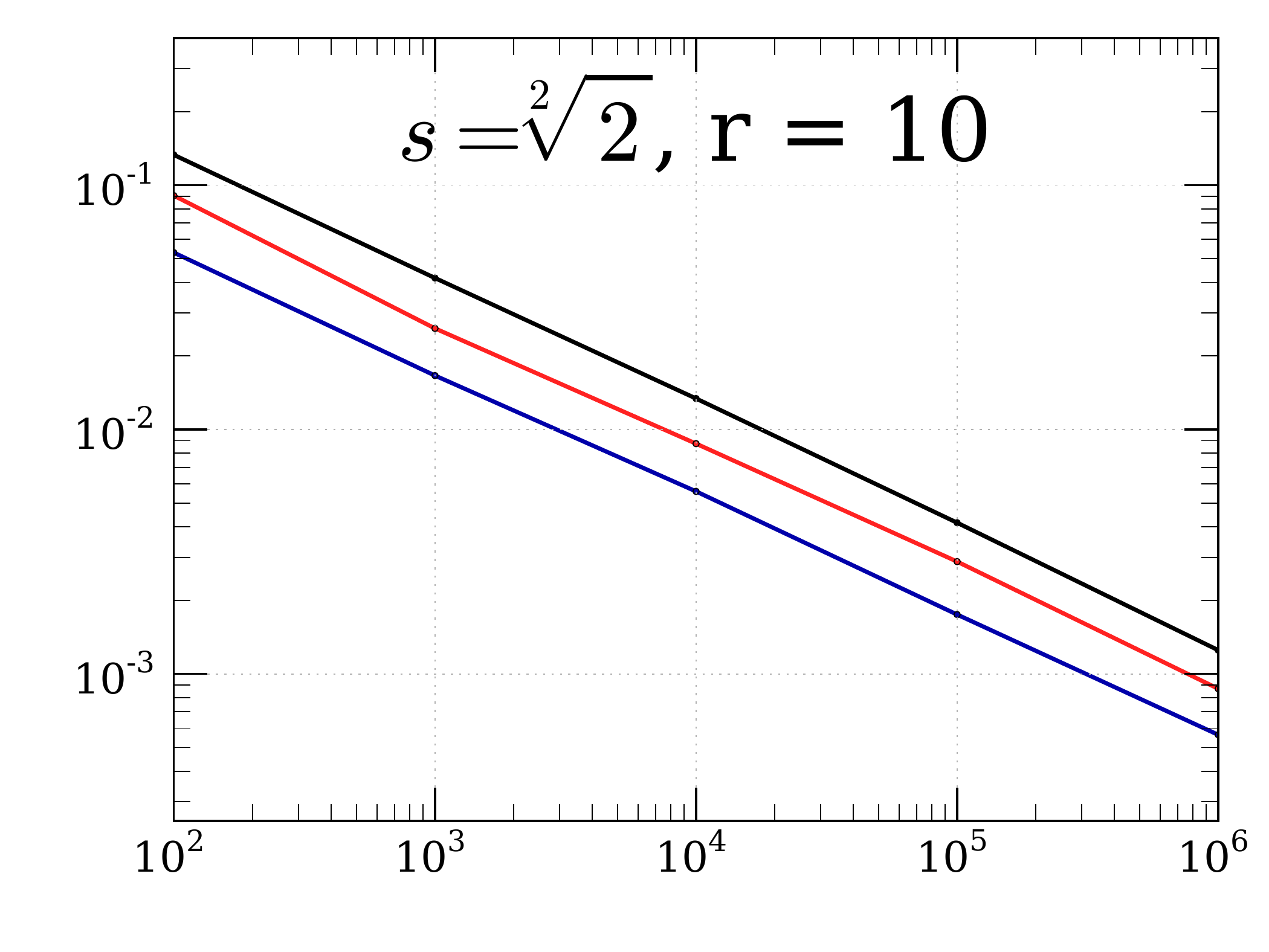}
    \includegraphics[height=\myspacing]{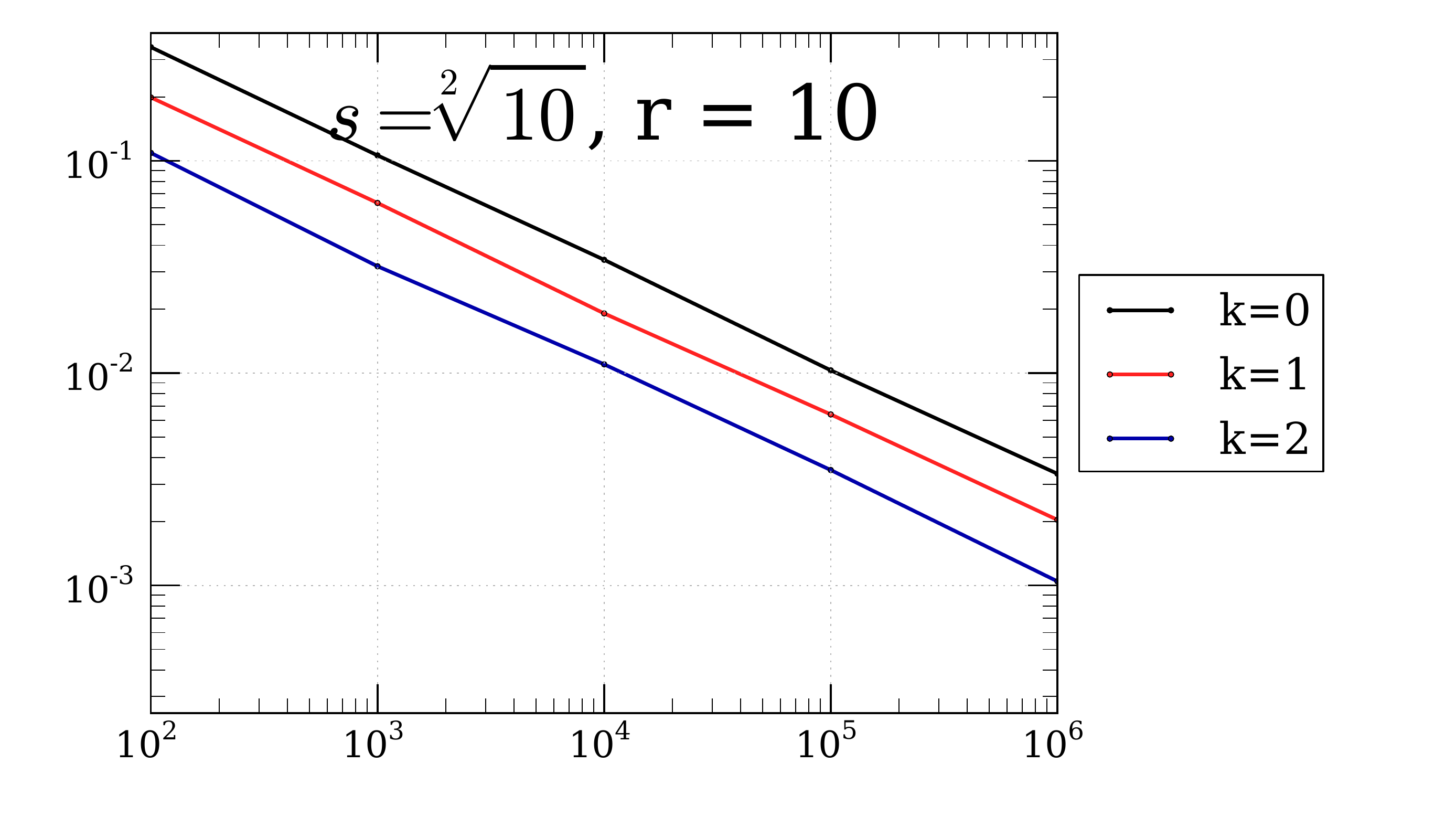}\\
$d=3$, varying rotation\\
    \includegraphics[height=\myspacing]{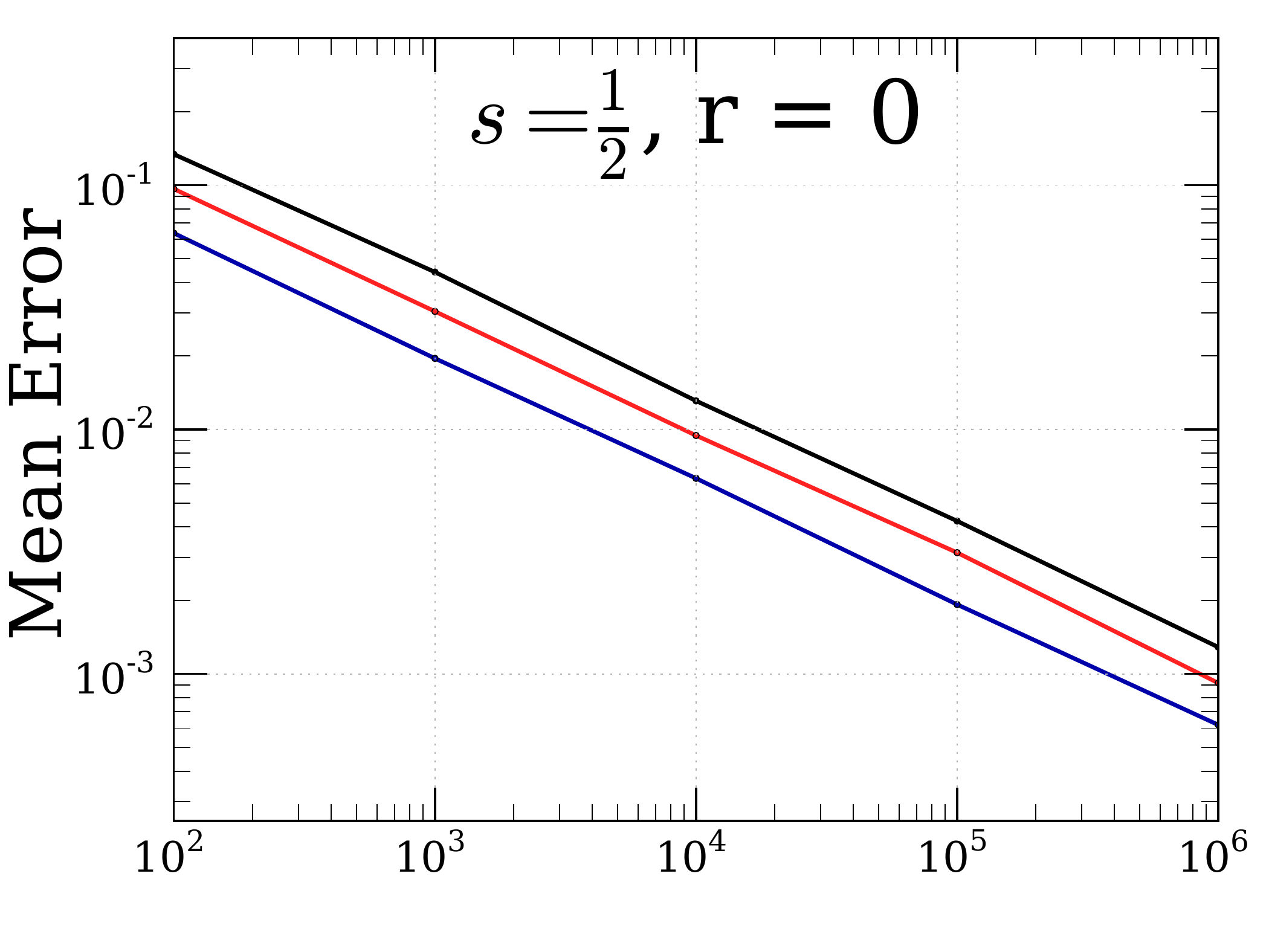}
    \includegraphics[height=\myspacing]{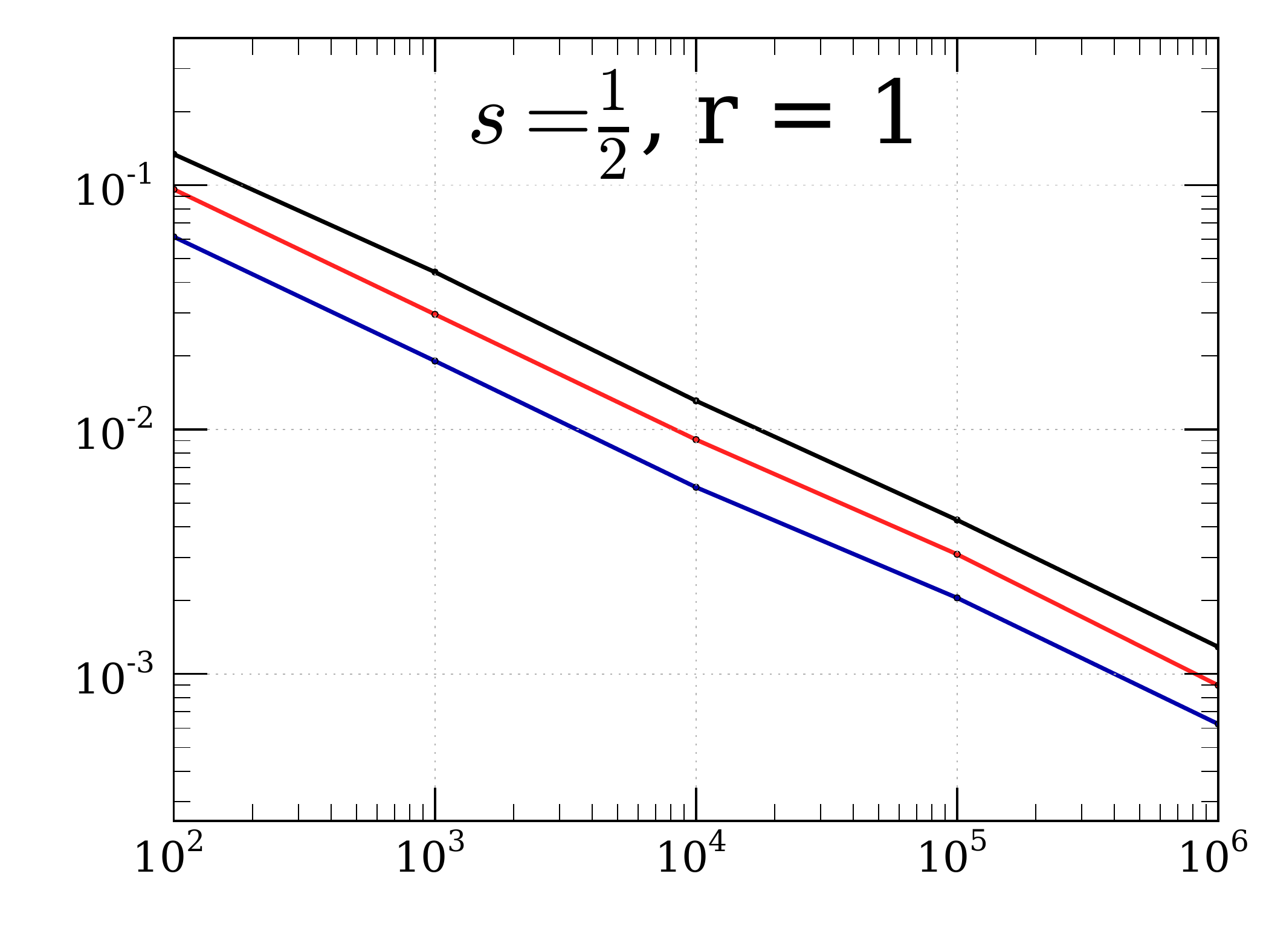}
    \includegraphics[height=\myspacing]{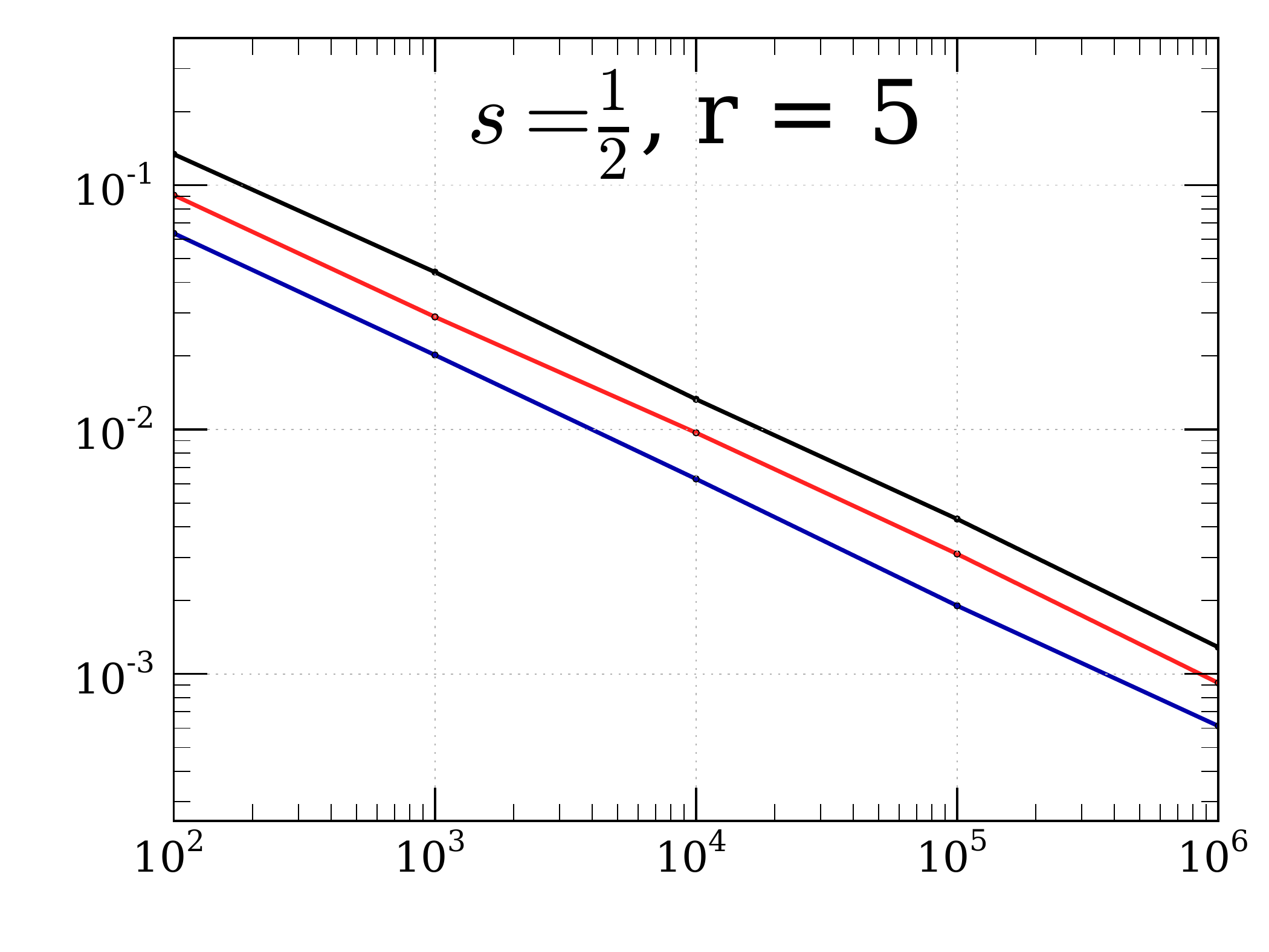}
    \includegraphics[height=\myspacing]{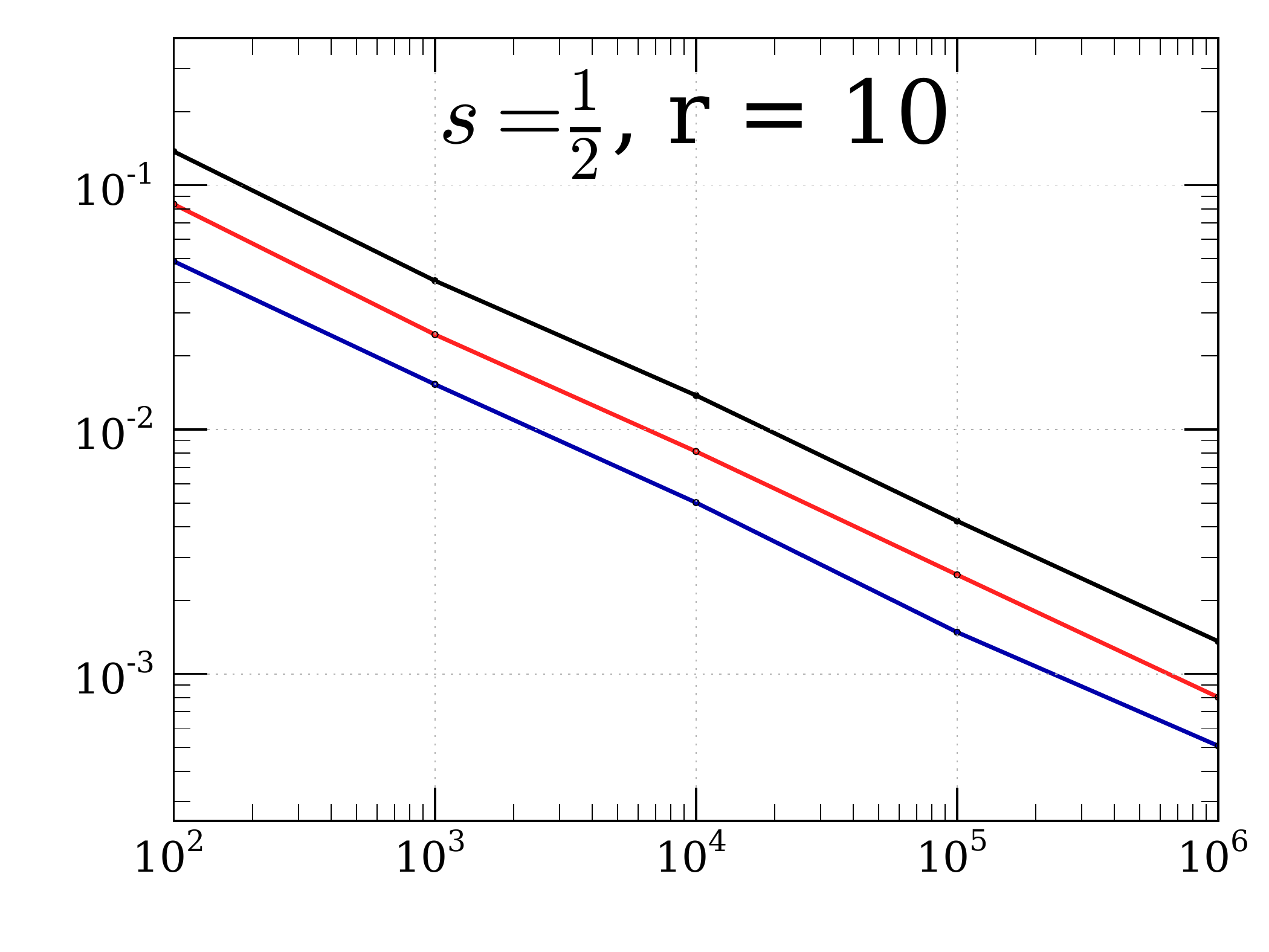}
    \includegraphics[height=\myspacing]{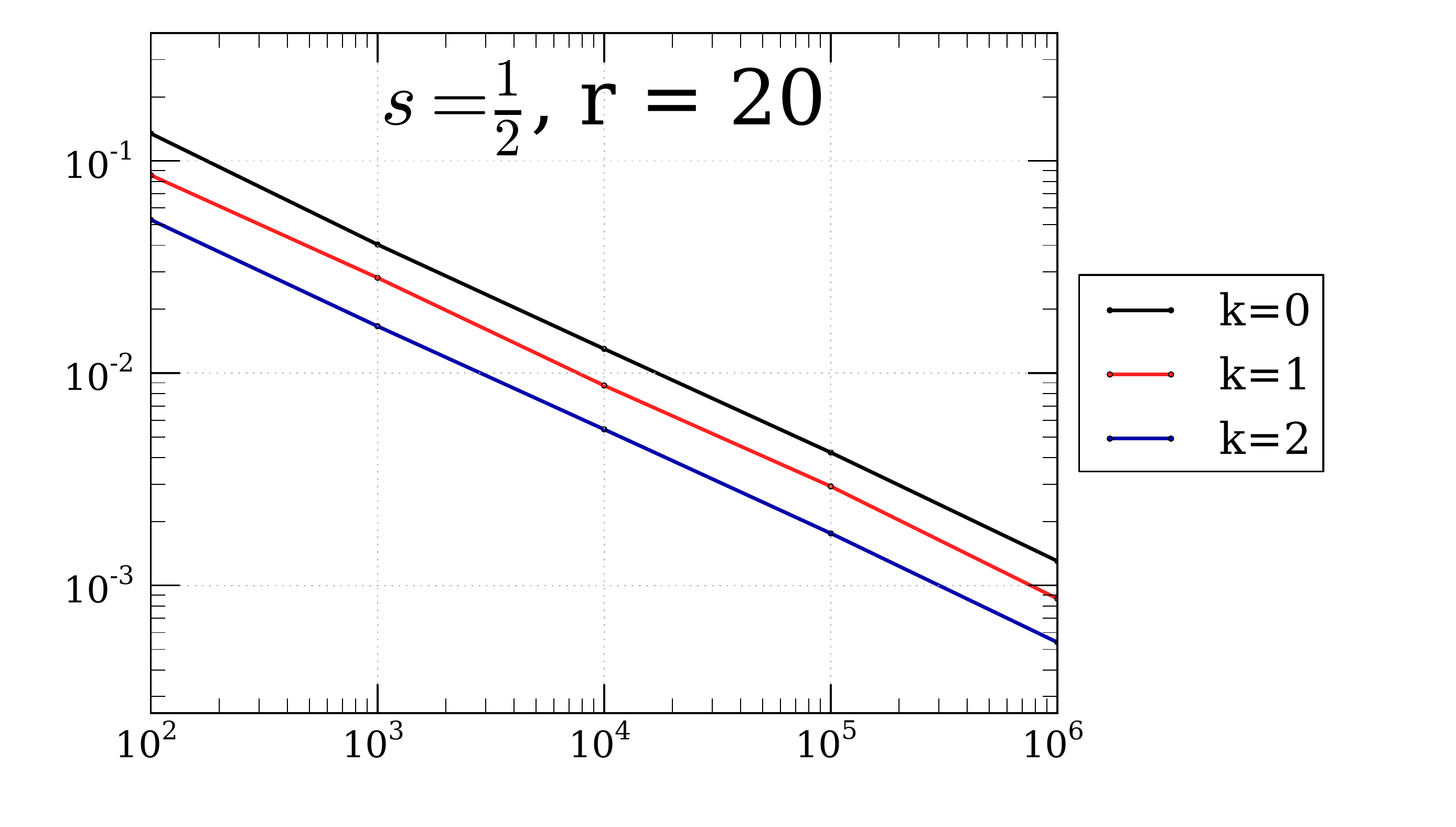}\\
$d=10$, varying squish\\
    \includegraphics[height=\myspacing]{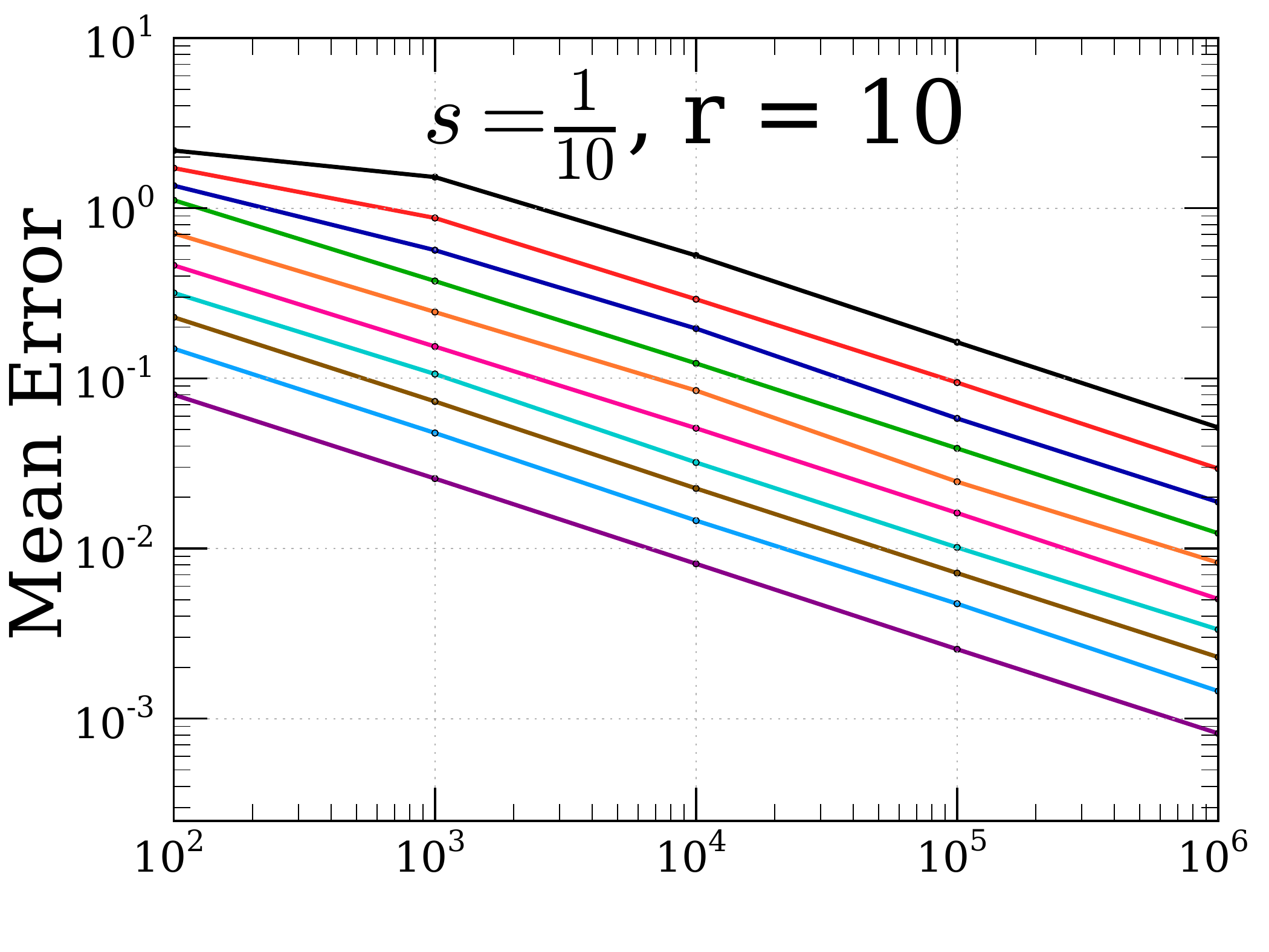}
    \includegraphics[height=\myspacing]{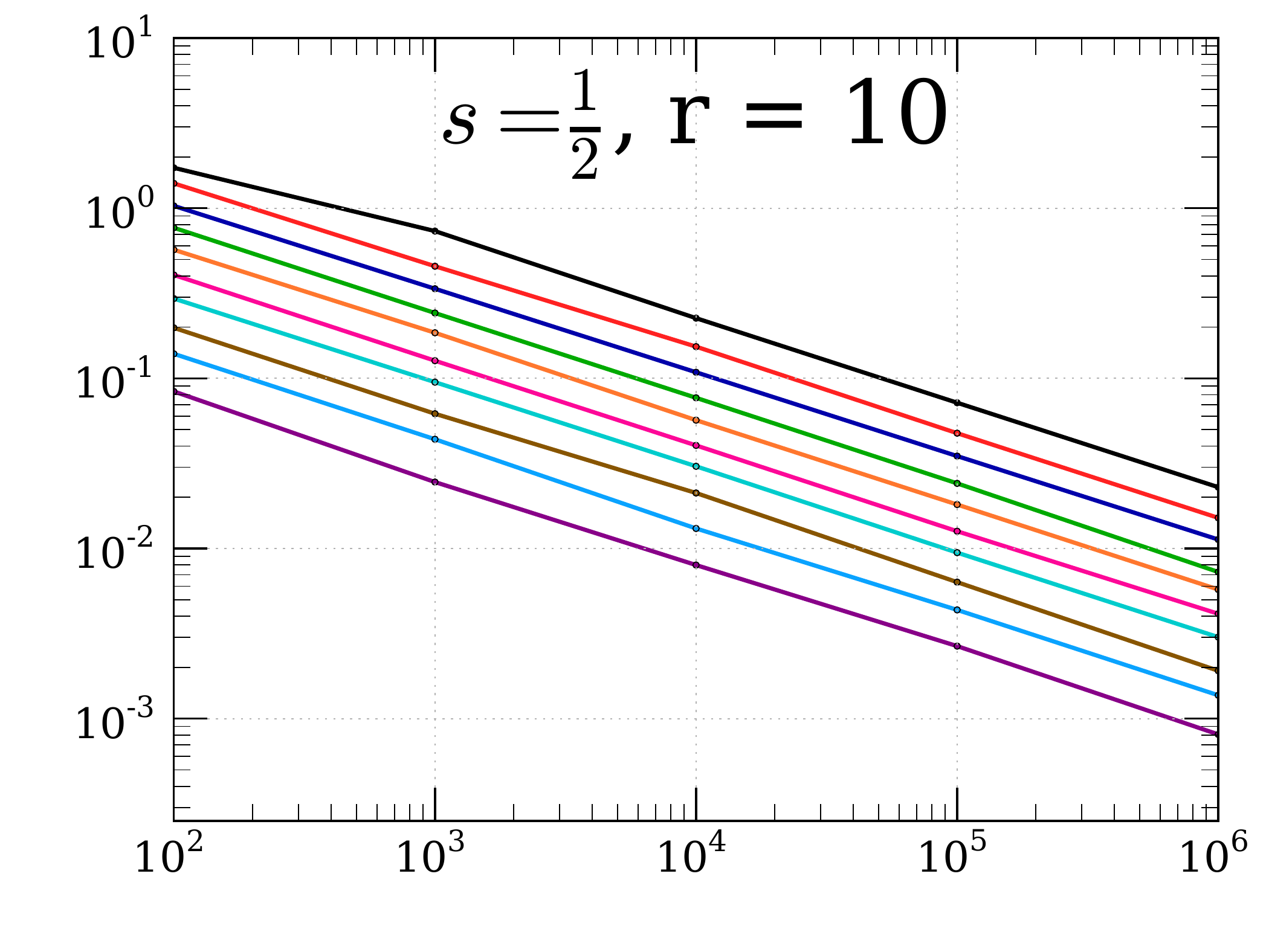}
    \includegraphics[height=\myspacing]{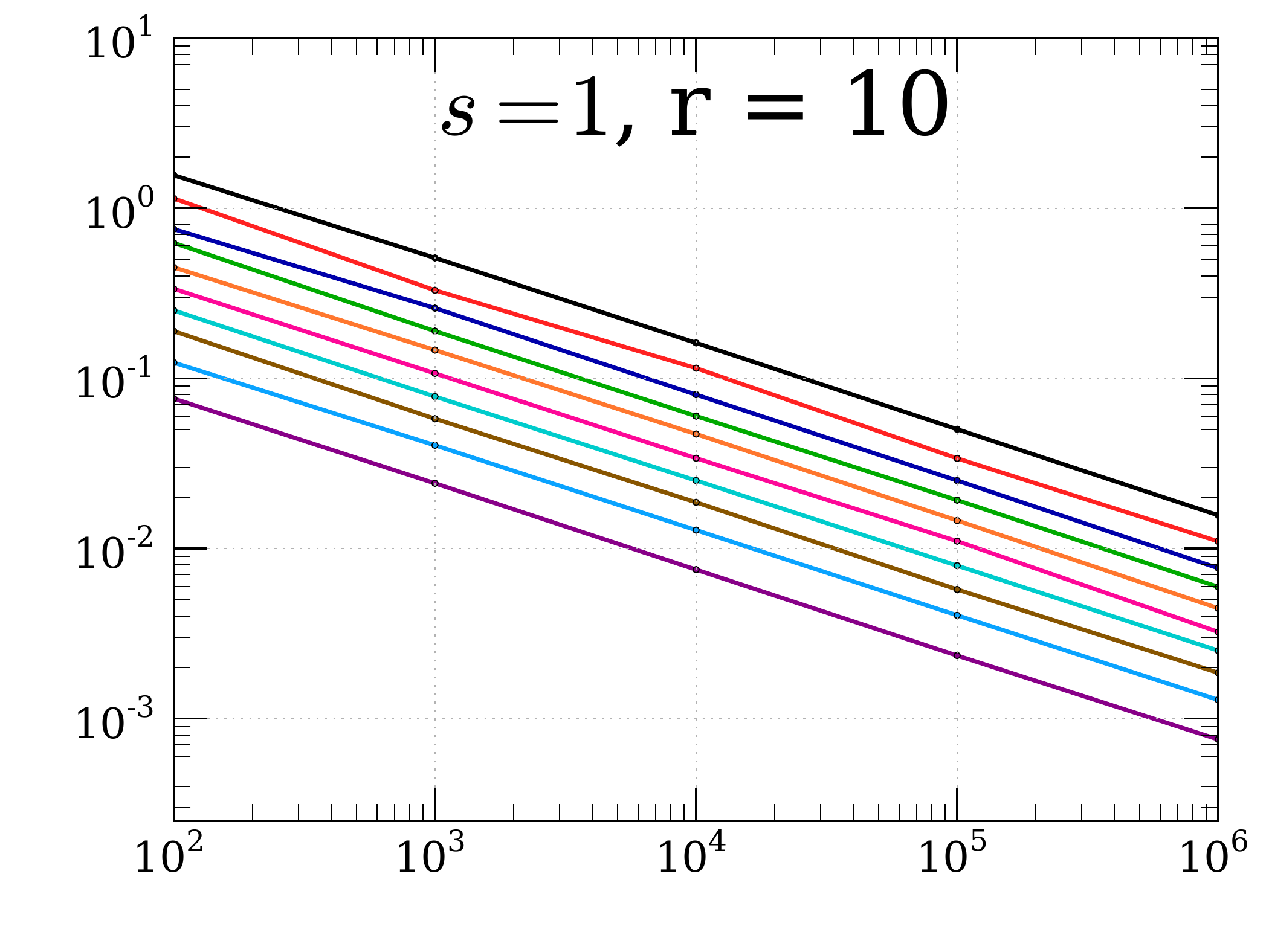}
    \includegraphics[height=\myspacing]{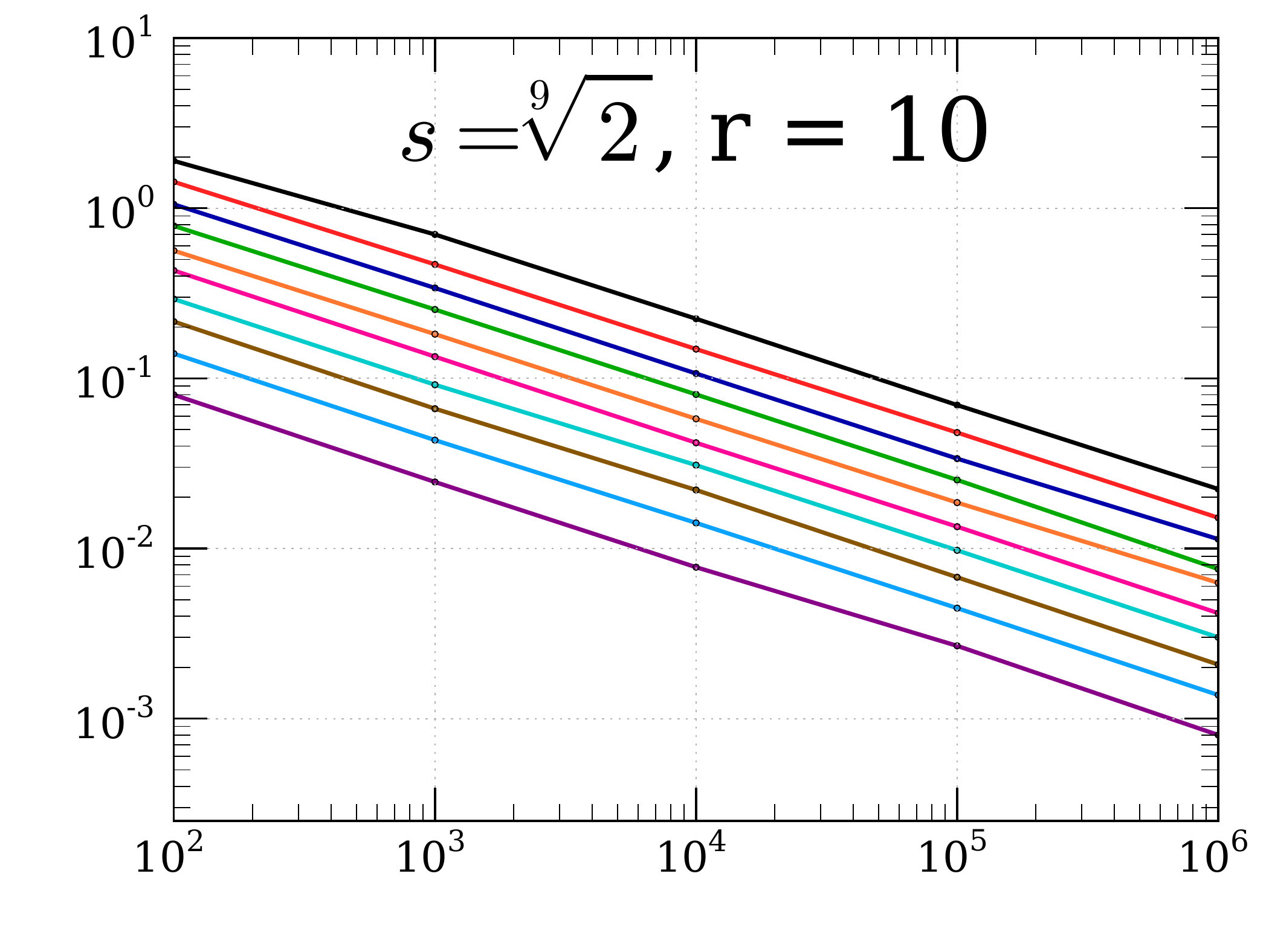}
    \includegraphics[height=\myspacing]{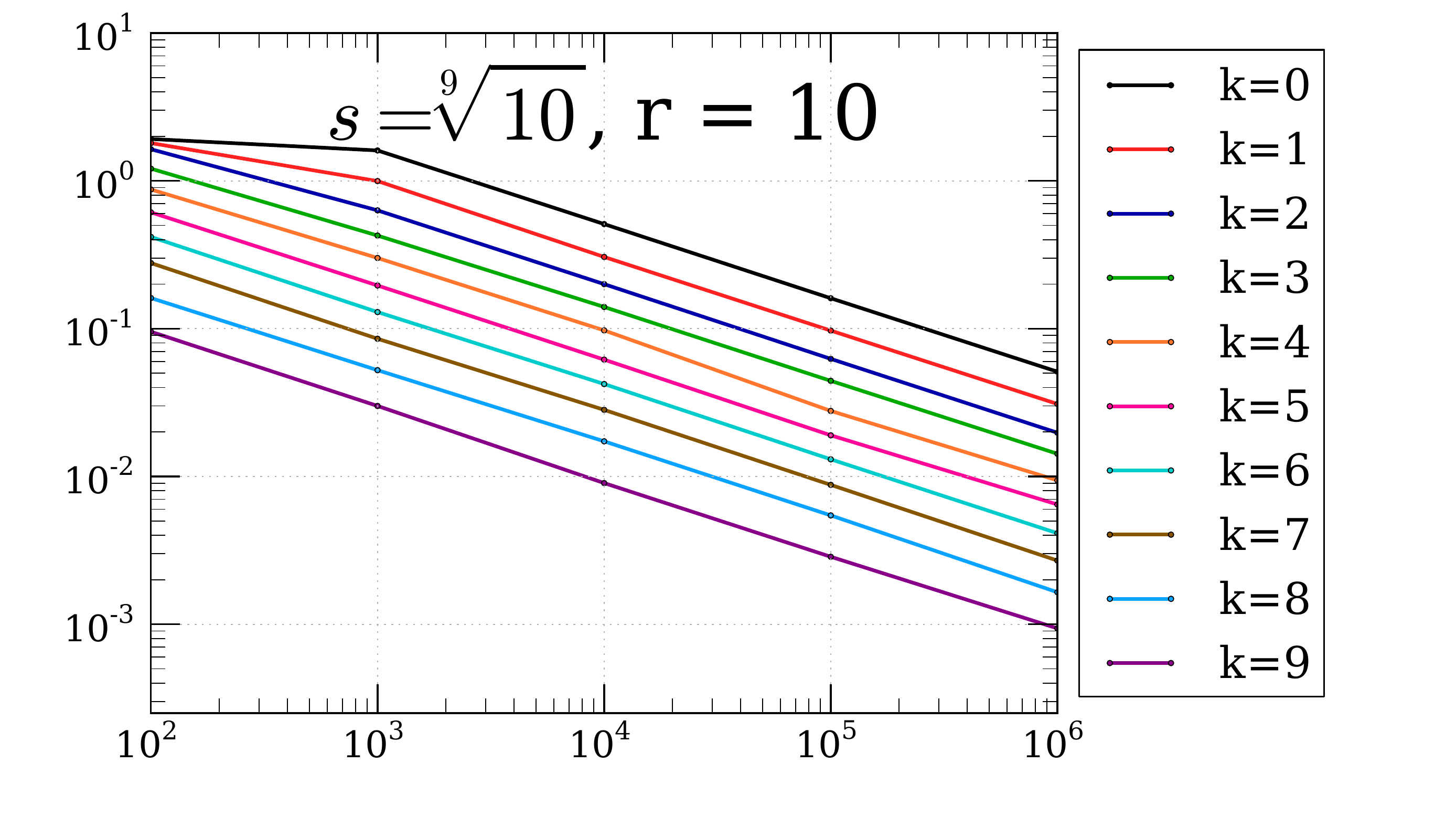}\\
$d=10$, varying rotation\\
    \includegraphics[height=\myspacing]{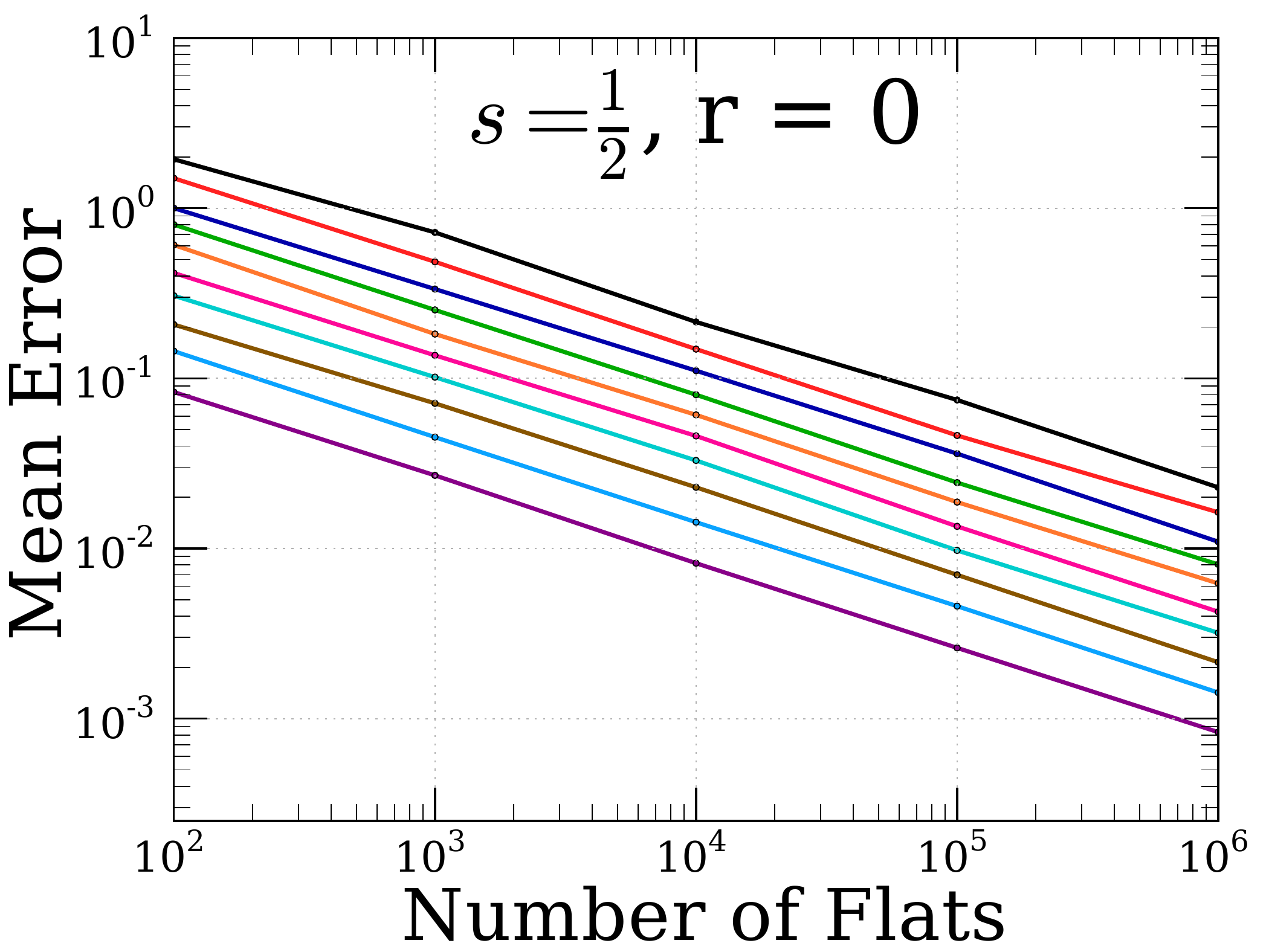}
    \includegraphics[height=\myspacing]{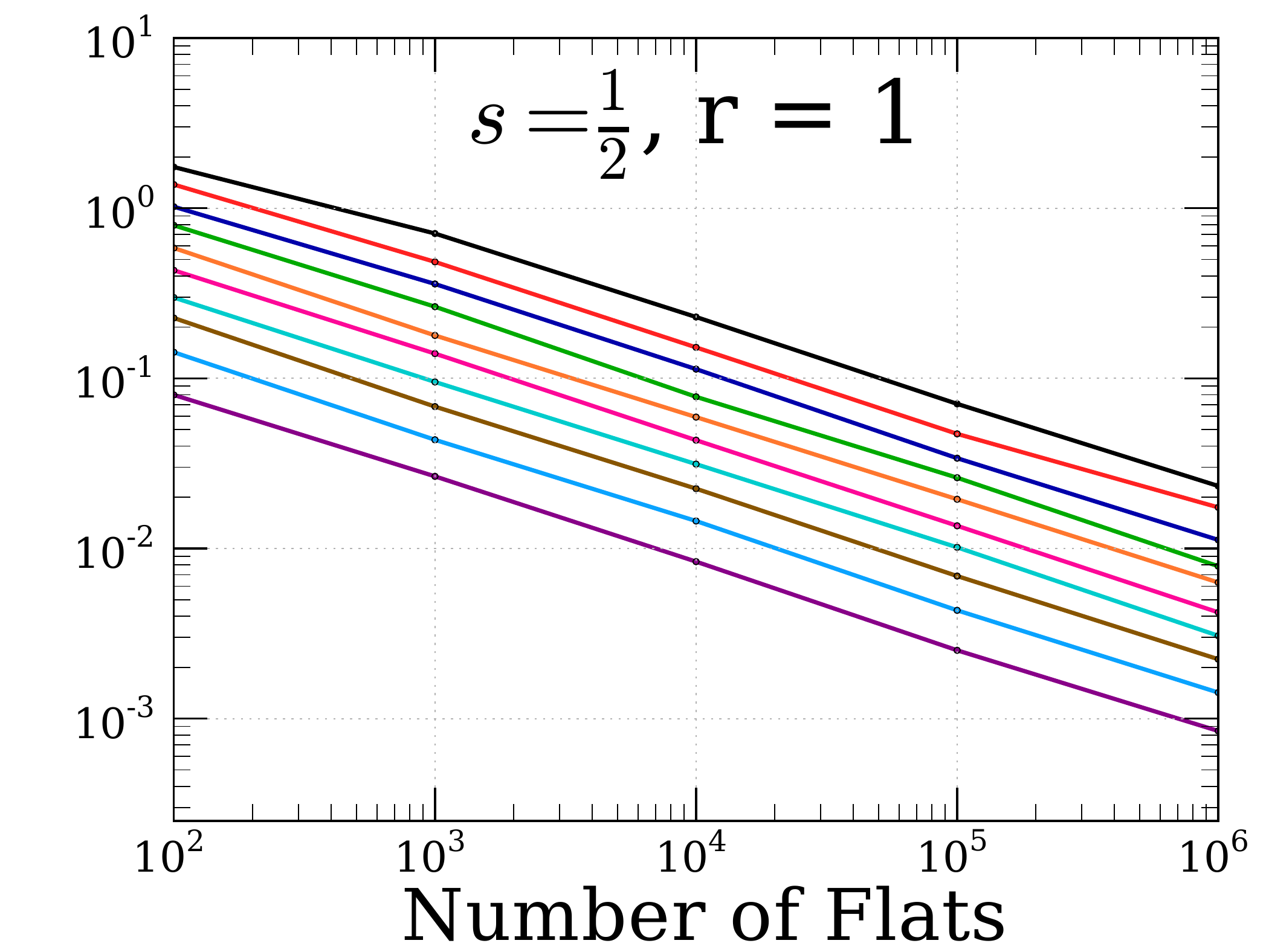}
    \includegraphics[height=\myspacing]{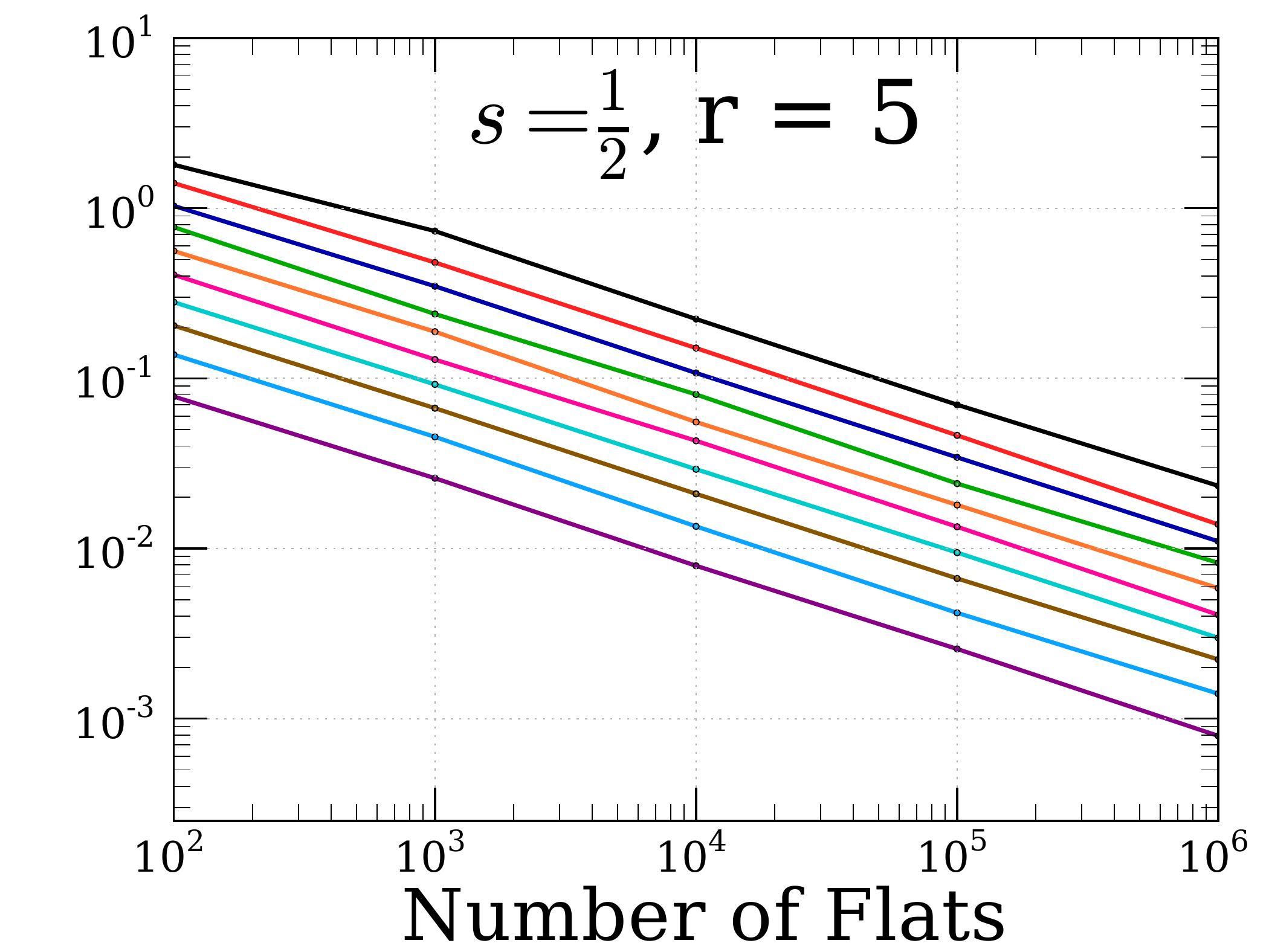}
    \includegraphics[height=\myspacing]{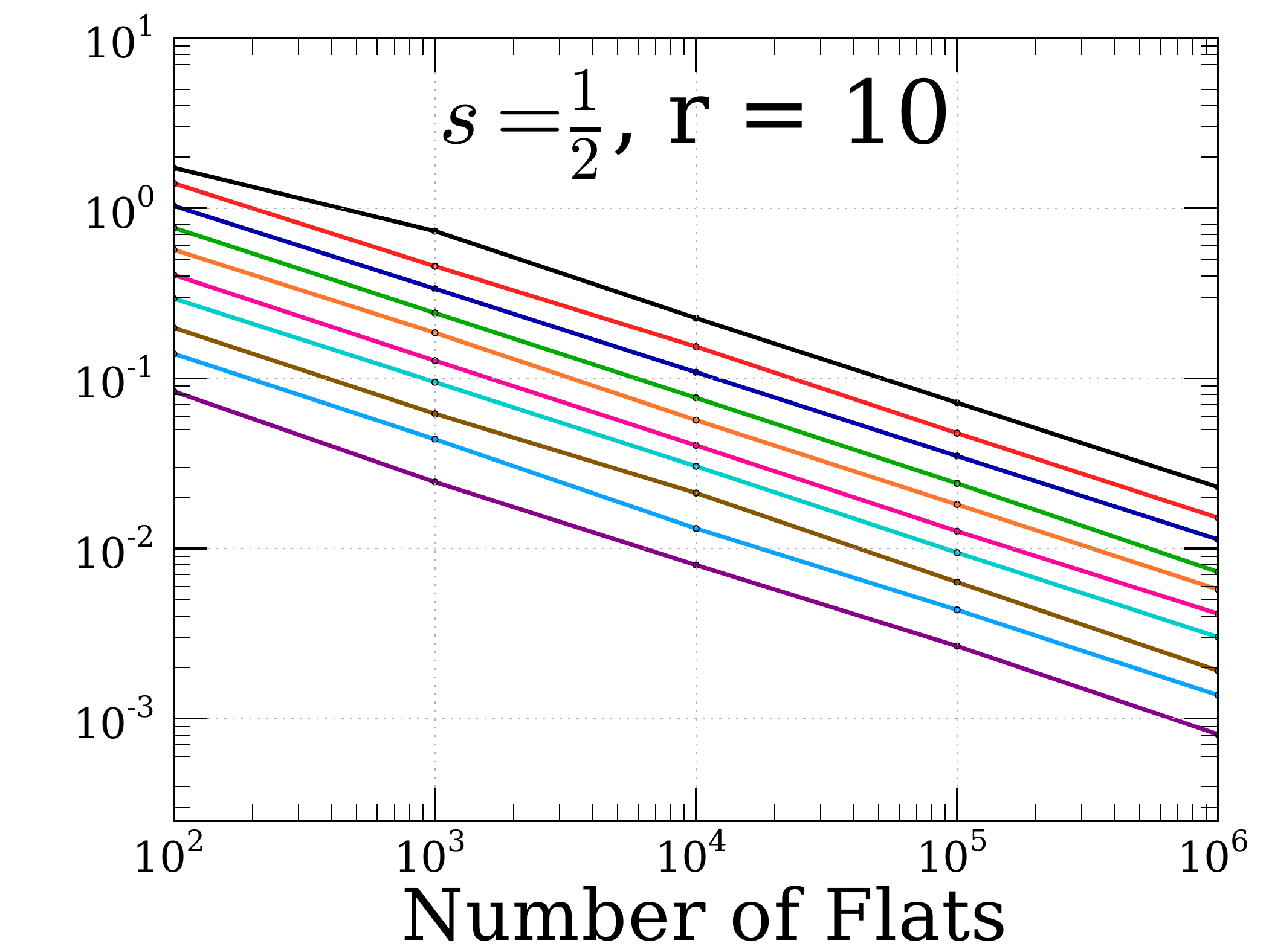}
    \includegraphics[height=\myspacing]{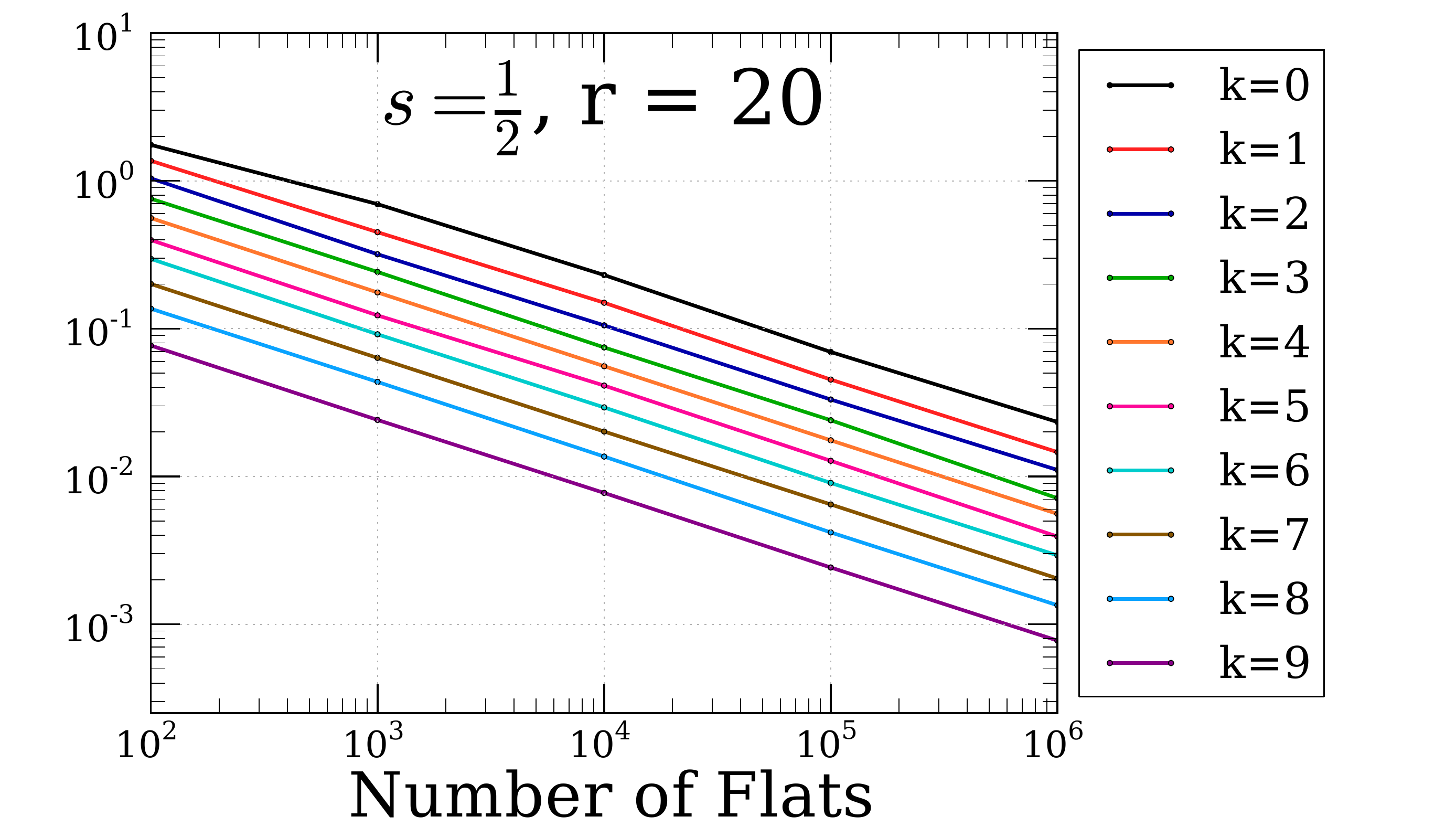}%\\
  \end{center}  
   \caption{Mean error of volume estimation, $| \textrm{mean} - \textrm{true} | / \textrm{true}$ by $n$. See also \tabref{tab:dart_experiments} and \figref{fig:histogram_estimates}.}
   \label{fig:mean_error}
\end{figure*}

\begin{figure*}[!ht]
  \begin{center} 
\fontsize{8}{8}\selectfont%fontsize, spacing after each line
$d=2$, varying squish\\
    \includegraphics[height=\myspacing]{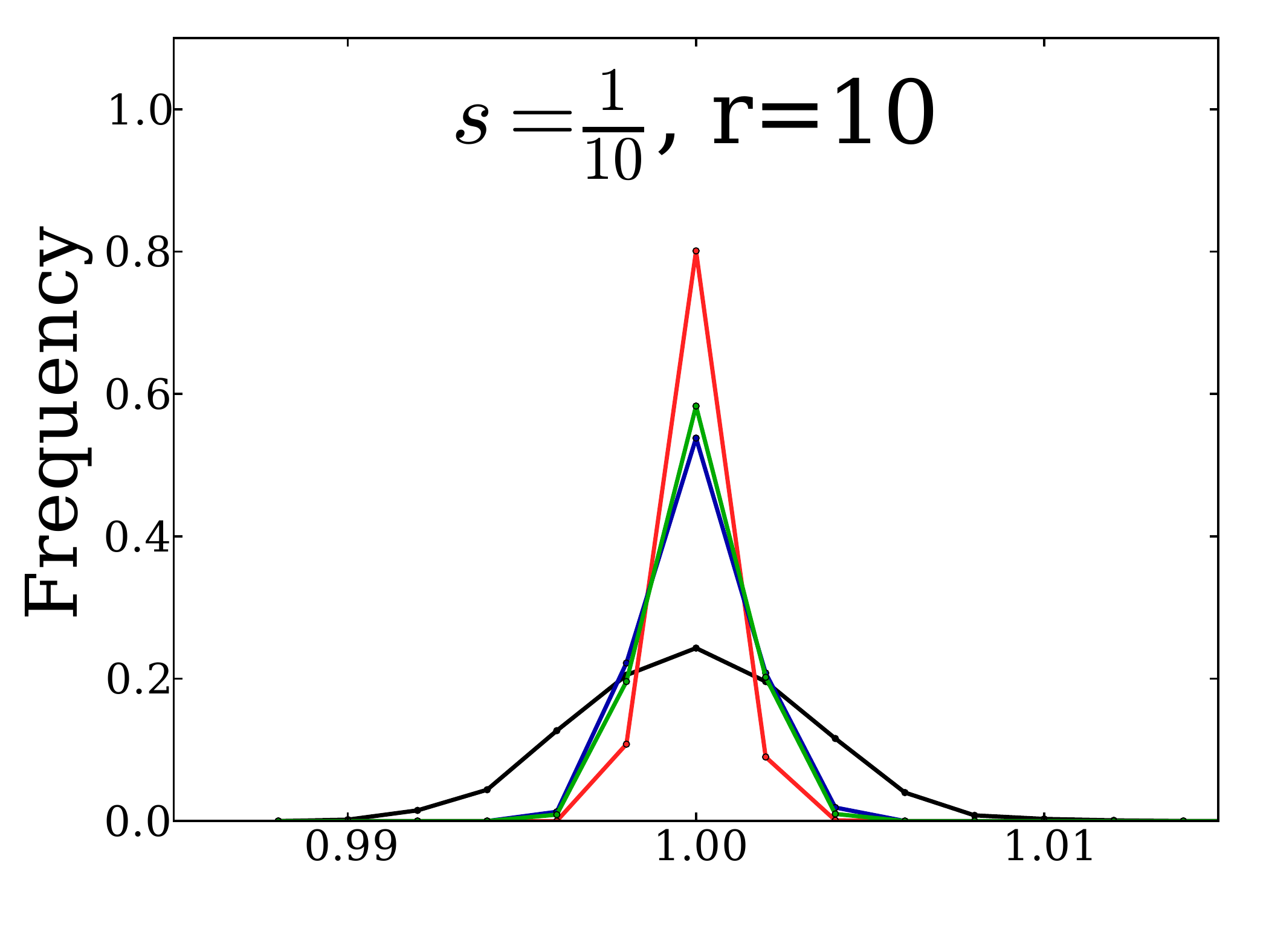}
    \includegraphics[height=\myspacing]{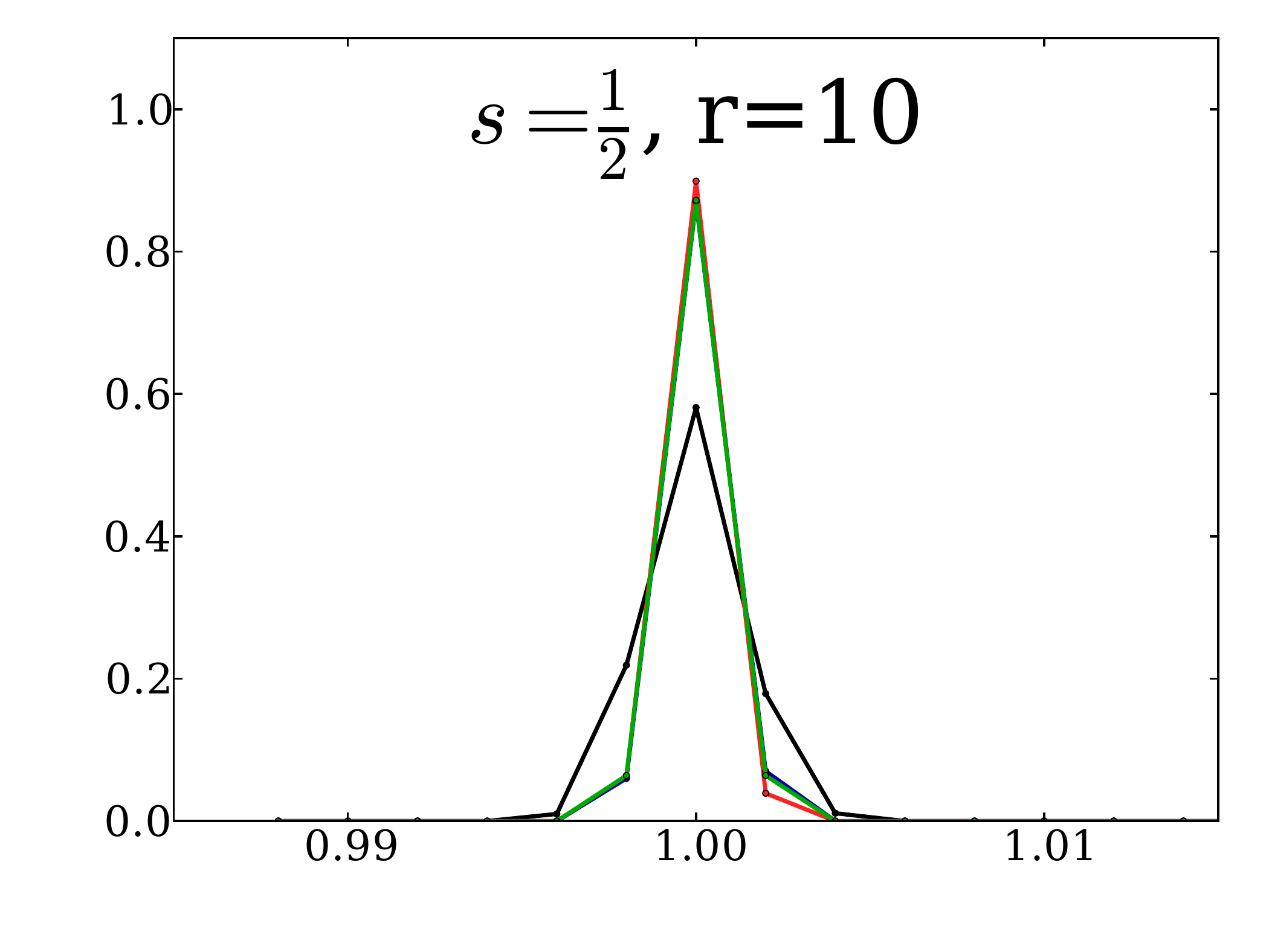}
    \includegraphics[height=\myspacing]{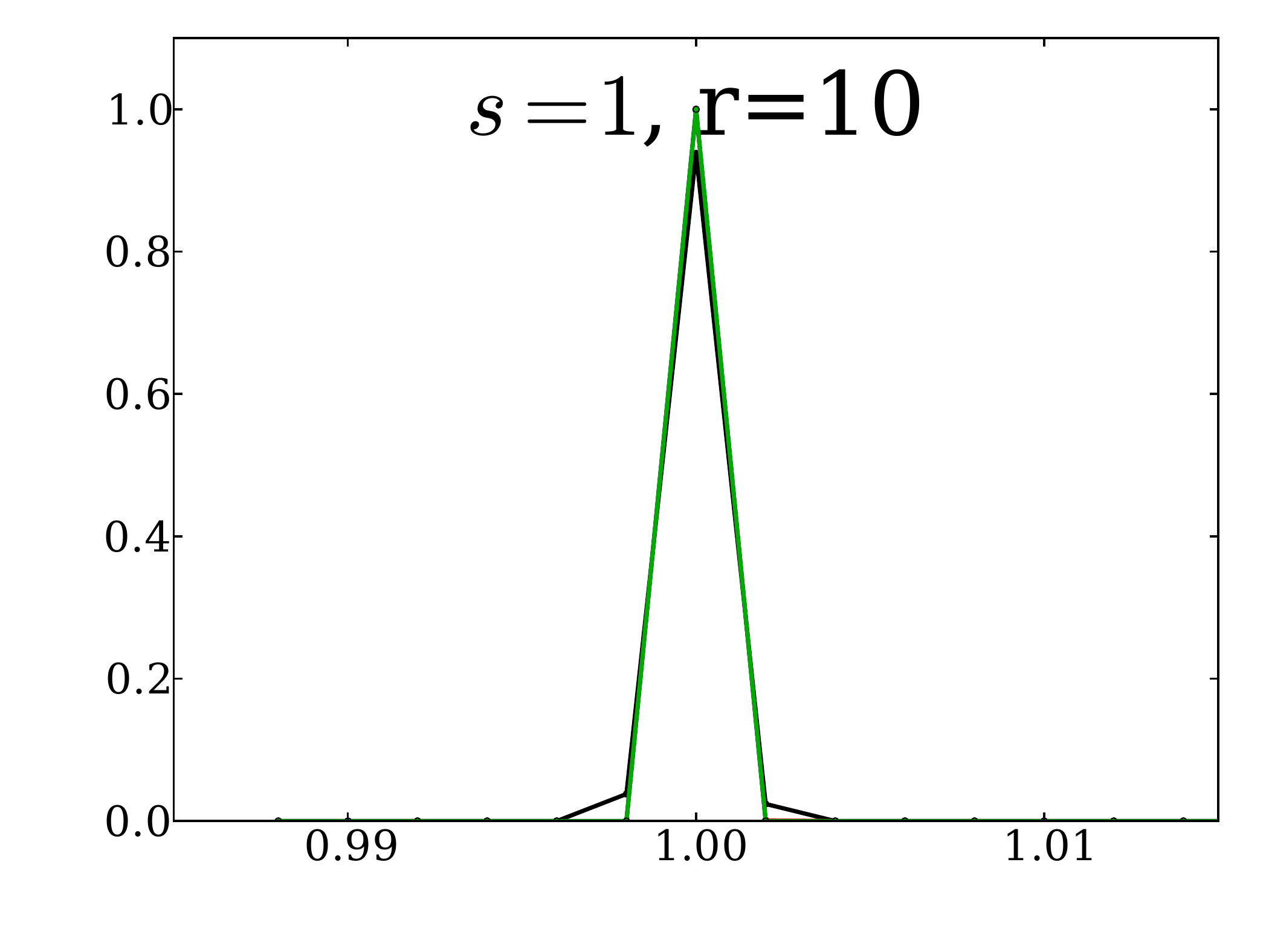}
    \includegraphics[height=\myspacing]{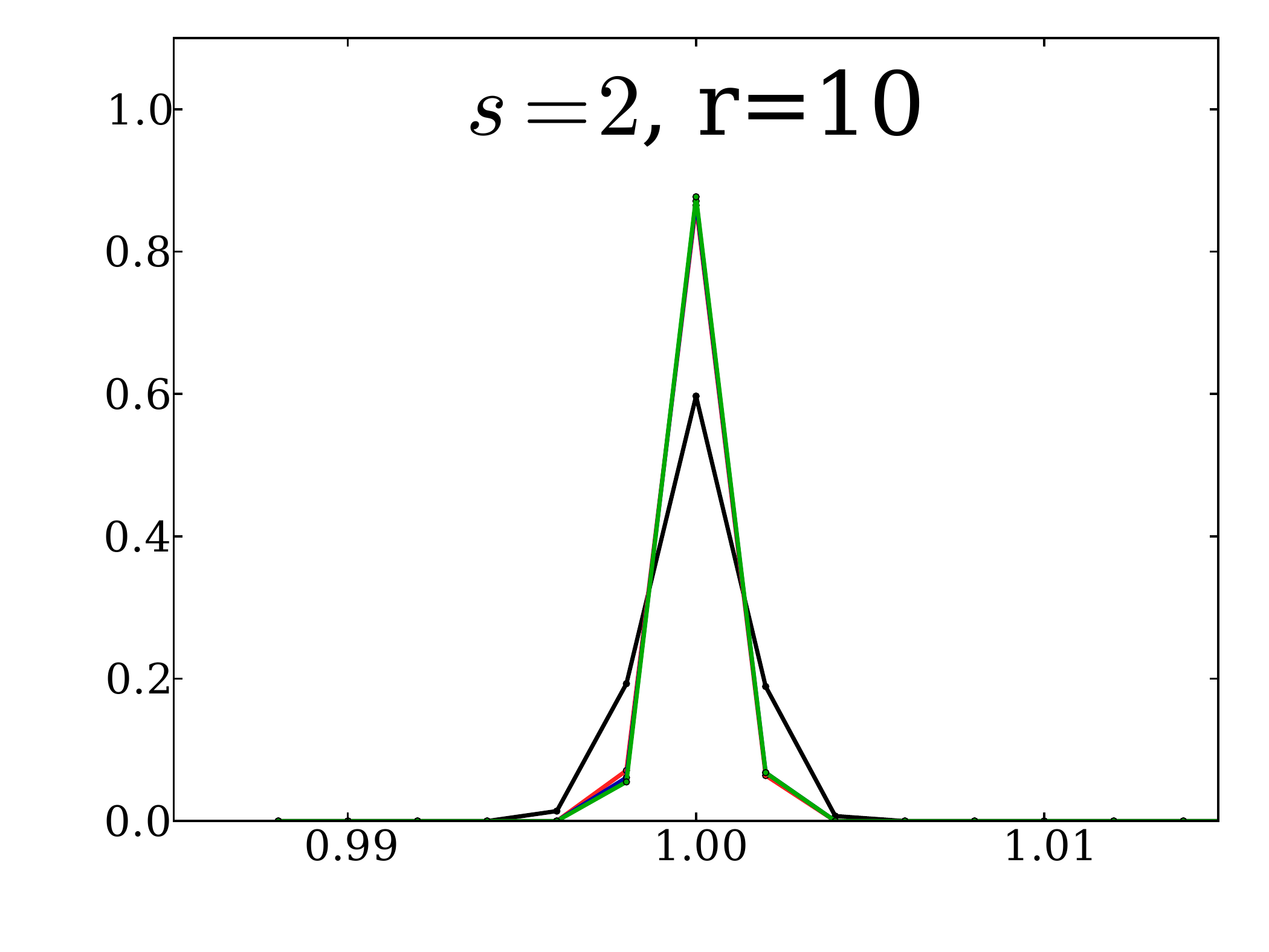}
    \includegraphics[height=\myspacing]{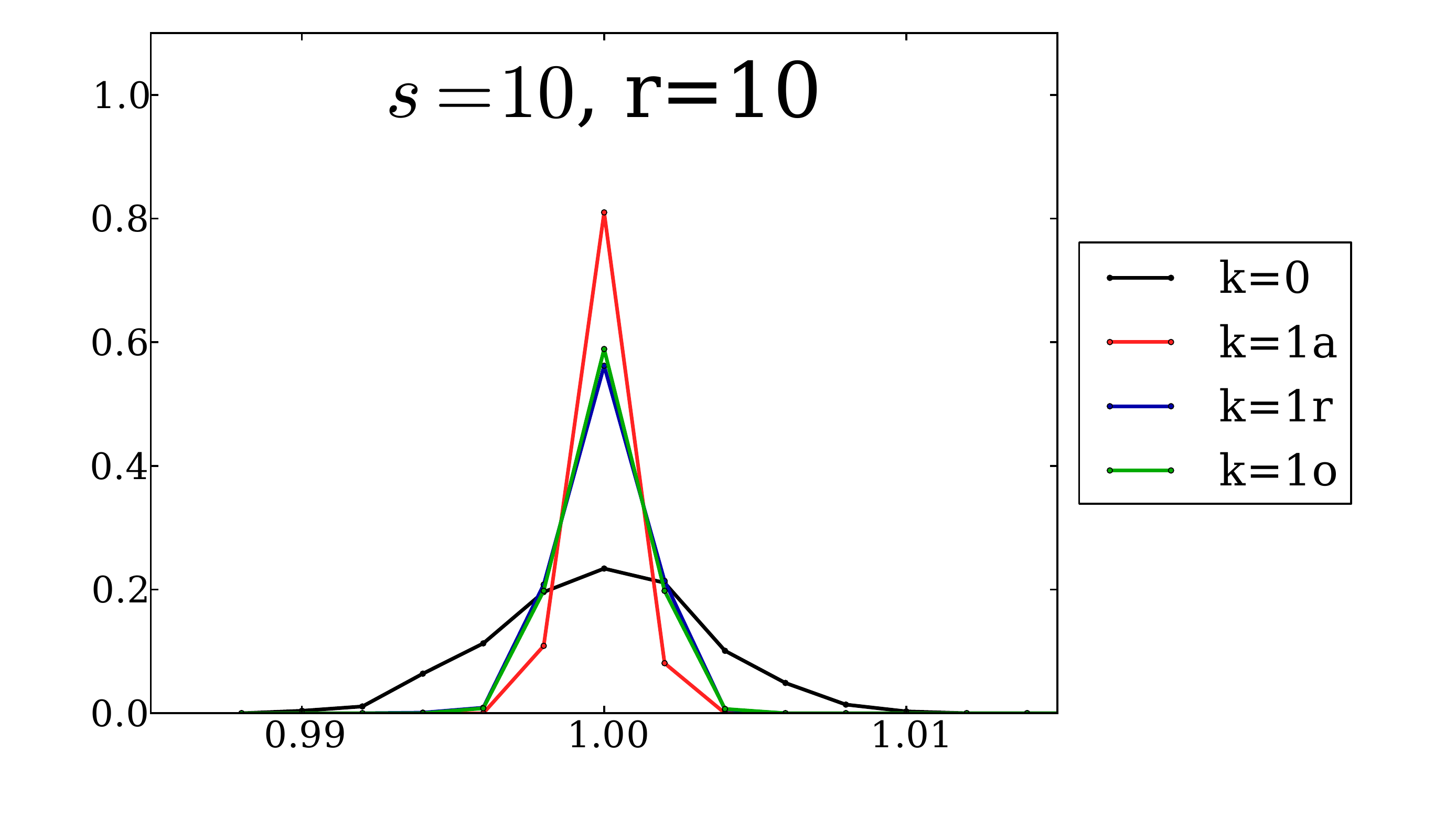}\\
 $d=2$, varying rotation\\    
    \includegraphics[height=\myspacing]{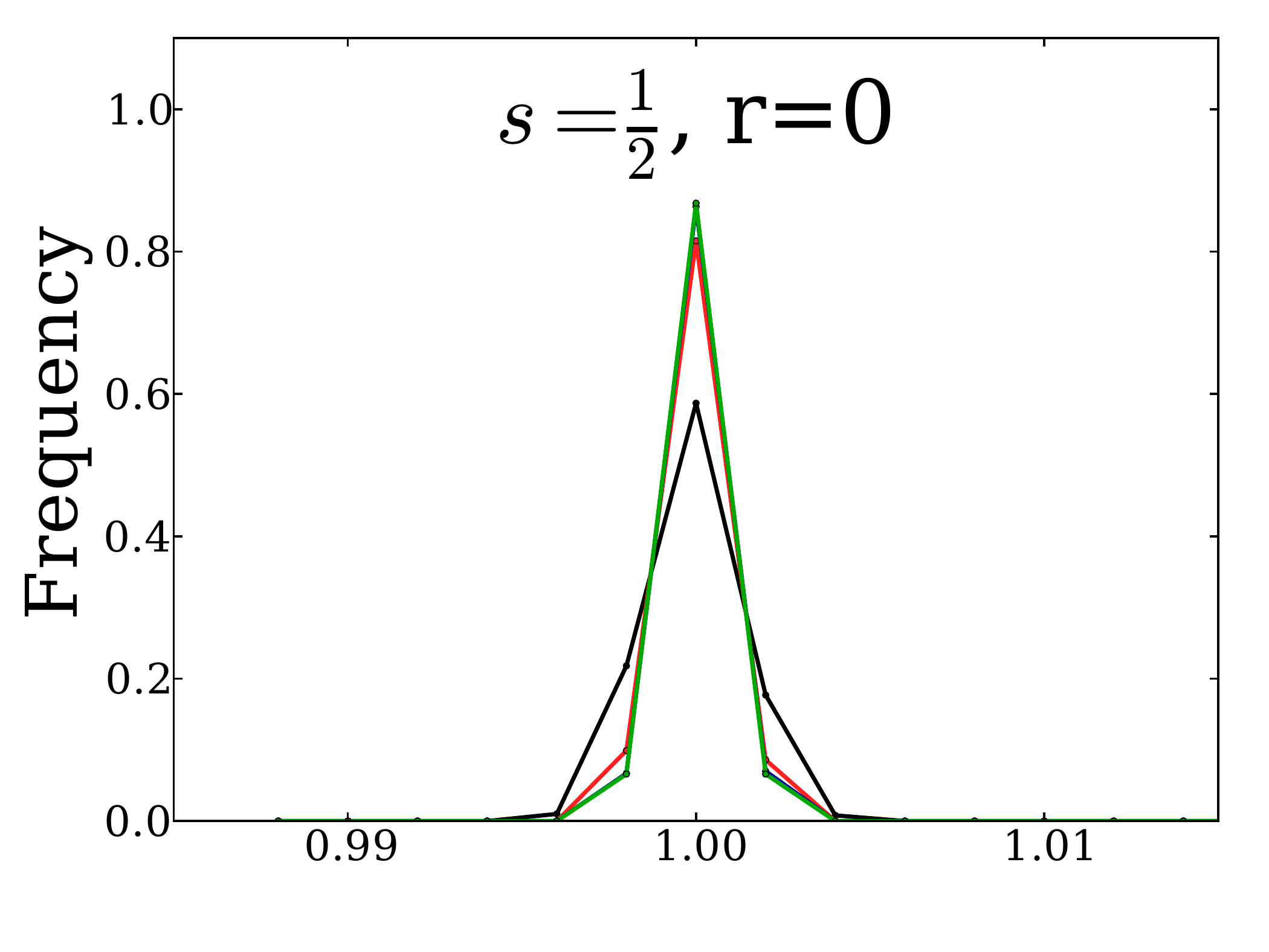}
    \includegraphics[height=\myspacing]{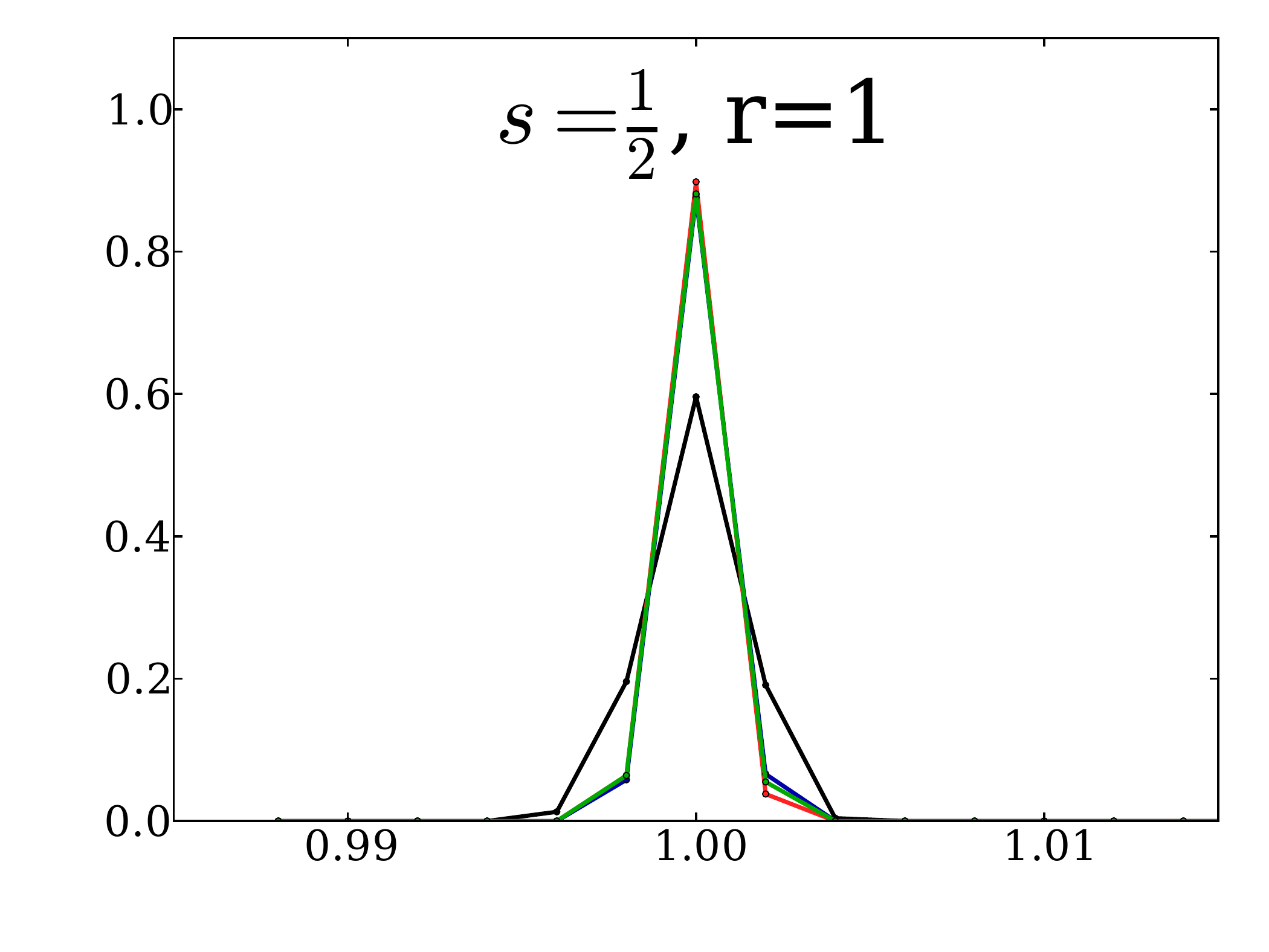}
    \includegraphics[height=\myspacing]{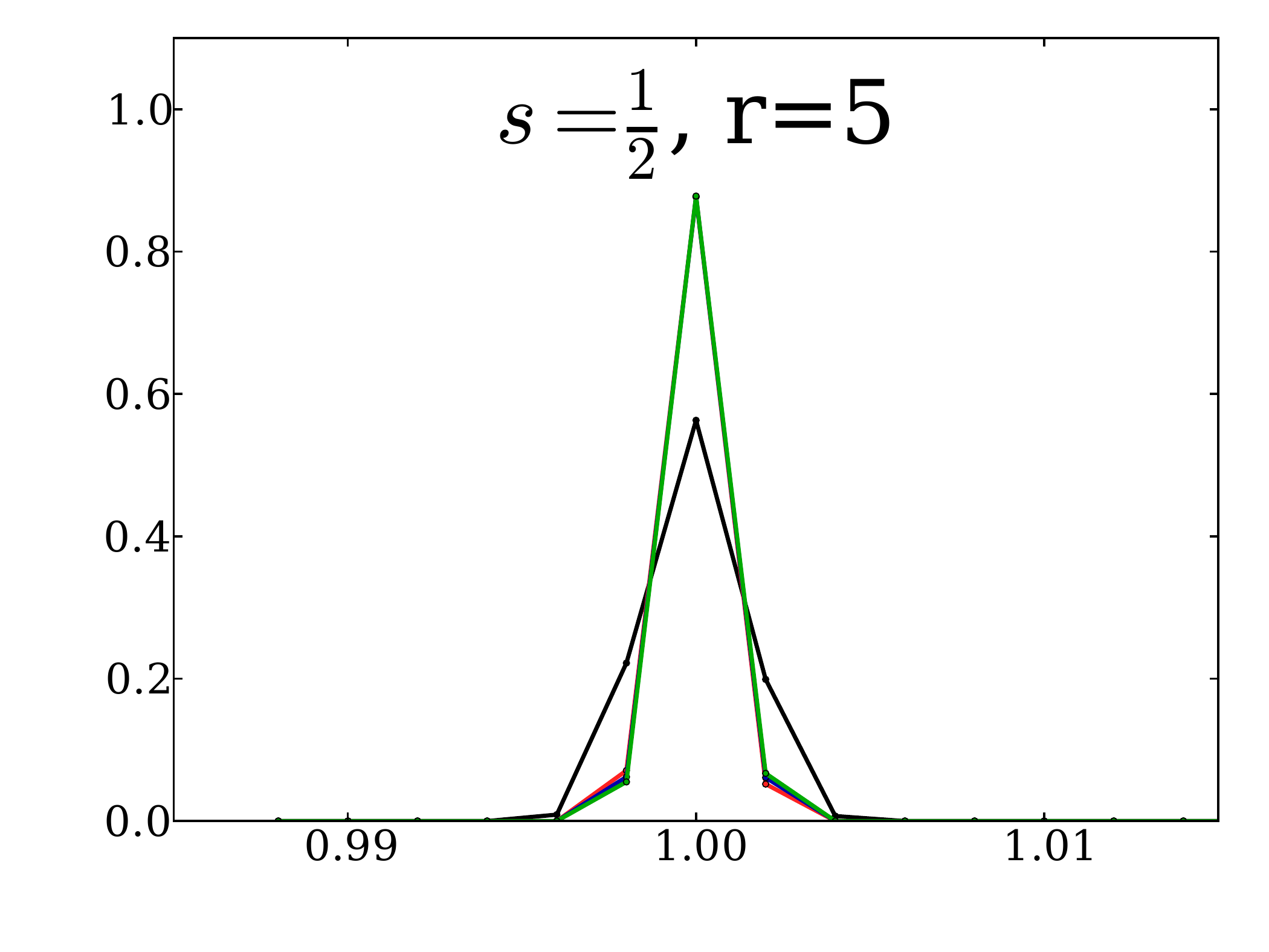}
    \includegraphics[height=\myspacing]{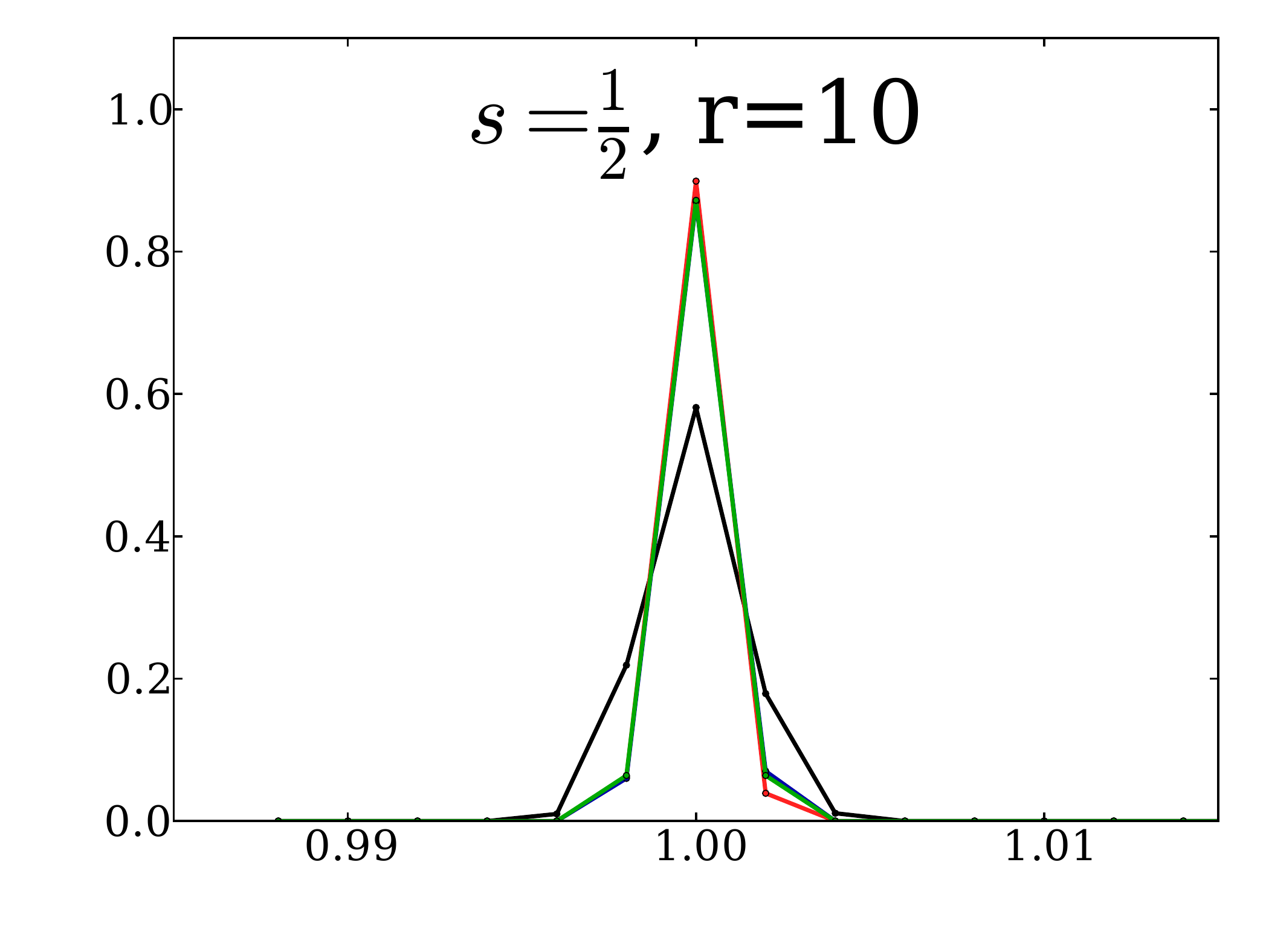}
    \includegraphics[height=\myspacing]{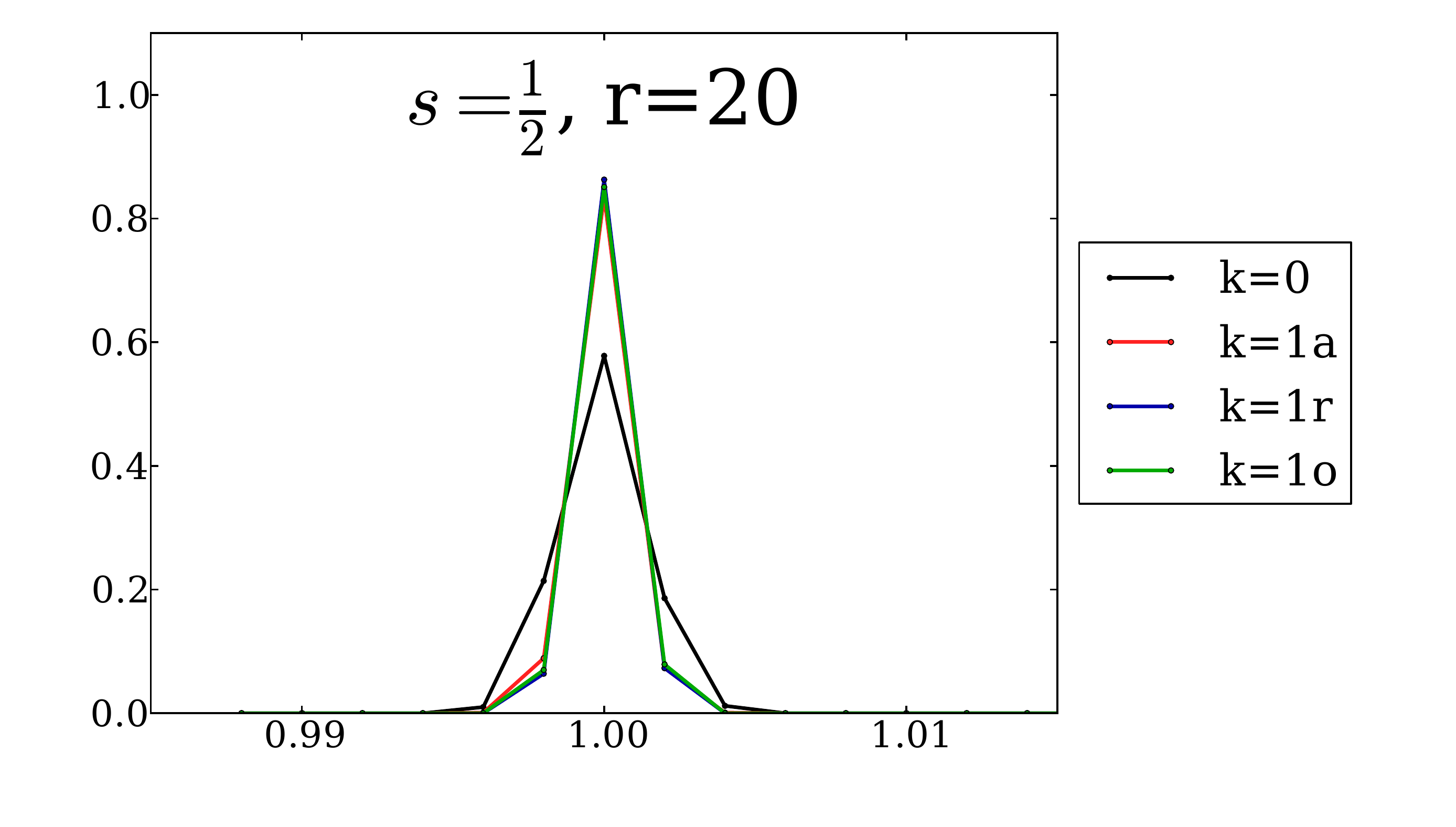}\\
$d=3$, varying squish\\
    \includegraphics[height=\myspacing]{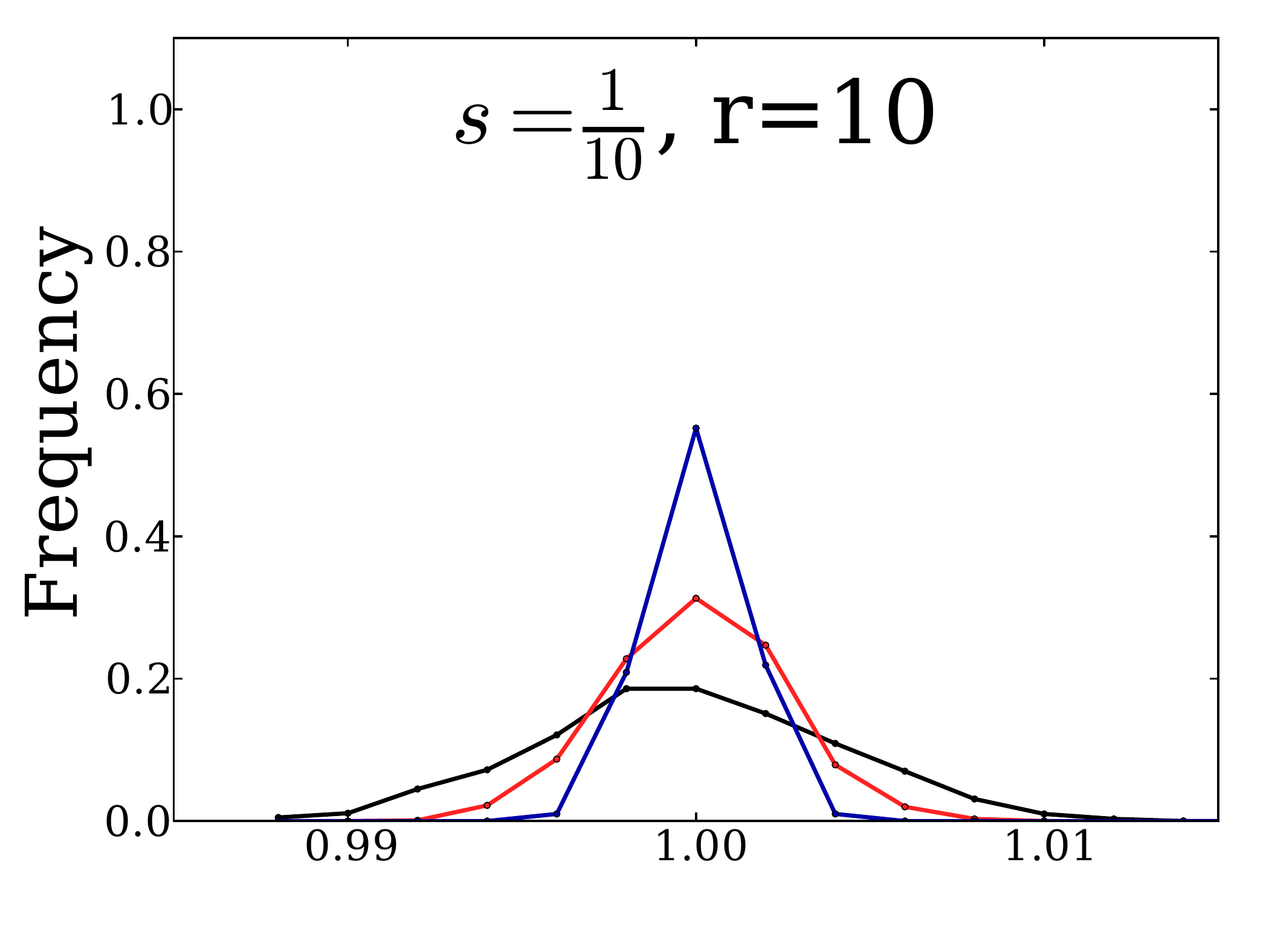}
    \includegraphics[height=\myspacing]{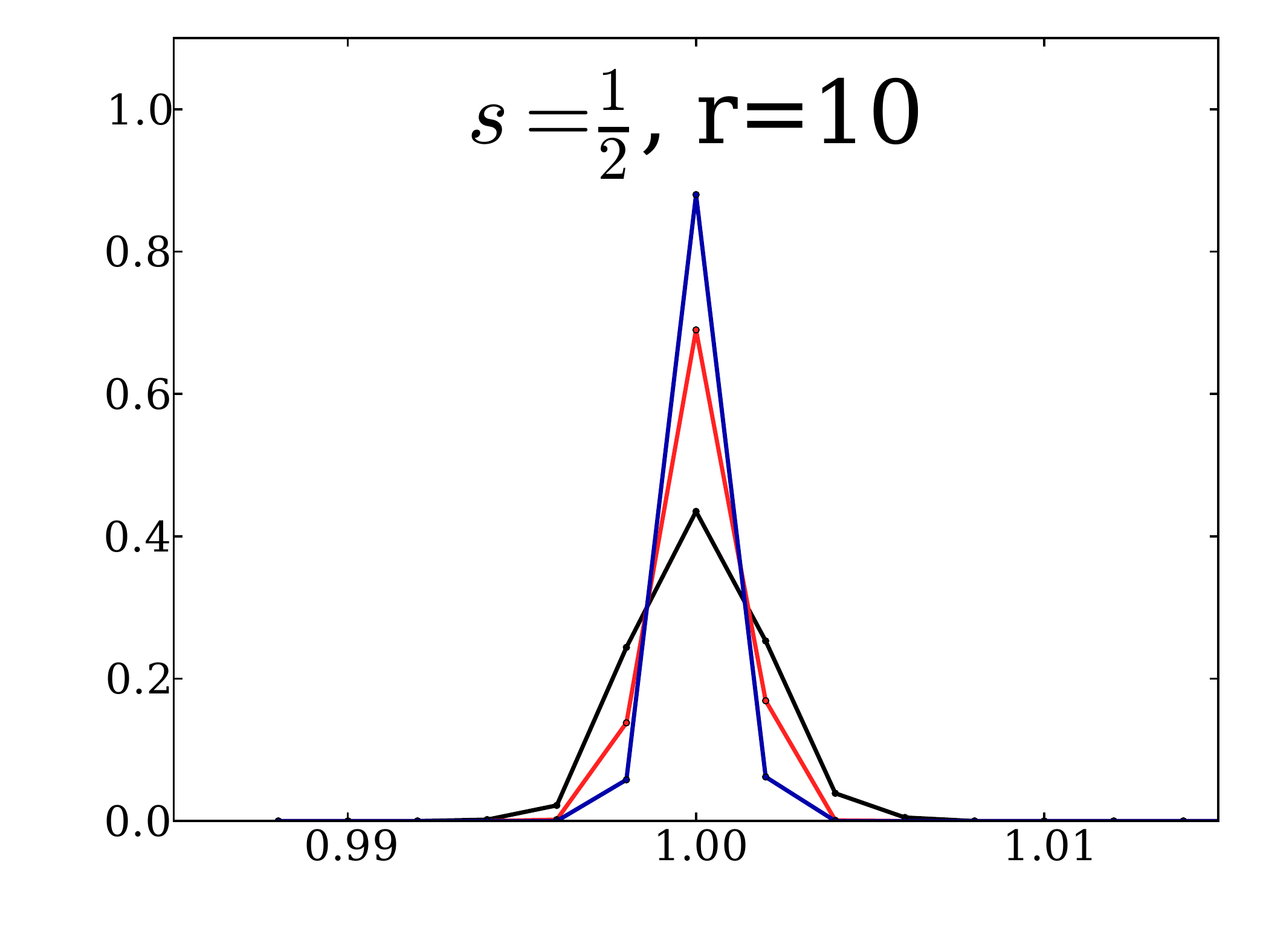}
    \includegraphics[height=\myspacing]{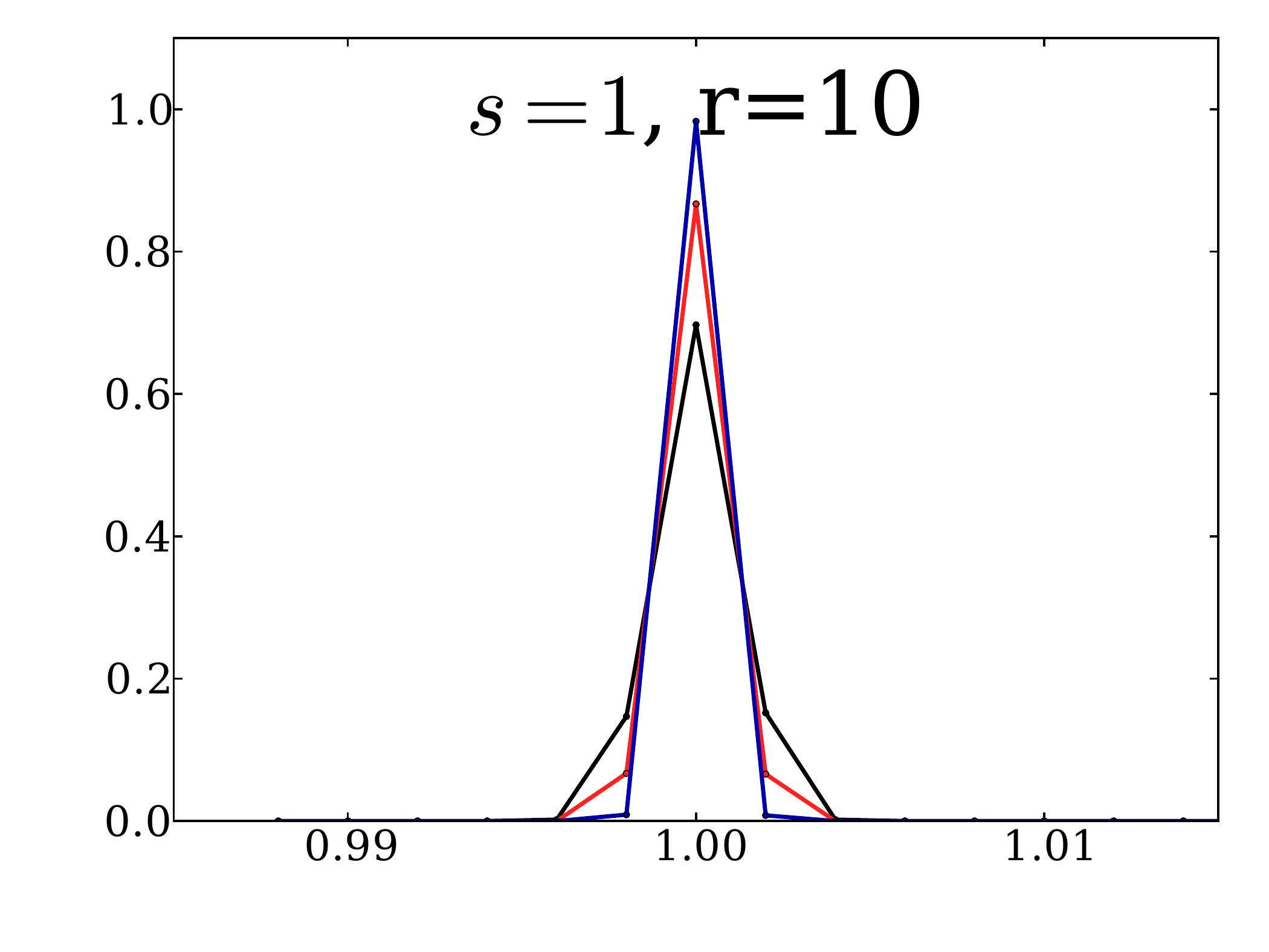}
    \includegraphics[height=\myspacing]{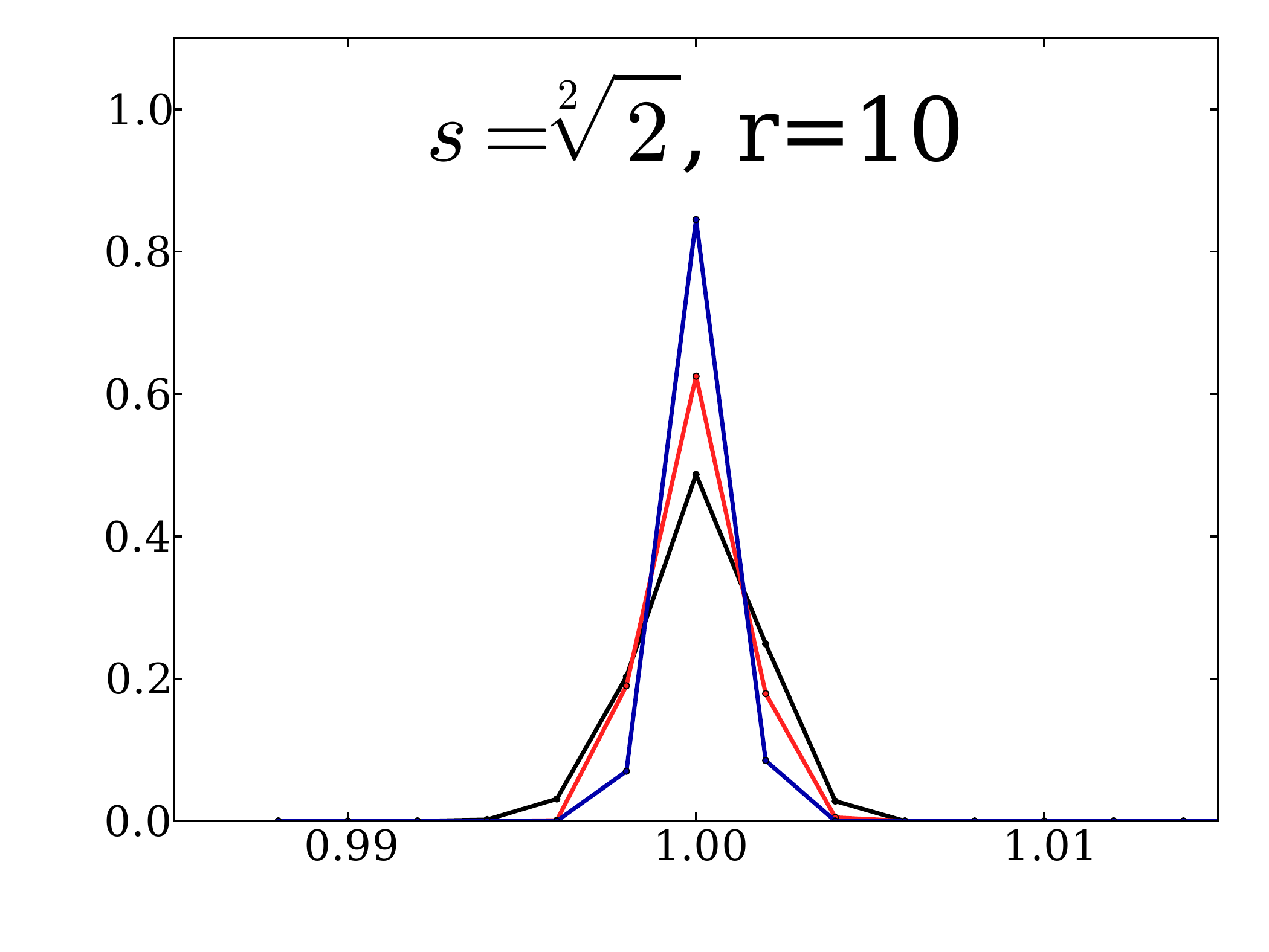}
    \includegraphics[height=\myspacing]{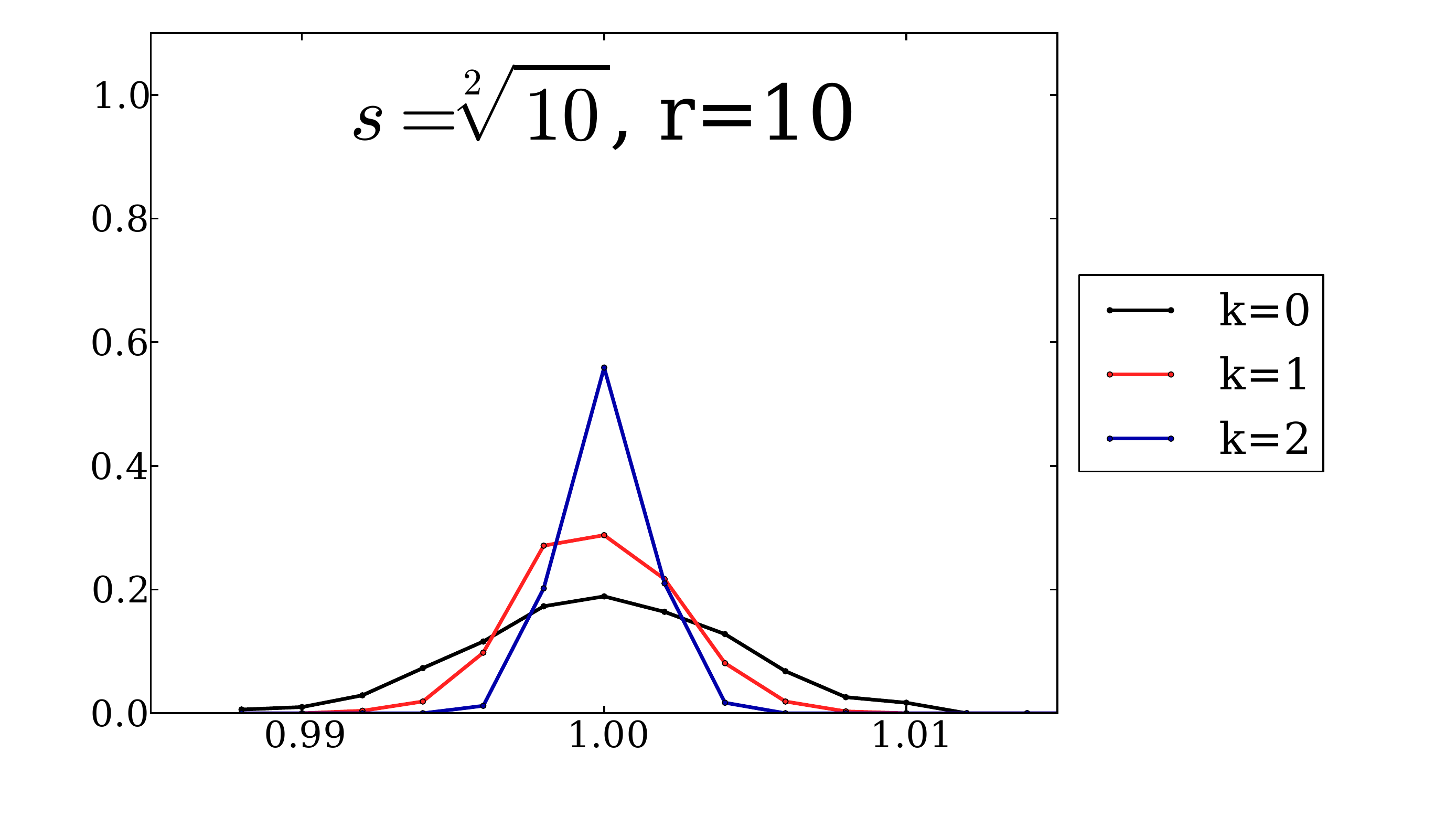}\\
$d=3$, varying rotation\\
    \includegraphics[height=\myspacing]{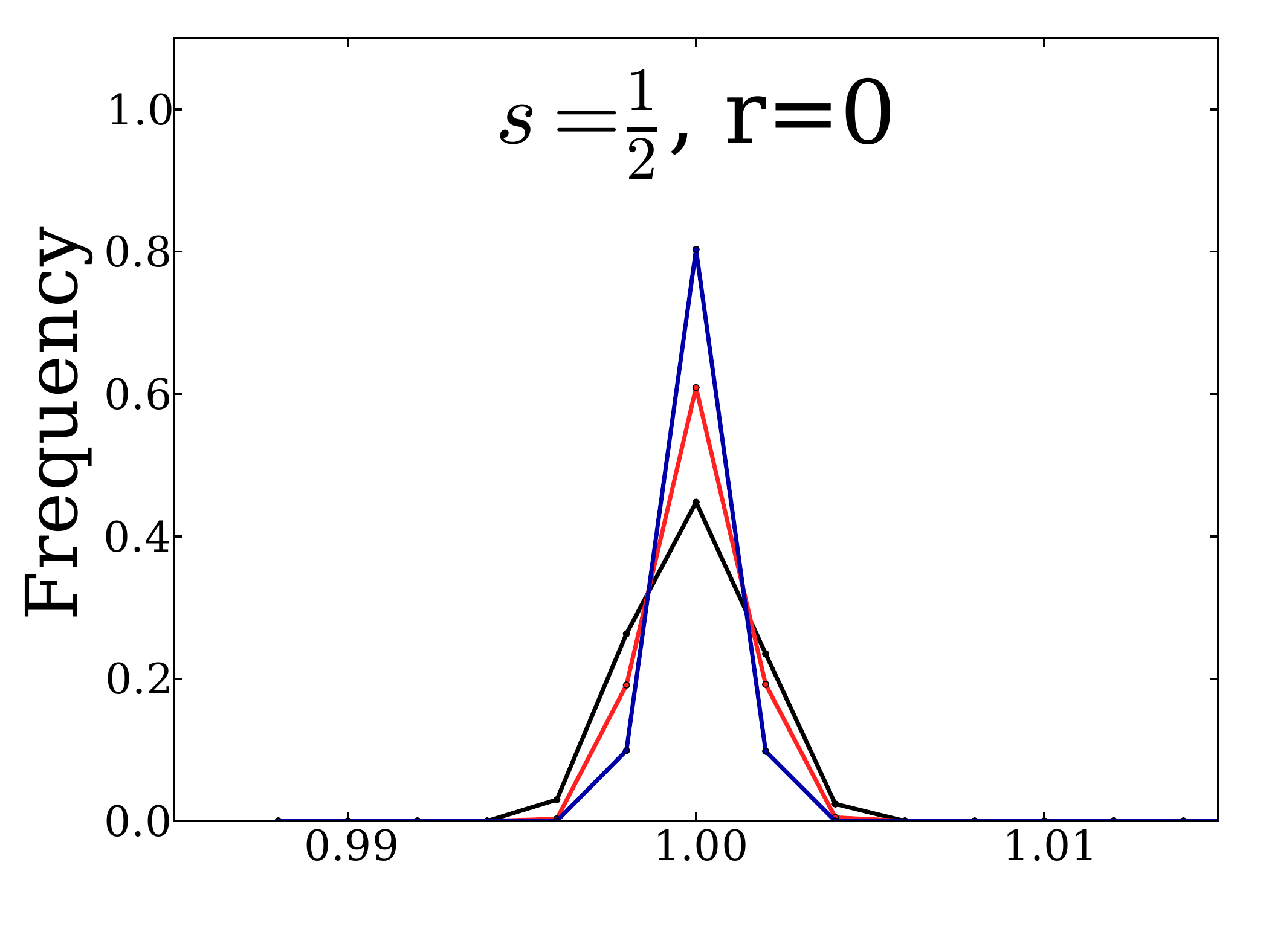}
    \includegraphics[height=\myspacing]{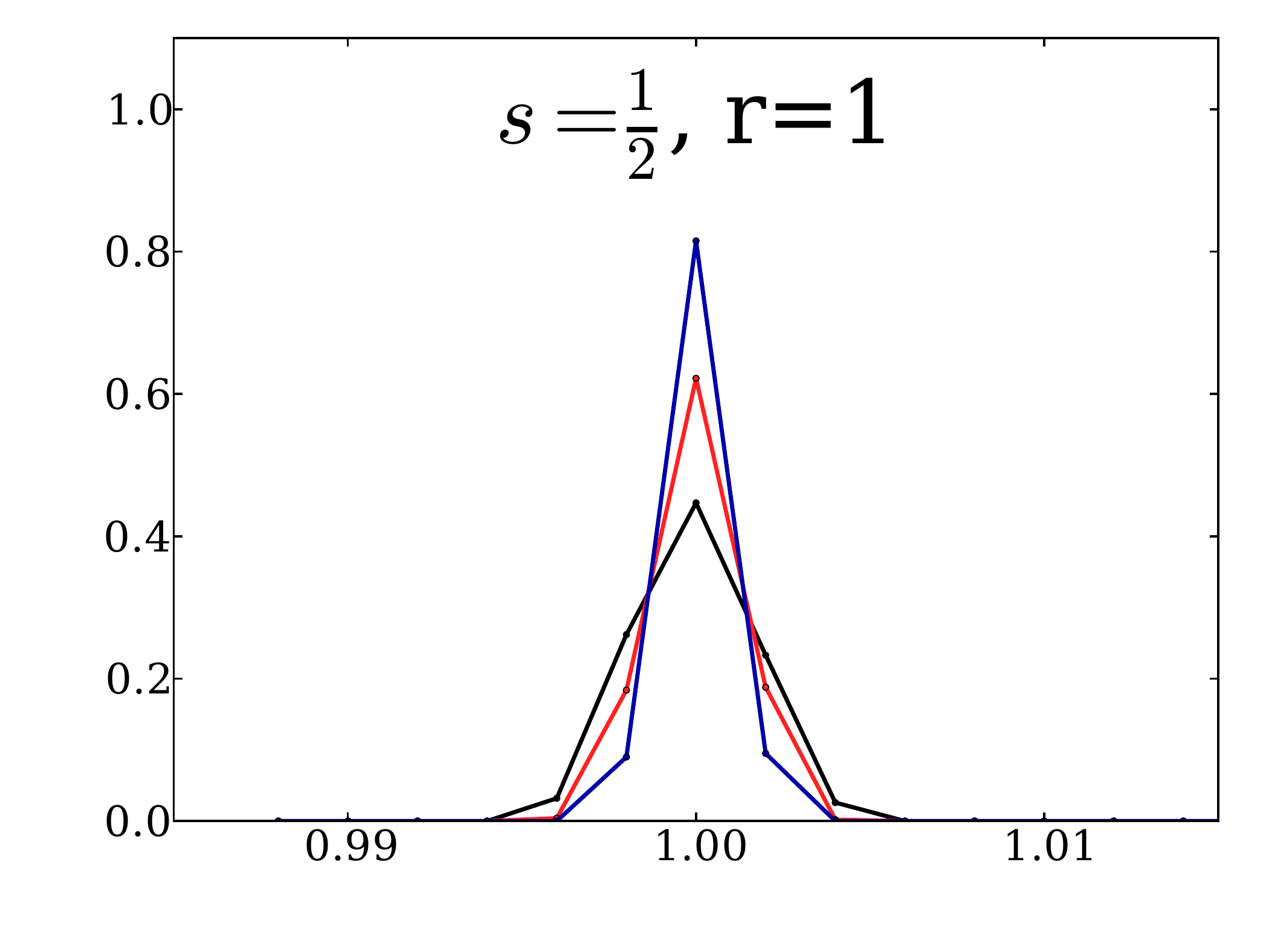}
    \includegraphics[height=\myspacing]{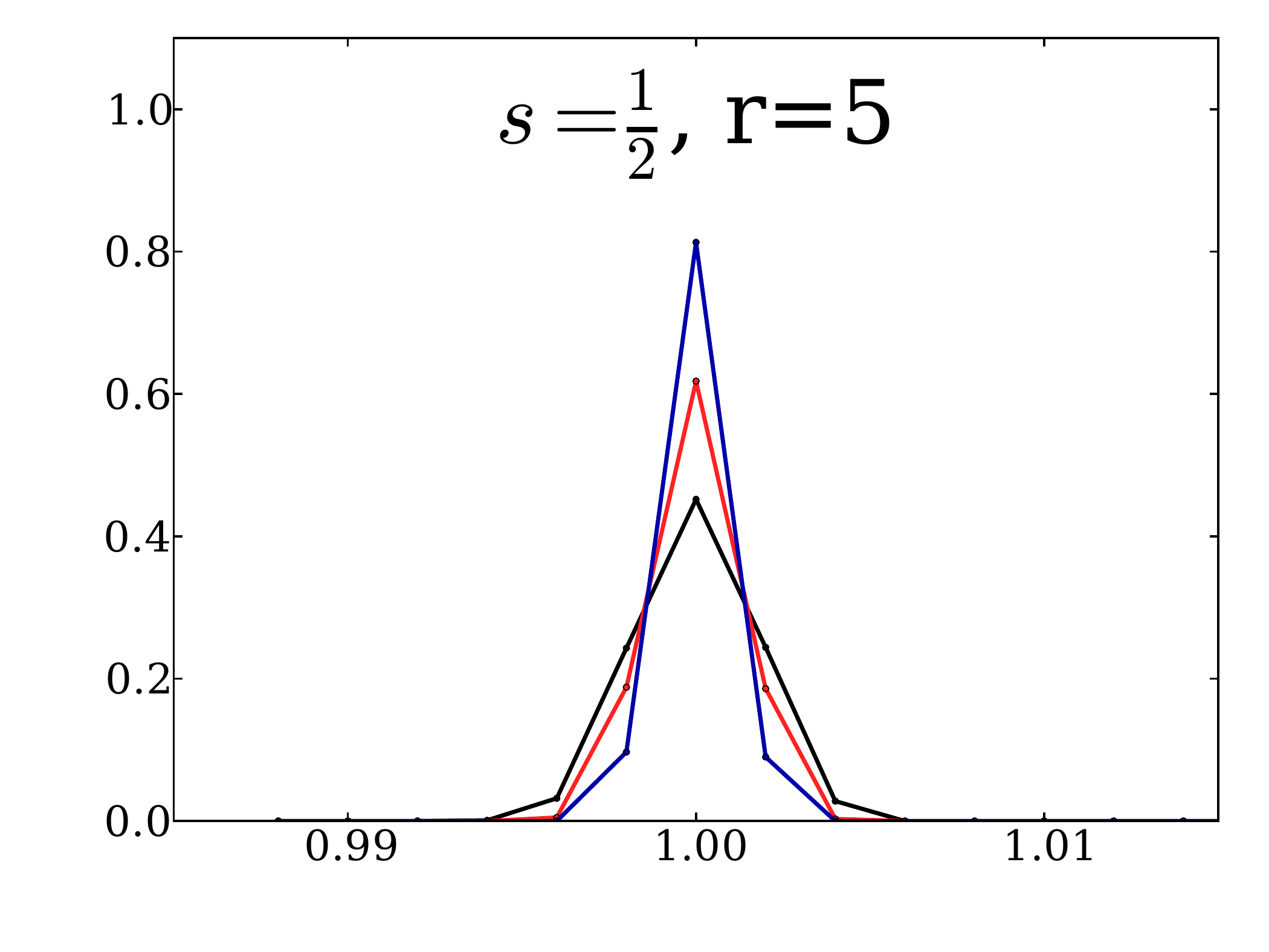}
    \includegraphics[height=\myspacing]{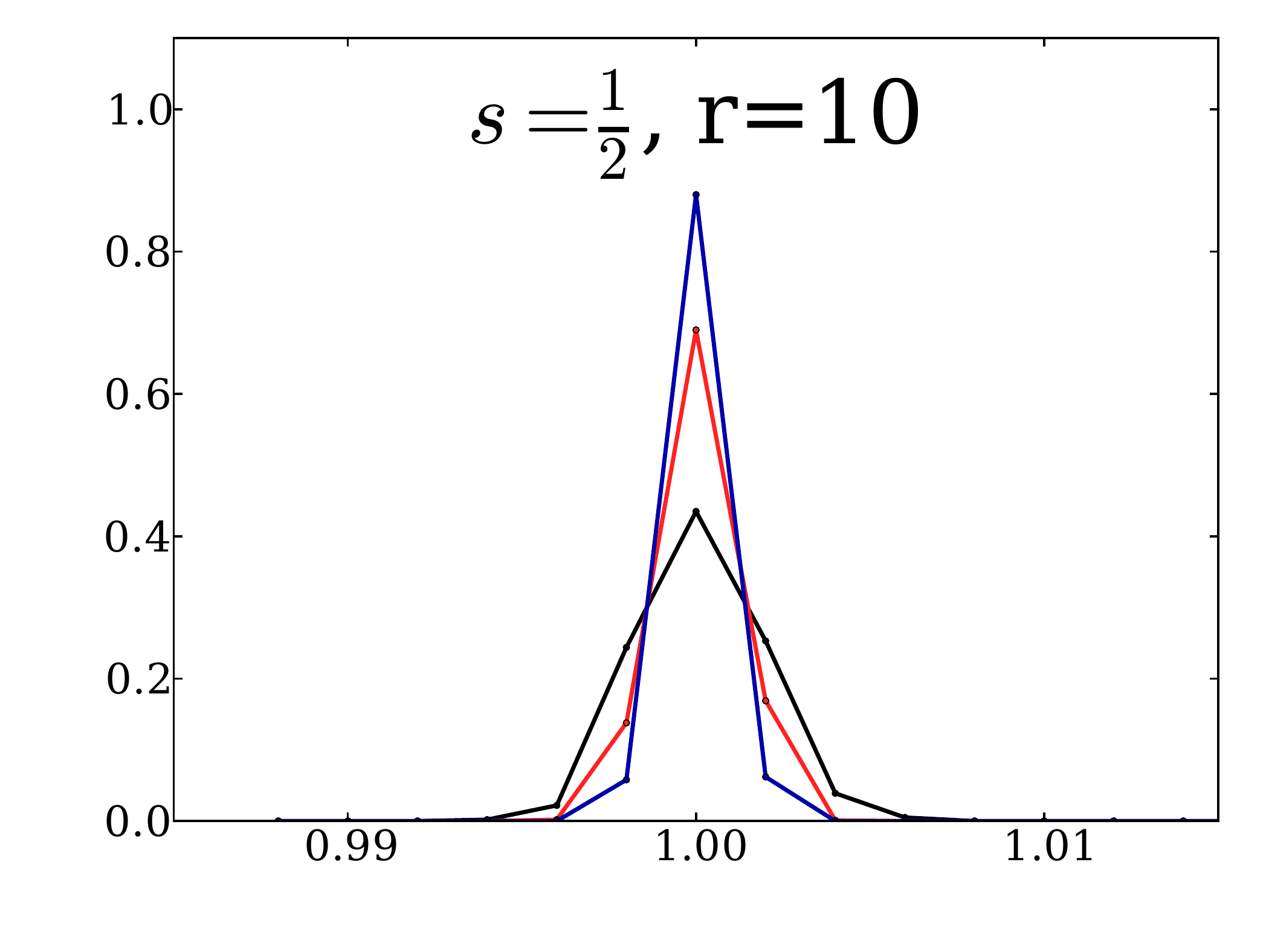}
    \includegraphics[height=\myspacing]{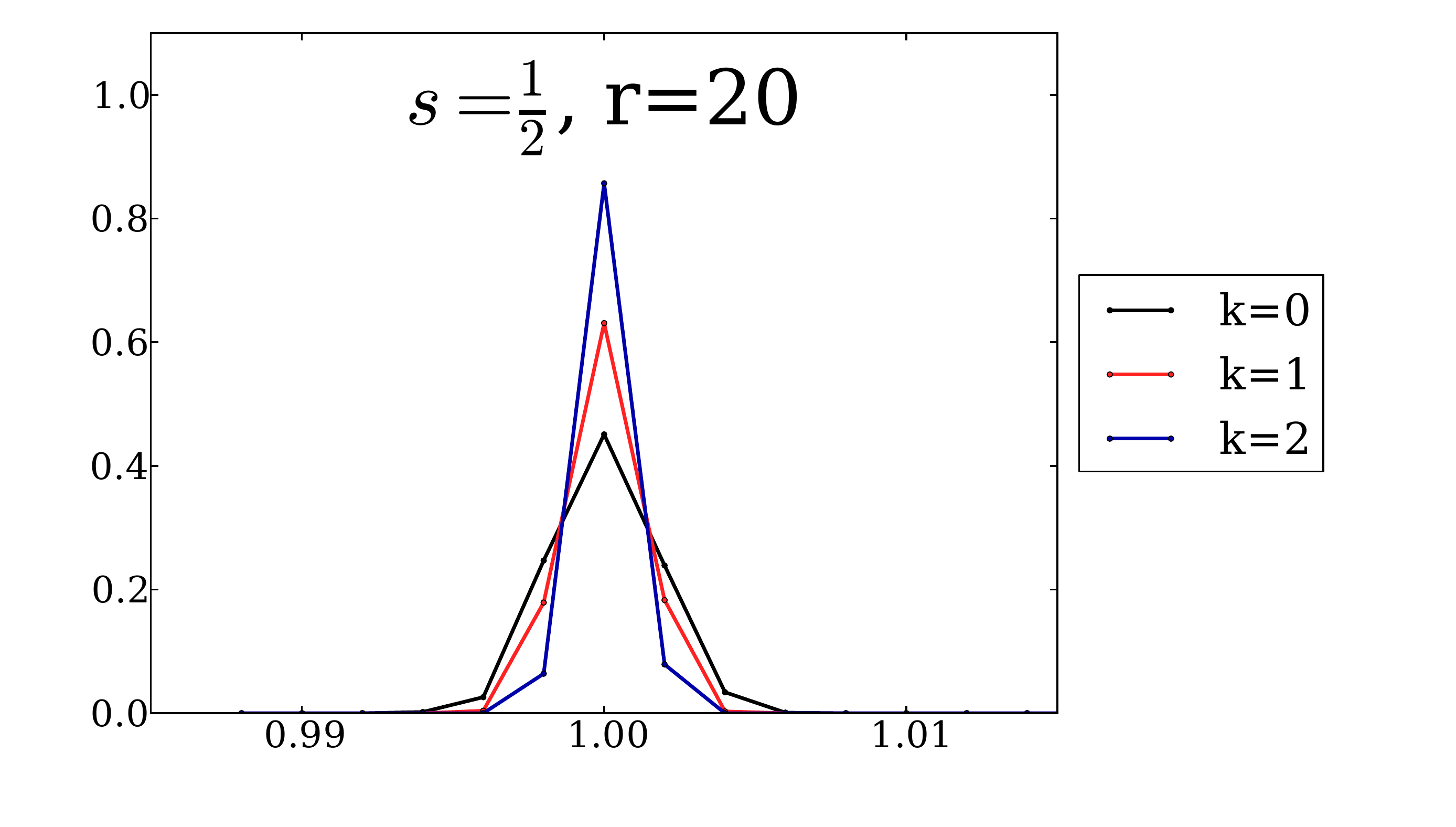}\\
$d=10$, varying squish\\
    \includegraphics[height=\myspacing]{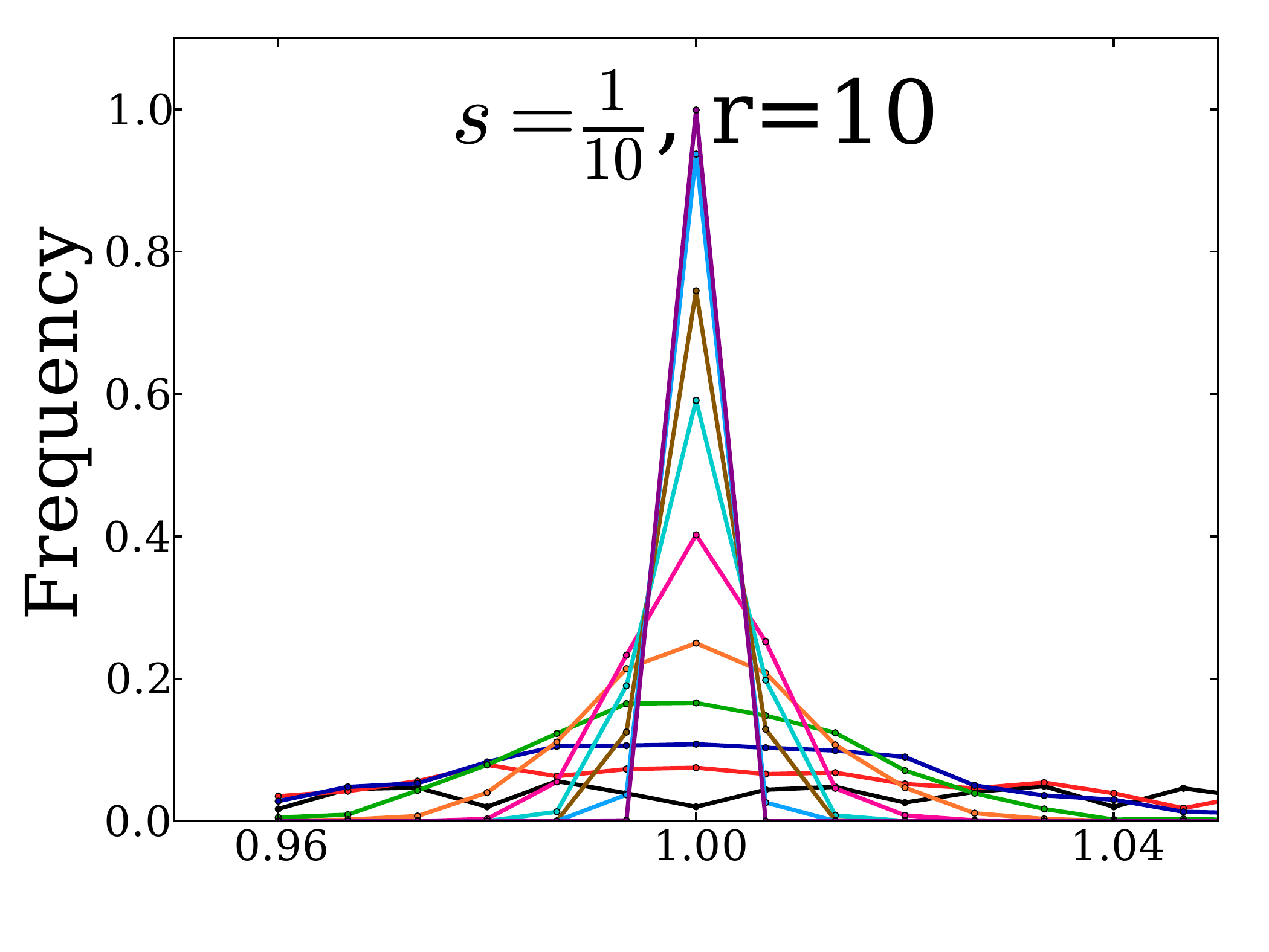}
    \includegraphics[height=\myspacing]{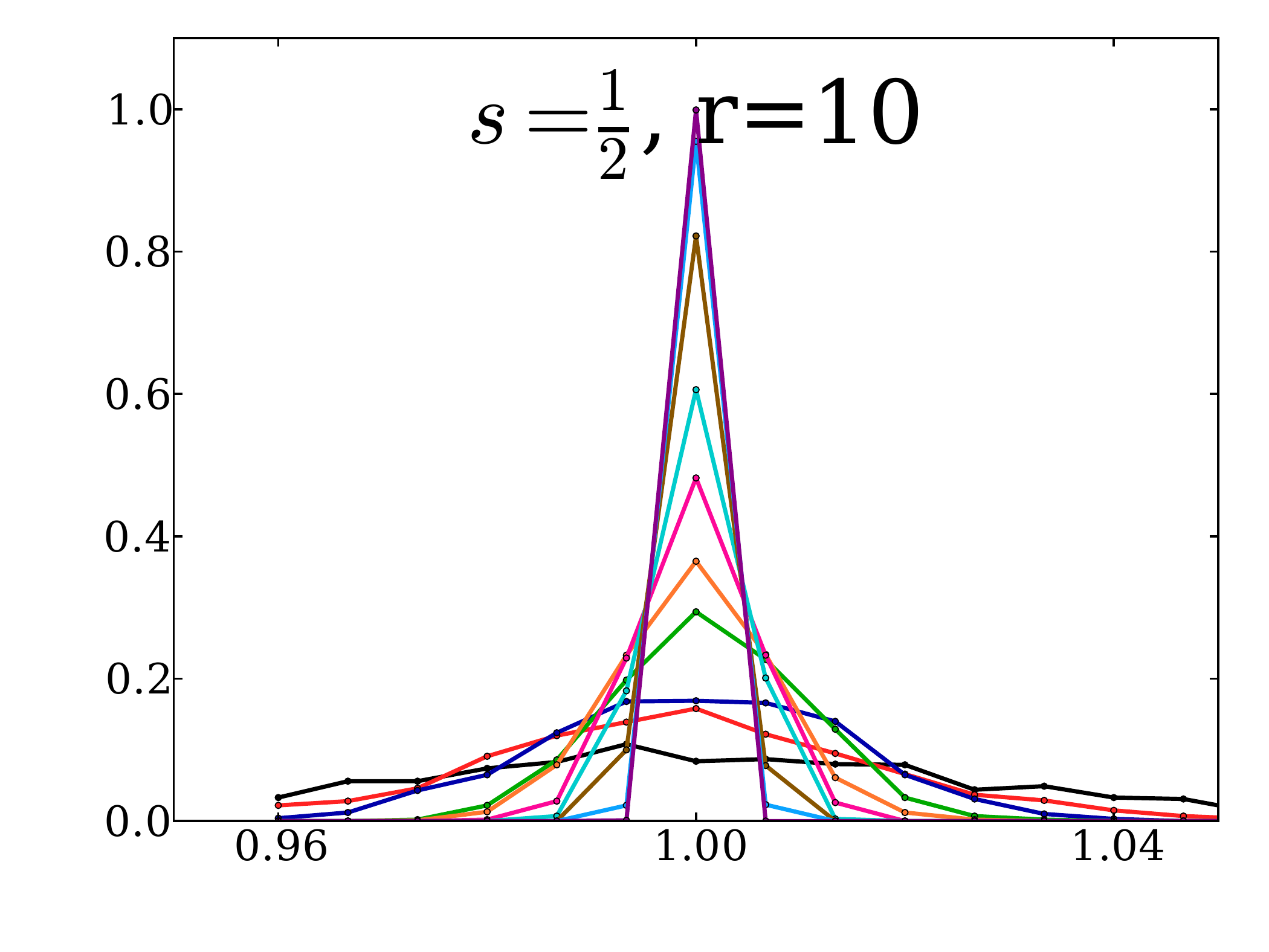}
    \includegraphics[height=\myspacing]{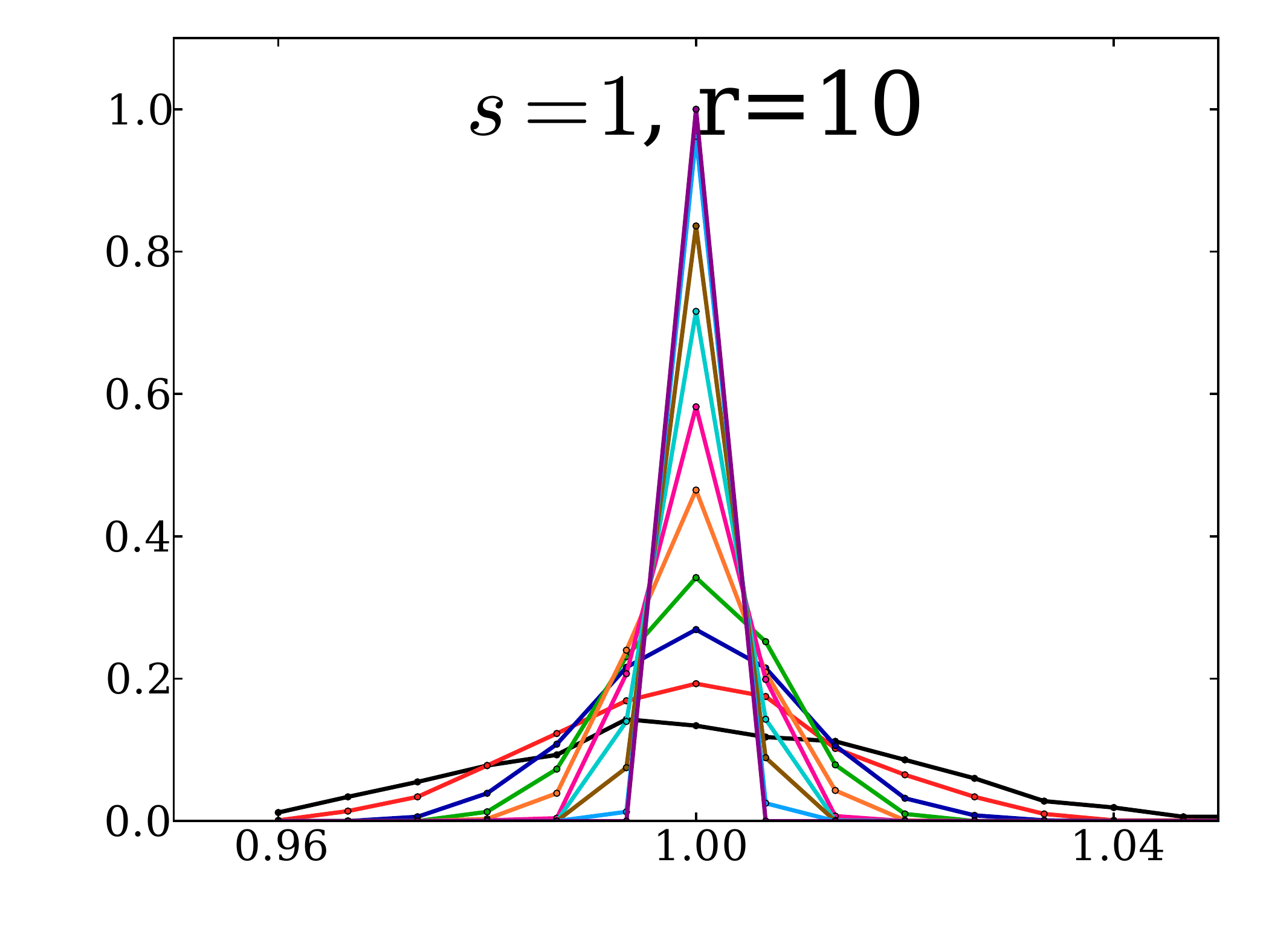}
    \includegraphics[height=\myspacing]{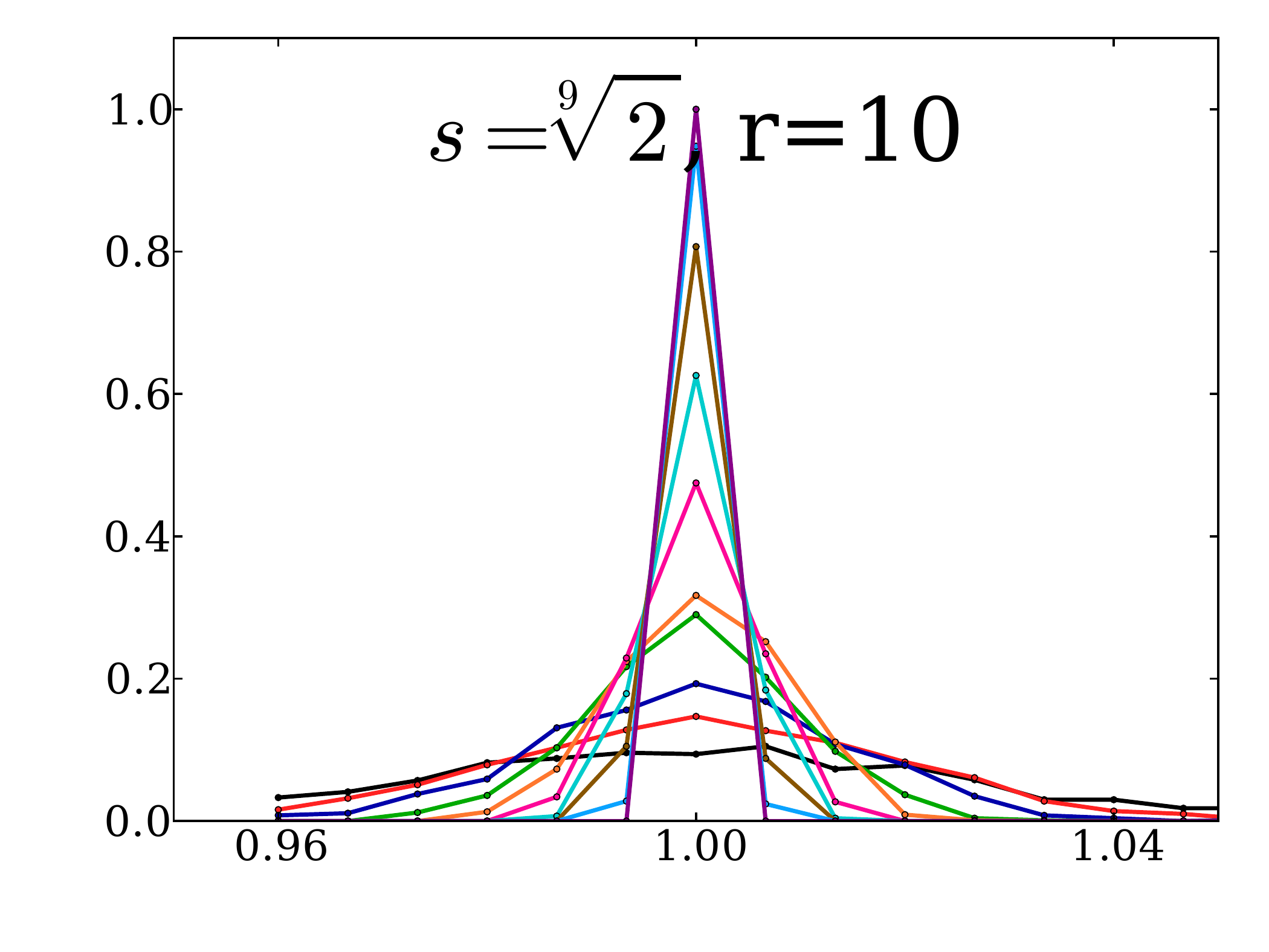}
    \includegraphics[height=\myspacing]{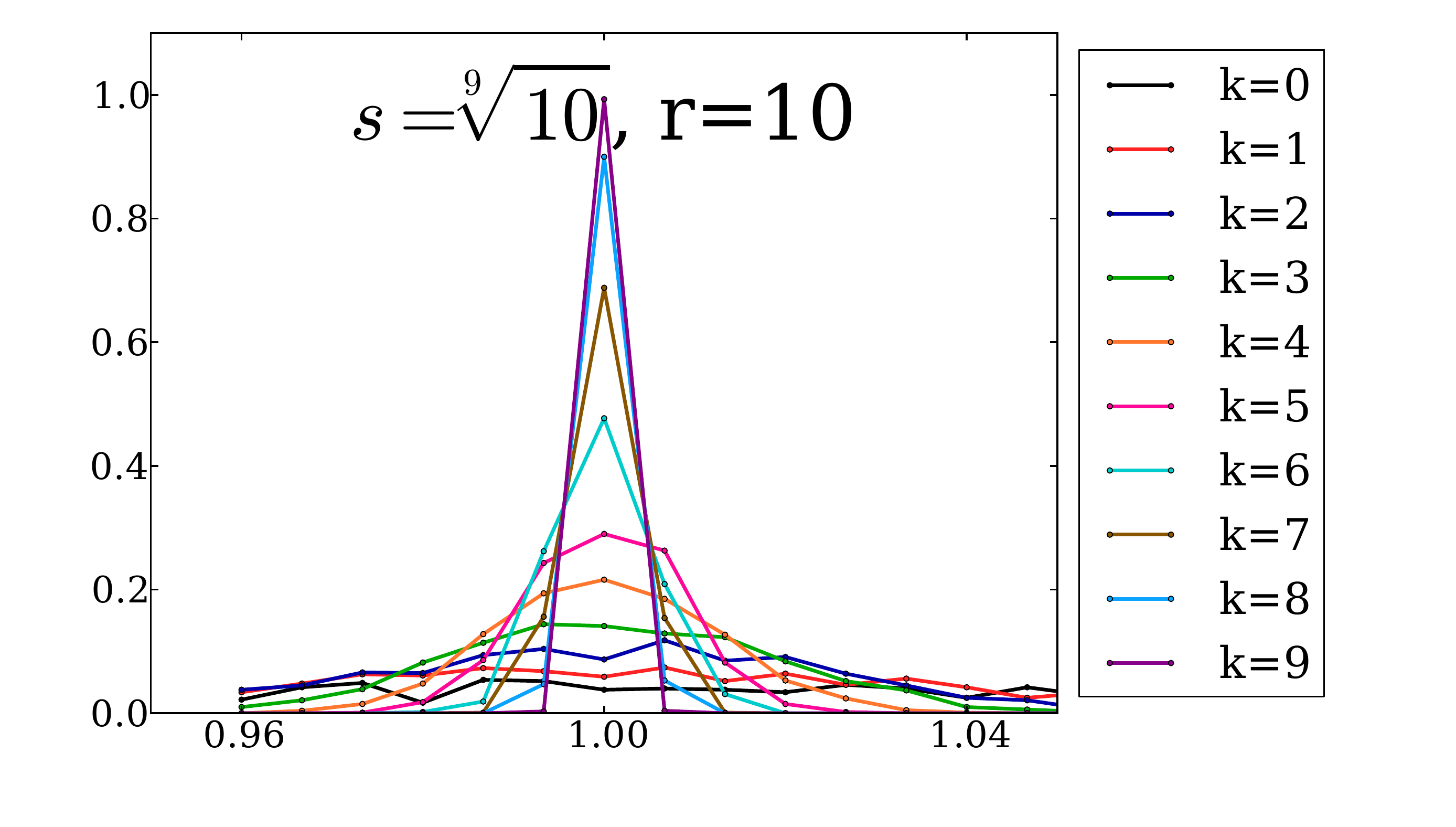} \\
$d=10$, varying rotation\\
    \includegraphics[height=\myspacing]{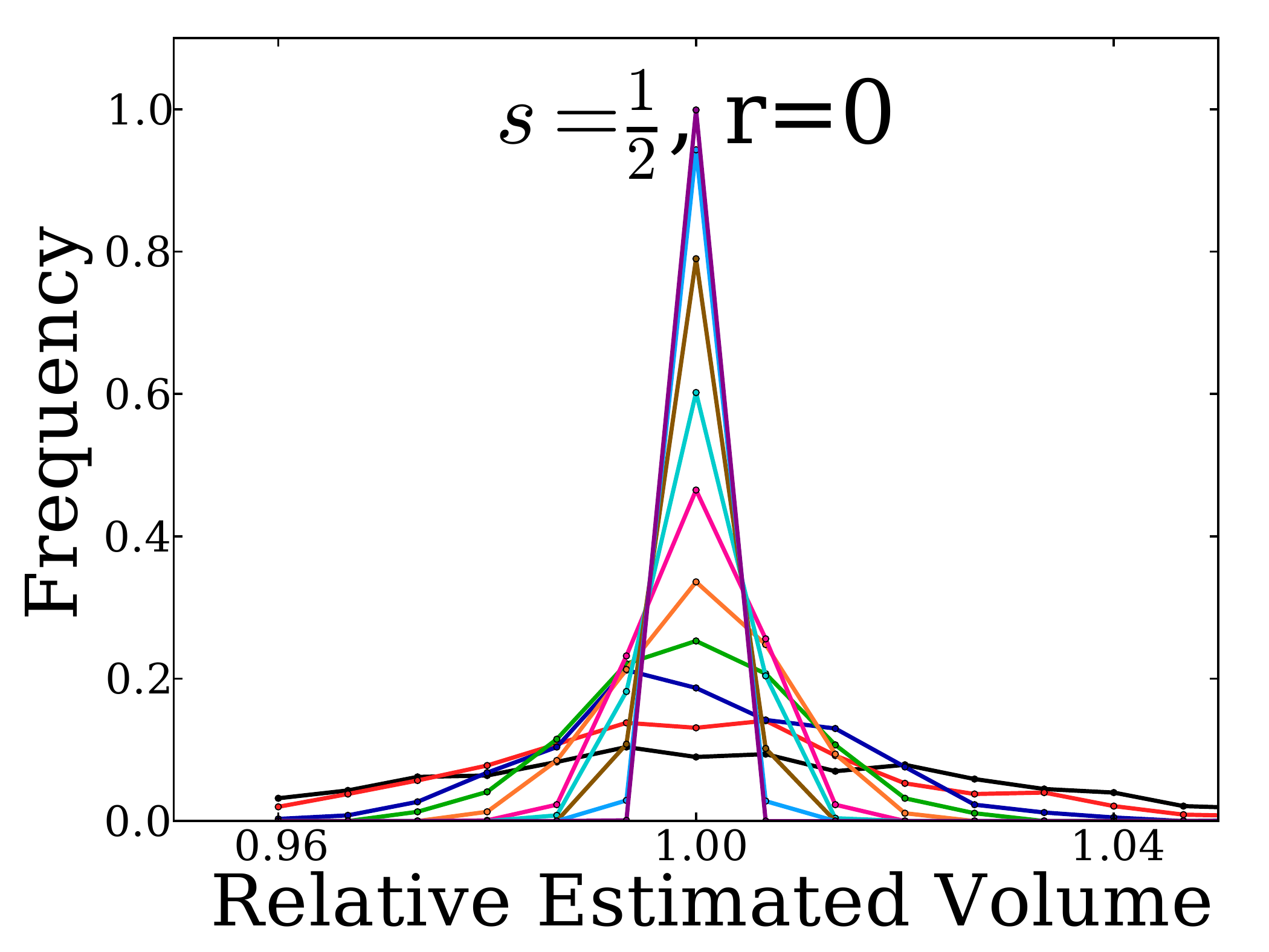}
    \includegraphics[height=\myspacing]{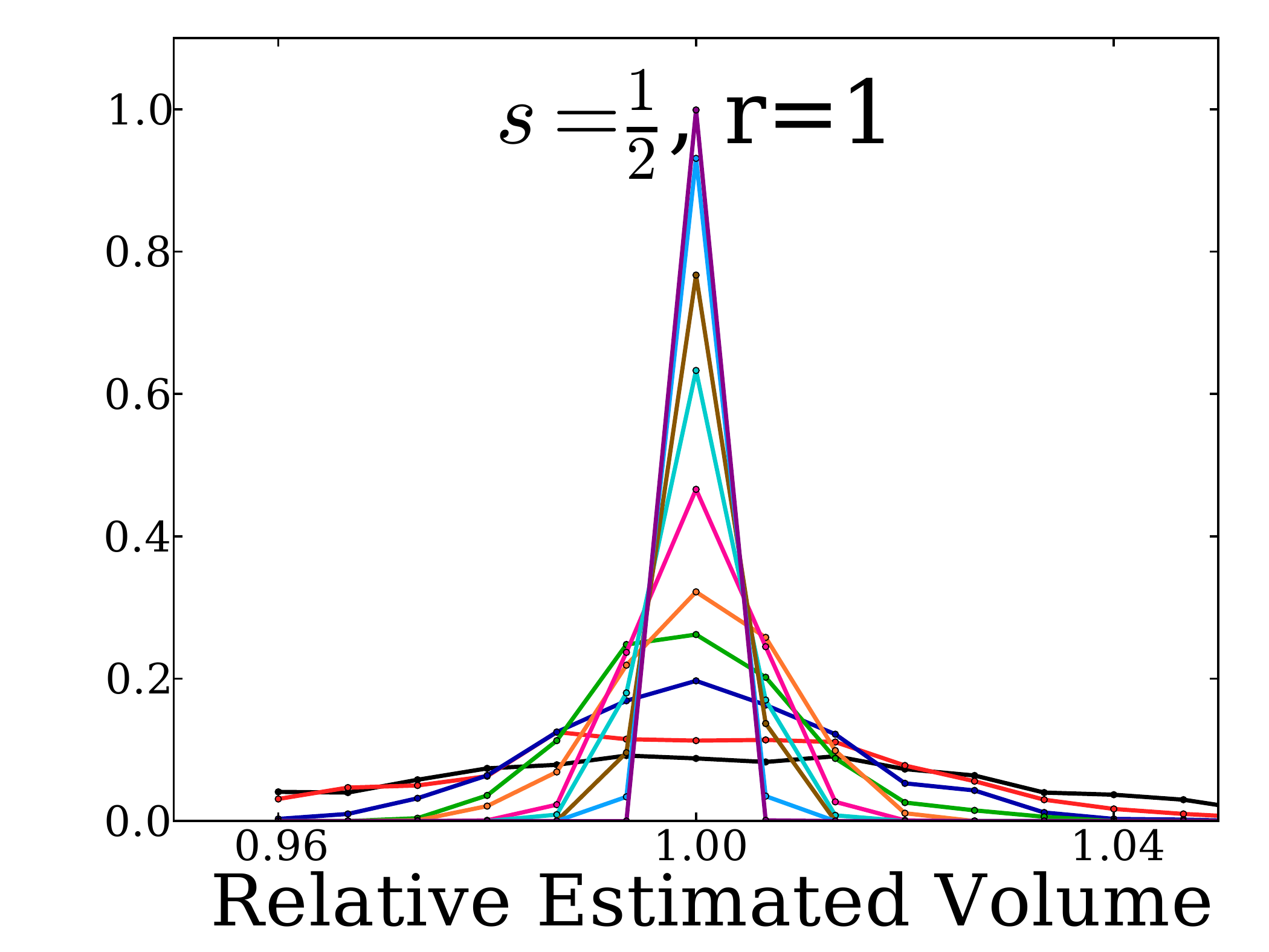}
    \includegraphics[height=\myspacing]{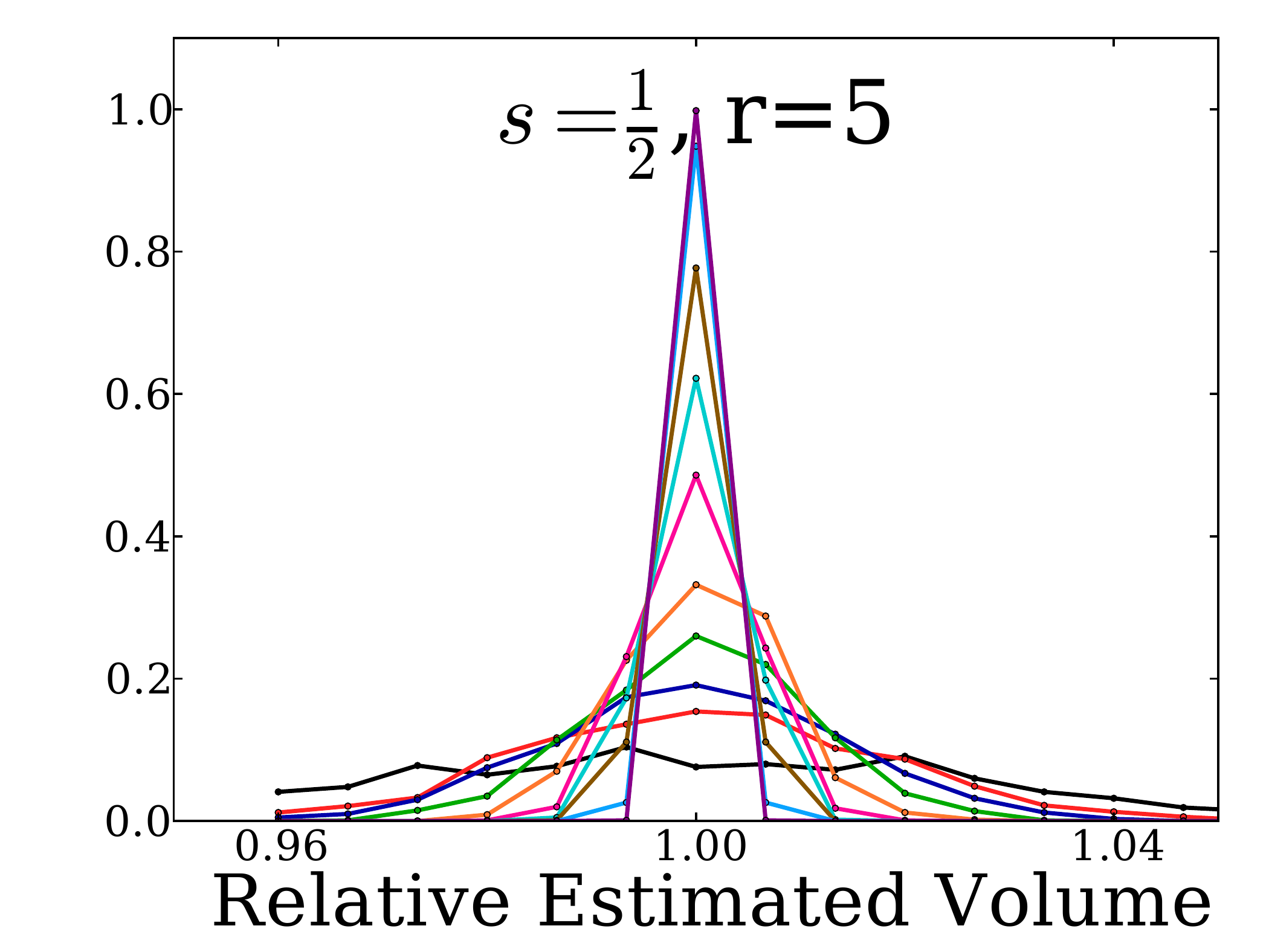}
    \includegraphics[height=\myspacing]{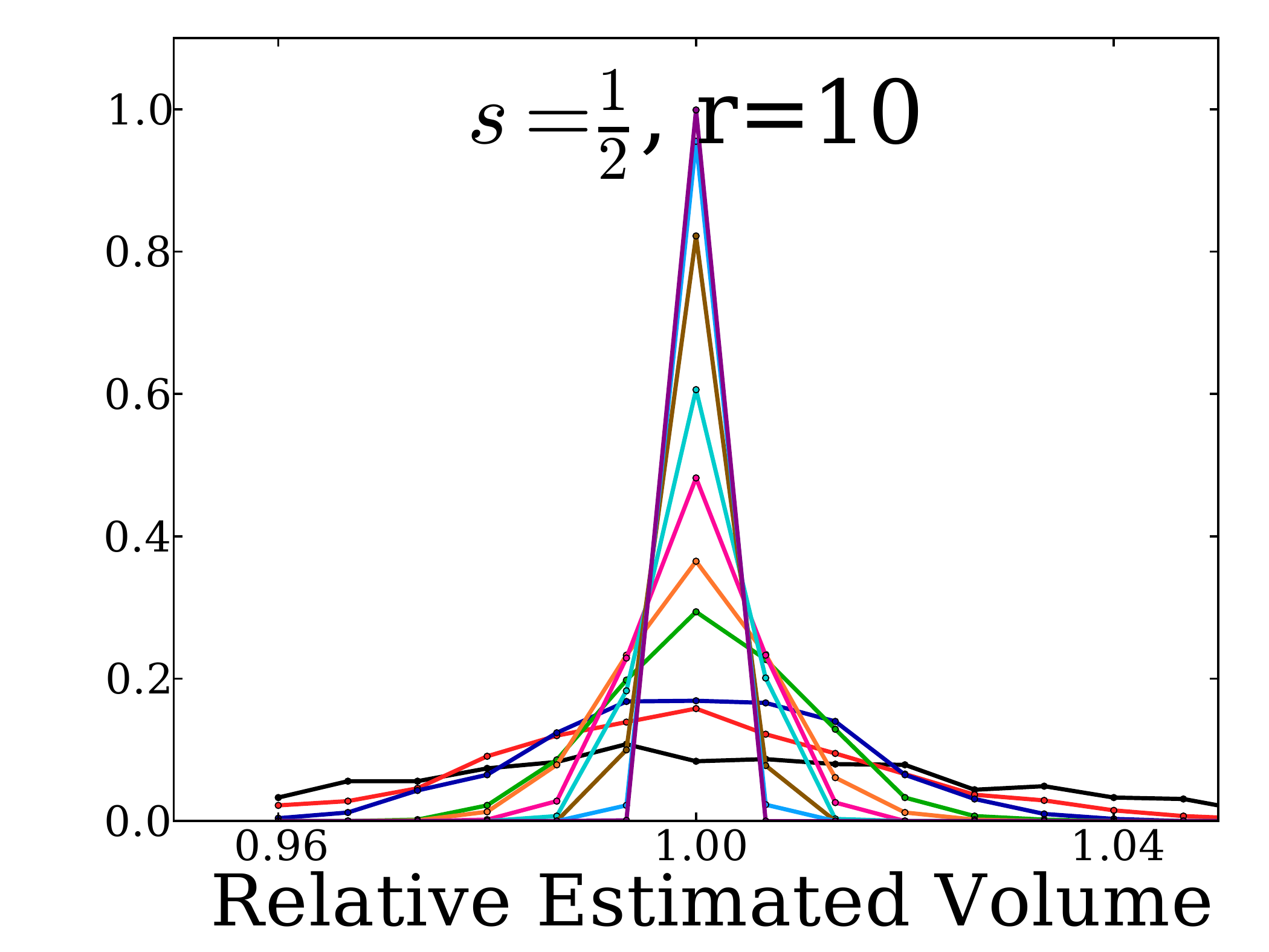}
    \includegraphics[height=\myspacing]{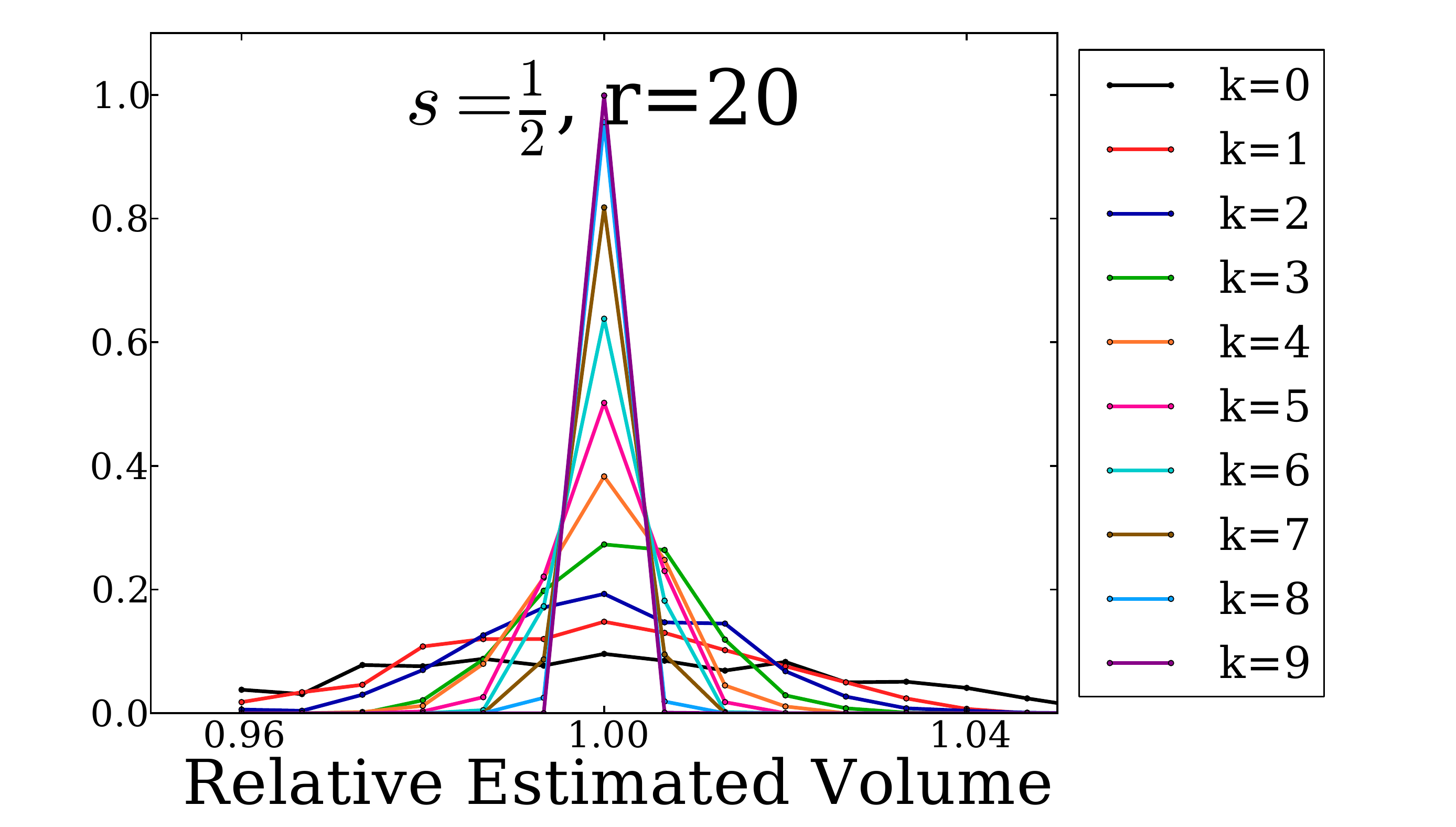}%\\
  \end{center}  
   \caption{Volume estimation histograms, estimate / true by frequency, for $n=10^6.$
   ``1a'' is axis-aligned darts, ``1r'' is random-angle lines, and ``1o'' is random-orientation darts, pairs of orthogonal flats. See also \tabref{tab:dart_experiments} and \figref{fig:mean_error}.}
   \label{fig:histogram_estimates}
\end{figure*}
%
%row1_3_10_0p100000.pdf
%row1_3_10_0p500000.pdf
%row1_3_10_1p000000.pdf
%row1_3_10_1p414214.pdf
%row1_3_10_3p162278.pdf
%row2_3_0_0p500000.pdf
%row2_3_10_0p500000.pdf
%row2_3_1_0p500000.pdf
%row2_3_20_0p500000.pdf
%row2_3_5_0p500000.pdf
%row3_10_10_0p100000.pdf
%row3_10_10_0p500000.pdf
%row3_10_10_1p000000.pdf
%row3_10_10_1p080060.pdf
%row3_10_10_1p291550.pdf
%row4_10_0_0p500000.pdf
%row4_10_10_0p500000.pdf
%row4_10_1_0p500000.pdf
%row4_10_20_0p500000.pdf
%row4_10_5_0p500000.pdf
%row5_2_10_0p100000.pdf
%row5_2_10_0p500000.pdf
%row5_2_10_10p000000.pdf
%row5_2_10_1p000000.pdf
%row5_2_10_2p000000.pdf
%row6_2_0_0p500000.pdf
%row6_2_10_0p500000.pdf
%row6_2_1_0p500000.pdf
%row6_2_20_0p500000.pdf
%row6_2_5_0p500000.pdf
%
%
%row1_hist_3_10_0p100000.pdf
%row1_hist_3_10_0p500000.pdf
%row1_hist_3_10_1p000000.pdf
%row1_hist_3_10_1p414214.pdf
%row1_hist_3_10_3p162278.pdf
%row2_hist_3_0_0p500000.pdf
%row2_hist_3_10_0p500000.pdf
%row2_hist_3_1_0p500000.pdf
%row2_hist_3_20_0p500000.pdf
%row2_hist_3_5_0p500000.pdf
%row3_hist_10_10_0p00000.pdf
%row3_hist_10_10_0p100000.pdf
%row3_hist_10_10_1p000000.pdf
%row3_hist_10_10_1p080060.pdf
%row3_hist_10_10_1p291550.pdf
%row4_hist_10_0_0p500000.pdf
%row4_hist_10_10_0p500000.pdf
%row4_hist_10_1_0p500000.pdf
%row4_hist_10_20_0p500000.pdf
%row4_hist_10_5_0p500000.pdf
%row5_hist_2_10_0p100000.pdf
%row5_hist_2_10_10p000000.pdf
%row5_hist_2_10_1p000000.pdf
%row5_hist_2_10_2p000000.pdf
%row6_hist_2_0_0p500000.pdf
%row6_hist_2_10_0p500000.pdf
%row6_hist_2_1_0p500000.pdf
%row6_hist_2_20_0p500000.pdf
%row6_hist_2_5_0p500000.pdf

\subsection{Results}
We plot the mean of the absolute value of the relative error, $| \textrm{mean} - \textrm{true} | / \textrm{true}$, 
vs.\ the number of flats $n$ in \figref{fig:mean_error}.  
We plot the histograms of the ratios of estimated / true volume for $n=10^6$ in \figref{fig:histogram_estimates}.
Each subfigure shows results for all $k$  for some combination of the other parameters.
We observe the following.

% 1/2 convergence
In \figref{fig:mean_error} the experimental slopes for all $k$ and $d$ are about $-1/2$, in agreement with theory (the $n$ to the power $-1/2$ in \equref{eqn:mcstderr}). 
The accuracy is insensitive to the orientation of the object.

In \figref{fig:histogram_estimates} the histograms are all sharply peaked at the true value.
This shows that the variations, the higher order moments of the estimates, are reasonable.

The major trend of these figures is that the accuracy of the estimates improves with $k$. 
Moreover, the larger the $k$, the smaller the variation of the estimate and the sharper the peak near the true value.

% ???
For small-volume objects, aligned darts are more accurate than unaligned darts.
This is illustrated by the red curves in the top-right and top-left subfigures in Figures~\ref{fig:mean_error} and 
\ref{fig:histogram_estimates}.
We were initially surprised by this,  but the explanation 
is that unaligned flats are shorter on average than aligned ones, because they might clip a corner of the square rather than being the side length of the square, so they more often miss the object. 
(They will be even shorter as the dimension of the space increases.)
For moderate-volume objects, the accuracy is about the same regardless of object orientation.
For unaligned flats, it appears that our $n$ was large enough that using pairs of orthogonal flats or independently-random individual flats does not make much difference.
Our conclusion is that aligned darts are universally better when the sample domain is a square.

% ???
The accuracy is primarily sensitive to the volume of the object and, secondarily, to the squish value.
Higher dimensional darts are better than lower dimensional ones, and the advantage is more pronounced for
small-volume objects.
This can been seen by considering the second-from-bottom row in Figures~\ref{fig:mean_error}. 
To see the volume dependence, note that the lines are closer together for $s=1$, and farther apart for larger and smaller squish factors. 
In low dimensions, 2 and 3, the smaller the volume of the object, the less accurate are all estimates, for all dimensions $k$. 
Moderate-dimensional darts in high dimensions, e.g.\ $d=10$ and $k=4$, also exhibit this trend.
However, 9-d darts in 10-d space have the same accuracy regardless of the volume or squish, because they are close to the dimension of the underlying space and they more fully span it. For example, a dart with $k=d$ always evaluates to the true value.
%(We suspect 9-d dart accuracy might degrade for extreme squish, say $s<10^{-6}$, and small $n$.)
That is, the advantage of higher-dimensional darts over lower-dimensional darts is greater for small objects in high dimensions, which is where we are advocating their use. 

A secondary phenomena is that the estimate is slightly more accurate for squish $s=1/10$ than for $s=\sqrt[9]{10}$, despite the  volume being the same.  This can be seen by careful examination of the height of the lines in the $d=10,$ varying squish row in 
\figref{fig:mean_error}.
For instance, the mean error is 10\% lower in the first column than the fifth for $n=10^4$ and $k=4$. 
This supports our intuition that darts are effective at hitting thin, coin-shaped regions. 
Since $\sqrt[9]{10}\approx1.3$, a 10-d object with this squish factor is actually roundish and not very needle-shaped. 
Preliminary experiments show that the accuracy gained by increasing $k$ is a complicated function of $s$ and the volume. High dimensional darts have more advantage over low dimensional ones for very small and sharp needle-shaped objects, compared to their advantage for coin-shaped objects.
% based on Scott's average symmetric proportion plots from  Anjul's high-squish data

\section{Conclusions}

%\subsection{Higher dimension \dart s}
In this work we introduce \dart{s} as a conception of higher-dimensional sampling.
We described a \dart{} framework for general dimension $k$, and then demonstrated efficiency and accuracy over three applications using  using $k=1$, and accuracy for one application using $k \ge 1$. In particular, we showed that darts produce accurate estimates of the volume of an object regardless of the dimension, orientation and aspect ratio of the object. 
Axis-aligned darts are universally preferable to unaligned ones for sampling square domains, and we expect this to extend to hyper-rectangles, e.g.\ bounding boxes. 

In principle, 
high-aspect-ratio (small-volume) objects are more efficiently sampled by higher-$k$ darts.
Demonstrating that efficiency for applications requires either analytical expressions, as in the toy sphere-volume problem in the Introduction; or efficient numerical techniques for evaluating the underlying function along higher dimensional flats. In future work we plan to explore a numerical technique using recursive sampling designs by dimension.

% keep conclusions in same order as the sections
For maximal Poisson-disk sampling,
line darts are helpful in getting close to maximality in high dimensions.
Below a given acceptable void ratio, they are more efficient than point darts. 
In terms of bounding the distance from a domain point to the nearest sample point, we are actually closer to maximality as the dimension increases. 
Any difference in the distribution of points produced by classical maximal Poisson-disk sampling and our line darts is small by the standard measures. Our line dart algorithm is efficient with respect to memory usage, which enables the production of larger samples in higher dimensions. %We produced a point cloud with 43,527 points in $d=30$ in less than three hours. 

For depth-of-field, using generalized \dart{}s gives us a high-quality noise-free image without aliasing. 
Although each 1-d dart requires more processing than a point sample, we only need
a few of them to render a high-quality image. This results in our \dart{} method 
outperforming point sampling. 
We suggest sampling over a higher-dimensional space to render both depth-of-field and motion blur in animations.
We suggest exploring weighted 1-d darts, where scene information determines which flats 
contribute more to a scene.

For uncertainty quantification, \dart\ Monte Carlo sampling can be more efficient than point sampling.
The key for all these applications is exploiting the problem structure to take advantage of what \dart{}s  provide.
%An optimization/root-finding method that leverages earlier 
%simulations to reduce the cost of later flats is needed to 
%realize this potential for general high-dimensional problems.

%
%\Dart{}s are an example of higher dimensional hyperplane sampling;
%however, there are a variety of other higher-dimensional sampling
%techniques, such as sampling along the path of a circle (or
%hypersphere) with varying radii, or sampling along a z-pattern. We
%plan to explore some of these for
%graphics applications and study what advantages, if any, we can discover.

%\todo{More general conclusions? }

%\todo{graphics conclusion:

%previous work and ours has proven that kd-dart-sampling is extremely
%beneficial for efficiently reducing noise while rendering complex visual
%effects. going forward, we would like to:

%1. implement on gpu to get real-time perf

%2. think about merging point sampling with line sampling to mitigate
%artifacts of both

%3. area sampling, path sampling, etc
%}

\anon{} % no acknowledgements
\known
{
  \section*{Acknowledgements}
%We are grateful to Gamito and Maddock for freely providing their software.
%
The Sandia authors thank Sandia's Validation and Verification program, and Computer Science Research Institute for supporting this work. 
The UC Davis authors thank the National Science Foundation (grant \#
CCF-1017399), NVIDIA and Intel Graduate Fellowships, the Intel Science
and Technology Center for Visual Computing, and Sandia LDRD award
\#13-0144 for supporting this work.

Sandia National Laboratories is a multi-program laboratory managed and operated by Sandia Corporation, a wholly owned subsidiary of Lockheed Martin Corporation, for the U.S. Department of Energy's National Nuclear Security Administration under contract DE-AC04-94AL85000.

}

\bibliographystyle{plain}
\bibliography{references}

\begin{thebibliography}{10}

\bibitem{Qhull}
C.~B. Barber, D.~P. Dobkin, and H.~T. Huhdanpaa.
\newblock The quickhull algorithm for convex hulls.
\newblock {\em ACM Transactions on Mathematical Software}, 22(4):469--483,
  1996.

\bibitem{Cook86}
Robert~L. Cook.
\newblock Stochastic sampling in computer graphics.
\newblock {\em ACM Transactions on Graphics}, 5(1):51--72, January 1986.

\bibitem{Dippe85}
Mark A.~Z. Dipp{\'{e}} and Erling~Henry Wold.
\newblock Antialiasing through stochastic sampling.
\newblock In {\em Computer Graphics (Proceedings of SIGGRAPH 85)}, pages
  69--78, July 1985.

\bibitem{ebeida_mitchell_vor}
Mohamed~S. Ebeida and Scott~A. Mitchell.
\newblock Uniform random {V}oronoi meshes.
\newblock In {\em 20th International Meshing Roundtable}, pages 258--275, 2011.

\bibitem{EbeidaMPSEurographics2012}
Mohamed~S. Ebeida, Scott~A. Mitchell, Anjul Patney, Andrew~A. Davidson, and
  John~D. Owens.
\newblock A simple algorithm for maximal {P}oisson-disk sampling in high
  dimensions.
\newblock {\em Computer Graphics Forum}, 31(2):785--794, May 2012.

\bibitem{Ebeida11}
Mohamed~S. Ebeida, Anjul Patney, Scott~A. Mitchell, Andrew Davidson, Patrick~M.
  Knupp, and John~D. Owens.
\newblock Efficient maximal {P}oisson-disk sampling.
\newblock {\em ACM Transactions on Graphics}, 30(4):49:1--49:12, July 2011.

\bibitem{Gamito09}
Manuel~N. Gamito and Steve~C. Maddock.
\newblock Accurate multidimensional {P}oisson-disk sampling.
\newblock {\em ACM Transactions on Graphics}, 29(1):8:1--8:19, December 2009.

\bibitem{glynn1989importance}
Peter~W. Glynn and Donald~L. Iglehart.
\newblock Importance sampling for stochastic simulations.
\newblock {\em Management Science}, 35(11):1367--1392, November 1989.

\bibitem{Gribel:2011:HQS}
Carl~Johan Gribel, Rasmus Barringer, and Tomas Akenine-M\"oller.
\newblock High-quality spatio-temporal rendering using semi-analytical
  visibility.
\newblock {\em ACM Transactions on Graphics}, 30:54:1--54:11, August 2011.

\bibitem{Gribel:2010:AMB}
Carl~Johan Gribel, Michael Doggett, and Tomas Akenine-M\"oller.
\newblock Analytical motion blur rasterization with compression.
\newblock In {\em Proceedings of High Performance Graphics}, pages 163--172,
  June 2010.

\bibitem{Haeberli:1990:TAB}
Paul~E. Haeberli and Kurt Akeley.
\newblock The accumulation buffer: Hardware support for high-quality rendering.
\newblock In {\em Computer Graphics (Proceedings of SIGGRAPH 90)}, pages
  309--318, August 1990.

\bibitem{Hall1873edp}
A.~Hall.
\newblock On an experimental determination of $\pi$.
\newblock {\em Messenger of Mathematics}, 2:113--114, 1873.

\bibitem{Hammon:2008:PPP}
Earl {Hammon Jr.}
\newblock Practical post-process depth of field.
\newblock In Hubert Nguyen, editor, {\em {GPU} Gems 3}, chapter~28, pages
  583--605. Addison-Wesley, 2008.

\bibitem{Jarosz:2011:ACT}
Wojciech Jarosz, Derek Nowrouzezahrai, Iman Sadeghi, and Henrik~Wann Jensen.
\newblock A comprehensive theory of volumetric radiance estimation using photon
  points and beams.
\newblock {\em ACM Transactions on Graphics}, 30(1):5:1--5:19, January 2011.

\bibitem{Jones11}
Thouis~R. Jones and David~R. Karger.
\newblock Linear-time {P}oisson-disk patterns.
\newblock {\em Journal of Graphics, GPU, and Game Tools}, 15(3):177--182, 2011.

\bibitem{Jones:2000:AWL}
Thouis~R. Jones and Ronald~N. Perry.
\newblock Antialiasing with line samples.
\newblock In {\em Proceedings of the Eurographics Workshop on Rendering
  Techniques 2000}, pages 197--206, London, June 2000. Springer-Verlag.

\bibitem{Lagae08}
Ares Lagae and Philip Dutr{\'{e}}.
\newblock A comparison of methods for generating {P}oisson disk distributions.
\newblock {\em Computer Graphics Forum}, 27(1):114--129, March 2008.

\bibitem{Lehtinen:2011:TLF}
Jaakko Lehtinen, Timo Aila, Jiawen Chen, Samuli Laine, and Fr{\'e}do Durand.
\newblock Temporal light field reconstruction for rendering distribution
  effects.
\newblock {\em ACM Transactions on Graphics}, 30(4):55:1--55:12, July 2011.

\bibitem{Liu:2006:QCM:1649589.1649888}
Yu-Shen Liu, Jun-Hai Yong, Hui Zhang, Dong-Ming Yan, and Jia-Guang Sun.
\newblock A quasi-{M}onte {C}arlo method for computing areas of point-sampled
  surfaces.
\newblock {\em Computer-Aided Design}, 38(1):55--68, January 2006.

\bibitem{Max:1990:ASD}
Nelson~L. Max.
\newblock Antialiasing scan-line data.
\newblock {\em IEEE Computer Graphics {\&} Applications}, 10(1):18--30, January
  1990.

\bibitem{McKay1979ctm}
M.~D. McKay, R.~J. Beckman, and W.~J. Conover.
\newblock A comparison of three methods for selecting values of input variables
  in the analysis of output from a computer code.
\newblock {\em Technometrics}, 21(2):239--245, May 1979.

\bibitem{Mitchell:1987:GAI}
Don~P. Mitchell.
\newblock Generating antialiased images at low sampling densities.
\newblock In {\em Computer Graphics (Proceedings of SIGGRAPH 87)}, pages
  65--72, July 1987.

\bibitem{lattice}
Gabriele Nebe and Neil Sloane.
\newblock A catalogue of lattices.
\newblock
  \url{http://www.math.rwth-aachen.de/~Gabriele.Nebe/LATTICES/index.html},
  2012.

\bibitem{Owen1992clt}
A.~B. Owen.
\newblock A central limit theorem for {L}atin hypercube sampling.
\newblock {\em Journal of the Royal Statistical Society. Series B
  (Methodological)}, 54(2):541--551, 1992.

\bibitem{Ramaley}
J.~F. Ramaley.
\newblock {B}uffon's noodle problem.
\newblock {\em The American Mathematical Monthly}, 76(8):916--918, October
  1969.

\bibitem{Rovira:2005:PSU:2386366.2386385}
J.~Rovira, P.~Wonka, F.~Castro, and M.~Sbert.
\newblock Point sampling with uniformly distributed lines.
\newblock In {\em Proceedings of the Second Eurographics / IEEE VGTC Conference
  on Point-Based Graphics}, SPBG'05, pages 109--118, Aire-la-Ville,
  Switzerland, Switzerland, 2005. Eurographics Association.

\bibitem{psa}
Thomas Schl\"omer.
\newblock {PSA} point set analysis.
\newblock \url{http://code.google.com/p/psa/}, 2011.

\bibitem{Spanier66}
Jerome Spanier.
\newblock Two pairs of families of estimators for transport problems.
\newblock {\em J. SIAM Appl. Math.}, 14:702--713, 1966.

\bibitem{Sun:2010:LSG}
Xin Sun, Kun Zhou, Stephen Lin, and Baining Guo.
\newblock Line space gathering for single scattering in large scenes.
\newblock {\em ACM Transactions on Graphics}, 29(4):54:1--54:8, July 2010.

\bibitem{Tzeng:2012:HPD}
Stanley Tzeng, Anjul Patney, Andrew Davidson, Mohamed~S. Ebeida, Scott~A.
  Mitchell, and John~D. Owens.
\newblock High-quality parallel depth-of-field using line samples.
\newblock In {\em Proceedings of High Performance Graphics}, pages 23--31, June
  2012.

\bibitem{Veach:1997:MLT}
Eric Veach and Leonidas~J. Guibas.
\newblock {M}etropolis light transport.
\newblock In {\em Proceedings of SIGGRAPH 97}, Computer Graphics Proceedings,
  Annual Conference Series, pages 65--76, August 1997.

\bibitem{Wei08}
Li-Yi Wei.
\newblock Parallel {P}oisson disk sampling.
\newblock {\em ACM Transactions on Graphics}, 27(3):20:1--20:9, August 2008.

\bibitem{White07}
Kenric~B. White, David Cline, and Parris~K. Egbert.
\newblock {P}oisson disk point sets by hierarchical dart throwing.
\newblock In {\em RT '07: Proceedings of the 2007 IEEE Symposium on Interactive
  Ray Tracing}, pages 129--132, September 2007.

\end{thebibliography}

%\received{January 2013}{continuity from SIGGRAPH-ASIA 2012 submission papers\_0450} %second field is accepted date

\end{document}